\RequirePackage{fix-cm} 
\documentclass[twoside,openright,titlepage,numbers=noenddot,headinclude,
                footinclude=true,cleardoublepage=empty,abstractoff, 
                BCOR=5mm,paper=a4,fontsize=10pt,
                ngerman,american,dottedtoc]{scrreprt}

\PassOptionsToPackage{utf8}{inputenc}
	\usepackage{inputenc}
\PassOptionsToPackage{eulerchapternumbers,listings,%
					 pdfspacing,
					 subfig,beramono,eulermath,parts}{classicthesis}                                        

\newcommand{\myTitle}{SQL Comprehension and Synthesis\xspace}

\newcommand{\myName}{George Rabeshi Obaido\xspace}

\newcommand{\myFaculty}{Put data here\xspace}

\newcommand{\myUni}{Put data here\xspace}

\newcommand{\myyear}{2020\xspace}
\newcommand{\myVersion}{version 4.2\xspace}
\newcommand{\mysponsor}{Council for Scientific and Industrial Research under the Department of Science and Technology in South Africa.\xspace}

\newcounter{dummy} 
\providecommand{\mLyX}{L\kern-.1667em\lower.25em\hbox{Y}\kern-.125emX\@}

	\usepackage{babel}   
\usepackage{csquotes}
\PassOptionsToPackage{fleqn}{amsmath}       
    \usepackage{amsmath}

\PassOptionsToPackage{T1}{fontenc} 
    \usepackage{fontenc}     
\usepackage{textcomp} 
\usepackage{scrhack} 
\usepackage{xspace} 
\usepackage{mparhack} 
\usepackage{fixltx2e} 
\PassOptionsToPackage{printonlyused,smaller}{acronym} 
    \usepackage{acronym} 
    
\usepackage{tabularx} 
\usepackage{caption}
\usepackage{subfig}  

\usepackage{listings} 
\PassOptionsToPackage{pdftex,hyperfootnotes=false,pdfpagelabels}{hyperref}
    \usepackage{hyperref}  
\PassOptionsToPackage{pdftex}{graphicx}
    \usepackage{graphicx}

\hypersetup{%
    colorlinks=true, linktocpage=true, pdfstartpage=3, pdfstartview=FitV,%
    breaklinks=true, pdfpagemode=UseNone, pageanchor=true, pdfpagemode=UseOutlines,%
    plainpages=false, bookmarksnumbered, bookmarksopen=true, bookmarksopenlevel=1,%
    hypertexnames=true, pdfhighlight=/O,
    urlcolor=webbrown, linkcolor=RoyalBlue, citecolor=webgreen, 
    pdftitle={\myTitle},%
    pdfauthor={\textcopyright\ \myName, \myUni, \myFaculty},%
    pdfsubject={},%
    pdfkeywords={},%
    pdfcreator={pdfLaTeX},%
    pdfproducer={LaTeX with hyperref and classicthesis}%
}   

\makeatletter
\@ifpackageloaded{babel}%
    {%
       \addto\extrasamerican{%
                }%
       \addto\extrasngerman{%
                }%
    }{\relax}
\makeatother

\newcommand\tabb[1][1.5cm]{\hspace*{#1}}
\newcommand{\inline}[2]{%
	\begin{tikzpicture}[baseline=(word.base), txt/.style={shape=rectangle, inner sep=0pt}]
	\node[txt] (word) {#1};
	\node[above] at (word.north) {\footnotesize{#2}};
	\end{tikzpicture}%
}

\usepackage{classicthesis} 
\usepackage[left=1.3in,right=1.3in,top=1in,bottom=1.15in]{geometry}
\usepackage{palatino,lettrine}
\usepackage[square]{natbib}
\usepackage{enumitem}
\usepackage{placeins}
\usepackage{mdframed,lipsum}
\usepackage{algorithm,algpseudocode}
\usepackage{pdfpages}
\usepackage{url}
\usepackage{float}
\usepackage{xcolor} 
\usepackage{paracol}
\usepackage{textcomp}
\usepackage{graphicx}
\usepackage{listings}
\usepackage{colortbl}
\usepackage{amsmath,amsthm, amssymb,amsfonts}
\usepackage{pifont}
\usepackage{nccmath}
\usepackage{tikz}
\usepackage{graphicx}
\usepackage{longtable}
\theoremstyle{definition}
\newtheorem{definition}{Definition}
\newtheorem*{remark}{Remark}
\newtheorem{exmp}{Example}[section]
\definecolor{ao(english)}{rgb}{0.0, 0.5, 0.0}
\definecolor{bostonuniversityred}{rgb}{0.8, 0.0, 0.0}
\definecolor{antiquebrass}{rgb}{0.8, 0.58, 0.46}
\definecolor{antiquefuchsia}{rgb}{0.57, 0.36, 0.51}
\definecolor{asparagus}{rgb}{0.53, 0.66, 0.42}
\definecolor{cornsilk}{rgb}{1.0, 0.97, 0.86}
\lstdefinestyle{Common}
{
	extendedchars=\true,
	language={[Visual]Basic},
	frame=single,
	rulecolor=\color{Red}
}

\lstdefinestyle{A}
{
	style=Common,
	basicstyle=\scriptsize\color{Black}\ttfamily,
	keywordstyle=\color{Orange},
	identifierstyle=\color{Cyan},
	stringstyle=\color{Red},
	commentstyle=\color{Green}
}

\lstdefinestyle{B}
{
	style=Common,
	backgroundcolor=\color{Black},
	basicstyle=\scriptsize\color{White}\ttfamily,
	keywordstyle=\color{Orange},
	identifierstyle=\color{Cyan},
	stringstyle=\color{Red},
	commentstyle=\color{Green}
}
\colorlet{punct}{red!60!black}
\definecolor{background}{HTML}{EEEEEE}
\definecolor{delim}{RGB}{20,105,176}
\colorlet{numb}{magenta!60!black}

\lstdefinelanguage{json}{
	basicstyle=\normalfont\ttfamily,
	numbers=left,
	numberstyle=\scriptsize,
	stepnumber=1,
	numbersep=8pt,
	showstringspaces=false,
	breaklines=true,
	keywordstyle=\color{blue},
	frame=lines,
	morekeywords={SELECT,UPDATE,FROM,WHERE, DELETE},
	backgroundcolor=\color{background},
	literate=
	*{0}{{{\color{numb}0}}}{1}
	{1}{{{\color{numb}1}}}{1}
	{2}{{{\color{numb}2}}}{1}
	{3}{{{\color{numb}3}}}{1}
	{4}{{{\color{numb}4}}}{1}
	{5}{{{\color{numb}5}}}{1}
	{6}{{{\color{numb}6}}}{1}
	{7}{{{\color{numb}7}}}{1}
	{8}{{{\color{numb}8}}}{1}
	{9}{{{\color{numb}9}}}{1}
	{:}{{{\color{punct}{:}}}}{1}
	{,}{{{\color{punct}{,}}}}{1}
	{\{}{{{\color{delim}{\{}}}}{1}
	{\}}{{{\color{delim}{\}}}}}{1}
	{[}{{{\color{delim}{[}}}}{1}
	{]}{{{\color{delim}{]}}}}{1},
}
\renewcommand\bibname{References}  

\usepackage{natbib} \input{natbib-add}
\bibpunct{[}{]}{;}{a}{}{;}  
\begin{document}
\frenchspacing
\raggedbottom
\selectlanguage{american} 
\pagenumbering{roman}
\pagestyle{plain}
\begin{titlepage}
    \begin{addmargin}[-1cm]{-2cm}
    \begin{center}
        \large  

%
\begingroup
\color{Black}\huge \bfseries \Huge{\myTitle} \\ \bigskip
\endgroup

        \includegraphics[width=6cm]{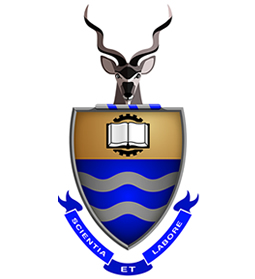} \\ \medskip
        \Large {George Rabeshi Obaido}\\ \medskip
\large{School of Computer Science and Applied Mathematics,\\University of the Witwatersrand, Johannesburg, South Africa.}\\[3.2cm]

A thesis submitted to the Faculty of Science,\\ in fulfilment of the requirements for the degree of \\ \textit{Doctor of Philosophy} in Computer Science.\\[4.2cm] 
\textit{}\\[0.2cm]


\textit{}\\[0.4cm]

\textit\textbf{Supervisors:\\Dr Hima Vadapalli \\Prof Abejide Ade-Ibijola}\\[1.2cm]

{\large 22 May 2020}\\[4cm]         

        \vfill                      

    \end{center}  
  \end{addmargin}       
\end{titlepage}   
\thispagestyle{empty}

\hfill

\vfill

\noindent\myName\\ 
\textit{\myTitle, \myyear} \\ 
\bigskip

\noindent Copyright \textcopyright\ University of the Witwatersrand, Johannesburg, South Africa.

\bigskip

\noindent\spacedlowsmallcaps{Supervisors}: \\
\noindent Dr Hima Vadapalli\footnote{School of Computer Science and Applied Mathematics, University of the Witwatersrand, Johannesburg, South Africa.} \\
\noindent Prof Abejide Ade-Ibijola\footnote{Formal Structures, Algorithms and Industrial Applications Research Cluster, School of Consumer Intelligence and Information Systems, University of Johannesburg, Johannesburg, South Africa.}

\medskip
\noindent\spacedlowsmallcaps{Supported by}: \\
\mysponsor\\

%
%

\thispagestyle{empty}
\refstepcounter{dummy}
\pdfbookmark[1]{Dedication}{Dedication}

\vspace*{3cm}

%

\begin{center}
    Dedicated to the\\ loving memory of my father, Mr Joseph Obaido.\\

\end{center}
\refstepcounter{dummy}
\pdfbookmark[0]{Declaration}{declaration}
\chapter*{Declaration}
\thispagestyle{empty}
I declare that this thesis is my original and unaided intellectual work. This thesis is submitted for the degree of Doctor of Philosophy in Computer Science at the University of the Witwatersrand, Johannesburg. This work has not been submitted to any other University, or for any other degree.\\
\bigskip
 
\bigskip 
\bigskip
\smallskip

\begin{flushright}

    \begin{tabular}{m{5cm}}
    	  \begin{center}
    		\includegraphics[scale=0.30]{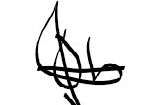}
    	\end{center}
        \\\hline
        \centering\myName \\\medskip
         May 2020.
    \end{tabular}
\end{flushright}

\pdfbookmark[1]{Abstract}{Abstract}
\begingroup
\let\clearpage\relax
\let\cleardoublepage\relax
\let\cleardoublepage\relax

\chapter*{Abstract}
\lettrine[lines=3,loversize=0.1] {S}\normalsize{tructured} Query Language (SQL) remains the standard language used in Relational Database Management Systems (RDBMSs), and has found applications in healthcare (patient registries), businesses (inventories, trend analysis), military, education, etc. Although SQL statements are English-like, the process of writing SQL queries is often problematic for non-technical end-users in the industry. Similarly, formulating and comprehending written queries can be confusing, especially for undergraduate students. One of the pivotal reasons given for these difficulties lies with the simple syntax of SQL, which is often misleading and hard to understand. An ideal solution is to present these two audiences: undergraduate students and non-technical end-users with learning and practice tools. These tools are mostly electronic, and can be used to aid their understanding, as well as enable them write correct SQL queries. This work proposes a new approach aimed at understanding and writing correct SQL queries using principles from Formal Language and Automata Theory. We present algorithms based on: regular expressions for the recognition of simple query constructs, context-free grammars for the recognition of nested queries and a jumping finite automaton for the synthesis of SQL queries from natural language descriptions. As proof of concept, these algorithms were further implemented into interactive software tools aimed at improving SQL comprehension. Evaluation of these tools showed that the majority of participants agreed that the tools were intuitive and aided their understanding of SQL queries. These tools should, therefore, find applications in aiding SQL comprehension at higher learning institutions and assist in the writing of correct queries in data-centered industries. 

\vfill

\endgroup			

\vfill
\pdfbookmark[1]{Publications}{publications}
\chapter*{Publications}

\lettrine[lines=3,loversize=0.1] {S}\normalsize{ome} portions of this thesis have been published in the following articles. These portions include figures, algorithms, equations and descriptions of concepts. These papers are:\\

\sloppypar
\begin{enumerate}[label={[\arabic*]}]
	\item Abejide Ade-Ibijola and George Obaido~(2017). \textit{S-NAR: Generating Narrations of SQL Queries using Regular Expressions}. In the ACM Proceedings of the South African Institute of Computer Scientists and Information Technologists (SAICSIT), pp 11-18, Bloemfontein, Free State. ISBN: 978-1-4503-5250-5. URL: \url{https://dl.acm.org/citation.cfm?doid=3129416.3129454}. [South Africa].
	\item  George Obaido, Abejide Ade-Ibijola, and Hima Vadapalli~(2018). \textit{Generating SQL Queries from Visual Specifications}. In Communications in Computer and Information Science (CCIS), vol. 963, pp 315–330, Springer, Cham, Switzerland. ISBN: 978-3-030-05813-5. URL: \url{https://link.springer.com/chapter/10.1007/978-3-030-05813-5_21}. [Switzerland].
	\item George Obaido, Abejide Ade-Ibijola, and Hima Vadapalli~(2019). \textit{Generating Narrations of Nested SQL Queries using Context-free Grammars}. In the proceedings of IEEE Conference on ICT and Society (ICTAS), pp 13:1–6, 6th to 9th March, Durban. URL: \url{https://ieeexplore.ieee.org/document/8703620}. [South Africa].
	\item George Obaido, Abejide Ade-Ibijola, and Hima Vadapalli~(2019). \textit{TalkSQL: A Tool for the Synthesis of SQL Queries from Verbal Specifications}. In the proceedings of the IEEE International Multidisciplinary Information Technology and Engineering Conference (IMITEC), pp 469–478, November 21st to 22nd, 2019. [South Africa].
	\item George Obaido, Abejide Ade-Ibijola, and Hima Vadapalli~(2019). \textit{Synthesis of SQL Queries from Narrations}. In the proceedings of the 6th IEEE International Conference on Soft Computing \& Machine Intelligence (ISCMI), pp 195–201, ISBN: 978-1-7281-4576-1, November 19th to 20th, 2019. [South Africa] --- \textit{\textbf{Best Presentation Award}}.\\
\bigskip

\noindent Other articles on related projects:
\item Abejide Ade-Ibijola and George Obaido~(2019). \textit{XNorthwind: Grammar-driven Synthesis of Large Datasets for DB Applications}. In International Journal of Computer Science (Scopus), Vol. 46, Issue 4, pp 541–551, International Association of Engineers (IAENG). URL: \url{http://www.iaeng.org/IJCS/issues_v46/issue_4/IJCS_46_4_05.pdf}. [Hong Kong].	
\end{enumerate}
\pdfbookmark[1]{Acknowledgments}{acknowledgments}


\bigskip

\begingroup
\let\clearpage\relax
\let\cleardoublepage\relax
\let\cleardoublepage\relax
\chapter*{Acknowledgments}
\lettrine[lines=3,loversize=0.1] {T}\normalsize{he} work presented in this thesis has been a life-changing experience and it would not have been possible without the support and guidance that I have received from many people. I, therefore, take this opportunity to extend my sincere gratitude and appreciation to all those who have made this PhD thesis possible.\\

Firstly, I would like to acknowledge my indebtedness and render my warmest thanks to my supervisors, Prof Abejide Ade-Ibijola and Dr Hima Vadapalli.  I am grateful to Prof Abejide for suggesting my topic as well as for introducing me to the exciting fields of Formal Language and Automata applications. He has taught me, both consciously and unconsciously, the elements of how good research should be done. He has spent countless hours editing my work in addition to constantly reminding me of how we needed to ‘flush’ accepted manuscripts; words which I will fondly remember. I will forever value all the time, advice, inspiration, encouragement and ideas that he has contributed to my PhD pursuit. The joy and enthusiasm that he has for research is truly remarkable and this has been a motivational factor throughout my PhD journey. Similarly, I would like to extend my appreciation and gratitude to Dr Hima for her constant encouragement, contribution and financial support towards the conferences that I was able to attend and at which the research papers found in this work were presented.\\

I am indebted to the Council for Scientific and Industrial Research - Department of Science and Technology of South Africa, under the Interbursary MSc/PhD bursary support programme for funding this research. I am also grateful to the Association for Computing Machinery, and Black in Artificial Intelligence, especially Rediet Abebe and Timnit Gebru, for providing travel grants to the United States of America and Canada respectively.\\

I would also like to extend my thankfulness to Prof Cajethan Iheka, Dr Sihlobosenkosi Mpofu, Dr Sarah Taylor, Dr Tarisai Rukuni, Abiodun Modupe, Michelle van Heerden and Dr Amos Anele and Dr Michael Ajayi for their words of encouragement during the writing of this thesis. My sincere appreciation goes out to Cynthia Lawrence, Haroon Motalib, Divashni Naicker and Jowilna Storm for offering me a part-time lecturing position at Pearson Institute of Higher Education, South Africa. In addition, I am grateful to my students, Laruschka Smit, Angelique Hassett, Jason Petrie, Rose Nnanna and Prisca Bifutuka, who have often assisted with crowdsourcing tasks, which has contributed to the completion of my research. \\ 

My appreciation also goes out to Jude Oyasor, Kehinde Aruleba, Blessing Ogbuokiri, Elvis and MaryAnn Umejiaku, Iyiola Ogunrekun, Michael Ayawei, Mustapha Oloko-Oba, Emmanuel Mudakiri, Segun and Bukola Adebayo, Dennis Uzizi, Oluwatobi Onawole, Izuchukwu Ohaegbu, Mbulelo Mthembu, Wainright Acquoi, Lance Krasner, Alan Michael, Ndivhuwo Nemudiviso and Senyo Cudjoe, who are my colleagues and friends for life, for making this journey exciting and fun.\\

Lastly and most importantly, I would like to thank my family for all their love and encouragement. I am thankful to my only surviving parent, my mother Margaret, who has constantly supported me in all of my academic pursuits. My appreciation goes to my in-laws, Melfie (mother), Rejoice (sister) and Whitney (niece) for constantly reminding me that I needed to rest throughout this journey. Many thanks to my siblings, Christiana, Anthony, Victoria, Sarah and Faith, for their unflinching support. And most of all, to my loving wife Thobekile, for her love and patience with me during those late nights spent in the research lab.

%

\endgroup

\pdfbookmark[1]{Preamble}{Preamble}
\begingroup
\let\clearpage\relax
\let\cleardoublepage\relax
\let\cleardoublepage\relax

\chapter*{Preamble}
\lettrine[lines=3,loversize=0.1] {T}\normalsize{his} PhD thesis investigates the problem of learning and writing correct SQL queries. It presents a number of approaches to the problem of SQL comprehension and synthesis. As a whole, it covers the application of the formal language and automata theory to the problem of SQL comprehension and synthesis. This preamble, by way of an introduction, provides the contributions and organisation of this thesis. It shows the list of related domains with keywords and highlights non-academic talks that have been presented on this work.\\

\noindent \textbf{Technical contributions.}\\
The technical contributions of this work are as follows:

\begin{enumerate}
	\item \textit{Formal Language and Automata Theory}: new formalisms for recognising SQL queries to generate \textit{narrations} using regular expressions (REs) and context-free grammars (CFGs), and a jumping finite automaton (JFA) for synthesising SQL queries from natural language specifications have been presented.
	
	\item \textit{Software Prototypes}: five software tools are presented. Firstly, \texttt{S-NAR} for assisting end-users to understand simple SQL queries~\citep{ade2017s}; secondly, a tool called the \texttt{SQL Narrator} for assisting end-users to understand nested SQL queries~ \citep{obaido2019generating}; thirdly, \texttt{Narrations-2-SQL} for translating natural language descriptions into SQL queries~\citep{obaido2019narrations}; fourthly, the \texttt{SQL Visualiser} that uses visual specifications to build a query~\citep{obaido2018generating}, and finally, \texttt{TalkSQL} aimed at assisting end-users to understand SQL queries using speech inputs~\citep{obaido2019talksql}.

	\item \textit{Evaluation of Prototypes}: The usefulness of these software tools was evaluated and the evaluation results are presented.
\end{enumerate}
\bigskip

\noindent \textbf{Thesis organisation.}\\
\noindent This thesis is organised into four parts. \autoref{part1} contains the introduction, definition of terms and a review of literature relevant to this study. \autoref{part2} and \autoref{part3} presents the major contributions of this work. \autoref{part4} evaluates the developed tools and presents the conclusions and future directions.
\bigskip

\noindent \textbf{Domain of research.}\\
The following categories shows the 2012 ACM\footnote{Association for Computing Machinery} CCS\footnote{Computing Classification System} that this research is based on.

\begin{itemize}
	\item Theory of computation, formal languages and automata theory, and grammars and context-free languages.
	
	\item Applied computing, computer-assisted instruction and interactive learning environment.

	\item Computers and education, computer and information science education and computer science education.
	
	\item Computing methodologies, artificial intelligence and natural language processing. \\
	
\end{itemize}

\noindent \textbf{Keywords.}\\
The keywords used in this research are as follows:
\begin{itemize}
	\item SQL comprehension, SQL query narration, SQL tutoring, intelligent tutoring system, regular languages, context-free grammar, jumping finite automaton, query by speech, visual specifications, verbal specifications, natural language processing, learning via visualisation, learning via narrations, language translation, relational database and synthesis of things.\\
\end{itemize}
\bigskip

\noindent \textbf{Non-academic Talks.}\\
Some ideas of this thesis was presented at the following event.

\begin{itemize}
	\item Doctoral Consortium of the ACM FAT*\footnote{Fairness, Accountability and Transparency} Conference in Atlanta, Georgia in the USA on the 29th January, 2019.
\end{itemize}
\endgroup			

\vfill

\pagestyle{scrheadings}

\refstepcounter{dummy}

\pdfbookmark[1]{\contentsname}{tableofcontents} 

\setcounter{tocdepth}{2} 

\setcounter{secnumdepth}{3} 

\manualmark
\markboth{\spacedlowsmallcaps{\contentsname}}{\spacedlowsmallcaps{\contentsname}}
\tableofcontents 
\automark[section]{chapter}
\renewcommand{\chaptermark}[1]{\markboth{\spacedlowsmallcaps{#1}}{\spacedlowsmallcaps{#1}}}
\renewcommand{\sectionmark}[1]{\markright{\thesection\enspace\spacedlowsmallcaps{#1}}}

\clearpage

\begingroup 
\let\clearpage\relax
\let\cleardoublepage\relax
\let\cleardoublepage\relax


\refstepcounter{dummy}
\pdfbookmark[1]{\listfigurename}{lof} 

\listoffigures

\vspace{8ex}
\newpage


\refstepcounter{dummy}
\pdfbookmark[1]{\listtablename}{lot} 

\listoftables
        
\vspace{8ex}
\newpage
    

\refstepcounter{dummy}
\pdfbookmark[1]{\lstlistlistingname}{lol} 

\lstlistoflistings 

\vspace{8ex}
\newpage
       

\refstepcounter{dummy}
\pdfbookmark[1]{Acronyms}{acronyms} 

\markboth{\spacedlowsmallcaps{Acronyms}}{\spacedlowsmallcaps{Acronyms}}

\chapter*{Acronyms}

\begin{acronym}[UML]
\acro{DRY}{Don't Repeat Yourself}
\acro{API}{Application Programming Interface}
\acro{UML}{Unified Modeling Language}
\end{acronym}  
\begin{longtable}{llll}
		\caption{List of Acronyms}\\
	AI & Artificial Intelligence & NLG & Natural Language Generation \\ 
	API & Application Programming Interface & NLIDB & Natural Language Interfaces to Database \\ 
	BI & Business Intelligence & NLP & Natural Language Processing \\ 
	BNF & Backus-Naur Form & NLQ & Natural Language Query \\ 
	CFG & Context-free Grammar & NLTK & Natural Language Toolkit \\ 
	CLT & Cognitive Load Theory & NLU & Natural Language Understanding \\ 
	CNN & Convolutional Neural Network & NMT & Neural Machine Translation \\ 
	CS & Computer Science & PoS & Part-of-Speech \\ 
	DB & Database & PBL & Problem Based Learning \\ 
	DBN & Deep Belief Network & QA & Question Answering  \\ 
	DDL & Data Definition Language & RDBMS & Relational Database Management System \\ 
	DML & Data Manipulation Language & RE & Regular Expression  \\ 
	EBNF & Extended Backus-Naur Form & RNN & Recurrent Neural Network  \\ 
	ERD & Entity Relationship Diagram & SFA & Syntax-free Approach  \\ 
	FLA & Formal Language and Automata Theory & SMT & Statistical Machine Translation \\ 
	GUI & Graphical User Interface & SPARQL & SPARQL Protocol and RDF Query Language \\ 
	HCI & Human-Computer Interaction & SQL & Structured Query Language \\
 	HMM & Hidden Markov Model & SSIS & SQL Server Integration Service \\ 
	 IBM & International Business Machines &	SVM & Support Vector Machine \\ 
	ITS & Intelligent Tutoring System & UD & Universal Dependency \\ 
	JFA & Jumping Finite Automaton & UI & User Interface \\ 
	MT & Machine Translation & WER & Word Error Rate \\ 
	\label{acronym}
\end{longtable}
\endgroup
\cleardoublepage\pagenumbering{arabic}
\cleardoublepage
\ctparttext{Structured Query Language (SQL) is the \emph{de facto} query language for most relational databases, containing commands used to access and manipulate data. Since released by \citet{codd1970relational}, SQL has been used by most database vendors for their products. Although SQL is highly declarative, many end-users encounter several challenges in writing correct queries. Such challenges have limited its use as the preferred database language of choice. This part of the thesis provides an introduction to what SQL is, describes several terms used, and reviews literature similar to this work.
	
\noindent This part contains three chapters. \autoref{ch:introduction} introduces and provides the context for this  
research. \autoref{ch:definitions} outlines several definitions that were used in this study. \autoref{ch:Background} presents the literature on SQL comprehension and other areas of study.
}
\part{Introduction and Background}\label{part1}
\chapter{Introduction}\label{ch:introduction}
\lettrine[lines=3,loversize=0.1]{F}\normalsize{or} the past three decades, Structured Query Language (SQL) has been the preferred database language for relational database management systems (RDBMSs)~\citep{kawash2014formulating,heller2019modify}. Since being adopted as an ANSI\footnote{American National Standards Institute} and ISO\footnote{International Organization for Standardization} standard, SQL has been widely used by most database vendors for their commercial products, such as IBM\footnote{International Business Machines} DB2, Microsoft SQL Server and Access, SAP HANA, Splunk DB, Teradata DB, etc~\citep{levene2012guided,bonham2014geographic,heller2019understand}. Similarly, many open-source RDBMSs have been introduced that support SQL, such as Oracle's MySQL, PostgreSQL, Mozilla's Firebird, MariaDB, Ingres, etc~\citep{soflano2015learning,heller2019understand}.

\begin{figure*}[h]
	\centering
	\includegraphics[width=200px]{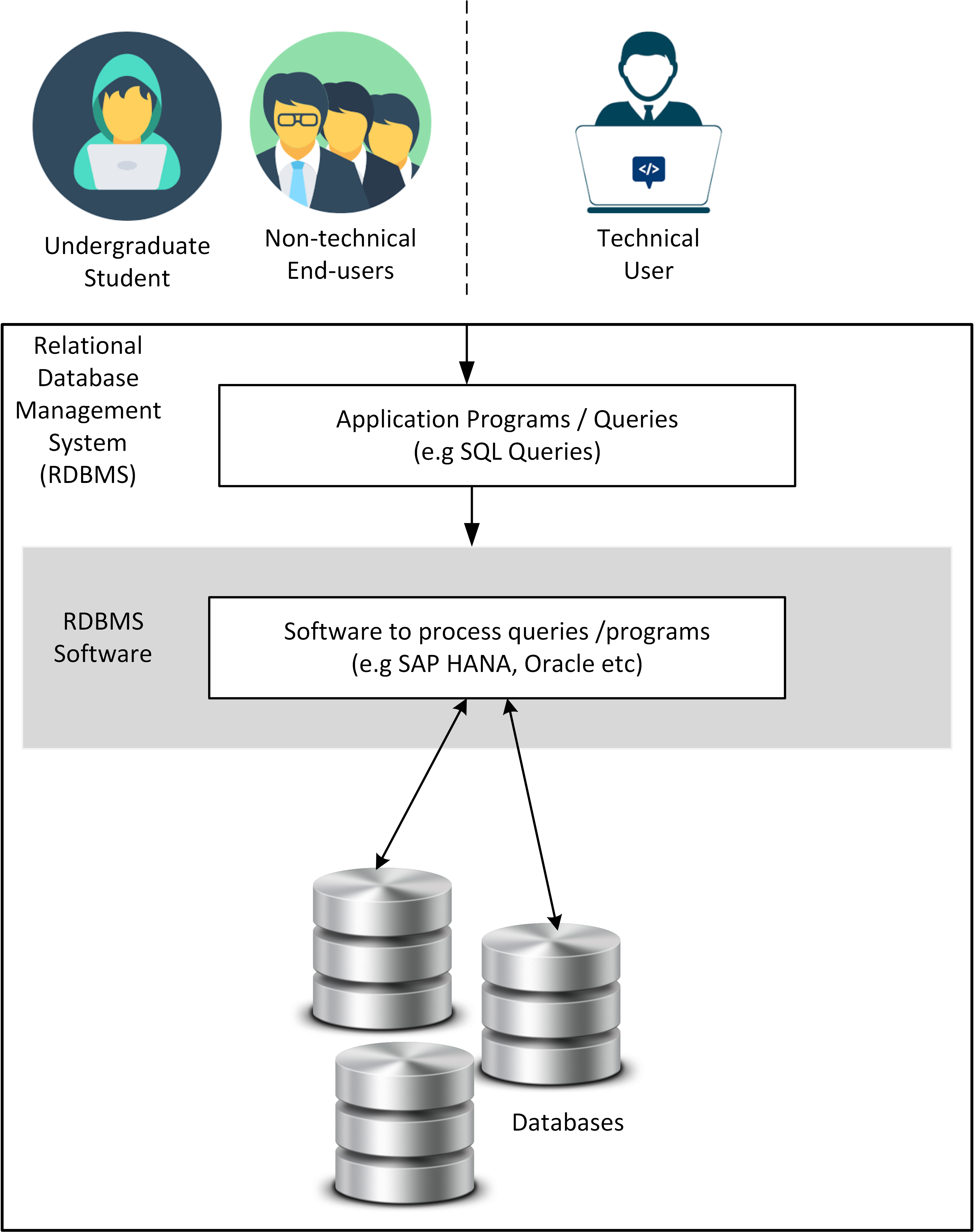}
	\caption[An interaction between users and a RDBMS ]{An interaction between users and a RDBMS}
	\label{fig:concept}
\end{figure*}

\bigskip
SQL has found many applications in academia and industry \citep{borodin2016development,chamberlin2012early,de2017relational}. In higher
learning institutions, SQL is taught as part of the introductory database course in the undergraduate curriculum \citep{silva2016sql}. Learning SQL is a pivotal skill that a Computer Science (CS) student ought to master as it is pertinent for an entry role in many diverse industries~\citep{cappel2002entry,sander2019integrating}. \citet{garner2015learning} suggest that even non-technical end-users, such as financial managers, stock brokers and controllers as well as HR managers in industry, should be able to write queries as part of their job functions. However, that may not always be the case. \autoref{fig:concept} shows the interaction between end-users and a RDBMS. \\

It is worth noting that SQL underpins a range of applications and programming languages to allow users to manipulate and retrieve information. These applications range from e-commerce, Internet of Things (IoT), commercial as well as open-source software. The following examples show some applications of SQL:

\begin{enumerate}
	\item A number of Extract Transform Load (ETL)\footnote{A data warehousing process for extracting data from different sources to fit into organisational needs} tools such as SQL Server Integration Services (SSIS), Skyvia and Informatica use SQL to communicate with databases.
	\item Programming languages such as Python and PHP usually embed SQL in their query strings when connecting to a database.
	\item Top Business intelligence (BI) tools such as the Microsoft Power BI and Tableau use SQL to create reports and charts while working with data.
	\item Most mobile application development frameworks: hybrid (Ionic, PhoneGap, Xamarin, etc) and native (Android Studio and Swift) support SQL in their engines; since many of these frameworks support SQLite.
	\item Big data analytic tools that support the RESTFUL API such as Elasticsearch and Sphinx execute SQL queries to produce results.
\end{enumerate}

\bigskip
Just like any other language, it is generally agreed that SQL is challenging for non-technical end-users and undergraduate students alike~\citep{bider2016yasqlt,grillenberger2012eledsql,folland2016visqlizer}. Many studies have identified the difficulties faced by non-technical end-users when writing SQL queries~\citep{garner2015learning,najar2016learning,atchariyachanvanich2017development}. These difficulties include: the burden of memorising database schemas, its declarative nature, and the naive perception that SQL is easy. \citet{li2014constructing} acknowledge that non-technical end-users struggle to comprehend queries written by technical users~\citep{li2014constructing}. Similarly, \citet{garner2015learning} noted that this is also true for undergraduate CS students. They emphasise that students struggle to learn SQL alongside a procedural or object-oriented programming language. Another study by \citet{ahadi2015quantitative} highlighted that simple and nested SQL queries are often problematic for students. Hence, it has become imperative for researchers to design learning and practice aids (mostly electronic) for these two audiences -- undergraduate students, and non-technical industry end-users -- to aid their comprehension of SQL. Moreover, tools that provide brief explanations of a query's functionality and visualisation of a query's output, can assist these users in understanding SQL~\citep{kokkalis2012logos,danaparamita2011queryviz}. In addition, tools with speech recognition capabilities can assist visually impaired learners in understanding SQL~\citep{berque2003case,mealin2012exploratory}.  

\bigskip
This work introduces “SQL Comprehension and Synthesis'', using approaches that allow users (students and non-technical end-users) to understand SQL queries and provide interfaces to write queries in clear English terms -- in a natural language. Such approaches abstract user-level queries and provide granularity of representation, identify syntax and semantic errors and provide possible solutions. Query comprehension involves tasks that make use of a mental process aimed at query reading and explanation~\citep{shneiderman1978improving,shneiderman2000creating}. Although query comprehension has been investigated as \textit{patterns} introduced to aid knowledge transfer by \citet{faroult2006art}, it was specifically targeted for experienced database developers to solve more complex query tasks. However, if a novice user cannot even construct simple queries, these patterns tend to offer no solution to the current challenge. Compared to the program comprehension domain that has gained popularity in recent years~\citep{storey2006theories,ade2014abstracting,ade2015introducing}, there are still many unexplored areas of research in the SQL comprehension domain.

\bigskip
Over the past decades, the program comprehension domain has recorded many successes. This is motivated by two classic cognitive theories, namely the \emph{top-down} and \textit{bottom-up} \citep{brooks1977towards,storey2006theories} approaches. The top-down approach focuses on how programs are perceived by programmers based on a series of hypotheses. The bottom-up approach uses tools to aid the comprehension process. This process was termed chunking, which shows obvious parts of code that a programmer may recognise. Together, these theories are regarded as the unified model~\citep{von1993program}. Within Computer Science Education, a systematic approach was introduced by \citet{fincher1999we} aimed at teaching programming without the syntax of the language. Such an approach is regarded as the Syntax-free Approach (SFA). The SFA has been applied in many program comprehension problems in elementary education~\citep{mannila2014computational}, middle-age education~\citep{grover2015designing}, and for undergraduate programs \citep{lahtinen2005study,ade2014abstracting,ade2015introducing,ade2016automatic}. Therefore, this was the approach employed in this research to assist students and non-technical end-users understand SQL queries.

\section{Problem Description}
This section describes the problems that were solved in this thesis. They are listed under the following categories.

\subsection{Generating Narratives for Simple Queries}
Many tools have been developed to assist end-users understand SQL queries \citep{cembalo2011savi,folland2016visqlizer}. The majority of these tools apply visualisation and interactive techniques to aid the comprehension of SQL queries. While these tools have shown to be effective at improving comprehension, many end-users struggle to understand the constructs and underlying logic behind SQL. Given this difficulty, it is important to granulate these queries into a form that is free from SQL syntax. With respect to this, we have answered these questions:
\begin{enumerate}
\item How else can we abstract queries so that they are easily understood by students and non-technical end-users?
\item Can we design a tool that addresses this problem?
\end{enumerate}
To answer these questions, we took a cue from the program comprehension domain. Within the programming pedagogy, \citet{fincher1999we} suggested that if teaching a programming language with the syntax affects the learning process, teaching without it would attempt to avoid it. In her words, this is “paradoxical”~\citep{fincher1999we}. The author suggested that the ideal way of teaching a programming language was the SFA, which abstracts the syntax of the language into a readable form. That is, translating programs back into syntax-free algorithms specified in natural language descriptions known as “narrations” \citep{ade2014abstracting}. Hence, we \textit{propose a tool that generates narratives for simple SQL queries using Regular Expressions (REs) for SQL comprehension}.

\subsection{Generating Narratives for Nested Queries}
A number of studies have identified that nested queries pose great difficulties for end-users \citep{woolf2010building,ahadi2016students}. In particular, end-users struggle to understand nested query constructs using the \texttt{SELECT} commands and \texttt{GROUPBY} clauses. An end-user who struggles to understand a simple SQL query will surely find nested queries challenging. Many tools have been built to handle simple queries, hence there is a growing demand for tools that aid nested SQL query comprehension~\citep{neumann2015unnesting}. It has become attractive for researchers to develop tools that handle these types of queries. We have answered the following questions:

\begin{enumerate}
	\item How can we abstract nested SQL queries so that users can understand them?
	\item Can we develop a tool for this?
\end{enumerate}

\noindent Given these questions, we \textit{extended the use of narrations to describing nested queries using a Context-free Grammar (CFG) for nested SQL query comprehension}. Since nested queries appear in more complex forms, REs are not suitable to handle this hitch as they are only useful for lexical analysis. 

\subsection{Synthesising SQL Queries from Narrations}
In the business sector, many professionals such as financial analysts, marketing executives, stock brokers, and mining experts use DB applications on a daily basis. These end-users often are unable to write correct queries. In most cases, they can clearly describe the intended task but lack the ability to formulate a correct query. As a result, they seek help from online sources and from technical experts to assist them with their query operation~\citep{yaghmazadeh2017sqlizer}. \citet{zhang2013automatically} identified two popular approaches to assist end-users in writing SQL queries: Graphical User Interfaces (GUIs) and the use of programming languages. It is worth noting that most RDBMSs are designed with GUI features that an end-user may struggle to find. Even so, writing SQL alongside a programming language requires good technical skills. Since end-users lack good programming skills, they may struggle to write correct SQL queries \citep{folland2016visqlizer, prior2014assesql}. An ideal approach would be to allow these users to express their requests in natural language. We have used the term “narrations” to describe a natural language. The following questions have been answered:

\begin{enumerate}
	\item How can we synthesise SQL queries from narrations?
	\item Can we describe an approach that automatically synthesises these queries?
	\item Can we develop a tool for this?
	\item What are end-users' perceptions of the tool?
\end{enumerate}
To answer this question, an approach that considers a context-sensitive language such as a natural language will suffice. Since natural languages are ambiguous, we \textit{propose the use of a Jumping Finite Automaton (JFA) to translate natural language descriptions into SQL queries}. 

\subsection{Generating SQL Queries using Visual Specifications}
Many RDBMSs contain query builders which are used to visualise a database schema through multiple mouse clicks~\citep{marcus2019neo}. Query builders allow a user to specify a query which retrieves multiple columns and table relations from a database. Whilst this approach provides many advantages, end-users need to remember that memorising a database schema may be difficult~\citep{garner2015learning}. In some cases, syntax errors from RDBMSs may be difficult for end-users to debug and the feedback received may not offer much help \citep{lavbivc2017recommender}. In addition, the SQL \texttt{SELECT} command is often problematic for learners \citep{sadiq2004sqlator}. In relation to this, we have answered the following questions:

\begin{enumerate}
	\item Which visualisation would assist users understanding of SQL queries?
	\item Does our approach require knowledge of SQL?	 
	\item Can we compare our approach with other existing tools?
\end{enumerate}

To answer these questions, we \textit{propose an image query visualiser that uses the drag and drop interactions to generate a SQL query}.

\subsection{Generating SQL Queries using Verbal Specifications}
Many Intelligent Tutoring Systems (ITSs) use speech commands to provide immediate and customised instruction to serve both educational and industrial needs~\citep{graesser2012intelligent}. Whilst these speech ITSs perform operations seamlessly, they only consider the SELECT statement, ignoring other query commands \citep{garner2015learning}. In addition, some speech ITSs fail to provide comprehensive feedback to a user. The following questions have been answered:

\begin{enumerate}
	\item Can we build a tool that supports speech inputs?
	\item Can visually impaired users use this tool?
\end{enumerate}
To answer these questions, we \textit{propose a speech-based system that takes speech inputs from a user, converts these into a query and provides speech, textual and visual feedback to the user}.

\section{Research Context}
\subsection{Aim and Objectives}
The aim of this research is to develop a new approach that makes it easier to assist non-technical end-users and students to understand SQL queries and also build software tools that tests this new approach. The objectives that make up the aim are to:

\begin{enumerate}
	\item \emph{aid the comprehension of queries through narrations}. The term “narrations’’ was first coined by \citet{ade2014abstracting} used to describe a textual approach aimed for novice program comprehension. This textual approach describes
	a program’s functionality free from a program's syntax, which is written in plain text. This idea is based on the SFA idea by \cite{fincher1999we}. In this work, we have extended the use of narrations to describe simple and nested queries. This will be based on REs and CFGs, two formal language techniques for generating languages. 
	\item \emph{translate natural language specifications of queries into standard SQL queries}. In most cases, non-technical end-users struggle to write correct SQL queries. This aspect of the research focuses on assisting non-technical end-users to write SQL queries. It allows end-users to specify their request using a natural language, which undergoes a number of transformations before a query is generated. This uses a JFA, an automata algorithm.
	\item \emph{build an interactive visualiser}. Visualisation has shown to improve the cognitive workload for understanding a concept~\citep{cembalo2011savi,mitrovic2016implementing}. The visualisation we have employed uses images depicting SQL operations to build queries. This approach uses the ‘drag and drop’ interactions for generating SQL queries.
	\item \emph{build a speech to SQL query synthesiser}. This is meant to assist users understanding of queries using speech inputs. This aspect also employs REs for the recognition of queries for feedback generation. 
	\item \emph{evaluate the impact of the comprehension aids proposed}. An online survey will be used to determine the effect of this tools.
\end{enumerate}

\subsection{Scope}
This research is based on the assumptions that:

\begin{enumerate}
	\item the scope is confined to the use of formal language techniques using REs and CFGs for the recognition of SQL queries, and an automata-based algorithm using a JFA for the automatic synthesis of SQL queries from natural language specifications.
	\item the focus will be on comprehending only SQL queries in simpler and nested forms.
	\item we only focus on comprehending and synthesising SQL queries for these end-users, namely: \textit{undergraduate students} in higher learning institutions and \textit{non-technical users} in the business sectors. 
\end{enumerate}

\subsection{Questions}
The questions of interest that have been answered in this research are as follows:
\begin{enumerate}
	\item \emph{can we build tools that can automatically narrate an SQL query?} -- yes, we answered this question in \autoref{part2}. In this part, we developed two tools; using REs to recognise simple SQL queries (in \autoref{ch:regular}) and a CFG for nested SQL query recognition (in \autoref{ch:cfg}). The resultant narrations were presented in these chapters.
	\item \emph{given good narrations, is it possible to come up with a valid SQL query?} -- yes, we showed that this was possible. We used a JFA to recognise natural language descriptions and used algorithms to generate a valid query. The idea is presented in \autoref{ch:nsql} under \autoref{part3}.
	\item \emph{can visually impaired users use these tools?} Definitely, they can use these tools. We showed that with the aid of a conversational tool in \autoref{ch:vsql} in \autoref{part3}, that these type of learners can use these tools to manipulate and access data from a database.
	\item \emph{can these tools find applications in both academic and business environments?} This question is answered in \autoref{ch:eproto} in \autoref{part4}. Participant feedback is also presented here.
\end{enumerate}

\section{Technical Contributions}
The contributions of this work are divided into three categories: formal techniques for the comprehension and synthesis of SQL queries, software prototypes of these techniques and evaluation of the prototypes.

\subsection{Formal Language and Automata Theory}
Formal language and automata theory (FLA) has been used in a wide spectrum of application areas. This research explored the ideas from this domain for SQL understanding. In \autoref{part2} and \autoref{part3}, we have:
\begin{enumerate}
	\item designed REs, a class of regular languages for the recognition of SQL query constructs,
	\item designed a CFG, a subset of irregular languages for the recognition of nested SQL queries, and
	\item used a JFA, an automata-based algorithm, to recognise natural language specifications.\\
\end{enumerate}

\subsection{Software Prototypes}
The FLA algorithms using REs, CFG and JFA were implemented into a number of software prototypes. The following prototypes are:
\begin{enumerate}
	\item \texttt{S-NAR}~\citep{ade2017s}: We developed a tool that uses REs which automatically generates a narration from a query in order to aid the comprehension of SQL queries.
	\item \texttt{SQL Narrator}~\citep{obaido2019generating}: This aspect is an improvement of \texttt{S-NAR} that describes the automatic generation of narrations from nested SQL queries using a CFG.
	\item \texttt{Narrations-2-SQL}~\citep{obaido2019narrations}: We presented a tool that uses a JFA for the recognition of natural language specifications of queries and translated them into a SQL query.
	\item \texttt{SQL Visualiser}~\citep{obaido2018generating}: We used visual specifications that represent SQL commands to build queries.
	\item \texttt{TalkSQL}~\citep{obaido2019talksql}: We presented a speech-based system that takes speech inputs from a user and uses REs to convert this into a query. 
\end{enumerate}

\subsection{Evaluation of Prototypes}
For each aspect of the study, we present the results of the evaluation of the tools that were developed in \autoref{part4}.
\begin{enumerate}
	\item \texttt{S-NAR}~\citep{ade2017s} was tested on 5000 queries and reported an accuracy of 96\%.
	\item \texttt{SQL Narrator}~\citep{obaido2019generating} reported that 98.1\%  agreed that the tool enabled them understand nested queries.
	\item \texttt{Narrations-2-SQL}~\citep{obaido2019narrations} showed that 96.9\% agreed the tool would be helpful to end-users.
	\item \texttt{SQL Visualiser}~\citep{obaido2018generating} reported that 92.16\% indicated that the tool aided their understanding of SQL syntax.
	\item \texttt{TalkSQL}~\citep{obaido2019talksql} concluded that 87.6\% acknowledged that the tool will help visually impaired learners to write correct queries using speech inputs.	
\end{enumerate}

\section{Thesis Organisation}

\begin{figure*}[h]
	\centering
	\includegraphics[width=280px]{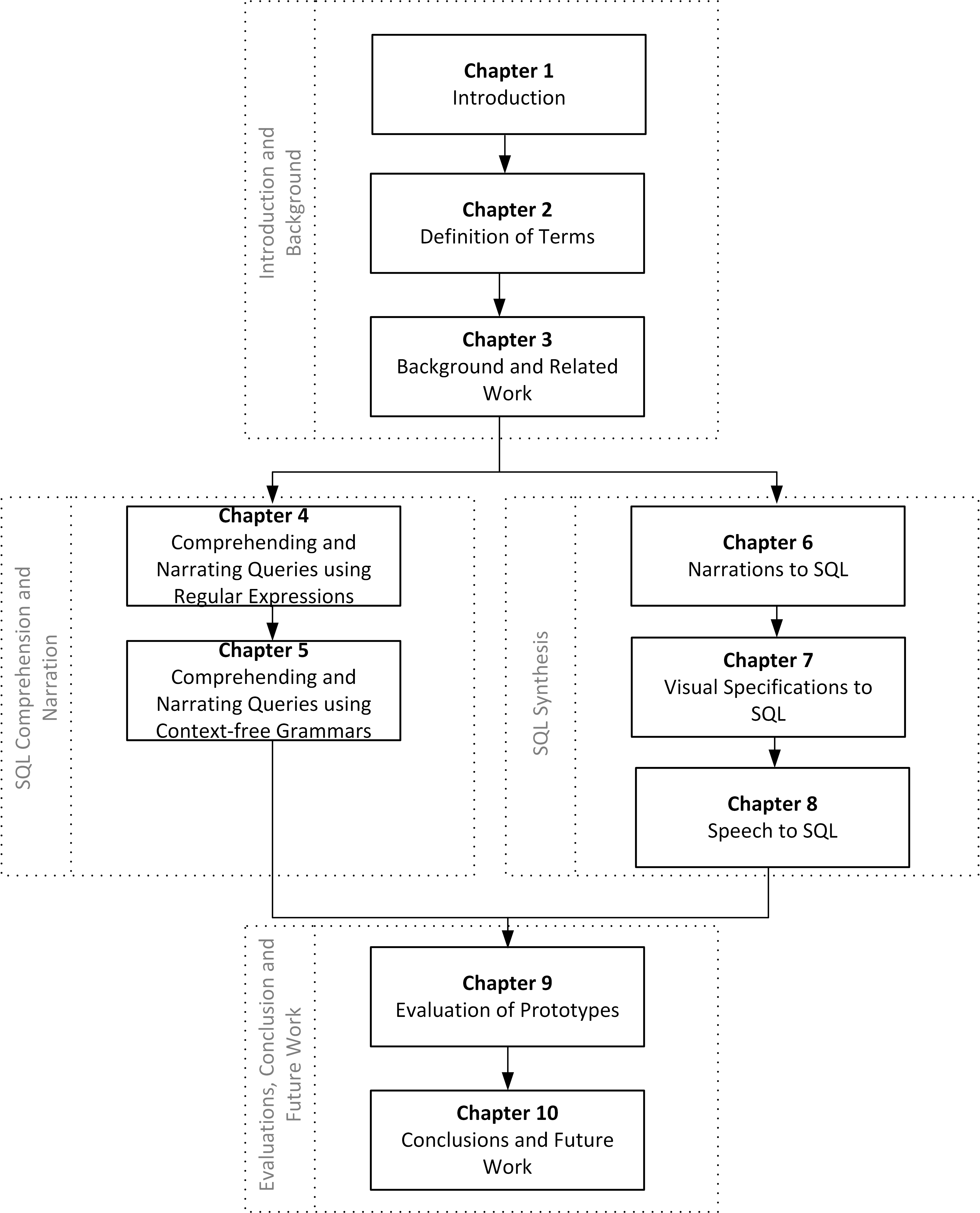}
	\caption[The organisation of this thesis]{The organisation of this thesis}
	\label{fig:structure}
\end{figure*}

\noindent This thesis is organised into parts with ten chapters as depicted in \autoref{fig:structure}. The description of Parts and Chapters are as follows.

\begin{description}
	\item[\autoref{part1}] contains two chapters. \autoref{ch:introduction} presents the introduction of this thesis and the context. \autoref{ch:definitions} highlights terms and definitions used in this work. \autoref{ch:Background} presents the literature reviewed for this work.
	\item[\autoref{part2}] focuses on the comprehension aspect using \textit{narrations} for aiding SQL comprehension. In \autoref{ch:regular}, narrations for simple queries using REs are introduced with detailed illustrations on how this will aid learning SQL queries for the first time. \autoref{ch:cfg} describes the generation of narrations using CFGs, aimed at assisting a user in understanding nested queries.
	\item[\autoref{part3}] describes the synthesis aspect of our work in three chapters. \autoref{ch:nsql} presents the use of a JFA to synthesise SQL queries from natural language specifications. This aspect shows the use of natural language descriptions to write SQL queries. \autoref{ch:vsql} introduces a visualiser to generate a query. The visualiser uses drag and drop interactions using visual specifications to generate a query. \autoref{ch:ssql} describes a speech to query synthesiser, targeted at assisting end-users to write correct SQL queries using speech inputs.
	\item[\autoref{part4}] presents the evaluation and concluding aspects of this thesis in two chapters. Chapter \ref{ch:eproto} provides the evaluation of the study. \autoref{ch:conclusion} presents the conclusion and future directions of this work. 
\end{description}
\chapter{Definition of Terms}\label{ch:definitions}
\lettrine[lines=3,loversize=0.1]{T}\normalsize{his} chapter presents the definitions of the terms used in this thesis. The terms are listed in different categories across the following areas: FLA, SQL comprehension and concepts, computational linguistics, and other terms.
\section{Formal Terms}
\begin{definition}[Lexical Analysis~\citep{grune2012modern}] This is the initial phase of a compiler where program texts are converted into a stream of tokens, white spaces and comments are removed. 
\end{definition}
\begin{definition} [Syntax Analysis~\citep{wilhelm2013compiler}] At this phase, the stream of tokens received at the lexical analysis phase is used to produce a tree-like data structure (parse tree or abstract syntax tree). This phase uses a CFG to construct the parse tree. This phase is also called parsing. 
\end{definition}	

\begin{definition}[Formal Language Theory~\citep{post1944recursively,perrin2003automata}] The field of formal language theory (FLT) has its root in mathematics. This became popular in 1956 when Noam Chomsky conducted an investigation into natural languages \citep{chomsky1956three,chomsky1959certain}. Since its inception, FLT has been applied in different domains such as switching circuits, neural networks, compiler designs (parsers), cryptography and computer graphics \citep{karhumaki2007automata,kari2013automata}. According to the \citet{chomsky1956three} hierarchy, formal languages are categorised into classes of increasing complexity such as regular languages, context-free languages, context-sensitive languages and recursive enumerable languages. Each of these languages is generated from grammars and can be defined by its respective types. For instance, regular languages are defined by regular grammars, etc. \autoref{fig:noam} presents Chomsky's hierarchy.

\begin{figure*}[htbp]
	\centering
	\includegraphics[width=200px]{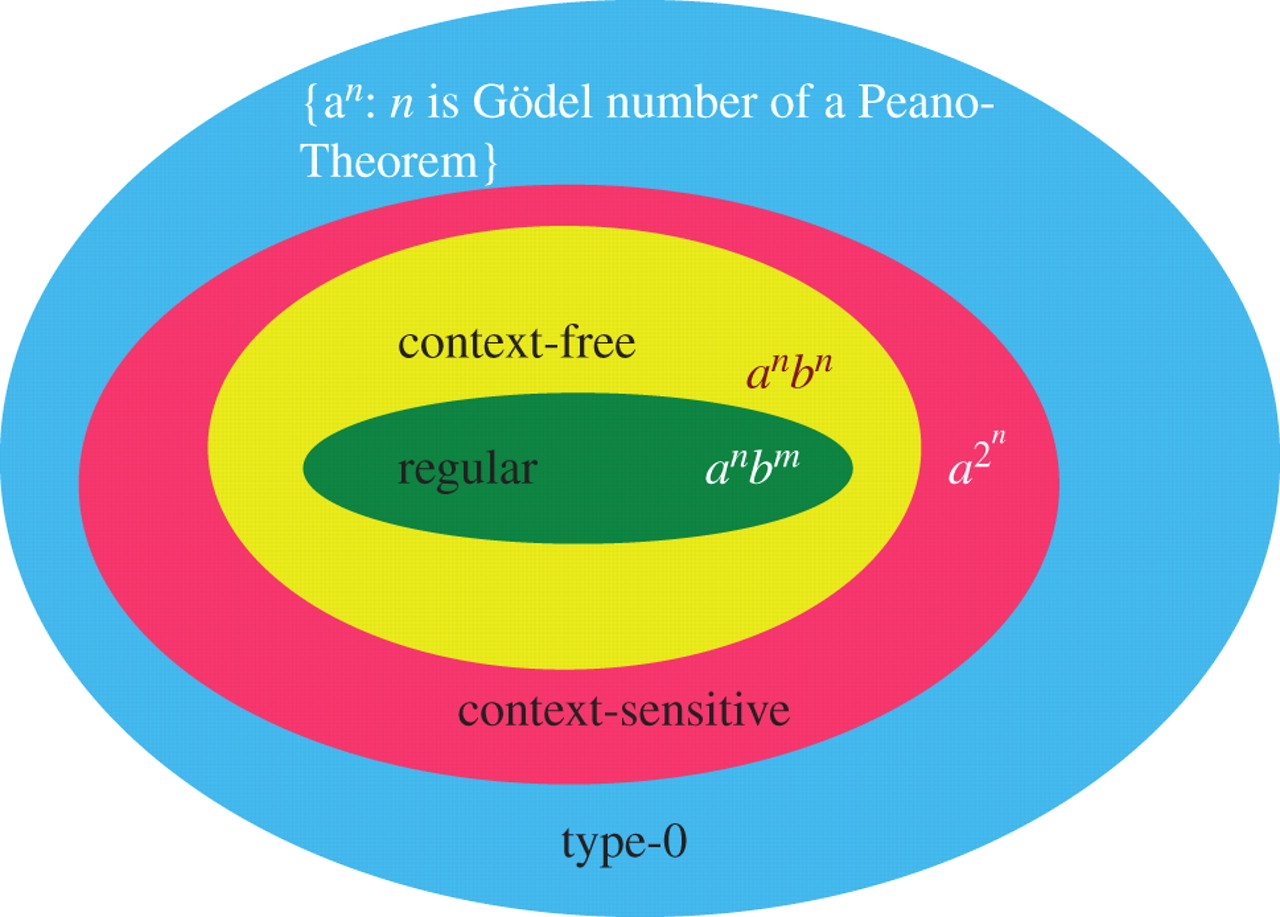}
	\caption[Chomsky's hierarchy of increasing complexity~\citep{jager2012formal}]{Chomsky's hierarchy of increasing complexity~\citep{jager2012formal}}
	\label{fig:noam}
\end{figure*}
\end{definition}
\begin{definition}[Basics~\citep{domosi2016chomsky,chomsky1956three}]In this section, we present some basic definitions. 
	\begin{itemize}
		\item An \textit{alphabet}, $\Sigma$, is a finite set or (collection) of symbols.
		\item A \textit{string} or (\textit{word}) is a finite sequence of zero or more symbols.
		\item A \textit{symbol} is an abstract entity or an item.
		\item A \textit{language}, ${L}$, is an infinite collection of strings over some alphabet $\Sigma$.
		\item The symbol $\Sigma^*$ is a set of finite or (non-empty) strings over $\Sigma$, where the symbol ($^*$) is known as the Kleene star.
		\item The symbol $|w|$ is the length of a string $w$.
		\item The symbol $\lambda$ or $\varepsilon$ is an empty string.
	\end{itemize}
\end{definition}
\begin{definition}[Regular Languages and Regular Expressions (REs)~\citep{ruohonen2009formal,ade2016finchan}]\label{regex2}
	Let \textit{R} denote the regular language over an alphabet $\Sigma$ or \textit{R$_\Sigma$}. The REs that follow are:
	
	\begin{enumerate}
		\item $\emptyset$ is in \textit{R}, representing the empty set.
		\item $\lambda$ is in \textit{R}, representing the empty string.
		\item For each symbol a,  the language $\lbrace a \rbrace$ in \textit{R}, representing the regular expression $a$.
		\item Let ${L_x}$ and ${L_y}$ is in \textit{R}, then, 
		\begin{itemize}
			\item ${L_x}\cup{L_y}$ is in \textit{R}, representing the union of both languages. 
			\item ${L_x}{L_y}$ is in \textit{R}, representing the concatenation of both languages. ${L_x}^{*}$ is in \textit{R}, denoting the Kleene star.
		\end{itemize}	
	\end{enumerate}
Regular languages have a wide range of applications from circuit design and text editing to pattern matching \citep{campeanu2003formal,yu2012regular}. Their widespread use is due to their highly expressive power. They have become well recognised pattern matching languages~\citep{yu2012regular}.
\end{definition}	
\begin{definition}[Context-free grammars (CFGs) \citep{ruohonen2009formal,ade2016finchan}]
	CFGs consist of four key components (or \textit{tuples}) such as $G$ = \{$N$, $\Sigma$, $P$, $S$\} where:
	
	\begin{enumerate}
		\item $N$ is a set of non-terminals (or lexicon) symbols.
		\item $\Sigma$ is a set of terminal symbols.
		\item $P$ is a set of production rules.
		\item $S$ is a start symbol, where $S$$\in$ $N$.
	\end{enumerate}	
\begin{remark}
Hence, a language generated by CFG can be repeatedly enumerable by applying production rules $P$, by starting with the start symbol $S$, then replacing the non-terminals $N$ with the corresponding production rules $P$ until all non-terminals $N$ have been reached. CFGs can be rewritten in Backus-Naur form (BNF), also denoted as (::=) as presented below, and have been applied to many real-world problems \citep{javed2009techniques,huang2014historical}. It is interesting to note that the SQL ISO 2003 uses CFGs for parsing \citep{schmitz2007approximating}. 
\end{remark}

\end{definition}	

\begin{definition}[Finite Automaton (FA) or Finite State Machine (FSM)~\citep{kupferman2018automata,genise2019homomorphic,meduna2017modern}] A FA or FSM is a 5-tuple where:
	\begin{enumerate}
		\item $Q$ is a finite set of states.
		\item $\Sigma$ is the finite input alphabet.
		\item $\delta:$ $Q$ $x$ $\Sigma$ $\longrightarrow$ $Q$ is the transition function.
		\item $s$ $\in$ $Q$ is the start state.
		\item $F$$\subseteq$$Q$ is the set of final (accept) state.
	\end{enumerate}	
	
A FA consists of several parts, which has a set of states and rules for moving from one state to another, depending on the input symbol. FAs are drawn with states as circles, start state indicated by the arrow pointed at it, accept or final states with a double circle and arrows going from one state to another as the transitions. This is shown in \autoref{fig:M1}. \\
\begin{figure}[h]
	\begin{center}
		\begin{tikzpicture}[scale=0.15]
		\tikzstyle{every node}+=[inner sep=0pt]
		\draw [black] (19.4,-23.4) circle (3);
		\draw (19.4,-23.4) node {$X$};
		\draw [black] (37,-23.4) circle (3);
		\draw (37,-23.4) node {$Y$};
		\draw [black] (37,-23.4) circle (2.4);
		\draw [black] (55.3,-23.4) circle (3);
		\draw (55.3,-23.4) node {$Z$};
		\draw [black] (11.7,-23.4) -- (16.4,-23.4);
		\fill [black] (16.4,-23.4) -- (15.6,-22.9) -- (15.6,-23.9);
		\draw [black] (18.077,-20.72) arc (234:-54:2.25);
		\draw (19.4,-16.15) node [above] {$a$};
		\fill [black] (20.72,-20.72) -- (21.6,-20.37) -- (20.79,-19.78);
		\draw [black] (21.827,-21.647) arc (119.25024:60.74976:13.043);
		\fill [black] (34.57,-21.65) -- (34.12,-20.82) -- (33.63,-21.69);
		\draw (28.2,-19.48) node [above] {$b$};
		\draw [black] (39.307,-21.493) arc (122.7854:57.2146:12.638);
		\fill [black] (52.99,-21.49) -- (52.59,-20.64) -- (52.05,-21.48);
		\draw (46.15,-18.98) node [above] {$a$};
		\draw [black] (53.289,-25.614) arc (-49.98957:-130.01043:11.104);
		\fill [black] (39.01,-25.61) -- (39.3,-26.51) -- (39.95,-25.75);
		\draw (46.15,-28.71) node [below] {$a,b$};
		\draw [black] (36.056,-20.565) arc (226.14669:-61.85331:2.25);
		\draw (38.25,-16.12) node [above] {$b$};
		\fill [black] (38.68,-20.93) -- (39.59,-20.7) -- (38.87,-20);
		\draw [black] (36.056,-20.565) arc (226.14669:-61.85331:2.25);
		\fill [black] (38.68,-20.93) -- (39.59,-20.7) -- (38.87,-20);
		\end{tikzpicture}
			\caption{A finite automaton, M, with three states} \label{fig:M1}
	\end{center}
\end{figure}
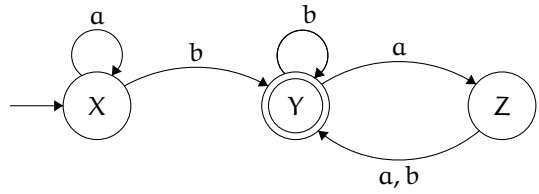	

The finite automaton, M, can be described formally by ($Q$, $\Sigma$, $\delta$, $X$, $Z$), where:
\begin{enumerate}
	\item $Q$ = \{X,Y,Z\} is a finite set of states.
	\item $\Sigma$ = \{a,b\} is the finite input alphabet.
	\item $\delta:$  is the transition function, represented by a transition table: 
\begin{displaymath}
\begin{array}{c|c|c|c}
{} & {} & \\ \hline
 & a & b \\\hline
X & X & Y \\
Y & Z & Y \\
Z & Y & Y \\
\end{array}
\end{displaymath}

	\item $X$ $\in$ $Q$ is the start state.
	\item $Z$$\subseteq$$Q$ is the set of final (accept) state.
\end{enumerate}
\begin{remark}
If the set of all strings (A) that the machine (M) accepts, then A is a language of machine M or $L(M)$ = $A$. A machine may accept several strings but always recognises only one language. If no string is accepted by a machine, M, it still recognises an empty, $\theta$, language. 
\end{remark}
\end{definition}
\begin{definition}[A General Jumping Finite Automaton or GJFA \citep{meduna2012jumping,kvrivka2015jumping,fernau2015jumping}]
		A GJFA is a quintuple such that $M$ = ($Q$, $\Sigma$, $R$, $s$, $F$) where:
		
		\begin{enumerate}
			\item $Q$ is a finite set of states.
			\item $\Sigma$ is the finite input alphabet.
			\item $R$ is the finite set of rules, where $py$ $\rightarrow$ $q$ ($p$,$q$ $\in$ $Q$, $y$$\in$$\Sigma$). 
			\item $s$$\in$$Q$ is the start state.
			\item $F$$\subseteq$$Q$ is the final state.
		\end{enumerate}		
If all rules $py \rightarrow  q$  $\in$ $R$ satisfy $\mid y\mid  \leq 1$, then M is a JFA. Also, L(M) is the language accepted by the automaton. A JFA is based on a Finite Automaton (FA). In a JFA, the input string is not read in a left to right manner. That is, if a symbol $M$ is read, it jumps continuously over a pool of information to an execution point. During computation, a symbol cannot be re-read once it has been read once.

\begin{remark}
	As a JFA, M is any string in  $\Sigma^*$$Q$$\Sigma^*$, which represents the binary jumping ${\curvearrowright}$. It satisfies the condition:
\begin{center}
$vpw$ $\curvearrowright$ $v^\prime qz^\prime$ $\Longleftrightarrow$ $\exists$ $p$$y$ $\rightarrow$ $q$ $\in$ R $\exists$ z $\in$ $\Sigma$*: w =  yz $\wedge vz = v^\prime z^\prime$.
\end{center}	

It is worth noting that a JFA can be used to represent a Context-sensitive Language (CSL) \citep{meduna2014regulated,meduna2017modern}.
\end{remark}

\end{definition}

\section{SQL Comprehension Terms}
\begin{definition}[Cognitive Workload~\citep{miklody2017maritime}] This is the level of measurable effort exerted by a brain in multiple cognitive tasks. Cognitive workload is reflected in the level of brain activity.
\end{definition}
\begin{definition}[Mental Model \citep{johnson2010mental,de2013interaction}] This is an abstract notion that builds psychological explanations of how something works. Mental models guide reasoning, behaviour and perception.
\end{definition}
\begin{definition}[Pedagogical Models~\citep{renaud2004teaching}] Pedagogical models (or pedagogical patterns) aim to find the best way of teaching for the purpose of sharing knowledge. Patterns have been taken up in many fields and each has a particular structure and vocabulary. For example, the medical science pattern is different from that in software engineering. The pedagogical models consist of:
	\begin{enumerate} 
	 \item Issues, which involves knowledge transfer of a particular type. 
	 \item Strategy, aimed at transfering knowledge in a particular manner. 
	 \item Implementation, which provides the delivery of content in the way specified by the strategy.
	\end{enumerate}
\end{definition}
\begin{definition}[Theory of Constructivism \citep{bruner1966toward}] This idea suggests that new ideas can be constructed based upon experiences. The theory shows how humans learn from their past experiences.
\end{definition}	
\begin{definition}[Learning Theories~\citep{pritchard2017ways}] These are sets of principles that explain how humans acquire, process and retain knowledge. These principles show how learners progress through the phases of learning.
\end{definition}	
\begin{definition}[Learning Taxonomy~\citep{adams2015bloom,sarfraz2017strategic,verenna2018role}] This classification shows the skills that educators set for their students to achieve. The taxonomy shows the level of cognition required for a course using a set of objectives. An example of a learning taxonomy is the Bloom's taxonomy of learning. This taxonomy shows the movement from basic (knowledge recall) to highest (evaluation) level of cognition. Other levels fall in between, which include comprehension, application, analysis and synthesis. The different taxonomies of learning discussed in this thesis are Bloom's Taxonomy, Anderson and Krathwol's Taxonomy, Gorman's Taxonomy and the CS Taxonomy.  
	
%
	
\end{definition}	
\begin{definition}[Problem-based Learning (PBL)~\citep{hmelo2004problem}] This is an instructional approach in which students learn from problem solving. In this method, there is no single correct answer, learners collaborate in groups to identify what is required to solve a problem. 
\end{definition}
\begin{definition}[Pedagogy Style~\citep{renaud2004teaching,mitrovic2003intelligent}] This is the principle that shows how a concept should be taught. The SQL pedagogy involves two methods, using instructor-led and electronic tools.
\end{definition}	
\begin{definition}[Syntax-free Approach (SFA)~\citep{pyott1991alex,fincher1999we,ade2014abstracting,ade2016automatic}] This approach is based on the principle of teaching novices how to program without its inherent syntax. The approach suggested that programming should be taught using clear English terms with the aim of improving program comprehension.
\end{definition}
\begin{definition}[Narrations~\citep{ade2014abstracting,ade2016automatic}] Narrations adopted the SFA approach in an attempt to aid program comprehension. This approach uses a high-level descriptions of programs written in plain English often longer than programs they describe. They are also called syntax-free textual algorithms. This approach was employed in this thesis in an attempt to assist users to understand SQL queries.
\end{definition}	

\section{SQL Concept Terms}
\begin{definition}[Relational Model~\citep{codd1970relational}] The relational model was developed to model data in the form of relations (tables).
\end{definition}

\begin{definition}[Relational Database Management System (RDBMS)~\citep{coronel2016database}] RDBMS enables users to create and maintain a relational database. Once the relational database is structured appropriately, it is referred to as \emph{normalisation}. According to \citet{batra2018history},  we define the following terminologies to better understand the RDBMS concept:
\begin{enumerate}
	\item Field: This is the smallest unit of information that consists of a column in a table. A field is also termed an \emph{attribute}.
	\item Record: The record consists of each row in a table. This is also regarded as a \emph{tuple}.
	\item Table or relation: In the relational model, every relation can be depicted as a table but not every table can be termed as some relation. Hence, a table is a collection of related data organised within a database.
	\item Database: This is an organised collection of related tables and data.
	\item Query: This is the composition of a table, presented in the form of a predefined SQL query.
\end{enumerate}
 
\end{definition}
\begin{definition}[Structured Query Language (SQL)~\citep{hogan2018practical}] The SQL is a standardised, de facto language for creating and maintaining a RDBMS. Every RDBMS engine supports SQL, which has made it the most comprehensive database language. It consists of statements that support queries, updates and data definitions (DML and DDL). We define these terms as:
\end{definition}
\begin{enumerate}
	\item Data Manipulation Language (DML): The DML consists of commands which are used to load, update and query a database. These consist of statements such as SELECT, INSERT, UPDATE and DELETE.
	\item Data Definition Language (DDL): The DDL defines commands for table creation, indexes and views. Popular commands in this category are CREATE, ALTER and DROP.
\end{enumerate}
\begin{definition}[Simple Queries~\citep{beaulieu2009learning}] These are queries that include operations for selection, mapping, built-in functions and simple Boolean values. 
	\begin{exmp}
		 An example of a simple query that displays all information from a Student table with a WHERE condition as presented in \autoref{lst:sq}.
	\end{exmp}
\begin{lstlisting}[language=SQL, upquote=true, showstringspaces=false, stringstyle=\color{violet}, basicstyle=\ttfamily, caption={A simple query}, label={lst:sq}, captionpos=t]
SELECT * 
FROM Student 
WHERE Name = "Steve";
\end{lstlisting}
\end{definition}

\begin{definition}[Nested Queries~\citep{beaulieu2009learning}] Nested queries include grouping, set operations and relational operators.
\begin{exmp}
A nested query shows nesting properties with multiple SELECT statements as seen in \autoref{lst:cq}.
\end{exmp}
\begin{lstlisting}[language=SQL, upquote=true, showstringspaces=false, stringstyle=\color{violet}, basicstyle=\ttfamily, caption={A nested query}, label={lst:cq}, captionpos=t]
SELECT lastname 
FROM Student
WHERE lastname IN (SELECT lastname
                    FROM records);
\end{lstlisting}
\end{definition}
\begin{definition}[Query Builders~\citep{ceballos2012criteria}] Query builders improve the understanding of a query using drag and drop functionality. Typically, a query builder is mostly used to create queries and filters.
\end{definition}
\begin{definition}[Intelligent Tutoring Systems (ITSs)~\citep{graesser2012intelligent}] ITSs are learning platforms that incorporate computational models to provide immediate and comprehensive feedback to a learner without requiring an instructor. ITSs evolved from Intelligent Computer-Aided Instruction (ICAI) in 1987 which attempted to produce human-like behaviour \citep{elsom1987intelligent}. Such activities are classified as 'good teaching'. 
\end{definition}	
\begin{definition}[End-users vs Student \citep{connolly2005database}] The end-users are the custodians of db applications, who maintain and use data from these applications to serve their information needs. In this work, end-users are classified as:
	\begin{enumerate}
\item Non-technical end-users. They are typically less knowledgeable of the RDBMS and SQL. The majority of these users are in diverse fields such as marketing, finance, mining, etc. Typically, they access their databases through special-purpose applications to speed up their processing task. For example, QuickBooks\footnote{https://quickbooks.intuit.com/}, a popular accounting software, is mostly used by end-users such as HR managers to carry out payroll tasks.
\item Technical users. Conversely, technical users are familiar with the features offered by the RDBMS. These users are very knowledgeable in SQL, and can write application programs to support their routine tasks. Examples of such users are database administrators, web programmers, data scientists, etc.
	\end{enumerate}

A student is an individual who is studying towards a degree. A student can be an undergraduate or a postgraduate. Here, our focus is on undergraduate students as they are studying SQL for the first time. Students and learners are used interchangeably in this thesis.
\begin{remark}
In this thesis, our focus is on these users, namely: non-technical end-users in industry, and undergraduate students in academic institutions. These users are our focus because they struggle to write and understand SQL queries.
\end{remark}
\end{definition}

\begin{definition}[Cognitive Models~\citep{al2013sql}] The cognitive models for SQL involves learning, understanding and remembering. These models require that learners should accumulate a range of contexts so that their formulation and translation skills are improved. 
	\begin{figure*}[h]
	\centering
	\includegraphics[width=400px]{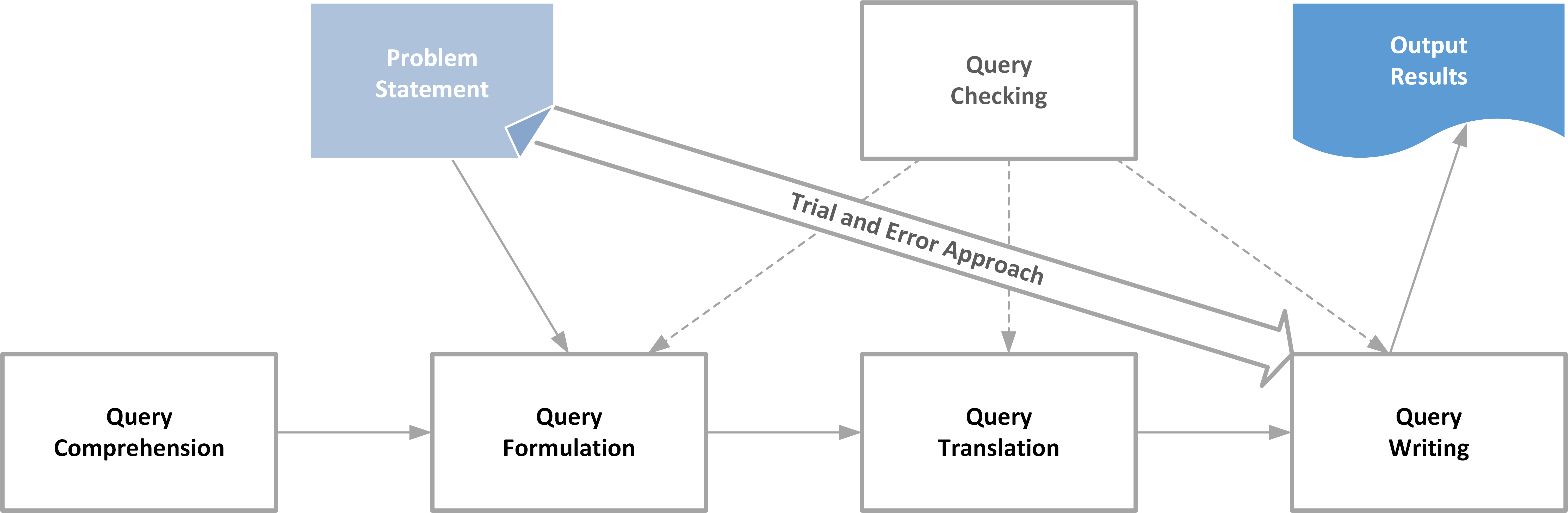}
	\caption[The cognitive models for learning queries~\citep{al2013sql}]{The cognitive models for learning queries~\citep{al2013sql}}
	\label{fig:cm}
\end{figure*}

\autoref{fig:cm} shows the cognitive models for query learning. The cognitive models are:
\begin{enumerate}
	\item Query comprehension is the stage that requires learners' skills of reading and understanding of SQL queries. This task involves reading and identifying the required query. 
	\item Query formulation is an approach that a learner is required to perform during problem solving. This approach requires the student to understand the context of the given problem.
	\item Query translation (synthesis) is a stage that requires the learner to express a query in clear English and write the related SQL query.
	\item Query writing is a factor that influences how a learner performs once a scenario is given. This phase requires that learners apply their knowledge of the SQL syntax for the provided scenario.
\end{enumerate}	
\end{definition}
\section{Computational Linguistics Terms}
\begin{definition}[Natural Language Processing (NLP)~\citep{nadkarni2011natural,manning1999foundations}] This is a branch of computational linguistics that explores how computers understand words written in human languages. NLP began in the 1950s and has been used in many applications, such as email-spam detection, text summarisation, question-answering (QA), and machine translation (MT), etc. 


 Generally, NLP is classified into Natural Language Understanding (NLU) and Natural Language Generation (NLG)~\citep{liddy2001natural}.
 
\end{definition} 
\begin{definition}[Natural Language Understanding (NLU)~\citep{sharma2019natural}] This concept helps computers to understand and interpret human languages, in speech forms. Most NLUs systems are based on statistical models, which are an important branch of NLP.
\end{definition}
\begin{definition}[Natural Language Generation (NLG)~\citep{staykova2014natural}] This process generates a natural language from non-natural language inputs. NLG is also regarded "translator", and classified as a sub-field in Computational Linguistics.
\end{definition}
\begin{definition}[Natural Language Interfaces to Databases (NLIDBs)~\citep{reinaldha2014natural,elsayed2015arabic,sharma2019natural}] NLIDBs are query interfaces, used to translate a natural language query (NLQ) in English into a database query. The first known NLIDB system was LUNAR \citep{woods1972lunar}, which was developed in the late sixties. Since then, there has been a continuous development of interfaces to assist end-users write queries.
\end{definition}
\section{Other Terms}
\begin{definition}[Visual Specifications~\citep{rojit2016visual}] Visual specifications are symbols used to represent features, which can be used to display some text or program. In the programming concept, visual specifications have been used to build and demonstrate a programming solution. Examples of visual specification tools are Scratch~\citep{resnick2009scratch}, Alice~\citep{dann2011learning} and Blockly~\citep{fraser2015ten}. These tools are generally called Block-based programming languages since they use “drag and drop” interactions to build a program. \autoref{fig:scratch} shows an example of Scratch with annotations describing program blocks.
	\begin{figure*}[h]
		\centering
		\includegraphics[width=200px]{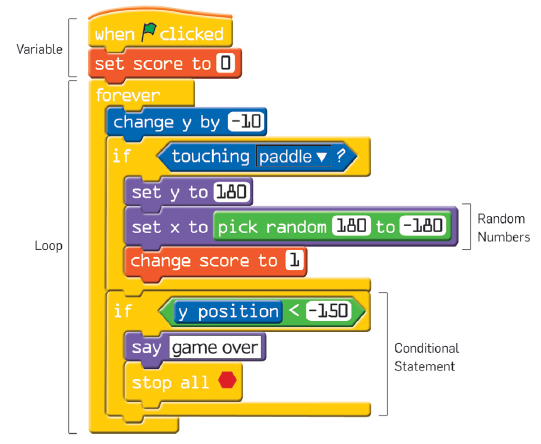}
		\caption[Example of Scratch showing program blocks~\citep{resnick2009scratch}]{Example of Scratch showing program blocks~\citep{resnick2009scratch}}
		\label{fig:scratch}
	\end{figure*}
\end{definition}	
\begin{definition}[Verbal Specifications~\citep{kantorowitz2014verbal}] Verbal specification is a process that clearly expresses the terminology of a domain where the subject matter is less understood. This process involves \textit{speech} forms. Verbal specifications have been applied in many scenarios where a concept is less well understood~\citep{kantorowitz2014verbal,meziane2008generating}. For example, in programming, to assist students produce programs from speech using an informal elicitation to allow users to translate speech into formal Unified Manipulation Language (UML) diagrams \citep{meziane2008generating}.
\end{definition}
\begin{definition}[Parsing~\citep{straka2017tokenizing}] Parsing is the transformation of a sequence of characters into a syntax tree. During parsing, a sequence of characters such as a sentence, are usually grouped into syntactic parts for recognition.  For example,"The dog barks" will be assigned to subject (“The dog”) and predicate (“barks”).
\end{definition}
\begin{definition}[Stopwords~\citep{saini2016continent}] This is a well-established method used to reduce noisy features in textual data. Stopword is based on the idea that words that are not relevant may help produce more accurate results for a classifier. This method is widely used in the NLP domain especially for document classification and information retrieval.
\end{definition}
\begin{definition}[Stemming~\citep{kotov2017nlp}] This is the process of reducing words to their base form, and has been applied to many computational linguistic problems. For example, the word “going”, would be reduced to “go” through stemming. In this case, the gerund (ing) in the word is removed from the base/root form.
\end{definition}
\section{Chapter Summary}
In this chapter, we have presented the definition of terms used in this thesis. We started by defining terms for formal language and its related languages, then we provided definitions for the SQL comprehension concepts. Next, we defined terms used in the computation linguistic domain, and other newer terms introduced in this work. \autoref{ch:Background} reviews the background and related work for this study. 

\chapter{Background and Related Work}\label{ch:Background} 
\lettrine[lines=3,loversize=0.1]{T}\normalsize{his} chapter reviews the literature related to SQL learning and other related approaches. We start by providing a general overview of SQL and a brief history, then we discuss the challenges associated with learning SQL. Next, we investigate existing learning theories and pedagogies. Furthermore, we highlight state-of-the-art tools used for teaching and learning SQL queries. Last, we discuss recent works in the NLP domain by describing existing methods and related tools.

\section{Introduction}\label{3.0}
It is a well established fact that SQL is predominantly used in the industry and taught as part of an introductory database course in the undergraduate curriculum~\citep{ahadi2015quantitative,bider2016yasqlt,grillenberger2012eledsql}. For an introductory database course, the basic SQL concepts that students are expected to master are clearly defined \citep{renaud2004teaching,mitrovic2016implementing}. More so, there are many useful resources available that are designed to assist students progress from basic to advanced levels of understanding SQL \citep{heller2019write,bergin2012pedagogical}. Many publications have presented different pedagogical approaches that address the relevant aspects of SQL that need to be covered by an educator \citep{renaud2004teaching,lavbivc2017recommender,coronel2016database,prior2014assesql}. \\

Extensive knowledge of SQL skills is vital for many organisations~\citep{liu2003examination}. Many studies have highlighted that sufficient knowledge of writing correct SQL queries remains a skill that most CS graduates have not mastered \citep{sander2019integrating,liu2003examination,mcgill2008critical}. A proficiency in SQL is highly sought after by industry employers, and is required for most entry level jobs in programming \citep{cappel2002entry}. A research study by \citet{verma2019investigation} showed that to take up programming roles, an individual must possess knowledge of SQL. Furthermore, \citet{chiang2012business} opined that a role in business analysis, data engineering and web development requires extensive knowledge of SQL. Despite the lingering demand for SQL skills, there is still the question of which specific pedagogy and learning style is preferred~\citep{sander2019integrating}. It is a well known fact that teaching and learning are connected processes that help to stimulate a learner \citep{menekse2019reflection}. \citet{mayer2016handbook} assert that learning is a process that connects new information to previous knowledge, hence it is imperative to assist learners develop their knowledge of a concept. To support learning, the mode of instruction needs to be designed to facilitate the learning process. Examples are: web-based presentations, educational games or using specialised software programs (or ITSs) \citep{collins2018rethinking}. Despite the usefulness of these instructional materials, students struggle to understand basic and complex SQL queries \citep{mitrovic2012fifteen,cembalo2011savi,chu2017cosette}. \\

Similarly, non-technical end-users in industry frequently use applications that rely on databases that store structured data \citep{sagiroglu2013big,hashem2015rise,de2017relational}. These applications make use of SQL queries to manipulate and retrieve data from these databases. \citet{wang2017interactive} opined that non-technical end-users would like to access databases and write SQL queries if they had the means to do so. Unfortunately, only technical specialists can accurately write correct SQL queries that extract useful information from these data sources \citep{elder2009natural}. As a result, non-technical end-users must rely on these specialists to generate reports of the query tasks for them. Such an approach can be time-consuming and laborious for an end-user \citep{zhang2013automatically,yaghmazadeh2017type,wang2017interactive}. A few questions come to mind, including but not limited to: What is SQL? Why is it so hard for students and non-technical end-users to understand SQL queries? What are the main causes of this and how can they be addressed? \\

This chapter continues by presenting a brief history of SQL in \autoref{3.1}. The challenges encountered when learning SQL are presented in \autoref{3.2}. The pedagogical approaches used for SQL are described in \autoref{3.3}. This is followed by cognitive theories of learning in \autoref{3.4}, and the state-of-the-art tools for SQL are presented in \autoref{3.5}. The NLP methods and similar areas are presented in \autoref{3.6}. The formal language and automata applications are discussed in \autoref{3.7}. The gaps noticed in literature that motivated this research are highlighted in \autoref{3.8}. This chapter concludes with \autoref{3.9}.

\section{Brief History of SQL}\label{3.1}   
The idea behind the relational model paved the way for \citet{codd1970relational} to release a language called the Structured English QUEry Language (or SEQUEL) to communicate with relational databases. SEQUEL was later renamed to SQL because SEQUEL had already been used for a hardware product \citep{levene2012guided,coronel2016database}. Since 1986, a joint effort by the ANSI and ISO adopted SQL as the default language for relational databases \citep{barrera2018big,heller2019understand}. Since then, SQL has undergone refinements in 1989, 1992, 1999, 2003, and 2006 \citep{coronel2016database}. The current version, SQL:2006, supports the object-oriented functionality, alongside with XML features for querying data, among other updates to the language. Most database vendors such as IBM, Oracle, Microsoft and Informix have continuously used SQL for their products. To date, SQL has remained the preferred language used to communicate with both commercial and open source database products \citep{harrison2015languages}. \\

Many of SQL statements are “English-like”, and generally are referred to as a declarative language \citep{de2017relational}. The English-like construction indicates that SQL statements are easier to learn and understand. That is, SQL statements resemble English language sentences in their formulation. In addition, SQL is widely considered as a non-procedural language by which its operation is specified (in this context, a result set) rather than a step-by-step computation~\citep{myalapalli2015appraisal}. As a result, a database engine decides the tasks and produces a result. Compared to a procedural language such as Java or C, a program can be broken into smaller chunks and a compiler can perform the desired computation in a step-by-step manner. Simply, a user is in control of what the program does. There are two variants of SQL commands: DDL and DML \citep{dekeyser2007computer}. The DDL commands allow users to manipulate data in a database, while the DML decides the commands for defining a database schema.

\section{Challenges of Learning SQL}\label{3.2}
Despite its simple syntax and highly declarative nature, learning SQL and its underlying concepts pose difficulties for students \citep{kleiner2013automated,jones2016desired} and non-technical end-users \citep{li2016understanding, soylu2016experiencing}. Several authors have tried to identify the reasons for the difficulties encountered by these groups of users while learning SQL. \citet{prior2003online} investigated why students experience difficulties in learning SQL. In this study, students were asked to submit a formal assignment on a piece of paper without having to practise against a relational database. The research highlighted that students perform badly in writing correct queries because most of them are only interested in passing the module rather than taking time to practise sufficiently. Furthermore, the study concluded that constructing and writing correct queries in SQL is a practical skill that cannot be gained without repeated practise.  \citet{mitrovic2016implementing} opined that learning SQL from a RDBMS often lead to learning challenges. The authors advised that errors generated from most RDBMS are not helpful to learners, because they are limited to the syntax of the RDBMS itself.\\

Writing DML and DDL expressions have shown to be problematic for learners \citep{dekeyser2007computer,seyed1993sql,qian2012designing}. \citet{sadiq2004sqlator} presented two reasons for the difficulties experienced while learning SQL. First, the straightforward syntax of the SQL \texttt{SELECT} command is often deceptive. To a learner, it might appear easy to learn, but the reverse always seems to be the case. Second, the declarative nature of SQL can be difficult for students to comprehend, especially if they are learning it alongside a procedural or object-oriented programming language. These difficulties were further discussed by \citet{dekeyser2007computer} and \citet{ahadi2015quantitative}. They argued that approaching programming problems in procedural or object-oriented programming languages requires learners to think in \textit{steps}, while SQL requires one to approach a problem in \textit{sets} rather than \textit{steps}. \\

Another difficulty faced in learning SQL is the burden of memorising database schemas, resulting in inaccurate solutions due to wrong attributes or table names specified \citep{mitrovic2003intelligent,mitrovic2016implementing,lavbivc2017recommender}. This difficulty often misleads the learner in understanding the underlying concept of SQL.  \citet{prior2014assesql} reported that apart from the burden of memorising database schemas, knowing when they are necessary in writing queries and how to execute them poses great difficulties, requiring consistent practice and effort from learners. Other reasons are that learners misunderstand the use of first-order logic and other basic concepts of SQL \citep{grust1999comprehend, ahadi2015quantitative, soylu2017ontology}. \\

\citet{kearns1997teaching} suggested that failing to understand basic SQL queries will lead to problems comprehending other concepts such as group and aggregation functions, joins, universal quantification and some set operations. In addition, \citet{ahadi2015quantitative} conducted a survey and revealed that a high percentage of students struggle to learn and write correct subqueries (nested and correlated) enclosed in balanced parentheses. The study agreed that only after students have fully grasped the early stages of learning simpler SQL queries, can they begin to understand nested queries as these require a procedural understanding of SQL. Also, recent studies by \citet{cagliero2018improving} and \citet{taipalus2018errors} agreed that students should first understand simpler SQL queries before being taught nested queries.\\

Similarly, non-technical end-users struggle to understand SQL queries written by technical experts \citep{ardito2014gestures,li2014constructing}. If a database administrator leaves an organisation, and the non-technical user is left to use an enterprise application without proper support and training, this can result in redundant reporting, which contains repetitive and unwanted data that makes it difficult to use in these environments, since the users do not understand queries. In most cases, enterprise applications that use SQL as a back-end are often employed by these users, which are often provided in a software documentation \citep{warnke2009technical}. These materials may contain terminologies that are difficult to understand and interpret. If the documentation is not properly designed with non-technical end-users in mind, this may pose serious difficulties.\\

The majority of these non-technical end-users work in diverse fields such as marketing, finance, mining, etc. Many of these users are faced with the challenge of retrieving vital information from their databases. In most cases, they can clearly specify the intended task, but lack the knowledge to write a correct SQL query. Thus, end-users often seek help from technical users or through online forums \citep{wang2017interactive,wang2017synthesizing}. Such a process can be time-consuming and frustrating~\citep{yaghmazadeh2017sqlizer}. In view of these challenges, we discuss the pedagogical patterns and learning strategies that have been proposed over the years.

\section{Pedagogical Models}\label{3.3}
Pedagogical models (or pedagogical patterns) and their use in teaching have been widely researched and debated over the years \citep{kotze2008don,schulte2010introduction}. Instructors apply different pedagogical models in their teaching curriculum. The father of patterns, \citet{alexander1977pattern}, defined patterns as ``each pattern described a problem that occurs repeatedly, used to describe the solution to a problem over a million times''. This quote suggests that patterns give designers the freedom of problem solving through many variations. \\

\citet{bergin2012pedagogical} describe pedagogical models as the detailed description of work carried out by an educator to communicate knowledge to others and to solve recurrent problems. \citet{renaud2004teaching} added that pedagogical models consist of three forms, namely: issues, strategies and implementation. The issues refer to the transfer of knowledge. The strategy aims to transfer knowledge in a particular way and implementation describes the materials used by the strategy. In a teaching and learning scenario, patterns offer a way of transferring knowledge. The intent is to capture and present a concept in a compact form to those that require the knowledge. We present the different pedagogical patterns as discussed in the literature.

\subsection{Models in Human-Computer Interaction}
The first human-computer interaction (HCI) patterns were based on user-centered system design with reference to Alexander's ideas~\citep{borchers2000interaction}. These patterns were discussed in the Common Ground~\citep{tidwell1999common}, Designing Interfaces~\citep{tidwell2010designing} and User Interface (UI) patterns and techniques~\citep{tidwell2002ui}.  HCI pattern was defined by \citet{dearden2006pattern} as a description of a proven solution for a user interface that takes place within a particular context. HCI patterns are also referred to as UI patterns that assist software developers to reuse best practices and avoid reinventing the wheel~\citep{seffah2015patterns}. The study showed that patterns are applicable to every software system and are widely independent of the tools that are used to develop those systems.\\

\citet{borchers2001patterns} discussed that HCI patterns should model design experience based on the field of architecture where pattern ideas were orginally conceived. The authors stressed that HCI patterns should focus exclusively on the non-technical end-user to embrace their potentials. Using patterns only for an expert user would limit their importance. \citet{van2003pattern} proposed a top-down approach of HCI patterns organisation into a scale of hierarchy from high-level design problems which are gradually unpacked into low-level design problems.

\subsection{Models in Education}
Just as patterns are employed in many fields to teach students about certain concepts, their application in education are numerous \citep{borrego2007development,vermunt2017learning,hansen2015democratizing}. \citet{al2013sql} described patterns in education as a technique to help educators transfer their experience in a manner that helps to achieve good teaching and learning. In this thesis, the author suggested that using patterns in a CS class will enhance problem solving skills.  Educational patterns describe successful practices within an educational context that includes methods, content and curriculum design \citep{winthrop2015wait}.\\

\citet{laurillard2013teaching} argued that the current teaching approach is changing and teachers are required to cope with a technological environment. In addition, the author stressed that teaching should be seen as a creative design profession. \cite{clayer2013patterns} proposed an engineering framework aimed at collecting different pedagogical designs from instructors. The result showed that the tool allowed instructors to convey their opinion in a self-expressive manner.

\subsection{Models in SQL}
In teaching SQL queries, pedagogical patterns employed either make use of traditional face-to-face instructor-led teaching \citep{mao2007effectiveness} or electronic aids \citep{de2007system,dollinger2010sql}. \citet{prior2004backwash} proposed three approaches to teaching SQL. The first aspect of this approach required that students be graded using a method that would help improve their learning skills. The idea was that giving students practicals and grading them would improve their query formulation skills. The second part of this approach considered improving SQL skills by using real-world software development, while the third process encourages students to build their SQL query skills by practising online. A similar study was conducted by \citet{abello2008learn}, who encouraged the automatic grading of students which would enhance and improve their SQL skills.\\

Another pedagogical pattern for teaching SQL was presented by \citet{renaud2004teaching}. The authors compared and contrasted two approaches to teaching in terms of mental models and cognition. The first approach exposed students to using tools to learn the SQL syntaxes, while the second approach required students to formulate queries on paper for weeks before they are exposed to tools. The results showed that if learners are taught using a particular paradigm, exposure to other methods would compromise their SQL query formulation skills. They insist that students need to grasp the basic concept of SQL before they are exposed to tools. \citet{caldeira2008teaching} conducted a similar study, which required that students understand SQL thoroughly, by reading and understanding how to write SQL scripts, before they are exposed to tools.\\

\citet{ahadi2016students} presented common semantics that instructors need to consider when teaching students to write SQL queries. They emphasised that a deeper understanding of these semantics would improve students' learning outcomes and proper writing of SQL queries.

\section{Learning Approaches}\label{3.4}
The study of how humans learn is not new. As the study of learning continues to expand, researchers have continually applied their ideas to this concept. Over the past few decades, learning theorists have engaged in extensive debates on how people learn \citep{mowrer1960learning,gredler1992learning,seligman1970generality}. Similarly, these theorists agreed that there is no clear definition of learning that is universally accepted. We provide some of the definitions in the literature. \citet{lachman1997learning} refers to learning as a behavioural change due to experience. \citet{de2013interaction} defined learning as an activity that involves acquiring and modifying knowledge, attitude, skills and behaviours. \\

\citet{schunk2012learning} identified three criteria for learning. First, learning involves change -- people learn by doing things differently; second, learning endures over time -- those changes are temporary and may not last forever; and, learning occurs through experience -- humans learn from past experience. \citet{de2013learning} argued that not all of these definitions are valid; a change in behaviour may not necessary imply learning. Furthermore, the study concluded that a change in behaviour is only caused by some experience in an individual and may not count as instances of learning. \citet{eberl2018organizational} retorted that some definitions need to specify that learning requires changes in a specific psychological mechanism to clearly make a distinction between a behavioural change and learning.\\

The act of learning is an active area of research in psychology, behavioural ecology, neuroscience, CS and many other disciplines~\citep{barron2015embracing}. In the CSs, students learn how to program in the early stages of their academic pursuit \citep{resnick2013learn,moreno2016code,hu2013using}. Learning how to code is considered a hot skill and most industry employers are in dire need of software programmers~\citep{luxton2018introductory}. At the early stages, most students are introduced to tools to enable them to improve and learn programming~\citep{moreno2016code}. Even so, students are taught SQL queries in introductory database courses, but learning how to write correct queries is problematic \citep{ahadi2015quantitative, soylu2017ontology}. It is important to know what modes of learning exist in order to ensure knowledge impacted by the instructor through instructional materials, is understood and learned. The next section describes the different learning strategies.

\subsection{Learning Theories}
Learning theories are conceptual frameworks that explain how humans acquire, retain and recall information~\citep{politis2008process}. An effective instructional design is important to ensure clarity, direction and focus throughout the learning process~\citep{mcleod2003learning}. \citet{schunk2012learning} noted that learners progress through different learning stages from novice to expert, and most learning theories share common instructional principles which aid the learner. That is, learning theories assist instructors in designing instructional contents to facilitate learning.  Many educational psychologists have identified different learning theories that explain how individuals learn by acquiring and organising knowledge~\citep{politis2008process,ertmer2013behaviorism,hung2001theories}. According to \citet{hung2001theories}, the four learning theories are behaviourism, cognitivism, constructivism and social constructivism. 

\subsubsection{Behaviourism}
Behaviourism was the first learning theory developed in the late 1800s and early 1900s~\citep{james2006assessment}. According to behaviourism, the goal is to derive an elementary law of learning and behaviour that can be extended to more complex scenarios. \citet{davey2017applications} described the behaviourist theory as a theory based on all behaviours which are acquired through conditioning. Behaviourists believe that conditioning occurs through the environment and our responses to environmental factors shape our actions~\citep{sheldon2011cognitive,done2018responsibilisation}. \citet{merriam2006learning} highlighted three assumptions of learning used in the behaviourist theory: 

\begin{enumerate}
	\item All behavioural-related tasks have little regard for a learner's cognition.
	\item Learning activity is influenced by environmental factors.
	\item Events formation and reinforcement form important components of the learning process.
\end{enumerate} 

These assumptions imply that learning is based on time-controlled events and environmental factors which bring about a change in behavioural responses. The strengths and weaknesses of the behaviourist theory were identified by \citet{engestrom2012whatever} and \citet{omar2018psychology}. In their works, the main strengths are that an individual is expected to behave in a certain way regardless of the circumstance. The weakness of this theory is that the mental cues that a learner receives may not match what was previously learnt.

\subsubsection{Cognitivism}
Cognitivism is based on the concept that learning is more important than responses to external factors such as the environment~\citep{duffy2013constructivism,harasim2017learning}. Cognitivism suggests that a reorganisation of experiences could lead to learners making sense of what they are taught, which can be termed learning \citep{mayer2009constructivism}. \citet{tobin2012constructivism} argued that each learner experiences things by generating their own rules and mental models. Hence, the adjustment of mental models paves the way for newer experiences. \citet{kalina2009cognitive} noted that before instructors can start designing instructional materials, they need to consider the student's learning point and allow them to create personal meaning before knowledge can be passed to them. Modern instructional methods are based on cognitive theories~\citep{harasim2017learning,kalina2009cognitive}. \\

The merit of cognitive theories is that learners are trained so that they are able to accomplish tasks on their own~\citep{omar2018psychology}. The weakness of this theory as emphasised by \citet{omar2018psychology} is that learners are forced to learn and accomplish tasks in a certain way. For example, in programming, learners can produce different working versions of a program, but some may be more efficient than others.

\subsubsection{Constructivism}
\citet{pritchard2017ways} described constructivism as the idea that learners can construct knowledge for themselves. Here, each learner constructs meaning as they learn, which is a vital part of the learning process. Constructivism was formalised by Jean Piaget, who suggested that through assimilation, knowledge can be constructed from experience~\citep{kalina2009cognitive}. Since constructivism describes how learning happens, it is not a particular pedagogy~\citep{hein1999constructivist}. It suggests how learners can use their previous experiences to understand instructional materials. Constructivist theory also places emphasis on mental processes of the learner. It is required that a learner utilises different cognitive processes for tasks. \citet{hein1999constructivist} listed some underlying principles that guide constructivism theory:
\begin{enumerate}
\item Learning is an active process that learners can use to construct meaning.
\item Learning involves constructing meaning: people learn to learn as they learn.
\item The action of constructing meaning is mental.
\item Learning is a social activity.
\item Motivation is essential to learning.
\end{enumerate}

A major advantage of this theory is that real-world situations are understood by relating them to past events~\citep{omar2018psychology}. Where conformity in thinking and actions are required, this learning theory may not be the best approach~\citep{omar2018psychology}.
\subsubsection{Social Constructivism}
The theory of social constructivism was developed by Lev Vygotsky in the 1930s and was the first one to reject the claim made by Piaget~\citep{hirtle1996social,pass2007constructivists}. \citet{chaiklin2003zone} interpreted Vygotsky's work to indicate that learning can be separated from its social interactions. \citet{powell2015evaluating} described the social constructivism theory of learning as an effective method that involves collaboration and social interaction. This type of learning is based on social interactions that are developed alongside personal critical thinking in the classroom. The study concluded that adding this form of learning alongside the constructivist approach would improve active participation in a classroom. \\

Social constructivists encourage learners to develop their own version of knowledge by learning from knowledgeable experts through social interactions \citep{hein1999constructivist}. This study also adds that knowledge is distributed in the community and can be built through engagements. The social constructivist approach assists young children to develop their knowledge by interacting with their immediate peers, adults and the physical world ~\citep{hurst2013impact}. Hence, from this viewpoint, learning is acquired and improved as the immediate community helps shape the knowledge.

\subsection{Learning Taxonomies}
Learning taxonomies are used to describe different aspects of learning behaviours in the form of a classification~\citep{horn2017taxonomy}.  This classification shows the skills that educators set for their students to achieve~\citep{sarfraz2017strategic}. The taxonomy shows the level of cognition required for a course using some set of objectives~\citep{verenna2018role}. Learning taxonomies usually move from basic to higher levels of cognition. This section presents some different learning taxonomies.
\subsubsection{Bloom's Taxonomy}
\begin{figure*}[h]
	\centering
	\includegraphics[width=280px]{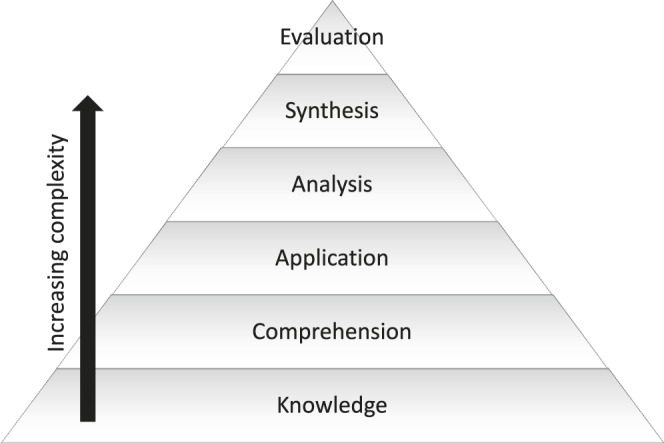}
	\caption[Bloom's taxonomy of learning~\citep{adams2015bloom}]{Bloom's taxonomy of learning~\citep{adams2015bloom}}
	\label{fig:bloom}
\end{figure*}

Bloom's taxonomy was devised in the 1950s, and was regarded as the stairway of learning that instructors use to enable students to reach a higher cognitive level~\citep{krathwohl2002revision,adams2015bloom}. Since then, Bloom's taxonomy has stood the test of time~\citep{sarfraz2017strategic}. The taxonomy divides the cognitive aspects of learning into six hierarchical levels with increasing complexity \citep{starr2008bloom}. The hierarchy ranges from the highest (analysis, synthesis and evaluation) to the lowest levels (knowledge, comprehension and application). Each cognitive level of Bloom's taxonomy assists the next level. The taxonomy can be used in almost all disciplines~\citep{sarfraz2017strategic}. \autoref{fig:bloom} shows the Bloom's taxonomy of learning.\\

\citet{crowe2008biology} applied Bloom's taxonomy in a biology study to assist instructors in creating instructional materials and successfully design questions that require students to apply their cognitive skills. The study showed that Bloom's taxonomy could help students to become successful biologists as this strategy can improve their learning skills. \citet{thompson2008bloom}, in their study, identified that many learning theorists believe that it was difficult to apply Bloom's taxonomy in introductory programming courses. In addition, the study reiterated that Bloom's taxonomy can be a very useful tool for the CSs discipline because it addresses cognitive processes that programmming courses require. Hence, it can be an invaluable tool for CS educators.

\subsubsection{Anderson and Krathwohl Taxonomy}
Bloom's taxonomy was revised to allow educators to understand and implement standards in their curriculum \citep{krathwohl2009taxonomy,forehand2010bloom}. This revised taxonomy is regarded as the \citet{krathwohl2009taxonomy} taxonomy. The taxonomy maps six well organised cognitive processes into knowledge levels and takes into consideration many of the criticisms levelled against Bloom's taxonomy. \\

\begin{figure*}[h!]
	\centering
	\includegraphics[width=250px]{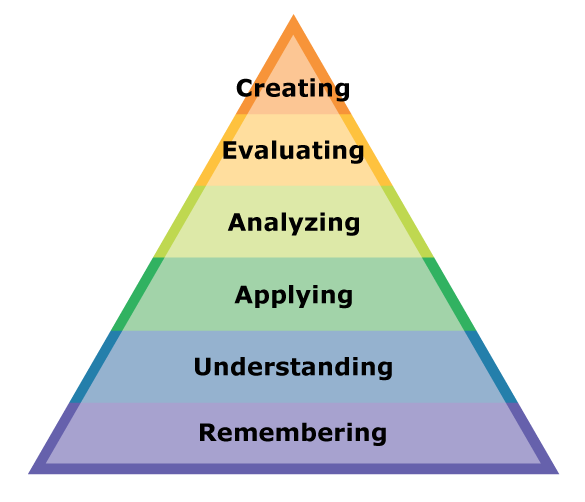}
	\caption[Anderson and Krathwohl's revised taxonomy~\citep{wilson2018anderson}]{The Anderson and Krathwohl's revised taxonomy~\citep{wilson2018anderson}}
	\label{fig:krathwol}
\end{figure*}

In comparison with Bloom's taxonomy, the Anderson and Krathwohl taxonomy is more comprehensive and has shown to assist instructors in designing instructional materials~\citep{pickard2007new}. In its formation, it reword nouns used in Bloom's taxonomy as verbs.  In this taxonomy, the `synthesis' phase of the Bloom's taxonomy was replaced with the `creating' phase at the top of the pyramid~\citet{thompson2008bloom}. Similarly, the lowest level of the Bloom's taxonomy called `knowledge' was replaced with `remembering'. Although Bloom's taxonomy is focused on the learning process in many forms, a disadvantage thereof is that it fails to indicate that learners must start at a lowest level before working up~\citep{churches2010bloom}. Instead, the learning process can be initiated at any point (context-free), and the learner can perform the learning task alongside. \autoref{fig:krathwol} presents the Anderson and Krathwohl's revised taxonomy. \\

\citet{jansen2009using} suggested that although the Anderson and Krawthwohl taxonomy was effective in investigating the cognitive learning processes of learners, developing questions for instructors is not straightforward.
 
\subsubsection{Gorman Taxonomy}
\begin{figure*}[h!]
	\centering
	\includegraphics[width=300px]{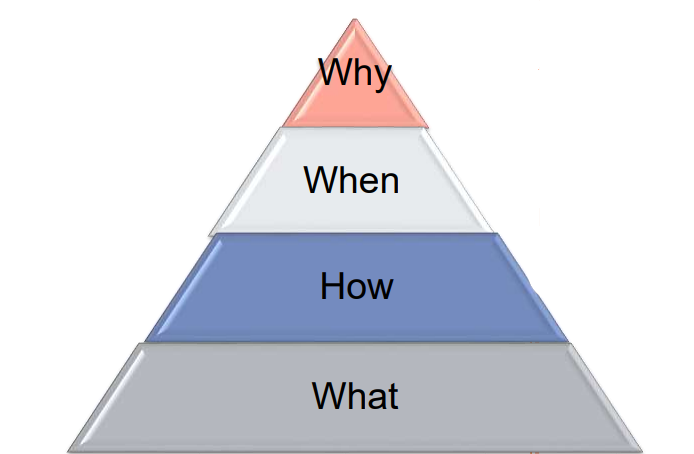}
	\caption[Gorman's taxonomy of learning~\citep{gorman2002types}]{Gorman's taxonomy of learning~\citep{gorman2002types}}
	\label{fig:gorman}
\end{figure*}

Michael Gorman proposed a taxonomy that consists of four simple learning levels~\citep{gorman2002types}. These levels shows how knowledge can be represented, whether implicitly or explicitly. \autoref{fig:gorman} includes the lower levels information (what) and skills (how) with higher levels of judgement (when) and wisdom (why). \citet{al2010sql} compared Bloom's and Gorman's learning taxonomies. The study indicated that the lowest level, `what' aligned with `knowledge and comprehension' in Bloom's taxonomy. In addition, the higher level word `why' matches with `evaluation. 

\citet{mohtashami2000application} noted that teaching the database concept with Gorman's strategy would require the first level to cover basic aspects such as entities, relations, etc. The second level should incorporate concepts such as ERDs\footnote{Entity-Relationship Diagrams} and SQL. The final level should investigate a problem-based approach to enable students use previous knowledge to tackle questions for the purpose of understanding the database concept.

\subsubsection{Computer Science Taxonomy}
Many educators have applied different taxonomies to the CS discipline \citep{maier2000we,johnson2006bloom,rutten2012learning}. \citet{johnson2006bloom} carried out a survey using Bloom's taxonomy to examine whether it is appropriate for the CS field. The survey showed that Bloom taxonomy can be helpful for educators developing instructional materials for their courses. In a similar study, \citet{lahtinen2007categorization} investigated Bloom's six cognitive activities for its use in CS education. The study revealed that the taxonomy was indeed useful to CS educators. \cite{bower2008taxonomy} presented a learning taxonomy that identifies different programming processes undertaken by students when learning programming. The taxonomy showed that students are encouraged to focus on tasks that foster memory retention. \\

%

\citet{shneiderman1978improving} introduced five learning tasks required for SQL learning. These tasks showed that learning SQL would require a learner to understand the syntax and semantics first before modifying queries written by oneself. \citet{renaud2009ideas} argued that before learning SQL, it is important that query construction skills are developed first. The study recommended Gorman's taxonomy, where students are required to learn the basic aspects first, before moving to a higher level.

\subsection{Learning Styles}\label{2.2}
As is widely discussed in the literature \citep{entwistle2015understanding,lowe2016sensory,adams2017failures}, individuals have unique ways of learning and processing information. The benefit derived from learning content and materials that match an individual's learning styles has been identified in the works of \citet{price2004individual} and \citet{fleming2011undergraduate}. This has also been identified in some computer-assisted learning systems \citep{truong2016integrating,sweta2016learner}. The terms ‘cognitive styles' and ‘learning styles' are used interchangeably in the literature. \citet{cassidy2004learning} defines an individual's cognitive style as a problem-solving, thinking, perceiving and remembering activity, while learning style indicates the application of cognitive style in a learning situation. \citet{soflano2015learning} defined learning style as an individual's choice and strategy for achieving learning objectives efficiently. In the domain of SQL, different learning approaches have been proposed. We discuss the different learning methods, such as PBL, learning from worked examples, learning through visualisation and learning from errors. 

\subsubsection{Problem-based Learning}\label{2.2.1}
\begin{figure*}[h]
	\centering
	\includegraphics[width=400px]{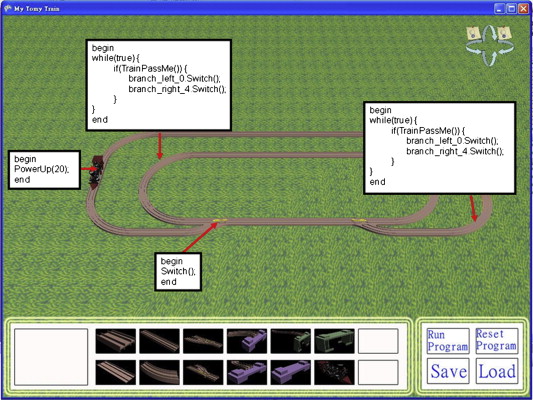}
	\caption[Problem-based task in a gaming scenario~\citep{liu2011effect}]{Problem-based task in a gaming scenario domain~\citep{liu2011effect}}
	\label{fig:gbl}
\end{figure*}

PBL is a constructivist\footnote{Constructivist - learning based on observation or through a scientific approach} approach in which learners learn through problem solving \citep{connolly2006constructivist}. In PBL, learning is based on acquiring and processing information that changes the knowledge previously acquired by a learner. In PBL, learning is achieved using problems to motivate students and there is a focus on student-centred activities. Using PBL, instructors act only as facilitators rather than as primary sources of knowledge \citep{hoic2009blended}. PBL is often applied to improve results in collaborative learning where students work in small groups \citep{thurley2008problem,azer2013cracks,chao2016exploring}. In collaborative learning, the task of finding the solutions to problems is shared among the group. \autoref{fig:gbl} describes a game-based problem task for a user, where the user has to solve a series of problems before accomplishing a specific task.\\

\citet{kreie2013database} explored the use of the PBL approach to improve database learning. In their study, students were asked to create a logical data model prototype of a database application. The prototype would require them to solve some data modelling errors which they would encounter. The researchers concluded that the PBL approach would challenge students to understand database concepts and improve their knowledge. \citet{ward2015teaching} extended the use of the PBL approach in teaching SQL through a game. The research presented a number of problem-solving skills for a learner in a typical gaming scenario. The researcher concluded that using the PBL approach, a learner would have a deeper interaction in solving problems in SQL.

\subsubsection{Learning from Worked Examples}\label{2.2.2}
\begin{figure*}[h]
	\centering
	\includegraphics[width=400px]{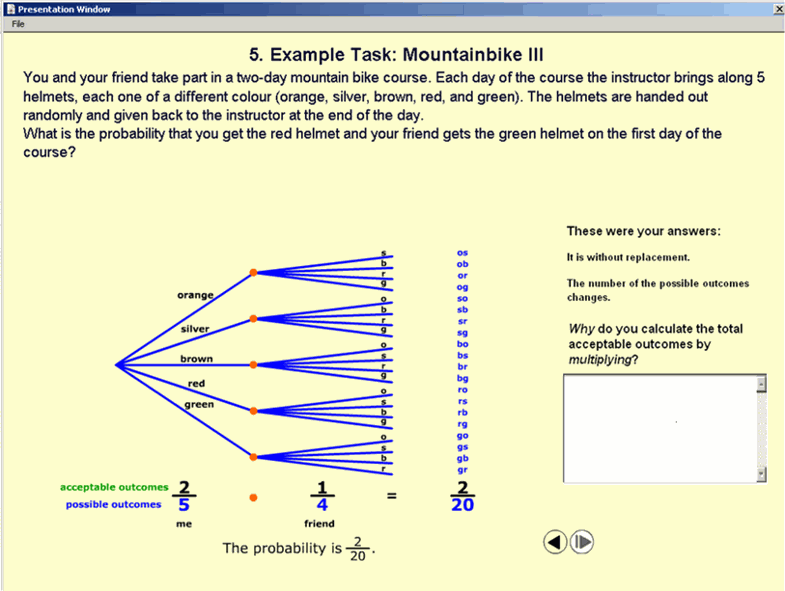}
	\caption[A worked example in the mathematics domain~\citep{berthold2009assisting}]{A worked example in the mathematics domain~\citep{berthold2009assisting}}
	\label{fig:math}
\end{figure*}

According to \citet{sweller2006worked}, learning from worked examples is the most effective learning strategy, especially when first learning a new domain. Worked examples are presented to students, followed by problem-solving techniques. This is done to ensure that they acquire adequate knowledge before being introduced to problems. Such an approach is ideal for novices, since examples help reduce cognitive workload and aid initial stage learning \citep{van2011effects,renkl2014toward}. \citet{najar2014adaptive} proposed the use of learning from worked examples ITS for teaching SQL. The study concluded that students' understanding increases significantly when presented with worked examples. \autoref{fig:math} shows an example of a probability worked example in mathematics, with the colours indicating how the answer is derived.\\

Another study of learning from worked examples was presented by \citet{chen2017much}. The study showed that in most cases, worked examples contain the right explanations for each and every step required. This helps novices gain the right information through the examples provided. The researchers concluded that learning from worked examples is ideal for novices rather than for advanced students.

\subsubsection{Learning through Visualisation}\label{2.2.3}
Learning through visualisation has proved useful in assisting novices understand a concept~\citep{watson2011learning,garner2003learning,kinchin2011visualising}. This technique has been used extensively in different application domains to present ideas \citep{ellis2010mastering,keim2008visual}. \citet{ellis2007taxonomy} defined visualisation as a systematic way of representing an abstract idea that facilitates human understanding. This abstract idea is usually designed in a way that is ``playful and aesthetically pleasing'', so that users can explore how to solve tasks. Visualisation can encourage active participation in learning and lead students' critical thought processes \citep{allenstein2008query}. \\

\begin{figure*}[h]
	\centering
	\includegraphics[width=400px]{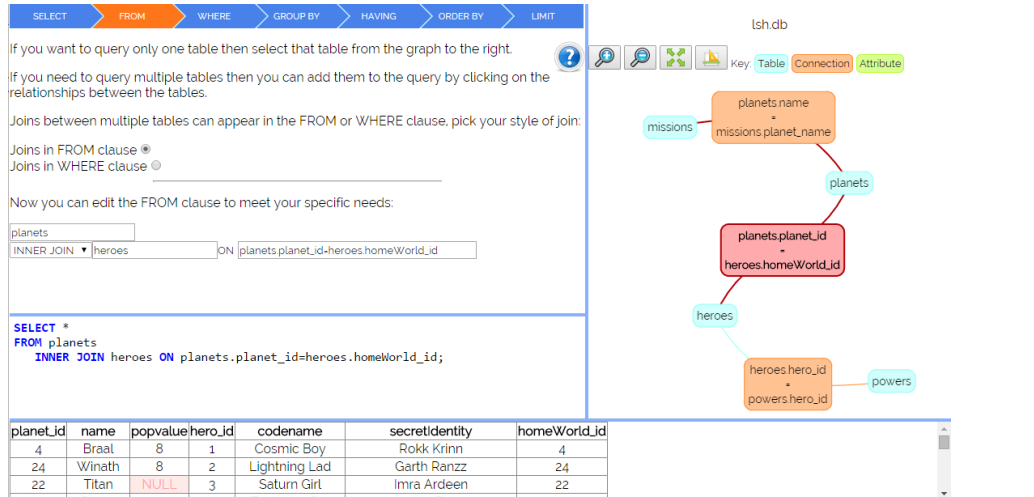}
	\caption[Sis, a visualisation tool~\citep{garner2015learning}]{Sis, a visualisation tool~\citep{garner2015learning}}
	\label{fig:visualiser}
\end{figure*}

A study conducted by \citet{kellems2016using} showed that visualisation can even help students with learning disabilities grasp information more easily. Their study showed that visual aids can also better meet the academic demands of students with autism spectrum disorder (ASD)\footnote{ A neurological and developmental disorder, which result in communication and interaction difficulties.}, since they do not require intensive training. A sample visualisation tool is presented in \autoref{fig:visualiser}. This example shows how query statements are interconnected.\\

In the area of program comprehension, visualisation has been explored to aid the understanding of programming~\citep{yassine2017serious,kinchin2011visualising}. \citet{lee2013drag} proposed the use of the `drag and drop' refactoring visualisation to assist programmers comprehend programs written in Java. Empirical evidence presented in their work showed that the approach was more efficient and less error-prone, and that it could help programmers comprehend programs easily. Studies conducted in block programming (a technique which represents programs as blocks and uses the drag and drop technique to generate a program) indicated that this type of visualisation increased student engagement and that it was effective in knowledge transfer \citep{malan2007scratch,rizvi2011cs0}. \\

A recent study with similar methodology, conducted on serious games, found that this approach paves the way for undergraduate students to learn programming \citep{yassine2017serious}. The benefit of this visual aid is that it enhances self-pacing, while its entertaining component attracts the attention of students and engages them in the learning process. It is worth noting that most SQL comprehension tools employ visualisation to  comprehend SQL queries and database schemas automatically either through textual or graphical representations \citep{sadiq2004sqlator, cembalo2011savi,folland2016visqlizer}. 

\subsubsection{Learning from Errors}
The concept ‘learning from errors' or LFE has been successfully applied in mathematics \citep{brodie2014learning}, physics \citep{grosse2007finding} and CS (especially in programming) \citep{shah2017qualitative}. A recent study by \citet{metcalfe2017learning} showed that errorful learning followed by constructive feedback is a vehicle to achieving purposeful learning. Other research conducted by \citet{metcalfe2017learning2} showed that learning from one's own errors and those of others is useful in helping to develop methods to improve students' learning. \\

\citet{mitrovic2012fifteen} applied the use of constraint-based modeling in SQL in dealing with errors. The study identified that fixing errors is a time-consuming process that requires a great deal of mental effort. Most importantly, people make errors because their procedural knowledge is poor. The study concluded that if one wants to learn from errors in SQL, one must first accumulate adequate declarative knowledge, which is later converted to procedural knowledge, a process requiring much practice. \citet{katz2016learning} extended LFE to the area of relational database modelling to examine the difficulties faced by students in understanding conceptual database modelling. The study showed that errors play a powerful role in database modelling, which encourages students to possess a deeper understanding of the course.

\subsubsection{Other Learning Styles}
While we have discussed only a few learning methods, other learning theories have been proposed that are used in computer-assisted learning. The learning theories, as discussed in the work of \cite{soflano2015learning} are:

\begin{enumerate}
	\item Kolb's learning style, a model developed by \citet{wolfe1984career} which is based on four elements: reflexive observation, concrete evidence, abstract conceptualisation and active experimentation.
	\item The VAK model, a model proposed by \citet{dunn1978teaching}, divides students into groups based on their learning preferences: visual, auditory or tactile.
	\item The Big-5 model, an approach by \citet{felicia2009profiling}, has five elements: openness, conscientiousness, extroversion, neuroticism and agreeableness.  
	\item Honey and Mumford's model, proposed by \citet{honey1992manual}, has four elements: activists, reflectors, theorists and pragmatists.
	\item The Felder-Silverman learning model, developed by \citet{felder1988learning}, which consists of four elements: perception, input, processing and organisation.
\end{enumerate}

\subsection{Cognitive and Mental Models}
Cognitive strategies that influence the effectiveness of teaching and learning are useful in education~\citep{mcdougle2016taking,lane2012cognitive}. Hence, cognitive models are representations of how humans gain knowledge. \citet{gentner2014mental} described mental models as concepts used to provide explanations for the purpose of understanding. These concepts may be from artifacts of technology, environmental factors, or tasks that need to be understood. They are mainly explanations that describe how novices understand a concept. This section provides some cognitive and mental models that have been used to support the understanding of SQL, as well as a number of cognitive theories discussed in the program comprehension domain.
\subsubsection{SQL Models}
\sloppypar
Many studies have investigated different cognitive theories used to support the learning of SQL~\citep{hatami2017towards,schlager1986cognitive,al2016framework}. SQL learning cognition, as defined by \citet{robins2003learning}, is the construction of schemas, which are organised chunks of related knowledge. The study further shows that learning either builds new schema or modifies existing schemas. Hence, to build cognitive processing of SQL, a mental model must be built to modify or construct knowledge in a schema. \\

\begin{figure*}[h!]
	\centering
	\includegraphics[width=400px]{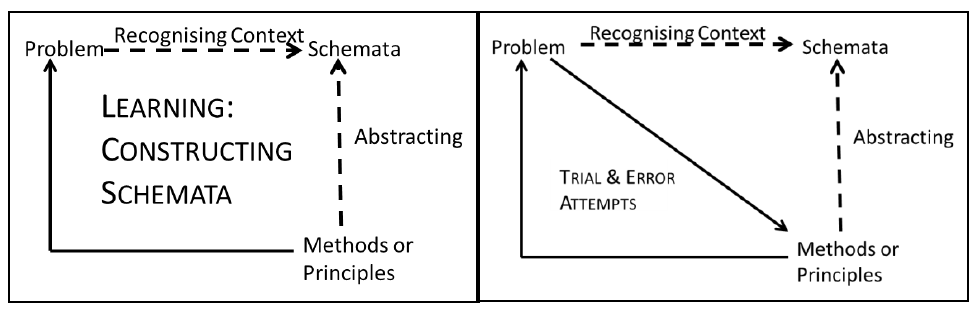}
	\caption[The schemata, and trial and error approaches~\citep{al2013sql,al2016framework}]{The schemata, and trial and error approaches~\citep{al2013sql,al2016framework}}
	\label{fig:scherror}
\end{figure*}

\citet{al2013sql} discussed that a complete understanding of SQL requires the novice to draw up mental models of the syntax and semantic concepts. The author noted that these concepts must be understood and their usage in a given scenario is very important to build good mental models of SQL. In addition, the study described four cognitive models for SQL to improve learning, understanding and remembering. These models are query comprehension, formulation, translation and writing to accumulate a range of contexts so that learning can be improved. In an extended study, \cite{al2016framework} argued that the schemata approach may not be the preferred solution. The study advised that knowledge of syntax and semantics is not sufficient to achieve mastery of SQL queries. The study added that a trial and error approach of solving schema problems may offer better cognitive processes for novices to achieve mastery of query problems. \autoref{fig:scherror} illustrates both approaches used to show mental processes and how problem solving is constructed.

\begin{figure*}[h!]
	\centering
	\includegraphics[width=350px]{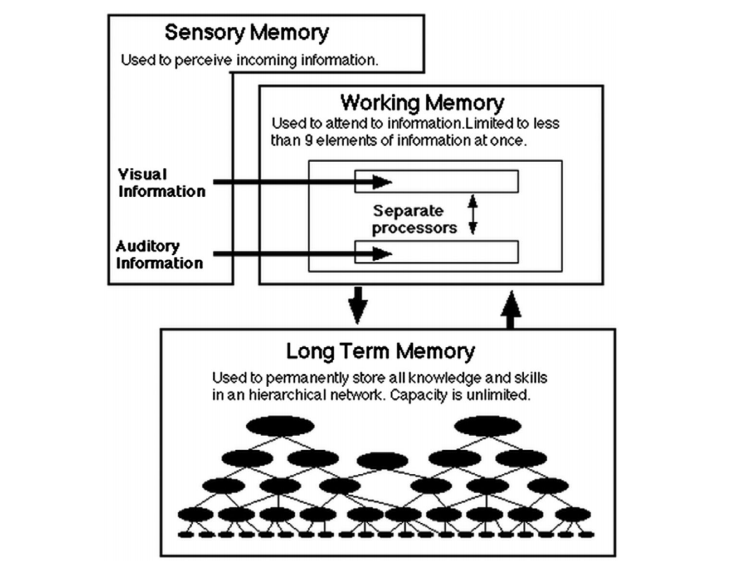}
	\caption[The CLT memory~\citep{sweller1988cognitive}]{The CLT memory~\citep{sweller1988cognitive}}
	\label{fig:memory}
\end{figure*}

\citet{mason2016applying} proposed a cognitive load theory (CLT) that addresses poor performances in an introductory database course, especially in the SQL concept. The study includes the human CLT model proposed by \citet{sweller1988cognitive} that focuses on long-term and working memory of knowledge storage as schema. The schema used in the study indicates cognition rather than database structure description. The study showed that schemas are retrieved from long-term memory and once they become complex, more information is used to expand the working memory. The human CLT concept is described in \autoref{fig:memory}. \\

\begin{figure*}[h!]
	\centering
	\includegraphics[width=350px]{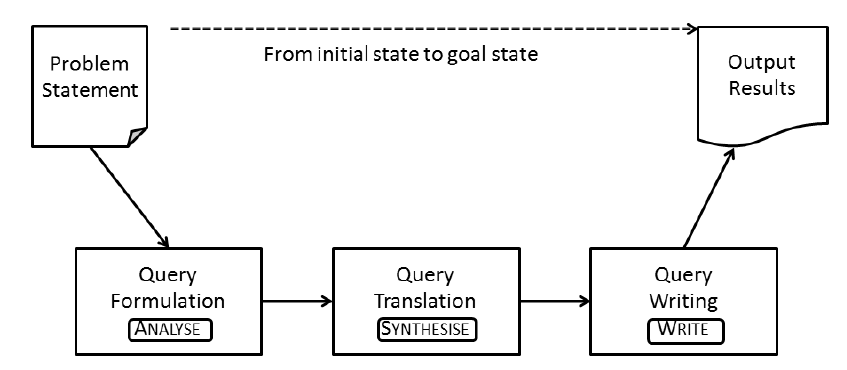}
	\caption[Problems to SQL generation ~\citep{almodel}]{Problems to SQL generation~\citep{almodel}}
	\label{fig:problems}
\end{figure*}

In recent studies, \citet{reisner1977use} described a study that generates and merges a set of lexical items and query templates for query generation. The study was followed by \citet{mannino2001database}, who proposed a two step approach that generates a query from problems and database representations. In addition, other studies expanded on the work of \citet{mannino2001database} by discussing three cognitive approaches to the SQL query generation \citep{siau2006cognitive,ogden1986implications}. The approaches are query formulation, query translation and query writing. The study concluded that combining these approaches will aid students in solving SQL problems. These approaches are summarised in \autoref{fig:problems}.

\subsubsection{Programming Models}
We present some of the programming cognitive models, as we take a cue from the SFA introduced by \citet{fincher1999we}. This approach was used in the current research. Program cognitive models were described as the representation of a program in a developer's memory~\citep{lemut2013cognitive,gulwani2015inductive,ade2016automatic,storey2005theories}. These theories were used to form a mental model useful to ensure that programs should be well understood. These models are described in this section.\\

\citet{schulte2010introduction} presented the \textit{top-down} approach as an assimilation process that applies knowledge of the program domain and maps this to the code itself. At this stage, the assimilation process is driven by hypothesis and used to convey some information about the program (Beacon). The top-down approach was first proposed by \citet{soloway1984empirical} to decompose programs into some levels of abstraction.\\

\citet{khuziakhmetov2016teaching} presented the \textit{bottom-up} program strategies based on developers reading code first and mentally grouping them. This strategy was referred to as the chunk model by \citet{pennington1987stimulus} and was used to group programs into a situation (or data-low) and program (or control-flow) model. Many learning theorists do not agree with either model (top-down or bottom-up), hence the \textit{opportunistic} approach was introduced to allow the reading of only terms that are necessary in a program~\citep{littman1986mental,storey2005theories}. \\

The \textit{knowledge-based} model, developed by \citet{letovsky1987cognitive}, that combined both top-down and bottom-up approaches. The study described three components: mental model, which is a memory representation, knowledge that contains plans and goals, and an assimilation process. The integration of all approaches led to the introduction of the \textit{integrated} model ~\citep{von1995program}. This model showed that switching between these models will improve the assimilation process of programmers.

\section{Comprehension Aids: State-of-the-Art Tools}\label{3.5}
In recent decades, tools to aid the comprehension of SQL have been proposed. Most of these tools employ visualisation to explain how a query interacts with a database, providing interactive examples of the basic concepts of SQL. Others offer solutions to problems faced while learning SQL queries \citep{brusilovsky2008open,cembalo2011savi,folland2016visqlizer}. In this section, we present some of the existing SQL tools from the earliest (that we know of) to the most recent that have been used in the comprehension of SQL.
\subsection{SQL Learning Tools}
\subsubsection{eSQL}
The \texttt{eSQL} system was proposed by \citet{kearns1997teaching} as an interactive learning system for students. It provides a step-by-step account of how a query result is determined, explaining each query step. For example, when a student writes a simple SQL query, \texttt{eSQL} provides a step-by-step formation of the resulting table. This ensures that even novices can easily grasp areas in SQL that are often found confusing. The authors stress that the dynamic stepwise mechanism provides a clear explanation of the underlying concepts of SQL, specifically allowing  students to visualise the behaviour of query operators, and is far superior to the traditional pen-and-paper explanation approach. \\

Although, the \texttt{eSQL} system provides more information to a user with its friendly user interface, it does not provide comprehensive feedback based on the user's solution, due to the lack of semantic analysis in its engine.
\subsubsection{WinRDBI}
\begin{figure*}[htbp]
	\centering
	\includegraphics[width=400px]{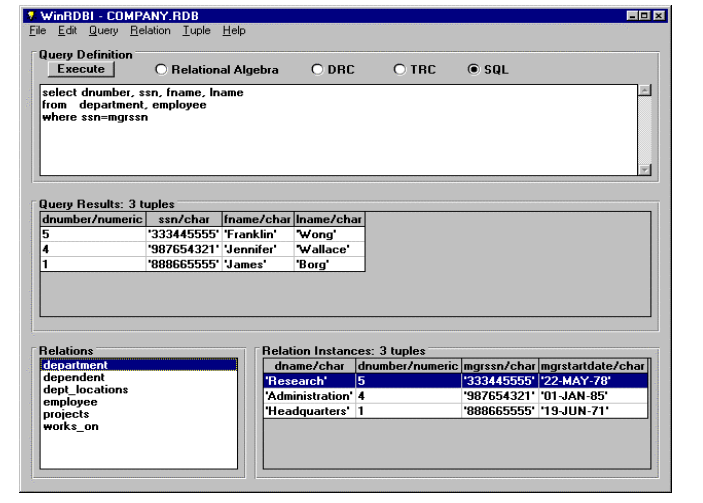}
	\caption[The user interface of the WinRDBI system \citep{dietrich1997winrdbi}]{The user interface of the WinRDBI system \citep{dietrich1997winrdbi}}
	\label{fig:winrdbi}
\end{figure*}

The \texttt{WinRDBI} system, also known as the Windows Relational DataBase Interpreter, is an educational tool that provides students with a friendly user interface to test their knowledge of SQL, relational algebra, and tuple relational calculus \citep{dietrich1997winrdbi}. The system is ideal for most introductory courses on database management systems and provides a platform to get immediate feedback by seeing the answers to the query specified. One major limitation of \texttt{WinRDBI} is its inability to provide comprehensive feedback, much as with the \texttt{eSQL} system. \autoref{fig:winrdbi} presents the \texttt{WinRDBI} system.

\subsubsection{SQL-Tutor}
\citet{mitrovic1998learning} developed the \texttt{SQL-Tutor}, as an intelligent teaching system (ITS), for teaching SQL, implemented on SUN workstations. \texttt{SQL-Tutor} focuses on the SQL SELECT query and uses semantic analysis to provide feedback on query solutions. It is based on the constraint-based modeling (CBM), an approach focused on identifying and providing feedback on errors \citep{mitrovic2003intelligent,mitrovic2016implementing}. An improved version of \texttt{SQL-Tutor} was introduced by \citet{mitrovic2003intelligent} known as \texttt{SQLT-Web}, it is a web version that addresses some of the shortcomings of the \texttt{SQL-Tutor}. 


\subsubsection{AsseSQL}
\texttt{AsseSQL} was developed by \citet{prior2003online} as an online assessment tool to test students' SQL formulation skills. It uses heuristics to evaluate whether a query entered is correct. The system is based on the SQL SELECT statement, and runs the submitted query on a test database. The query specified by a user is based on a question asked. \texttt{AsseSQL} compares the output of the query with the question. The goal of \texttt{AsseSQL} is to provide a deep learning experience for students. Although \texttt{AsseSQL} was successful in providing feedback and grading options, it is vulnerable to SQL injection attacks -- an attempt to make unauthorised changes to a database \citep{dekeyser2007computer}. An improved \texttt{AsseSQL} was introduced by \cite{prior2004backwash} to evaluate users' perception of the system. 

\subsubsection{SQLator}
The \texttt{SQLator} is a web-based interactive tool presented by \citet{sadiq2004sqlator} at the University of Queensland for learning SQL. It uses heuristics as its engine to evaluate the correctness of formulated query. Much like the \texttt{AsseSQL}, it supports assessments and grading to queries submitted by students, but does not provide suggestions or hints to query formulation. \texttt{SQLator} has three main components, namely: \textit{web application} -- this is used for providing access to users, \textit{engine} -- implemented as a Microsoft COM\footnote{Component Object Model} object for providing query formulation and evaluation, and \textit{databases} -- used for storing user data. Primarily, it supports only the SQL SELECT statement and judges if a proposed solution in SQL, corresponds to an English statement.


\subsubsection{SAVI}
\texttt{SAVI}, also known as SQL Advanced Visualisation, was created by \citet{cembalo2011savi} as a system to aid the teaching and understanding of the semantics of SQL. It uses reversible animations to explain how query operators can transform data from a database. \texttt{SAVI} was implemented using the Google Web Toolkit framework written in Java, which allows Internet applications to be executed in any browser. The motive for this system was to help users overcome problems related to SQL and improve mental visualisation of query concepts. Although \texttt{SAVI} is efficient when using tables with a limited number of rows, it performs poorly with large tables. 


\subsubsection{SiS}
\texttt{SiS} is an acronym for \textit{SQL in Steps}. \texttt{SiS} is an online learning platform which allows students to learn and build SQL queries in a series of steps \citep{garner2015learning}. The goal of \texttt{SiS} is to improve the way in which users learn the SQL SELECT query by building a series of steps in the form of graphs. Hence, \texttt{SiS} is focused on the SQL SELECT statement because it is identified as an area of difficulty for students \citep{prior2003online, sadiq2004sqlator, de2007system}. 


\subsubsection{VisQlizer}
\citet{folland2016visqlizer} proposed \texttt{VisQlizer} as a learning tool to help students create a mental model of the underlying concepts of SQL. \texttt{VisQlizer} uses animations and decomposition to aid comprehension. The researcher concluded that visualisation contributes to a better learning experience for students when coupled with traditional lectures and textbooks. 


\subsubsection{QueryViz}
The query visualisation system (or QueryViz) was designed by \citet{danaparamita2011queryviz}. It reduces the workload needed to understand queries. \texttt{QueryViz} uses visual constructs to address issues students face with nested queries. As a learning tool, \texttt{QueryViz} allows novices to intuitively familiarise themselves with the logical patterns behind the SQL syntax using the visualisation. 


\subsubsection{YASQLT}
Yet Another SQL Tutor ( or \texttt{YASQLT}) is an automated assessment tool for SQL developed by \citet{bider2016yasqlt} that teaches the introductory aspect of SQL queries to novice students. \texttt{YASQLT} focuses on the SQL SELECT and CREATE VIEW statements. The goal of \texttt{YASQLT} is for checking the result of a query and addressing common errors made by novices while learning SQL. Lastly, the survey conducted with students using \texttt{YASQLT} showed that it was helpful in aiding their comprehension of SQL queries. 


\subsubsection{OITS}
The Oracle Intelligent Tutoring System (OITS) is a tutoring system created by \citet{aldahdooh2017development}. \texttt{OITS} automatically generates problems related to SQL queries to be solved by students. \texttt{OITS} consists of four basic components. These components are: Expert Module -- responsible for identifying errors, Student Module -- highlights the problem solving steps, Tutoring Module -- keeps record of the student's progress, and UI module -- integrates multi-media applications to aid learning. Empirical evaluation was conducted on OITS and the outcome concluded that the tool was friendly and easy to use. 


\subsubsection{COSETTE}
\texttt{COSETTE} is an automated prover for SQL, developed by \citet{chu2017cosette}, that can determine the semantic equivalence between two SQL queries. The main goal behind \texttt{COSETTE} is to determine if two SQL queries are semantically equivalent. \texttt{COSETTE} works with the SQL SELECT statements. In its metric, not all queries are supported. The study suggested that COSETTE can be used in a variety of real-world applications such as semantic caching, automatic grading and verifying the correctness of RDBMS rewrite rules. 

%

\subsection{More Comprehension Aids}\label{2.9}
In the previous section, we discussed and presented tools for aiding the comprehension of SQL. However, we cannot discuss all of them comprehensively. Other closely related tools developed to aid SQL are:
\begin{enumerate}
	\item SQLify -- A SQL teaching and assessment tool, developed by \citet{de2006students}, intended to offer a richer learning experience and provide comprehensive feedback to students.
	\item LEARN-SQL -- A learning environment for automatic rating of notions of SQL, presented by \citet{abello2008learn}, that allows online assessment and learning in an interactive manner.
	\item eledSQL -- A web-based SQL learning tool proposed by \citet{grillenberger2012eledsql}, suitable for teaching SQL queries to novices.
	\item SCYTHE -- A web-based query-by-example system proposed by \citet{wang2017interactive} that synthesises SQL queries from I/O examples.
	\item SQL tester -- An online practice aid developed by \citet{kleerekoper2018sql} for assisting students to learn SQL queries, and providing immediate feedback.
	\item SQL-to-text -- A deep learning model using the graph-to-sequence approach proposed by \citet{xu2018sql}, which uses a graph encoder to generate SQL query to textual explanation.
	\item GeoSQL Journey -- A game-based tool designed by \citet{sandoz2018geosql} to stimulate student interest and simplify SQL learning.
	\item RSQLG -- A practice aid proposed by \citet{julavanich2019rsqlg}, aimed at providing a hands-on environment to stimulate student's interest of learning SQL.
\end{enumerate}
In this section, we have presented some SQL comprehension tools. The next section introduces NLP and its techniques.

\section{Natural Language Processing and Techniques}\label{3.6}
NLP has been applied in many fields such as CS, linguistics and cognitive science. In the next sections, we present the history and techniques used in NLP.
\subsection{Brief History of NLP}
NLP originated in the 1940s, just after World War II~\citep{chowdhury2003natural,jackson2007natural}. During this period, MT was the first language translation technique. At this time, people started to realise that a machine can be created to carry out this kind of task automatically. This phase was criticised due to primitive computing resources available~\citep{hutchins1986machine,hutchins1992introduction}. This was the era of punch cards and batch processing that required no suitable higher-level language. The predominant language at this time was the assembly language, systems developed used dictionary-lookup for word re-ordering in a target language, which produced poor results~\citep{bateman2003natural}.  NLP researchers realised that this task was more difficult than expected, hence there was a growing need to improve the theory of language~\citep{lehnert2014strategies}. \citet{bates1995models} discussed that the first application of NLP was abandoned due to the complexity of getting computers to map one natural language (NL) into another. Thus, the study showed it was difficult to map one string to another.\\

From the 1950s to the late 1970s, speech and language processing was split into two paradigms called \textit{symbolic} and \textit{stochastic}~\citep{jurafsky2000speech}. The symbolic approach led to the work of \citet{chomsky1956three}, who introduced the idea of generative grammars and other formal methods. These methods were published in a book, \textit{Syntactic Structures}, in 1957~\citep{chomsky2002syntactic}.  In addition, many linguists and computer scientists started developing parsing algorithms: top-down, bottom-up and then dynamic programming~\citep{roark1999efficient,kumar1988cdp}. The development of parsing algorithms led to the earliest parsing system, which was the Zelig Harris's Transformation and Discourse Analyst Project (TDAP)~\citep{heinroth2012introducing}. At this point, the linguistic field gained further insights on how to revive MT. \\ 

In 1956, the second aspect of the symbolic paradigm introduced the field of Artificial Intelligence (AI)~\citep{shanmuganathan2016artificial,mira2008symbols}. This field was based on the logic theorist perspective and the general problem solver model that created simple NLU systems. These NLU systems are based on pattern matching and keyword search, that were mostly used on QA systems. They use a knowledge-based approach that encodes knowledge and they are capable of producing answers for a provided question. The systems retrieve answers to questions from a database. Initially, LUNAR~\citep{woods1978semantics} and SHRDLU~\citep{winograd1973procedural} were the first NLIDBs that used this approach. The stochastic approach led to the development of the speech recognition engine, in particular, the Hidden Markov Model (HMM) began to show positive signs~\citep{juang1991hidden,sonnhammer1998hidden}. This introduced works on speech recognition and synthesis.\\

From the 1980s to the late 1990s, NLP systems became accessible, and areas such as semantic-oriented processing tasks were built on NL systems~\citep{sag2002multiword,ng1997corpus}. By the end of the 1980s, statistical approaches were showing progress, which complemented some of the significant NLP problems already addressed by the symbolic approaches~\citep{manning1999foundations}. It became apparent that the NLP field was expanding. Most statistical methods such as probabilistic parsing were combined in machine learning to derive both syntactic rules and their probabilities~\citep{hale2001probabilistic}. Even shallow processing methods such as finite state parsing and surface patterns were used in practical tasks~\citep{grefenstette2000recognizing,roche1997finite}. These methods were extended to dialogue structures in conversational systems especially in gaming technologies. Similarly, algorithms for discourse processing, reference resolution, parsing and part-of-speech (PoS) tagging began to incorporate probabilistic measures into their methods~\citep{ng2004chinese,nieuwland2007you}. There was an increase in the computer memory and speed of processing systems, which allowed for commercial implementation of subareas of NLP such as speech recognition. Grammar and spelling algorithms began to apply augmented alternative communication~\citep{garcia1992main,myers1998brief}. All these contributed to the rise of the Web with needs for language-based information retrieval and extraction methods~\citep{srinivasan2002speech}. These methods extended to the growth of linguistic resources such as WordNet~\citep{miller1995wordnet}, British National Corpus~\citep{leech1992100} and test tools such as the Penn Treebank Tagset~\citep{marcus1994penn} was developed.\\

The early 2000s saw the development of the first neural language model proposed by Yoshua Bengio and his team~\citep{bengio2003neural,bengio2008adaptive}. The neural network describes an artificial neural network, an aspect of the deep learning model that moves data in one direction through hidden nodes and extends this to an output node. More neural approaches have been developed, and extended to problems related to NLP~\citep{collobert2008unified,collobert2011natural}. Notably, many computational problems such as word sense disambiguation, and part-of-speech identification have become standards, which are used throughout NLP~\citep{warner2012detecting,li2010topic}. Currently, popular NLP powered tools such as Apple's Siri, Amazon's Alexa and Google's Home have been developed, which are used on most portable devices ~\citep{krusche201850,kugler2019being}.

\subsection{Classification of NLP}
\begin{figure*}[h]
	\centering
	\includegraphics[width=350px]{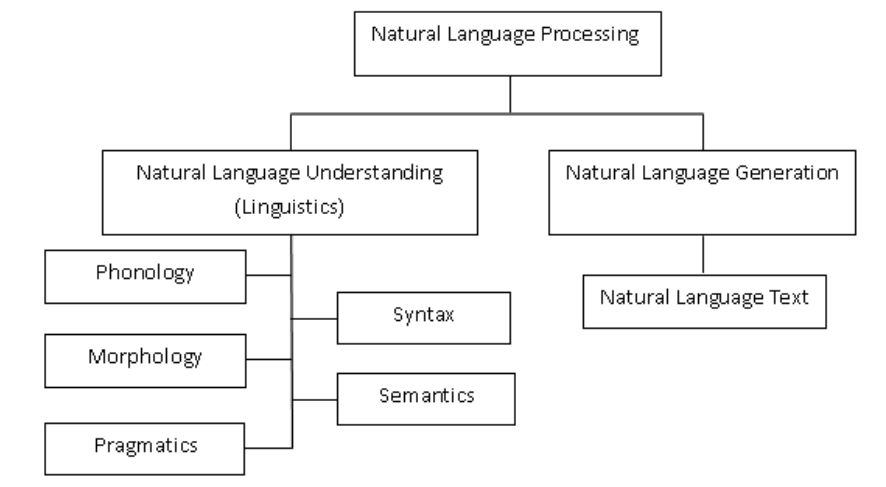}
	\caption[Classification of NLP~\citep{khurana2017natural}]{Classification of NLP~\citep{khurana2017natural}}
	\label{fig:nlpclassification}
\end{figure*}

NLP is a combination of AI and Linguistic methods, and aims to make computers understand human languages. According to \citet{manning2014natural}, NLP is classified into NLU and NLG. \citet{jusoh2012techniques} described NLU as a representation that deals with modeling human reading comprehension tasks which parses and translates inputs according to NL principles. Typically, this uses NLP algorithms to reduce human speech into structured forms. NLU consists of components such as phonology, morphology, pragmatics, syntax and semantics. \citet{reiter2000building} described NLG as a technique where texts are generated from human languages using computer-accessible data. This process understands texts in natural form such as English from  non-linguistic representation of information~\citep{manning2014natural}. These terms are illustrated in \autoref{fig:nlpclassification}. We describe each of these terms.

\begin{description}
\item[Phonology] The field of phonology deals with the study of sounds and how they are used in languages~\citep{gussenhoven2017understanding}. Phonology is applied to virtually all languages, and is predominantly used in the linguistic domain. In the computational discipline, phonology refers to the application of computational techniques to the processing of phonological information~\citep{bird2002computational}. \citet{trask2004dictionary} described that phonology tells us what sounds are contained within a language and shows what happens if they are combined into words. The study further described that phonology and phonetics are two sub-disciplines in linguistics and highlighted that speech organs and muscles are involved in the different aspects of a language. The human phonological features differ with varying frequencies according to their supralaryngeal vocal tract. This is depicted in \autoref{fig:voice}.

\begin{figure*}[h]
	\centering
	\includegraphics[width=250px]{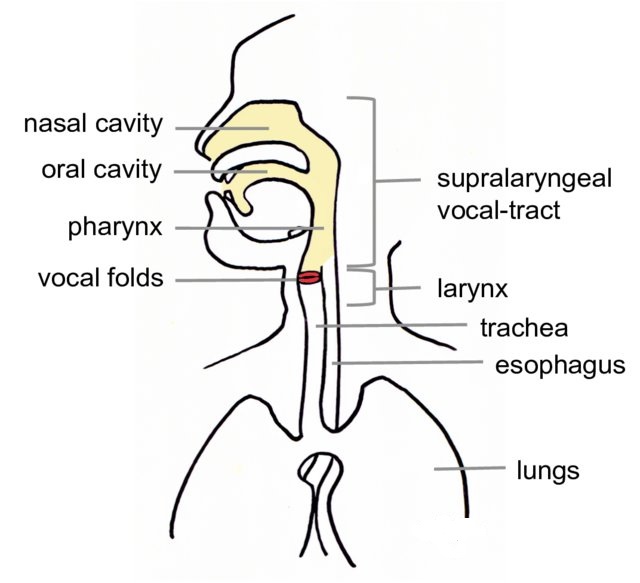}
	\caption[The human vocal apparatus~\citep{pisanski2014human}]{The human vocal apparatus~\citep{pisanski2014human}}
	\label{fig:voice}
\end{figure*}

\item[Morphology] The morphology of a language describes how words are put together to form a grammar~\citep{ritchie1992computational}. Words are an essential part of linguistics, and constitute an integral part of mental grammar~\citep{twain2013morphology}. A native speaker of a language is expected to know thousands of words. \citet{nordlinger2019morphology} noted that without words, it will be difficult to convey thoughts through a language or understand others' thoughts. Hence, a native speaker of English knows how to segment sounds of words in his or her lexicon. In NLP, morphology is important for lemmatisation and parsing. They are useful in many applications~\citep{twain2013morphology,nguyen2016funguild}. For example, morphology has been applied in lexical databases such as WordNet~\citep{miller1995wordnet} and Arabic Ontology~\citep{el2018generative}, and statistical parsing tools such as SPRML~\citep{tsarfaty2010statistical} and FUNGuild~\citep{nguyen2016funguild}.

\item[Pragmatics] Pragmatics is a branch of linguistics that deals with how contexts can be converted into meanings through a language~\citep{thomas2014meaning}. Pragmatics has its roots in sociology, anthropology and philosophy~\citep{morris1970pragmatic}. This study was influenced by the work of \citet{peirce1902logic}, who described three systems of signs (semiotics) namely, syntax, semantics and pragmatics. The study defined \textit{syntax} as the formal relation of signs, \textit{semantics} as the relation of signs to what they denote, and identified that \textit{pragmatics} is used to describe signs with relations to users and interpreters. Many NLP researchers discussed that, to produce effective NLP systems, it is important to understand the pragmatics of a natural language \citep{cruse2011meaning,ward2016semantics}. \citet{cherpas1992natural} stressed that semantics blends well with pragmatic theory, as it enables the recognition of human conversations, and how they can manipulate each other. In addition, the history of their interaction can shape future utterances.

\item[Syntax] \citet{covington1994natural} described syntax as the set of rules that governs the formation of words into phrases or sequences of well-formed words (sentences). The goal of syntax is to relate morphological components to semantic constituents~\citep{chakrabarti2004breaking,dale2000handbook}. In other words, syntax involves word tokens and structure. In syntactical forms, words combine in a way which mirrors the expected meaning. For example, \textit{Peter loves Mary might indicate something different from Mary loves John}. There is usually ambiguity involved at the syntatic phase~\citep{manning2010introduction}. To represent these forms, syntax trees are usually used to denote the syntactical analysis phase~\citep{van2005generator,hewitt2019structural}. In NLP, syntax is regarded as a lower level stage and involves the following activities~\citep{nadkarni2011natural,roomnatural,gill2019overview}:

\begin{enumerate}
	\item Lemmatisation: This process involves reducing words to their base form, known as \textit{lemma}.
	\item Morphological Segmentation: This process breaks words into morphemes. For example, the English word, ``horses'' contains two segments (horse) + (s). This is a word stem and its suffix.
	\item Word Segmentation: This is the process of dividing strings in a language into component words.
	\item Normalisation: This process involves the categorisation of words or tokens into a standard format.
	\item Stemming: This is the process of reducing words within the same stem to their root form.
	\item PoS Tagging: This is the process of parts of speech identification for each word. For example, the PoS for ``Johannesburg'' is a noun.
	\item Parsing: This is the grammatical analysis of a given sentence into a syntax tree.
	
\end{enumerate}

\item[Semantics] \citet{berant2014semantic} define semantics as a concept that conforms to meaning. The process of checking if words form sensible sets of instructions is known as \textit{semantic analysis}~\citep{nasukawa2003sentiment}. At the semantic analysis phase, this process relates the syntactic structures from granular levels of phrases, clauses to the language independent meaning~\citep{cambria2014jumping,sun2017review}. This has been applied in many domains such as: Medicine~\citep{pons2016natural}, Affective Computing~\citep{cambria2016affective} and Game theory~\citep{ryan2015we}, etc.  In NLP, if a language is to be understood by a computer, it must go through syntactic and semantic analysis phases~\citep{seuren2017semantic}. The upper level activities of NLP as described by \citet{nadkarni2011natural,khurana2017natural} are: 

\begin{enumerate}
	\item Named Entity Recognition: This task involves determining the pre-defined categories of named entities such as names of organisation, persons, locations, etc.
	\item Word Sense Disambiguation: This task determines the meaning of an ambiguous word within a context. 
	\end{enumerate}
\end{description}
\subsection{NLP Algorithms}
Since the inception of NLP, different algorithms have been described to work with tools in this field. \citet{duh2018bayesian} describe NLP algorithms as useful in multiple language variations. In this section, we review the NLP algorithms that have been proposed over the years.

\subsubsection{Naïve Bayes}
Naïve Bayes (NB) algorithm has emerged as one of most efficient and effective text classification techniques used for data mining and machine learning tasks~\citep{gao2018privacy,li2018differentially,xu2018bayesian}. This classification technique was developed by Thomas Bayes in 1763, hence its name~\citep{wang2016classification}. It was regarded as \textit{naïve} because it assumes features that are used by a machine learning model which are independent of each other. Although currently dubbed as `the punching bag' of newer classifiers by machine learning theorists, it remains widely used for classification of texts, which is fast and easier to implement~\citep{xu2018bayesian,mohammed2017normalised}. Most spam filters include NB in their commercial and open-source projects. In addition, it has been applied in many real-world classification problems in medicinal diagnosis, sentiment analysis, weather predictions, etc ~\citep{xu2018bayesian,wood2019private,kwon2019solar}. \\

\citet{li2018differentially} described NB as a probabilistic classifier that uses the Bayes theorem which can substitute logistic regression models, and can also be used to formulate dependency or independent variables. Furthermore, Bayes theorem is also referred to as a posterior probability, used for an event ~\citep{zhang2019research}.\\

%
%
%
%
%
%

In a comparative study, \citet{keogh2006naive} showed that NB was much faster to train, and insensitive to irrelevant features in data. A major flaw identified in the study showed that NB assumes independence on more linear features than on non-linear features. Furthermore, \citet{xu2018bayesian} discussed that the NB's classical counterparts such as the Hidden Markov Models (HMMs), Support Vector Machine (SVM) and principal component analysis (PCA) were much better for text classification compared to the NB.

\subsubsection{Support Vector Machine}
The SVM is a widely text supervised model proposed by \citet{boser2003training} that has been applied to numerous real-world problems. SVM belongs to a family of linear classifiers used for classification and regression tasks~\citep{ben2010user}. As a technique, it is regarded as one of the kernel methods that maximise predictive accuracy while avoiding over-fit to data~\citep{ben2010user}. These method provides two advantages: they have the ability to generate non-linear boundaries and they allow a classifier to use data with no obvious fixed space representation. This has been applied in protein synthesis, Deoxyribonucleic Acid (DNA) sequences and bio-informatics problems~\citep{shawe2004kernel}. \\

\begin{figure*}[h]
	\centering
	\includegraphics[width=300px]{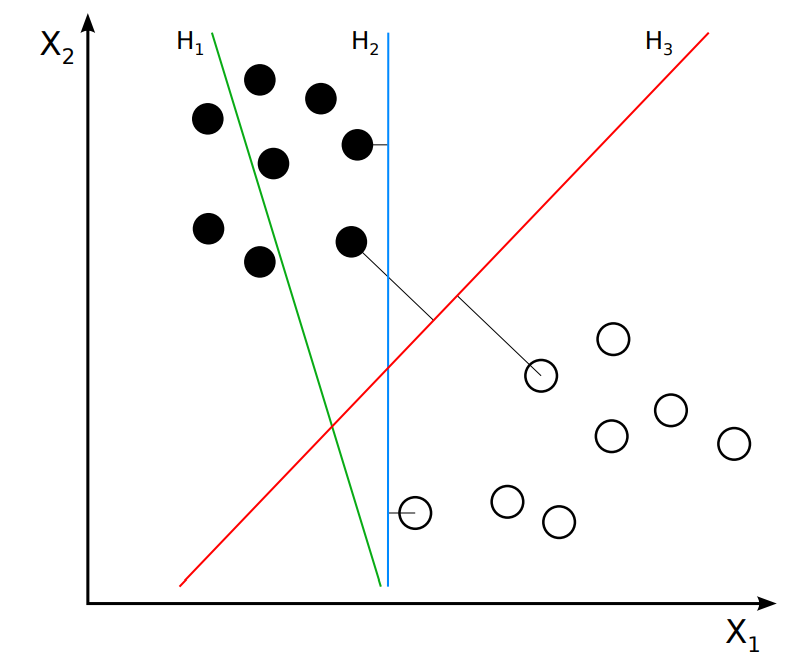}
	\caption[Support Vector Machine: Separating Hyperplanes~\citep{klein2006wits}]{Support Vector Machine: Separating Hyperplanes~\citep{klein2006wits}}
	\label{fig:svm}
\end{figure*}

One simple rule about SVMs is that they create a line (or a hyperplane) that separates data into classes, which can be applied to many classification problems ~\citep{evgeniou2001workshop,tang2013deep}. \citet{klein2006wits} explained the hyperplane using an illustration. In \autoref{fig:svm}, the first hyperplane, $H_1$, is indicated as a black circle which incorrectly classifies data points. The second and third hyperplanes, $H_2$ and $H_3$, correctly classify data points.\\

\citet{auria2008support} identified the strength and weaknesses of SVM. Notably, SVMs are good with structured and unstructured data e.g. text, trees and images. One major drawback of SVM is that it takes longer to train a large dataset. \citet{goldberg2008splitsvm} noted that despite the challenges faced with SVMs, they contribute immensely to solving problems in the NLP domain. 

\subsubsection{Hidden Markov Model}
One of the most popular methods used in machine learning for sequence modelling of speech and protein is the HMM~\citep{tobon2011hidden}. Since introduced by the Russian mathematician, Andrey Andreyevich Markov, as the \textit{theory of stochastic Markov processes}, it has contributed immensely to solving many real-world problems~\citep{van2010activity,narasimhan2016bcftools,fu2016dynamic}. Almost all modern speech recognition systems are based on HMM. Although, its framework has not changed drastically, it has evolved even further with more sophisticated features~\citep{gales2008application}.  \citet{bahdanau2016end} described the HMM as a double stochastic model that is based on intrinsic variability of spectral features and a statistical modeling framework that ensures consistency in detecting spoken languages. These stochastic processes are characterised by states and transitions probabilities. The first process, states, are not visible, hence it is considered to be \textit{hidden}. The second process, transitions, attempts to produce state-dependent probability distributions~\citep{kouemou2011history}.\\


\citet{keselj2009speech} described the HMM as based on the augmentation of the Markov chain, which shows the probabilities of sequences of random variables. The study further explained that for a NLP scenario, HMM shows observed events and hidden event (PoS tags) that are considered in a probabilistic model. Furthermore, \citet{awad2015efficient} explained that for given observations, HMM is used to describe the solution to a problem in a state-sequence determination manner and through model training means. 

\subsubsection{Deep Learning}\label{deeplearn}
\begin{figure*}[h]
	\centering
	\includegraphics[width=400px]{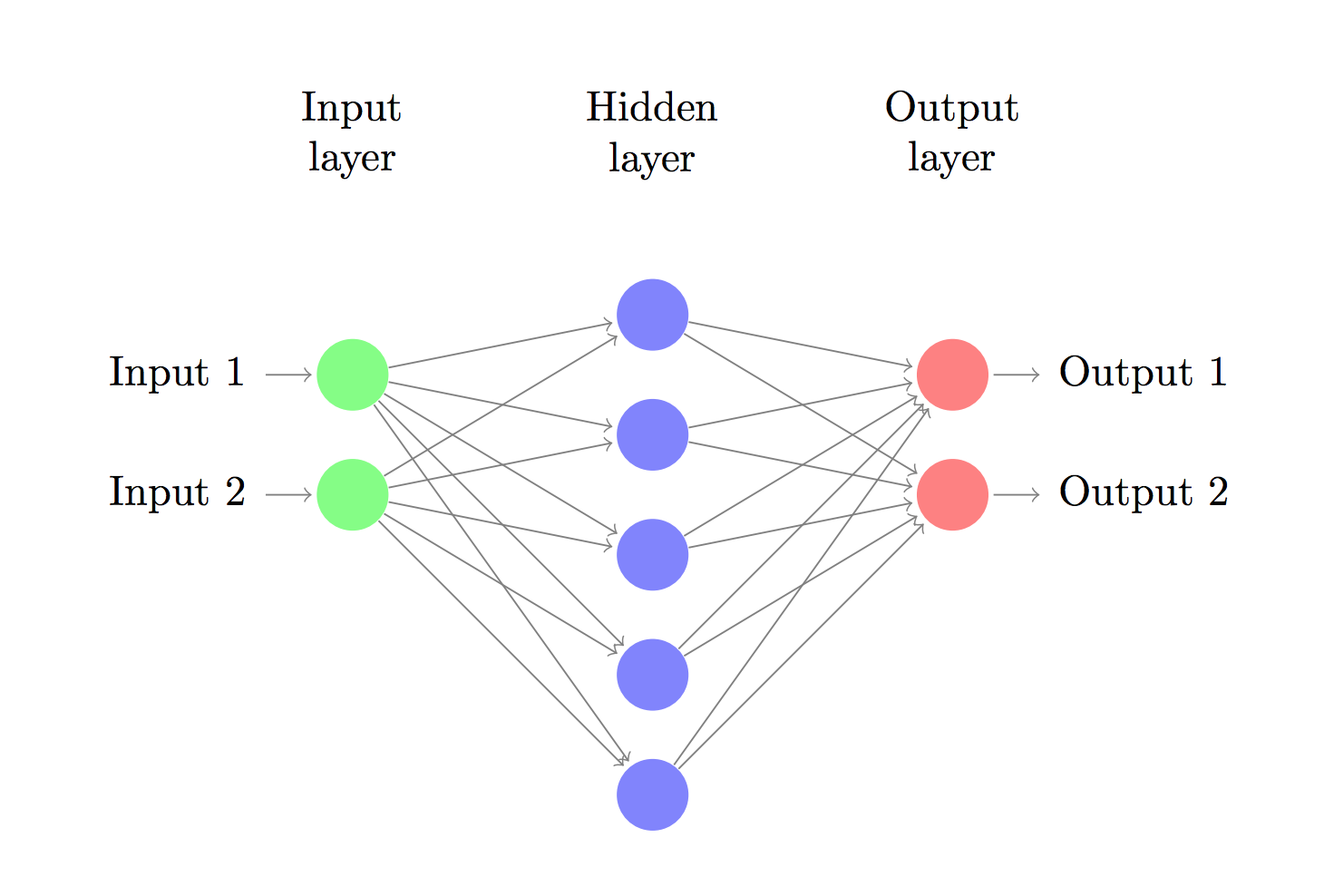}
	\caption[A neural network~\citep{liu2017survey}]{A neural network~\citep{liu2017survey}}
	\label{fig:snn}
\end{figure*}

The era of deep learning methods began in the 20\textsuperscript{th} century after the performance of traditional learning became less satisfactory to process human information forms in speech and vision~\citep{ohlsson2011deep,sun2017revisiting}. The deep learning methods were originated by Geoffrey Hinton in 2006, when he proposed the Deep Belief Network~(DBN), a deep learning structured architecture~\citep{lecun2015deep}. Since the DBN, there have been rapid developments of other deep learning techniques with significant impacts on information processing in the medical, engineering, and even the educational fields~\citep{ching2018opportunities,lim2016teachers}. Predominantly, the deep learning methods have contributed immensely to the NLP~\citep{socher2012deep,deng2014deep,bian2014knowledge}, speech recognition~\citep{hinton2012deep,amodei2016deep} and computer vision~\citep{kendall2017uncertainties,voulodimos2018deep}. After the DBN network, Artificial Neural Networks (ANNs) became popular and were an active area of research, utilising neurons to produce real-valued activation. These neurons receive inputs from external sources, and assign weights and biases during training. Then, it produces an output. This is illustrated in \autoref{fig:snn}. The diagram shows the input layer in the leftmost layer, the hidden layer indicates that it is neither an input nor an output medium. The output is the rightmost layer or output neurons. \\

Comparable to the ANNs are the Convolutional Neural Networks (CNNs) and Recurrent Neural Networks (RNNs). They are comprised of neurons that receive an input and perform a scaler operation to produce an output~\citep{goodfellow2016deep}. CNNs are mostly used for pattern recognition within images and have recorded major successes~\citep{khosravi2018deep,sturmfels2018domain}. RNNs are popularly used for sequential tasks such as speech and language (texts)~\citep{di2016artificial,goodfellow2016deep}. RNNs are powerful networks and training them requires backpropagation which exhausts a great deal of computational power. Together, these neural networks have performed well on various NLP tasks such as named-entity recognition, MT, phrase detection and language modeling~\citep{manning2015computational,zheng2013deep}. One major contributing factor of these neural networks is their ability to perform tasks without time-intensive engineering processes. This has led to an important concept in NLP, known as Word Embedding~\citep{weston2012deep}. In addition, a survey study conducted by \citet{kepuska2017comparing} showed that due to the deep learning method integrated in the Google Speech recognition engine, it improved its word error rate (WER) from 23\% in 2013 to 8\% in 2015. The study showed that Google's speech engine was significantly better compared to the Microsoft Speech API and Sphinx-4 engine. 
 
\subsection{Popular NLP Libraries}
NLP libraries are mostly used by researchers to extract information from texts~\citep{opennlp2011apache,manning2014stanford,zhang2019riscoper}. These libraries handle a wide range of NLP tasks such as topic modelling, text classification, sentiment analysis, POS tagging and many more. In this section, the libraries used for NLP tasks are presented.

\subsubsection{Apache OpenNLP}
The Apache OpenNLP\footnote{https://opennlp.apache.org/} was written in Java, as a free, open-source machine learning tool used for core NLP tasks, such as named entity recognition, parsing, sentence segmentation, and POS tagging~\citep{baldridge2005opennlp,opennlp2011apache}. Developers use the OpenNLP interfaces, provided by means of an API\footnote{Application Programming Interface} to implement these NLP tasks. The OpenNLP library uses a maximum entropy to build advanced text processing services~\citep{dandapat2007part,tratz2007pnnl,bilgin2019study}. The maximum entropy framework is based on the principle of making assumptions based on constraints imposed on training data, relationships between data features and expected outcome. In addition, the maximum entropy is used to recognise different entities such as locations, organisations, dates and persons. The OpenNLP has been applied to solve many problems in different domains. We describe them in no particular order.\\

\citet{rodrigues2018nlpport} used the OpenNLP tool for NLP tasks specified in the Portuguese language. In this study, sentences were split into tokens, which are a combination of words. These words are further analysed to broaden the result. The study reported that while using OpenNLP, many tools under-performed in some language constructs within the Portuguese context. During an analysis of the forum posts on the dark web, \citet{park2016temporal} applied the OpenNLP for sentiment scores generation. The study used the OpenNLP tool tokeniser feature to create an array, which is then used to find all parts-of-speech in each post, and generate frequencies of each noun in the post. In the biomedical field, \citet{zhang2019riscoper} developed a tool called RNA Interactome Scoper (RIscoper) that uses the OpenNLP for sentence segmentation which extracts contents from articles. According to the study, the developed tool will save time required for drafting of literature reviews and data organisation in databases. In addition, the study emphasised that the tool would be useful for bioinformaticians and experimental biologists.


\subsubsection{Natural Language Toolkit}
The Natural Language Toolkit\footnote{http://www.nltk.org/} (NLTK) provides a set of tools, released under an open-source licence for performing different NLP tasks~\citep{loper2002nltk,bird2008multidisciplinary,bird2009natural}. NLTK was developed in Python and contains libraries that supports statistical and symbolic NLP. Python was selected as the implementation language because its syntax and semantics have good string handling functionality and provide a shallow learning curve~\citep{bird2008multidisciplinary}. This toolkit has comprehensive documentation, including tutorial guides that contain the processing tasks it supports. Predominantly, NLTK is well suited to users learning or conducting research in NLP including related areas such as machine learning, cognitive science and information retrieval~\citep{klein2006computational}. \\

\citet{bird2009natural} described that before NLTK was developed, the following goals were kept in mind. The first goal is \textit{simplicity}, an intuitive framework that could give users a practical knowledge of NLP was created. Second, there was a need to ensure \textit{consistency} with interfaces that provide a uniform framework. Third, there is a need to provide a structure that allows new modules to be \textit{extensible}. Finally, the study showed that some components can be used \textit{independently} of other components. \citet{lobur2011using} highlighted the uses of NLTK, which includes chunk parsing, assignments, as well as advanced tasks such as word sense disambiguation and morphological analysis. 

%

\subsubsection{Stanford CoreNLP}
The Stanford CoreNLP\footnote{https://stanfordnlp.github.io/CoreNLP/} toolkit is an extensive annotation pipeline framework, developed in Java, that provides features to most NLP tasks from tokenisation, named-entity recognition through to coreference and basic dependencies~\citep{corenlp2016suite,hirschberg2015advances,angeli2014stanford}. \citet{manning2014stanford} described the toolkit as a combination of multiple components, each with their APIs tied together to function as a custom glue unit (or code). Before using the engine, raw text is inserted into some annotated object, which then undergoes various NLP processing tasks and the resulting feedback is provided by means of an annotated or plain text format. We describe some of the recent works that have extended the Stanford CoreNLP into other languages. \\


For recognising lexicons within the Chinese language, \citet{peng2015concrete} developed a tool called \textit{CONCRETE}, an NLP pipeline built on a number of open-source tools. This pipeline extended the CoreNLP framework to recognise Chinese language within a broader context. The pipeline supported word segmentation, named entity recognition, parsing and PoS tagging. \citet{bondielli2018corenlp} developed an extension of the Stanford CoreNLP toolkit based on the Universal Dependency (UD) framework into a tool, called the \texttt{CoreNLP-it}. This tool was developed for the recognition of the Italian language as a set of customisable classes. This study reported that the tool was UD compliant, catered for multi-word token representation and provided an extensible framework to support other languages. \citet{andreeva2018intellectual} extended the CoreNLP toolkit to recognise texts in the Russian language. This tool was able to determine parts of speech in the Russian language and this proved effective in this context.  

\subsubsection{spaCy}
The spaCy\footnote{https://spacy.io/} NLP engine was developed as an open-source Python toolkit, designed to work on large-scale commercial information extraction tasks~\citep{al2017choosing,bocklisch2017rasa,srinivasa2018natural}. The current version, spaCy v2.x, was developed in Python/Cython in 2017, which possesses an accuracy of 92.6\%, making it the fastest syntactic parser in the world. Till now, spaCy is the fastest NLP toolkit in the world with regards to language processing tasks, especially when compared to NLP libraries that have been developed~\citep{al2017choosing}. spaCy supports almost all NLP tasks from dependency parsing, tokenisation to POS tagging, and it works well with most deep learning libraries such as scikit-learn, TensorFlow, PyTorch~\citep{jangid2018aspect,goyal2018deep}.  We describe recent works that have applied the spaCy toolkit to their tools for various NLP tasks. \\


\citet{bocklisch2017rasa} built a tool called \texttt{Rasa}, an open-source Python framework, used to build conversational systems using the spaCy toolkit for NLP tasks. For NLU tasks, the spaCy toolkit was used to perform tokenisation and POS tagging. \citet{kejriwal2017flagit} extended the use of the spaCy framework into an open-source tool called \texttt{FlagIt}, a system for mining problems in the sex trafficking domain. The system has been integrated into a domain-specific search platform used by over 200 law enforcement agencies to minimise the problem of human trafficking. The study noted that spaCy was effective in the inherent complex NLP tasks used by \texttt{FlagIt}. In a study on grammatical error corrections, \citet{naplava2019cuni} developed a tool called the \texttt{CUNI} system that applies restricted, unrestricted and low-resource tracks trained using the Wikipedia source to resolve errors in a sentence. \texttt{CUNI} applied spaCy to correctly tokenise sentences. The study showed that spaCy was effective in this task.

\subsection{Applications of NLP}
NLP is one of the most important technologies of the current information age~\citep{klein2017opennmt,nakazawa2006example,jurafsky2012natural}. As humans communicate their thoughts in a language, this has paved the way for numerous applications of NLP: language translation, web search, advertisements, spam detection, text categorisation, and QA. In this section, a few NLP application areas are discussed. 

\subsubsection{Machine Translation}
\citet{sokolov2016learning} described MT as an important application area of NLP that deals with the automatic translation of speech or text from one human language into another. \citet{hirschberg2015advances} emphasised that MT is the most substantial way in which computers could facilitate human to human communication. MT is rated as the most difficult field in NLP, because it relates to human lives, and as such, it had become a million dollar affair~\citep{alsohybe2017machine}. \\

Over the past decades, MT has been an active area of research for linguists, computer scientists and engineers~\citep{klein2017opennmt,nakazawa2006example,costa2016character}. It is interesting to note that the first of numerous applications of computers was the MT. This was studied intensively in the late 1950s~\citep{hirschberg2015advances}. This era saw the development of the Statistical Machine Translation (SMT) and Rule-based Machine Translation (RMT). In the early 1990s, the MT field was transformed at the bilingual Canadian Parliament proceedings when there was parallel text translation of English and French sentences~\citep{jurafsky2012natural}. Since then, there has been improvement in language translation in the MT field, and currently, the field is in a state of flux with hybrid solutions, falling short of precision and accuracy of human translators~\citep{dale2019nlp}.\\

\citet{koehn2007moses} presented an open-source toolkit called \texttt{Moses} for SMT that consists of components for data preprocessing, language model training and result translation. This toolkit integrates well with NLP/speech processing tools with varying confidence in a consistent and flexible framework. The study concluded that \texttt{Moses} will be of immense value to the MT community. The advancement of the deep neural network has paved the way for Neural Machine Translation (NMT), which is a recently proposed approach to the MT field~\citep{koehn2017six,artetxe2017unsupervised,zoph2016transfer}. Unlike the SMT approach, NMT builds on a single neural network that maximises translation performance. In addition, the SMT field is problematic because most translation systems are specifically trained within a particular domain~\citep{farajian2017multi}. Thus, it might perform poorly in a different domain.  With the introduction of NMT, state-of-the-art systems perform better in English-French translation tasks~\citep{bahdanau2014neural}. 

%

\subsubsection{Question Answering}
QA is an active area of research and a specialised type of information retrieval (IR) or information extraction (IE) aimed at returning answers to queries presented in the form of a natural language~\citep{ong2009measurement,belinkov2015vectorslu,cantador2011second}. \citet{athenikos2010biomedical} discussed that the next generation of search engines will utilise the capabilities of QA. The history of QA systems dates back to the late 1960s and early 1970s when a major surge of research activities was seen within the IR/IE community~\citep{athenikos2010biomedical}. In 1999, this surge led to the establishment of the QA Track in the famous Text REtrieval Conference evaluations~\citep{voorhees1999text}. Since then, a number of techniques has been developed for answer generation for three questions types such as list, factoid and definitions~\citep{athenikos2010biomedical}. To support QA systems, numerous large datasets have been developed over the years. These datasets consist of questions posed by crowdworkers\footnote{a method that involves volunteers to accomplish a specific task} that are used to train QA systems, where the answer is a segment of text within the content. Examples are \texttt{SQuAD}~\citep{rajpurkar2016squad}, \texttt{HotpotQA}~\citep{yang2018hotpotqa}, \texttt{DAWQAS}~\citep{ismail2018dawqas}, \texttt{30M Factoid questions}~\citep{serban2016generating} and many others. \\



Generally, QA systems consist of three processing methods, namely: question, documents and answering phases~\citep{hirschman2001natural}. They are built with semantic knowledge throughout the QA process, in order to derive correct answers to questions. The semantic information is obtained from questions and ontological resources may be used to improve the performance of the QA system. According to a review study by \citet{athenikos2010biomedical}, QAs were classified into semantic-based QA systems, inference-based, and logic-based. The study concluded that QA systems will continue to grow and help users better utilise the evolving nature of information.

\subsubsection{Spam Detection}
There is a no single, accepted definition of spam, although this problem has gained prominence since the 1990s~\citep{van2010filtering,chandra2014survey,hayati2010definition}. \citet{iedemska2014tricks} described spamming as an activity perpetuated by cybercriminals, which is used to generate income to the tune of millions of dollars. Conversely, spam may contain unsolicited advertising contents~\citep{broadhurst2018malware}. Spamming activities are conducted using platforms such as websites reviews~\citep{lin2014towards,ghai2019spam,saini2019multi}, social media~\citep{barber2018similar,sharaff2016comparative}, opinion mining~\citep{rayana2015collective,chen2015opinion} and email~\citep{idris2015combined,seyyedi2018estimator} to swindle users. Email spamming appears the most popular amongst these activities~\citep{christina2010study}. This reduces human productivity, wastes bandwidth and storage, and has exceeded legitimate emails which are sent over the Internet~\citep{shue2009spamology}. \\

%

Over the past few years, numerous studies have been proposed to detect spam. We present the recent studies in no particular order. Using NLP algorithms, \citet{ezpeleta2016does} proposed the use of the Bayesian filtering classifier to detect unsolicited emails, which are major threats affecting millions of users per day. The research achieved an accuracy of 99.21\% exceeding those proposed by machine learning algorithms. Similarly, \citet{maguluri2019adaptive} extended the use of Bayesian classification to email spam detection. The study categorised email messages as either \textit{spam} or \textit{non-spam} and noted that spam could be an enormous problem for private and public organisations. Classification techniques such as SVM and Deep Neural Networks have also proved effective to efficiently identify spam in emails~\citep{torabi2015efficient,scholkopf2002learning,agarwal2016spam}. \\

\citet{kumaresan2017mail} applied the SVM technique to detect spam emails, showing a better accuracy over other techniques due to the small data size. The study reported an accuracy of 97.235\% when compared with an existing approach. \citet{roy2019deep} extended the deep networks using CNN and the Long Short Term Memory (LSTM) to detect spam in a social network such as Twitter. The study incorporated semantic databases such as WordNet and ConceptNet to improve semantic information representation. Furthermore, this improved the accuracy and F1-score of the result.

\subsubsection{Dialogue Systems}
Recent advances in NLP have led to multiple applications of dialogue systems, which have significantly eased tasks in medicine, online shopping, technical support, etc~\citep{deng2013recent,serban2016building,bowden2019data}. These dialogue systems provide either \textit{speech} or \textit{type-written} features, or both~\citep{mctear2002spoken,glas2012interaction}. Speech dialogue systems are interactive platforms used by humans to communicate with a computer with the intention of achieving a specific objective~\citep{serban2016building}. For example, a user may request all hotels that are based in a particular location from a chatbot\footnote{an AI-powered system designed to simulate conversation with human users}. The chatbot takes the request and provides immediate feedback to the user. With the growing rate of AI, numerous companies are beginning to design powerful goal-oriented spoken dialogue systems. Examples of such spoken dialogue systems are Google's Assistant, Apple's Siri, Amazon's Alexa and Microsoft's Cortana~\citep{lopez2017alexa,hoy2018alexa}. \citet{chen2017deep} described the components required of any speech dialogue system. These components are automatic speech recognition for recognition of human speech into text, NLU framed into speech identification, dialogue management using backend providers and NLG for generating texts based on some linguistic methods. \\


Nowadays, a number of NLP libraries have been developed, which are used by dialogue systems for communication. Examples of these libraries are OpenDial~\citep{lison2016opendial}, AllenNLP~\citep{gardner2018allennlp}, ParlAI~\citep{miller2017parlai}. General frameworks such as Stanford CoreNLP and spaCy have been used for the development of dialogue systems~\citep{burtsev2018deeppavlov,li2016deep}.\\

The origin of dialogue systems can be found in tools such as \texttt{ELIZA}~\citep{weizenbaum1966eliza} designed to allow humans to interact with computers in free-forms, specified in natural language. Similarly, other tools such as \texttt{Parry}~\citep{colby1975artificial} and \texttt{Alice}~\citep{wallace2009anatomy} were developed to allow users interact in a conversational manner. Although these tools were able to perform tasks seamlessly, they were not intelligent enough and lacked the ability to keeping conversations evenly~\citep{shum2018eliza}. Hence, they only work well in constrained environments. Since then, many researchers have developed smart solutions on dialogue systems that can handle complex tasks. \citet{huang2015guardian} developed a spoken dialogue system called \texttt{Guardian} that significantly can improve interaction in a cost-effective manner. The tool combines expert and non-expert processes that uses a Web API to scale up interactions, and can be embedded in other dialogue systems through its API features. Similarly, a tool called LS-SDS was developed by \citet{papangelis2017ld} as an advanced, complex, interactive interface that leverages on Linked Data to improve the linking of entities to user's input and data sources. \citet{goel2019hyst} proposed a hybrid approach that uses a trainable neural network coupled with a NLU and dialogue tracking to achieve a high accuracy. This hybrid approach helps to facilitate interaction between a user and a computer system.

\subsubsection{Text Categorisation}
In the data-intensive application fields such as banking, universities and funding agencies, unstructured data remains a serious problem~\citep{tang2016bayesian,selvi2017text}. One way of tackling this problem is to present the data in a format that can be used in these fields~\citep{al2019arabic}. This process is regarded as text categorisation (or text classification). \citet{dumais1998inductive} described text categorisation as an application of NLP, that deals with the assignment of natural language texts to one or more categories based on their linguistic contents. \citet{sebastiani2005text} explained that text categorisation involves the task of automatically sorting document sets into a set of categories. This area of NLP has received prominence in the last 10 years, and many researchers and software developers are deploying applications using this approach. There are well-known text classification methods such as language detection, sentiment analysis and topic labeling~\citep{stein2019analysis,huang2014text,chaturvedi2018distinguishing}. These methods have been used to solve numerous real-world problems. \\

\begin{figure*}[h!]
	\centering
	\includegraphics[width=300px]{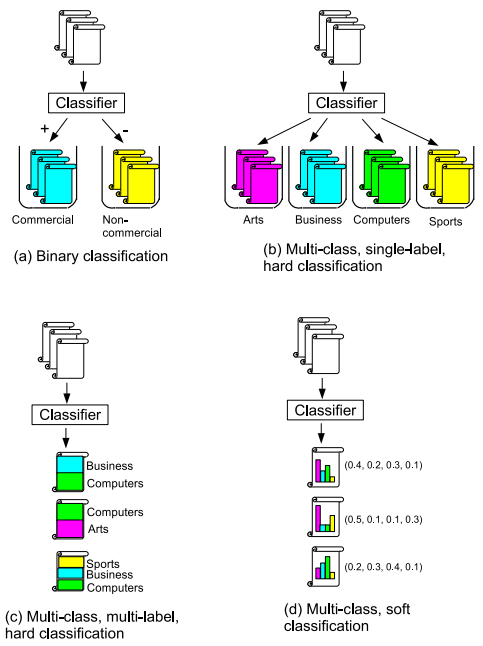}
	\caption[Text classification methods for web documents~\citep{qi2009web}]{Text classification methods for web documents~\citep{qi2009web}}
	\label{fig:page}
\end{figure*}

For a classification of web page content, \citet{qi2009web} described state-of-the-art practices as essential to many tasks such as information retrieval, web directory maintenance and data crawling. The study showed that classification can improve web search quality. The four classification techniques as described in \autoref{fig:page}, show how they can used to resolve search engine spams and improve search results in websites. For clinical data classification problem, \citet{bui2014learning} used the regular expression approach alongside the SVM classifier for text classification of clinical datasets. The study reported that using these two classifiers improved clinical text classification performance and showed that this hybrid approach was significantly better than using the SVM classifier alone. \citet{hughes2017medical} presented an approach for the classification of clinical text at the sentence level. The method used the CNN approach to train the health information dataset and indicated that the approach superseded previous approaches that have been developed by about 15\%. In a similar approach, \citet{chen2017deep} used a CNN model to classify radiology reports with accuracy reaching 99\%. The study reported that this approach may be used for large-scale applications by annotating texts in medical imaging reports. 

\subsection{Classification of Natural Language Interfaces to Databases}
\citet{mony2014overview} described NLIDBs as systems which provide easy access to databases using a natural language, without requiring a user to write in a query languages such as SQL, Prolog and Lisp. These systems are mostly used by non-technical end-users in fields such as banking, medical, engineering, mining, etc~\citep{yuan2019criteria2query,gantayat2019goal,kapetanios2008natural}. To use NLIDBs, \citet{wudaru2019question} emphasised that users are required to communicate with these systems in a natural language such as English. Other languages specified in Arabic~\citep{elsayed2015arabic} and French~\citep{etzioni2002high} have been proposed in the past. The development of NLIDB systems started in the 1960s~\citep{nihalani2011natural,sujatha2012survey}. Initially, tools such as \texttt{BASEBALL}~\citep{green1961baseball} were developed using the baseball league played in the US as a test case. \texttt{BASEBALL} provided answers related to location, dates, etc. This was followed by \texttt{LUNAR}~\citep{woods1973progress}, a NLIDB system that was developed from the Apollo lunar exploration that provided information about soil samples. Other systems that followed were \texttt{RENDEZVOUS}~\citep{codd1974seven}, \texttt{LADDER}~\citep{sacerdoti1977language} and Chat-80~\citep{warren1982efficient}. All these systems produced good results, but had a limited repository of information related to other domains~\citep{mishra2016survey,papantoniou2019cs563}. Similarly, these systems used \textit{hard-wired} knowledge and were dependent upon a limited application area, which was a major disadvantage. This has been improved by newer NLIDBs that cater for domain independence and multiple different databases. NLIDBs are classified into different categories. These categories show the knowledge-base used by NLIDBs generate answers to natural language questions~\citep{affolter2019comparative,pazos2013natural}. This section presents these categories of NLIDBs.

\subsubsection{Keyword-based}
The goal of keyword-based systems is to match keywords against a meta-data~\citep{shah2013nlkbidb}. In this approach, the systems attempt to retrieve keywords from an input sentence and convert the equivalent into SQL queries. The following are some of the recent keyword-based tools:

\begin{description}
\item[SINA] \citet{shekarpour2015sina} developed \texttt{SINA} as an online, scalable keyword search system used for transforming natural language questions into SPARQL Protocol and RDF Query Language (SPARQL) queries. In its engine, it uses the HMM to establish the most likely NLQ from different datasets. To reduce questions into keywords, \texttt{SINA} uses the tokenisation, lemmatisation and stop word removal methods, which are segmented and processed, before the generated SPARQL query is displayed to the user. A major weakness of \texttt{SINA} is that it reduces the number of answerable questions because it translates only to conjunctive SPARQL queries.


\item[Aqqu] A template-based system called \texttt{Aqqu} was proposed by \citet{bast2015more} that uses keywords to translate natural language questions to their matching SPARQL query. The tool uses POS-tagging for entity matching, generates the set of sequences of words, computes the list of all entities from the knowledge base and provide scores for each of the entities matched. \texttt{Aqqu}'s strength lies in the identification of relationships between entities. 


\item[DeepEye] The \texttt{DeepEye} tool was created by \citet{qin2018deepeye} that attempts to use keywords to create visualisations from user's queries in natural language. The idea behind this tool is to use a keyword query and a dataset, to generate all possible visualisation. The study described the framework of \texttt{DeepEye} and explained that it crawls, stores and provides good visualisation from multiple different sources.

%

\item[SpatialNLI] \citet{li2019spatialnli} presented a NLI tool called \texttt{SpatialNLI} that translates a natural language question into a structured SQL query that is executable by a DBMS. This tool learns from the keywords of the spatial comprehension model that takes a NLQ and captures the spatial-specific semantics of it. In addition, the tool uses a sequence-to-sequence approach, which is a deep learning concept to capture semantic meaning of a question. The accuracy reported by the tool indicates 90.7\%. The authors claimed that \texttt{SpatialNLI} outperforms other popular state-of-the-art methods.


\end{description}

\subsubsection{Pattern-based}
The pattern-based systems are an extension of the keyword-based approach~\citep{affolter2019comparative}. This approach answers more complex questions and uses patterns to generate SQL queries from a natural language. Some of the pattern-based NLIDB systems are:

\begin{description}
	\item[DBPal] \citet{utama2018end} developed \texttt{DBPal} as a novel data exploration tool that leverages advances in the deep model to provide a robust platform. This platform uses a pattern-based, deep learning model to translate NLQ into SQL queries. The tool attempts to support users in phrasing questions without knowing query features and database schema. \texttt{DBPal} uses the Paraphrase Database (PPDB) database that contains millions of paraphrases specified in 16 languages and uses this as an index that maps words to paraphrases for any given NL query.
	
	
	\item[NLQ/A] The \texttt{NLQ/A} tool was designed as an NLI interface to query a knowledge graph \citep{zheng2017natural}. This tool was developed for end-users struggling to understand query languages such as SPARQL and SQL. In addition, it conducted experiments over the QALD dataset and showed that the approach was effective as it surpasses previous state-of-the-art tools with regards to recall and precision. 

	\item[FANDA] The \texttt{FANDA} was developed by \citet{liu2019fanda} as a NLI that uses a FollowUp dataset to generate a SQL query. The tool was targeted for novices struggling with SQL query, and intends to allow them write their request in a free-form specified in a natural language. \texttt{FANDA} employs a ranking approach specified in sentence patterns with a weakly supervised learning method to transfer across multiple different domains.
	
\end{description}

\subsubsection{Parsing-based}
In parsing-based systems, the tools parse the input natural language question and use the information structure to generate a query~\citep{affolter2019comparative}. In most cases, they use a dependency parser to handle any glitches. These systems use more advanced heuristics compared to the keyword-based and pattern-based approaches.
\begin{description}
	\item[ATHENA] \citet{saha2016athena} developed an ontolog-driven system called ATHENA that enables users to write queries in natural language, which is then translated into a SQL query. ATHENA uses a two-stage approach where the input NLQ is first translated into an ontology, which is then translated into SQL. With ATHENA, the user is not expected to know how to write a query language such as SQL. The study concluded that ATHENA was used on three different open-source databases, and attained an impressive precision over them. 


	\item[NaLIR] The NALIR system translates a correct English language sentence into a SQL query and evaluates the query against a RDBMS~\citep{li2014constructing}. The system consists of three parts: dependency parser for understanding the NL query linguistically, parse tree node mapper for node identification in the parse tree, and an interactive communicator that explains how the queries are to be processed. 
	
	\item[BioSmart] \citet{jamil2017knowledge} presented the BioSmart tool that uses a syntactic classification that computes natural language sentence into several classes that fits into predefined syntactic templates, which is then interpreted to generate a SQL query. The generative process takes a natural language sentence by its NLP interface and maps this into a predefined sentence or query template. Next, the query mapper transforms the query into a logical query for the structural ontology which identifies the table, analysis tool and generates the query.
	
	\item[BELA] The BELA tool was designed by \citet{walter2012evaluation} as a QA system that processes natural language questions over linked data to generate a SPARQL query. Similarly, this parses the natural language questions and produces a set of query templates before generating the query. When compared to other systems, BELA attempts to reduce computation time and increases its user-friendliness for end-users to write correct SPARQL queries easily.
	
	\item[MaNaLa] \citet{giordani2008mapping} developed \texttt{MaNaLa} as a novel approach that exploits database meta-data and semantically maps natural language into SQL. The study showed that \texttt{MaNaLa} uses a machine learning algorithm that maps a dataset of natural language questions and SQL queries by their syntactic structures. 
\end{description}

\subsubsection{Grammar-based}
Grammar-based systems use a set of rules represented as \textit{grammar} that defines how the natural language questions can be used to generate a SQL query\citep{affolter2019comparative,song2015tr}. This supports end-users who are less knowledgeable about SQL to enable them to write correct queries. This section highlights a few grammar-based systems.

\begin{description}
	\item[Asknow] \citet{dubey2016asknow} developed the \texttt{Asknow} system where users can write their queries in English to a target database engine. The questions are first normalised into syntactic forms, before they are translated into SPARQL queries. In addition, the system is sufficiently adapted to query paraphrasing that enables it to use its grammar for the normalisation of a query which follows a syntactic process. 
	
	
	 \item[GFMed] The \texttt{GFMed} NLI system was proposed by \citet{marginean2017question} for biomedical linked data that applies a grammatical framework (GF)\footnote{https://www.grammaticalframework.org/} that translates natural language queries into SPARQL queries. To generate natural language into SPARQL, \texttt{GFMed} uses GF for syntactic and morphological processes and aggregates the path before using RE operators for recognition. The study highlighted that \texttt{GFMed} can cater for a separate language apart from English for users to ask questions.
	 
	 \item[SQLizer] \citet{yaghmazadeh2017sqlizer} presented \texttt{SQLizer} that uses a CFG for semantic parsing of natural language questions into SQL queries. \texttt{SQLizer} is an end-to-end system, which is fully automated to work with any database without needing additional customisation from the end-user. Due to natural language ambiguities, \texttt{SQLizer} adopts type-directed synthesis and repair techniques to generate a query, which is fully automated and non-database agnostic. In addition, \texttt{SQLizer} performed better when compared to the SQLizer system during an evaluation with three databases such as MAS~\citep{sinha2015overview}, IMDB~\citep{lu2012dataset} and YELP~\citep{huang2014improving}. Similarly, these datasets were used to evaluate NALIR~\citep{li2014nalir}.
     

\item[ln2SQL] The \texttt{ln2SQL} tool was initially designed for another engine called \texttt{fr2SQL} to convert natural language in French into SQL~\citep{couderc2015fr2sql}. In addition, this Python-based tool considers only the SELECT query command to alter a database using the French language. To parse the NL, \texttt{ln2SQL} uses a \textit{treetagger} according to the Parts of Speech (PoS) tagging approach to filter words in a sentence. This treetagger is based on the spaCy NLP framework. The filtered words are then extracted and mapped into keywords which are used to generate a query. The generated query is used to retrieve rows from a table in a database. The study showed that \texttt{ln2SQL} can be used to support multiple databases. 
\end{description}
\subsubsection{Speech-based}
Recently, speech-based NLIDBs have been introduced to ease conversations between humans and db applications without needing to typeset a request in natural language~\citep{serban2016building}. A popular example is the Microsoft's Cortana~\citep{lopez2017alexa,hoy2018alexa} In this section, we present a few of these tools.

\begin{description}
\item[EchoQuery] \citet{lyons2016making} built \texttt{EchoQuery} as a conversational system that uses speech commands to query a database system. The study concluded that \texttt{EchoQuery} was easy and flexible to use. Furthermore, the study was evaluated in \cite{utama2017voice}. The evaluation was conducted using two baselines, such as template-based and rule-based approach to map a semantic tree to a SQL. The result of the evaluation showed that \texttt{EchoQuery} performed at more accurately when compared with two exisiting NLIDBs.

\item[SpeakQL] \cite{chandarana2017speakql} presented an end-to-end speech driven interface used to convert NL into SQL queries. The authors combined four approaches to build the \texttt{SpeakQL} engine. The first of these processes was a state-of-the-art Automated Speech Recognition (ASR) technology that processes spoken SQL query into a transcribed output. Second, a transcribed output is processed to obtain a syntactically correct SQL statement that considers keywords, characters and literals using a CFG. Third, the literals are mapped to attribute names and values, then, a visual output is presented to the user. The study concluded that \texttt{SpeakQL} is friendlier, more interactive and significantly faster than other NLIDBs.

\item[Cyrus] \cite{godinez2018meet} developed a mobile speech assistant for the iOS\footnote{A mobile operating system developed by Apple Inc} platform that supports large query classes on a test database. To parse a natural language into SQL, \texttt{Cyrus} uses a speech recogniser to convert a user's speech into text transcription. Also, it uses a \textit{linguistic} tagger to parse the texts. In addition, it allows students to choose a sample database, perform operations on this database, and produce a result. The study showed that \texttt{Cyrus} was able to map simple NL statements into SQL successfully, and plans to improve its engine to accommodate complex queries.

\end{description}

\section{Formal Language and Automata Applications}\label{3.7}
In this section, we present some of the applications of FLA. First, we highlight some tools that have used REs, then we discuss CFG applications, and finally, we highlight a number of JFA applications.
\subsection{REs Applications}
This section contains some of the tools that have used REs. 

\begin{description}
	\sloppypar
	\item[Information Retrieval] \citet{li2008regular} used REs for an algorithm called \texttt{ReLIE} to retrieve specific information from a corpus. The study compared \texttt{ReLIE} with a machine-learning extractor known as Conditional Random Field (CRF) algorithm. The study showed that ReLIE was more effective at extraction tasks compared to its counterparts.
	\item[Clinical Applications] A software tool by \citet{turchin2006using} was designed to identify and extract blood pressure values from clinical records. The study concluded that REs provide an alternative approach for abstracting data elements in multiple clinical applications if a general purpose NLP software is not available. A similar study used REs to find patterns in genes~\citep{sharmila2018chronological}.
	\item[Computer Science Education] \texttt{NOPRON} was designed by~\citet{ade2014abstracting} to aid the comprehension of novice programs. This tool translates these programs into detailed textual descriptions using REs. 
	\item[Security Applications] \citet{xie2008spamming} designed a framework called \texttt{AutoRE}, that detects spam mail using REs. The study concluded that this approach significantly reduced the false positive rate in the result. The study was extended to a tool named \texttt{BotGraph}~\citep{zhao2009botgraph}, developed to detect botnet spamming targeted at most email providers.
\end{description}

\subsection{CFGs Applications}
CFGs have been applied to many diverse tools in different domains. In this section, we present some of the tools that have used CFGs. 
\begin{description}
	\item[Program Synthesiser] A program synthesiser that uses CFGs to automatically generate procedural programs in Python was developed by \citet{ade2018syntactic}. The study emphasised that CFGs can be extended to automatically generate programs in other procedural programming languages.
	\item[Financial Chat Analyser] A tool developed by \citet{ade2016finchan} that uses CFGs for the automatic comprehension and summarisation of financial chats retrieved from the Instant Bloomberg messaging application.
	\item[Fuzzy System] A tool using CFGs to improve the elicitation of linguistic information in decision making was designed by \citet{rodriguez2016position}. This study showed that CFGs were used to provide a formal approach to the building linguistic expressions in fuzzy systems.
	\item[Lyrics Generator] Designed by \citet{pudaruth2014automated}, this uses CFGs rules and statistical constraints to automatically generate song lyrics. 
	\item[Profile Synthesiser] A tool that uses a variation of CFGs in the automatic generation of hypothetical social media profiles  \citep{ade2017synthesis}. The study concluded that CFGs might be extended to similar problems in the health and social media domains.\\
\end{description}

\subsection{JFA Applications}
A number of studies have considered JFA for natural language abstraction. Here, we present a few of these studies.

\begin{description}
	\item [Automata-like Systems] \citet{cienciala2014towards} introduced Automaton-like P Colonies systems (or APCol systems) as formal methods that include JFA in membrane and distributed systems.
	\item [Game Theory] \citet{maubert2013jumping} proposed the use of JFA for uniform strategies in game theory applications. \citet{bozzelli2015uniform} extended the study for winning conditions and module-checking scenarios using a JFA.
	\item [Frequently Asked Questions] \citet{nnamdijfa2018} proposed the use of JFA for the abstraction of FAQs\footnote{Frequently Asked Questions} in natural language for information retrieval tasks. The study extended this approach into a QA system (chatbot) that allows the comprehension of customer queries.  
	\item [Tweet Comprehension] \citet{stephenjfa2019} presented a tool called an Automata-Aided Tweet Comprehension (ATC) using a JFA for the automatic comprehension of tweets. The study reported that JFA was effective for this task.
\end{description}

\section{The Gap}\label{3.8}
This section highlights the outstanding areas of research, given the background of works that have been investigated. While there are many studies that have proposed tools that aid the comprehension of SQL, there is still a persistent need for new/improved methods to:

\begin{enumerate}
	\item recognise using a formal approach such as REs for SQL queries that may seem confusing for users,
	\item seek to recognise nested SQL queries that exist in different forms cascaded with balanced parentheses using another formal approach using a CFG,
	\item create a tool that implements the synthesis of speech to SQL query,
	\item generate SQL queries using a visualiser that applies visual specifications to build the queries, and
	\item develop an approach to parse natural language into SQL query using a JFA. NLP is an area of AI, which introduces a higher complexity when SQL queries are synthesised from natural language.
\end{enumerate}

\section{Chapter Summary}\label{3.9}
In this chapter, the background related to SQL was presented. First, the challenges faced when learning and writing SQL queries were presented. Second, a number of pedagogical patterns and learning approaches were presented. Next, different state-of-the-art tools for the teaching and learning of SQL were highlighted.  Then, a review of NLP and related areas was conducted. The last section enumerated the gaps that motivated this research. \autoref{part2} introduces two of the methods using \textit{narrations} that were proposed for this research.

\cleardoublepage
\ctparttext{Teaching queries in plain English eliminating syntactic barriers, is the goal of achieving SQL comprehension. This approach to teaching was first proposed by \citet{fincher1999we} and regarded as the SFA. To aid program comprehension, this approach to teaching was termed \emph{narrations} by \citet{ade2014abstracting}. This approach helps to abstract a program without first considering the syntax of the language. Since SQL have less control structures, we have extended this approach of teaching for SQL for the first time, ignoring the syntax that describes unique sets of rules and guidelines of the language. The technique described was used for the automatic generation of explanations for simple queries using REs. In addition, to aid the understanding of nested queries, a CFG was used to recognise nested queries in an attempt to improve comprehension for these types of queries.\\

\noindent This part comprises of two chapters. \autoref{ch:regular} presents a formal technique, using REs to recognise simple SQL queries and describes the tool designed. In \autoref{ch:cfg}, a CFG was described for the recognition of nested queries and a tool was designed for the automatic generation of narrations for nested queries.
}

\part{SQL Comprehension and Narration}\label{part2}
\chapter{Comprehending and Narrating Queries using Regular Expressions }\label{ch:regular} 

\lettrine[lines=3,loversize=0.1]{T}\normalsize{he} previous chapter outlined the background related to this research. This chapter presents a formal approach using \textit{REs} for the recognition of different SQL query constructs. This was built into a tool called \texttt{S-NAR}, that automatically translates the recognised SQL constructs into textual explanations (i.e. narrations). \texttt{S-NAR} was tested with 5000 queries and some performance results were presented.

\section{Introduction}
REs are powerful methods for text processing which have played significant roles for many CS applications~\citep{gogte2016hare,harden2017regular,ganty2019regular}. These applications range from operating systems and search engines to text editors, etc. In many programming languages, REs are inbuilt as standard libraries and present in the syntax of others~\citep{cappers2017exploring}. Newer applications of REs are found in Logstash~\citep{hamilton2018scada}, Elasticsearch~\citep{zamfir2019systems} and Grep~\citep{ganty2019regular} for finding patterns in texts. In this study, we have used REs for the recognition of SQL constructs. The formal definition of REs have been provided in \autoref{ch:definitions}.

\begin{figure*}[h!]
	\centering
	\includegraphics[width=0.9\linewidth]{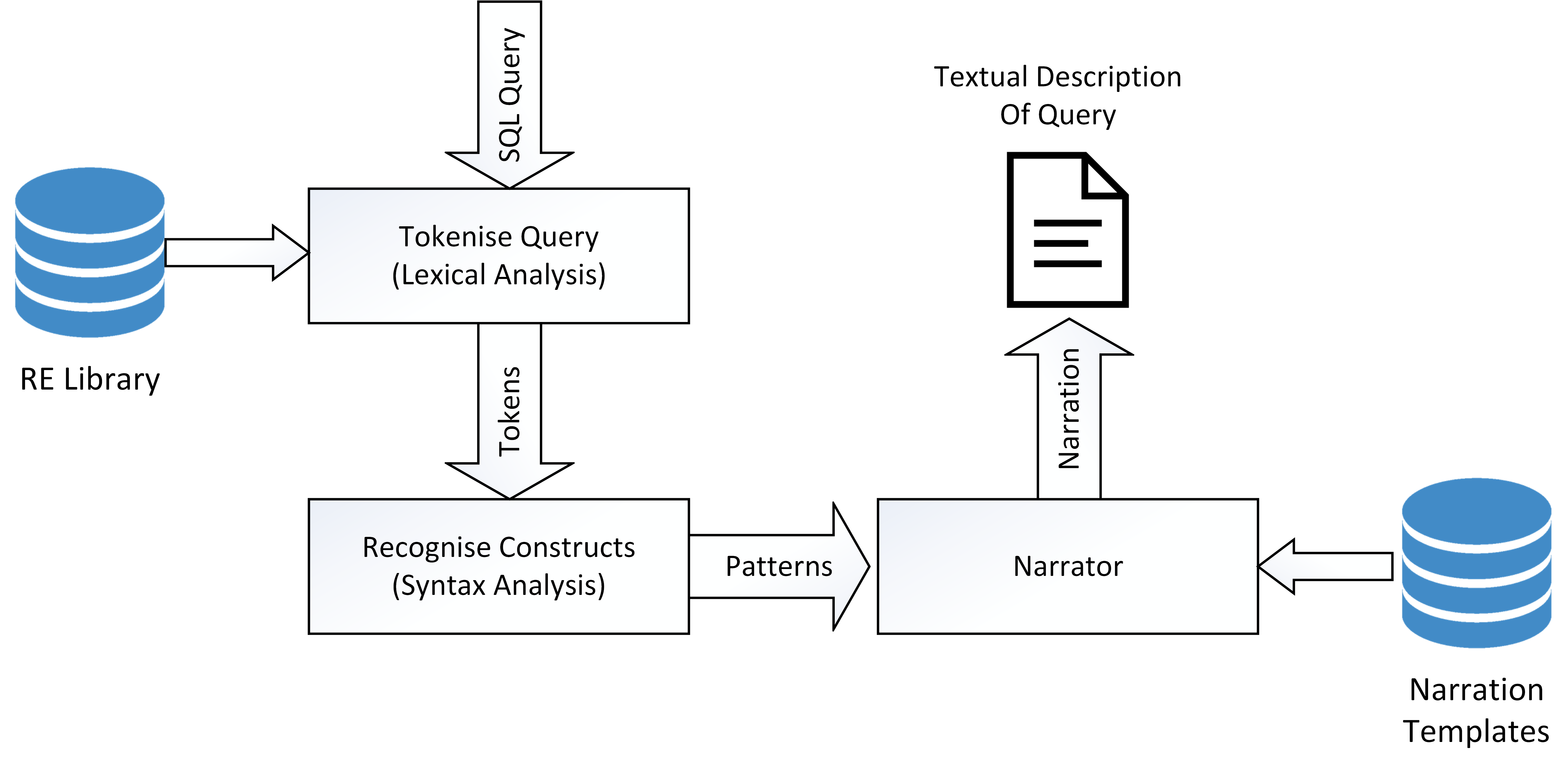}
	\caption{SQL query narration process}
	\label{fig:narration}
\end{figure*}

In this chapter, we propose a way to aid SQL comprehension via the automatic generation of narrations. Narrations are textual descriptions of queries that are presented in plain English, eliminating high level language and syntactic barriers often faced by novices. This style of teaching high-level languages was termed the SFA by \citet{fincher1999we} and has been shown to aid comprehension of novice programs~\citep{ade2014abstracting,ade2016automatic}. In this work, we have extended this approach to SQL queries for the first time and since SQL queries generally have lesser complex control structures (with no loops, nested constructs, etc), a recogniser based on the class of regular languages was used to parse and generate narrations from SQL queries. In \autoref{fig:narration}, we show how this technique works on SQL queries. The query is first tokenised\footnote{Here, \textit{Normalisation} is assumed to be a sub-stage of tokenisation.}. The tokens are grouped into syntactic parts at a recognition stage. The recognised sentential forms are then passed to a narrator. The narrator contains pre-defined templates that convert patterns of query into textual descriptions. Finally, the resulting narration is displayed to the novice/learner.

\section{Translating Queries into Narrations}	
Narrations have been applied in program comprehension \citep{ade2014abstracting,ade2016automatic}, with no application to scripting languages such as SQL. In this section we describe how we have translated SQL queries to narrations. Narrations (as previously generated from programs) are step-wise descriptions of programming instructions and often longer than programs they represent. Narrations are different from comments because they do not provide much semantics as often used in programs --- and are referred to as syntax-free textual algorithms~\citep{ade2014abstracting}. For SQL queries, we present a sample narration. In \autoref{lst:label5create}, we show a simple SQL query for creating a database table with five fields, while \autoref{algo:example} shows its narration. \\

\begin{lstlisting}[language=SQL,escapechar=@, morekeywords={DATABASE, SCHEMA},frame=bt,numbers=none, label={lst:label5create}, caption= SQL query to create a table with five fields] {Create a table with five fields ; StudentID, Lastname, Firstname, Address and City}
CREATE TABLE student_record(
StudentID int,
LastName varchar(255),
FirstName varchar(255),
Address varchar(255),
City varchar(255) 
);
\end{lstlisting}

\begin{algorithm}
	\caption{SQL query to create a table with six fields}
	\label{algo:example}
	\textit{This query creates a table named student\_record and declares StudentID as an integer, Lastname as an alphanumeric entry of at most 255 character, Firstname as an alphanumeric entry of at most 255 characters, Address as an alphanumeric entry of at most 255 characters and City as an alphanumeric entry of at most 255 characters.}
\end{algorithm}

In this work, we have developed a tool that takes SQL queries and generates narrations similar to \autoref{algo:example}. The next section of this work presents REs for the recognition of the syntax of SQL, a stage before narration generation.

\section{Regular Expressions for SQL Abstraction}\label{citeresql}
\begin{figure*}[h!]
	\centering
	\includegraphics[width=1.0\linewidth]{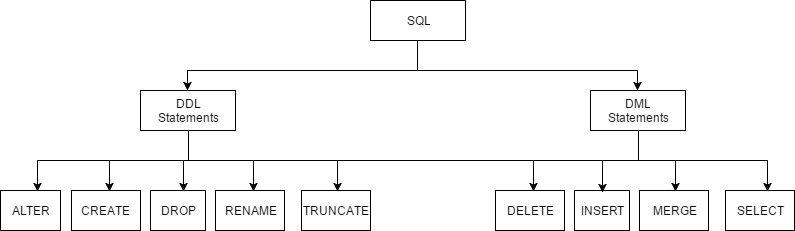}
	\caption{Categories of SQL commands}
	\label{fig:diagram}
\end{figure*}

This section presents REs for recognising the syntactic components of queries, from characters to tokens and the various distinguished commands. First, we introduce a hierarchical diagram in \autoref{fig:diagram} that describes the different categories of SQL statements. This helps us in structuring the granularity level of the REs to be designed to recognise the queries. In \autoref{fig:diagram}, we show how SQL can be broadly broken down into two categories of statements --- DDL and DML. The DDL contains five major types of commands that are used to define/redefine database structures (such as Tables, Views, etc.) in the memory, while the DML is used to perform record-changing operations on the data committed to already existing objects. The relation in \autoref{fig:diagram} is used in structuring the REs used in parsing queries prior to narration composition.\\

In \autoref{tab:lexemes}, we present the basic lexemes in queries (across DML and DDL queries) that form a building block for the syntactic structures in higher levels of granularity. These include: specifications for identifiers (specified in similar fashion as in programming languages), the end of line delimiter, white spaces, commas, brackets, operators and values. Other streams of characters are also presented at this level such as: list of quoted and unquoted string/integer values that are separated by commas. These values often appear in many queries. \\

\begin{table*}[htb!] 
	\scriptsize
	\begin{center}
		\caption{Queries at the Lowest Granularity Level}
		\label{tab:lexemes}
		\begin{tabular}
			{ p{3cm} | p{3cm} |  p{6cm} }
			\hline 
			\textbf{Token} & \textbf{Abbreviation} & \textbf{RE (.Net)} \\
			\hline  
			Identifier &	ident & \texttt{{\raggedright [A-Za-z\_][A-Za-z0-9\_]* }} \\ \hline
			Number &	number & \texttt{{\raggedright [1-9][0-9]* }} \\ \hline
			Semi colon & semi\_colon & \texttt{{\raggedright \textbackslash ;} } \\ \hline
			One or more white spaces & n\_spc & \texttt{{\raggedright \textbackslash s+}} \\ \hline
			Zero or more white spaces & spc & \texttt{{\raggedright \textbackslash s* }} \\ \hline
			All   & all & \texttt{{\raggedright \textbackslash * }} \\ \hline
			Assignment & ass\_sym & \texttt{{\raggedright \textbackslash=}} \\ \hline
			Bracket open & bra\_open & \texttt{{\raggedright \textbackslash ( }} \\ \hline
			Bracket close & bra\_close & \texttt{{\raggedright \textbackslash ) }} \\ \hline
			Comma & comma & \texttt{{\raggedright \textbackslash , }} \\ \hline
			Greater than & greater\_than & \texttt{{\raggedright(\textbackslash>)}} \\ \hline
			Less than & less\_than & \texttt{{\raggedright (\textbackslash <) }} \\ \hline
			Not equal to & not\_equal\_to & \texttt{{\raggedright(\textbackslash!\textbackslash=)}} \\ \hline
			Not greater than & not\_greater\_than & \texttt{{\raggedright(\textbackslash!\textbackslash>)}} \\ \hline
			Not less than & not\_less\_than & \texttt{{\raggedright(\textbackslash!\textbackslash<)}} \\ \hline
			Greater than or equal to & greater\_than\_equal & \texttt{{\raggedright(\textbackslash>\textbackslash=)}} \\ \hline
			Less than or equal to & less\_than\_equal & \texttt{{\raggedright(\textbackslash<\textbackslash=)}} \\ \hline
			Logical operators & log\_op &\texttt{(AND|OR|ANY|LIKE|NOT|BETWEEN|EXISTS)}\\\hline
			Arithmetic operators & ari\_op & \texttt{(\textbackslash+|\textbackslash-|\textbackslash*|\textbackslash /|\textbackslash \%)} \\ \hline
			Comparison operators & comp\_op &\texttt{(\textbackslash=|\textbackslash!\textbackslash=|\textbackslash<|\textbackslash>|\textbackslash!\textbackslash<|\textbackslash!\textbackslash>|\textbackslash<\textbackslash=|\textbackslash>\textbackslash=)} \\ \hline
			Integer value & int\_val&\texttt{(\textbackslash-?\textbackslash d+)[+-]?} \\ \hline	
			Varchar value& varchar\_val& \texttt{(A-Za-z\_)*} \\ \hline	
			Boolean value& bool\_val& \texttt{(true|false)} \\ \hline			
			Float value& flot\_val&\texttt{(\textbackslash d\textbackslash d?\textbackslash.\textbackslash d\textbackslash d?)} \\ \hline
			Data type & datatype & \texttt{(int|varchar|bool|float)} \\ \hline
			Ident separated by comma value & ident\_sep\_by\_comma & \texttt{((bra\_open)(ident)(spc)(comma)(spc))*
			((bra\_open)(ident)(bra\_close))} \\ \hline					
			Value in quote & val\_in\_quote &\texttt{((\textbackslash')(bra\_open)(ident)|(number) | (flot\_val)|(n\_spc)| (comma)(bra\_close))+ (\textbackslash'))}\\ \hline
			List of values in quote & list\_of\_vals\_in\_quote & \texttt{(bra\_open)(val\_in\_quote) (spc)(comma)(spc)(bra\_close)*
			(val\_in\_quote))}\\ \hline
			List of values separated by comma & list\_vals\_sep\_comma & \texttt{(bra\_open)(spc)}\texttt{(list\_of\_vals\_in\_quote)
			(spc)(bra\_close)} \\ \hline
			Ident equals value & ident\_equal\_val & \texttt{(ident)(comp\_op)(val\_in\_quote)(comma)
			*(ident)(comp\_op)(val\_in\_quote)})	\\ \hline
		\end{tabular}	
	\end{center}
\end{table*}
We proceed and present REs for DDL statements in \autoref{tab:sqlddl}. The ALTER construct allows for renaming a table or options to add, drop or modify columns of the table. Hence, we have two expressions in \autoref{tab:sqlddl}. The DROP RE recognises statements for deleting a database or a table. The RENAME, TRUNCATE and CREATE REs are similarly specified. The REs for recognising DML constructs are shown in \autoref{table:dml}, together with the different variations of the SELECT command.

\begin{table*}[!htb]
	\scriptsize
	\begin{center}
		\caption{DDL statement building blocks}\label{tab:sqlddl}
		\begin{tabular}
			{ p{3cm} | p{3cm} |  p{6cm} }
			\hline
			{\textbf{Statement}} & {\textbf{Abbreviation}} & {\textbf{RE (.Net)}} \\
			\hline
			ALTER & alter & \parbox{270pt}{\texttt{(ALTER)(\_n\_spc)(TABLE)(n\_spc)(ident)\\ 
			(n\_spc)(RENAME)(n\_spc)(TO)(n\_spc)\\
			(ident)(semi\_colon)\\
			(ALTER)(n\_spc)(ident)(n\_spc)\\
			(ADD|DROP|MODIFY)(n\_spc)(COLUMN)\\
			(n\_spc)(ident)(n\_spc)\\
			(datatype)(semi\_colon)}}\\ \hline
			DROP & drop & \texttt{(DROP)(n\_spc)(DATABASE|TABLE)(n\_spc)(IF)
			(n\_spc)(EXISTS)(n\_spc) 
			(ident\_sep\_by\_comma)
			(semi\_colon)} \\ \hline
			RENAME & rename & \parbox{270pt}{\texttt{(RENAME)(n\_spc)(TABLE)(n\_spc)(ident)\\
			(n\_spc)(TO)(n\_spc)(ident)(semi-colon)}} \\ \hline
			TRUNCATE & truncate & \texttt{(TRUNCATE)(n\_spc)(TABLE)(n\_spc)(ident)
			(semi-colon)} \\ \hline	
			CREATE & create & \texttt{(CREATE)(n\_spc)(DATABASE|TABLE)(n\_spc)
			(ident)(bra\_open)(ident\_sep\_by\_comma)
			(datatype)(semi\_colon)}\\ \hline			
		\end{tabular}
	\end{center}
\end{table*}

\begin{table*}[!htb]
	\scriptsize
	\begin{center}
		\caption{DML statement building blocks}	\label{table:dml}
		\begin{tabular}
			{ p{3cm} | p{3cm} |  p{6cm} }
			\hline
			{\textbf{Statement}} & {\textbf{Abbreviation}} & {\textbf{RE (.Net)}} \\
			\hline
			DELETE & delete & \texttt{\raggedright (DELETE)(n\_spc)(FROM)(n\_spc)(ident)(n\_spc)
			(WHERE)(n\_spc)(ident)(comp\_op)
			(val\_in\_quote)(semi\_colon)} \\ \hline
			
			INSERT & insert & \parbox{270pt}{\texttt{\raggedright (INSERT)(n\_spc)(INTO)(n\_spc)(ident)\\
			(n\_spc)(bra\_open)(n\_spc)\\
			(ident\_sep\_by\_comma)(spc)\\
			(VALUES)(spc)\\
			(list\_of\_values\_sep\_by\_comma)(semi\_colon)}}\\ \hline
			
			SELECT & select & \texttt{{\raggedright (SELECT)(n\_spc)(ident\_sep\_by\_comma|all)
			(n\_spc)(FROM)(n\_spc)
			(ident)(semi\_colon)}} \\ \hline
			
			SELECT DISTINCT & distinct & \texttt{{\raggedright (SELECT)(n\_spc)(DISTINCT)(n\_spc)\\
			(ident\_sep\_by\_comma)(n\_spc)(FROM)\\
			(n\_spc)(ident)(semi\_colon)}} \\ \hline
			
			SELECT WHERE &where & \parbox{270pt}{\texttt{{\raggedright (SELECT)(n\_spc)(ident\_sep\_by\_comma|all)
			(n\_spc)(FROM)(n\_spc)(ident)\\
			(n\_spc)(WHERE)(n\_spc)(ident)\\
			(comp\_op)((\textbackslash')(ident)(\textbackslash')|\\
			(number))(n\_spc)(AND|OR)(n\_spc)(ident)\\
			(comp\_op)((\textbackslash')(ident)(\textbackslash')|\\
			(number))(semi\_colon)}}} \\ \hline
			
			SELECT WHERE\_IN & where\_in & \parbox{270pt}{\texttt{{\raggedright(SELECT)(n\_spc)(ident\_sep\_by\_comma|all)\\
			(n\_spc)(FROM)(n\_spc)(ident)(n\_spc)(WHERE)\\
			(n\_spc)(ident)(n\_spc)(IN)\\
			(n\_spc)(list\_of\_values\_sep\_by\_comma)\\
			(semi\_colon)}}} \\ \hline
			
			SELECT WHERE\_BETWEEN & where\_between & \parbox{270pt}{\texttt{{\raggedright(SELECT)(n\_spc)(ident\_sep\_by\_comma|all)\\
			(n\_spc)(FROM)(n\_spc)(ident)(n\_spc)(WHERE)\\
			(n\_spc)(ident)(n\_spc)(BETWEEN)(n\_spc)\\
			((\textbackslash')(ident)(\textbackslash')|\\
			(number))(AND)(n\_spc)((\textbackslash')\\
			(ident)(\textbackslash')|(number))\\
			(semi\_colon)}}} \\ \hline
			
			SELECT WHERE\_LIKE & where\_like & \parbox{270pt}{\texttt{{\raggedright(SELECT)(n\_spc)(ident\_sep\_by\_comma|all)\\
			(n\_spc)(FROM)(n\_spc)(ident)(n\_spc)\\
			(WHERE)(n\_spc)(ident)(n\_spc)(LIKE)\\
			(n\_spc)((\textbackslash')(ident)\\
			(\textbackslash')|(number))(semi\_colon)}}} \\ \hline
			
			SELECT ORDER & order\_by & \parbox{270pt}{\texttt{{\raggedright (SELECT)(n\_spc)(ident\_sep\_by\_comma|all)\\
			(n\_spc)(FROM)(n\_spc)(ident)(n\_spc)\\
			(ORDER BY)(n\_spc)(ident)(n\_spc)\\
			(ASC|DESC)(semi\_colon)}}} \\ \hline
			
			SELECT GROUPBY & group\_by & \parbox{270pt}{\texttt{{\raggedright (SELECT)(n\_spc)(ident\_sep\_by\_comma|all)\\
			(n\_spc)(FROM)(n\_spc)(ident)(n\_spc)\\
			(GROUP BY)(n\_spc)(ident)(semi\_colon)}}} \\ \hline
			
			SELECT HAVINGCOUNT & having\_count & \parbox{270pt}{\texttt{{\raggedright (SELECT)(n\_spc)(COUNT)(ident\_sep\_by\_comma)\\
			(n\_spc)(FROM)(n\_spc)(ident)(n\_spc)\\
			(GROUP BY)(n\_spc)(ident)(n\_spc)\\
			(HAVING COUNT)(n\_spc)(ident)(comp\_op)((\textbackslash')\\
			(ident)(\textbackslash')|(number))\\
			(semi\_colon)}}} \\ \hline
			
		\end{tabular}
		
	\end{center}
\end{table*}
\section{Introducing S-NAR}
\begin{figure*}[h!]
	\centering
	\includegraphics[width=420px]{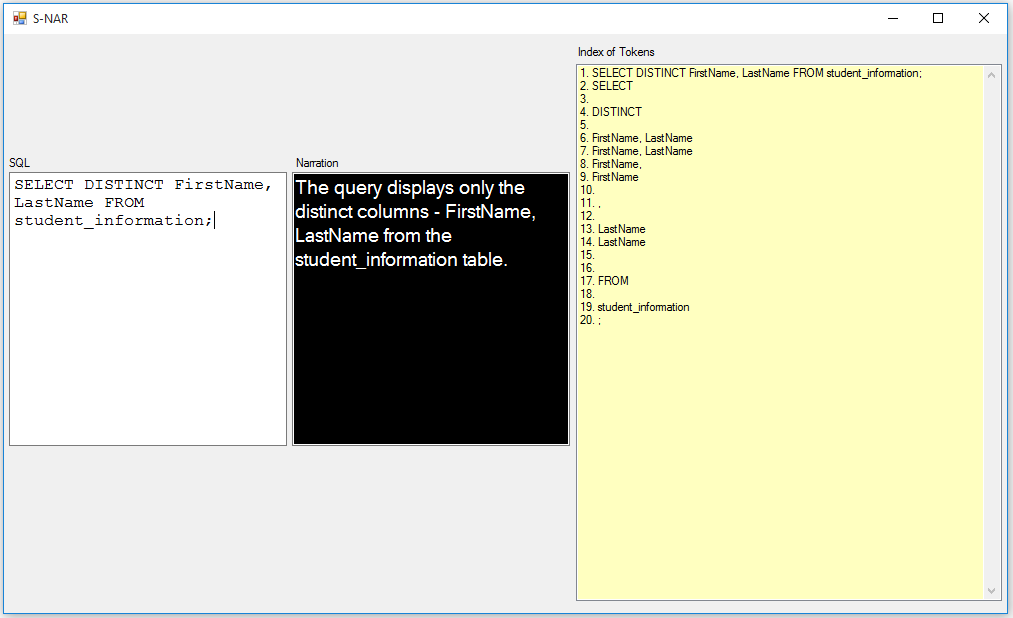}
	\caption{Screenshot of S-NAR during the narration of a SELECT query}
	\label{fig:select}
\end{figure*}

We have implemented \texttt{S-NAR} as a desktop application (\autoref{fig:select}) using the regular expression library provided in the .NET framework~\citep{Microsoftregex}. We tested \texttt{S-NAR} with a dataset of 5000 SQL queries. \texttt{S-NAR} successfully narrated 4824 queries presented to it (about 96.48\%). We noted that all failed instances were queries that had balanced parenthesis in them. This is because the language of balanced parenthesis is \textit{nonregular}, hence, REs did not suffice in those instances. A parser based on a CFG will sufficient to handle this hitch. The remainder of this section shows and discusses some results from \texttt{S-NAR}. \\

\subsection{Implementation and Results}
In this section, we present results obtained from the narration of some queries during the testing stage of \texttt{S-NAR}. \\

\subsection{DDL Statements}
\texttt{S-NAR} was successful in narrating DDL statements. \autoref{lst:label1} shows an ALTER statement and the generated narration for this statement is shown in \autoref{alter}. \texttt{S-NAR} removes all the technical terms such as DROP, MODIFY, etc., and presents an intuitive summary. English words that are usually intuitive, such as ALTER, are left in the narration generated by \texttt{S-NAR}. \\

\begin{lstlisting}[language=SQL,escapechar=@, morekeywords={DATABASE, SCHEMA, DROP, MODIFY},frame=bt, numbers=none, label={lst:label1}, caption={ALTER statement query}]
ALTER TABLE supplier ADD COLUMN 
supplier_name varchar(255);

ALTER TABLE supplier DROP COLUMN 
supplier_name varchar(255);

ALTER TABLE supplier MODIFY COLUMN 
supplier_name varchar(255);
\end{lstlisting}

\begin{algorithm}
	\caption{Narration of the ALTER statement}\label{alter}
	\begin{algorithmic}[1]
		
		\State \emph{This query alters the supplier table by adding a new column called supplier\_name that allows alphanumeric entry with at most 255 characters}
		
		\State 	\emph{This query alters the supplier table by removing a new column called supplier\_name that allows alphanumeric entry with at most 255 characters}
		
		\State 	\emph{This query alters the supplier table by modifying a new column called supplier\_name that allows alphanumeric entry with at most 255 characters}		
	\end{algorithmic}
\end{algorithm}

The CREATE statement in \autoref{lst:createdd} is narrated in \autoref{create}. In this case, \texttt{S-NAR} abstracted the \texttt{varchar} data type, and referred to it as \textit{alphanumeric} --- in order to aid comprehension. \autoref{lst:drop} is a DROP statement that is narrated in \autoref{drop} --- here the DROP keyword is abstracted as ERASE for ease of understanding. \\

\begin{lstlisting}[language=SQL,escapechar=@, morekeywords={DATABASE, SCHEMA, DROP, MODIFY},frame=bt, numbers=none, label={lst:createdd}, caption={CREATE statement query }]
CREATE DATABASE student_db;
CREATE TABLE student_information (
ID int,
Firstname varchar(255),
Lastname varchar(255),
Gender varchar(50),
Address varchar(255) 
);
\end{lstlisting}

\begin{algorithm}
	\caption{Narration of the CREATE statement}\label{create}
	\begin{algorithmic}[1]
		\State \textit{This query creates a database named student\_db}
		\State \textit{This query creates a table named student\_information, and declares ID as an integer, Firstname as an alphanumeric entry of almost 255 character, Lastname as an alphanumeric entry of at most 255 characters, Gender as an alphanumeric entry of at most 255 character, Address as an alphanumeric entry of at most 255 characters.}	
	\end{algorithmic}
\end{algorithm}


\begin{lstlisting}[language=SQL,escapechar=@, morekeywords={DATABASE, SCHEMA, DROP, MODIFY},frame=bt, numbers=none, label={lst:drop}, caption={DROP statement query }]
DROP DATABASE IF EXISTS student_information;
\end{lstlisting}

\begin{algorithm}
	\caption{Narration of the DROP statement}\label{drop}
	\begin{algorithmic}[1]
		\State \textit{This query erases the student\_information database from the computer memory given that it previously exists}	
	\end{algorithmic}
\end{algorithm}

\autoref{lst:rename} and \autoref{lst:truncate} describe two queries for renaming and truncating a table. These queries are narrated in \autoref{algol:rename} and \autoref{algo:truncate} respectively. 

\begin{lstlisting}[language=SQL,escapechar=@, morekeywords={DATABASE, SCHEMA, DROP, MODIFY, RENAME,TO},frame=bt, numbers=none, label={lst:rename}, caption={RENAME statement query }]
RENAME TABLE student_record TO student_information;
\end{lstlisting}

\begin{algorithm}
	\caption{Narration of the RENAME statement}\label{algol:rename}
	\begin{algorithmic}[1]
		\State \textit{This query renames the student\_record table to student\_information}	
	\end{algorithmic}
\end{algorithm}

\begin{lstlisting}[language=SQL,escapechar=@, morekeywords={TRUNCATE},frame=bt, numbers=none, label={lst:truncate}, caption={TRUNCATE statement query }]
TRUNCATE TABLE student_information;
\end{lstlisting}

\begin{algorithm}
	\caption{Narration of the TRUNCATE statement}\label{algo:truncate}
	\begin{algorithmic}[1]
		\State \textit{This query empties the contents from the student\_information table}	
	\end{algorithmic}
\end{algorithm}

\subsection{DML Statements}
\texttt{S-NAR} was successful in narrating DML statements. \autoref{lst:label6} shows the DELETE statement and its narrations. Here, the equals sign is abstracted to ``is'' as depicted in \autoref{algo:Deleted}. The INSERT statement in \autoref{lst:label7} is narrated in \autoref{insert} with the keyword INSERT abstracted as ADD. The SELECT statement follows a similar narration pattern with its different variations (sometimes, having optional WHERE clause, DISTINCT and COUNT keywords, etc.) shown in \autoref{select}.\\

\begin{lstlisting}[language=SQL,escapechar=@, morekeywords={TRUNCATE},frame=bt, numbers=none, label={lst:label6}, caption={DELETE statement query }]
DELETE FROM student_information WHERE 
student_firstname='peter';
\end{lstlisting}

\begin{algorithm}
	\caption{Narration of the DELETE statement}\label{algo:Deleted}
	\begin{algorithmic}[1]
		\State \textit{This query removes from the student\_information table where the student\_firstname is peter}	
	\end{algorithmic}
\end{algorithm}

\vspace{0.45in}
\begin{lstlisting}[language=SQL,escapechar=@, morekeywords={TRUNCATE},frame=bt, numbers=none, label={lst:label7}, caption={INSERT statement query }]
INSERT INTO student_information (FirstName, 
LastName, Address, City, PostalCode, Country)
VALUES ('Peter', 
'Tom', 
'21 claim street', 
'Rivonia', 
'2001', 
'South Africa');
\end{lstlisting}

\begin{algorithm}
	\caption{Narration of the INSERT statement}\label{insert}
	\begin{algorithmic}[1]
		\State \textit{This query adds into the student\_information table into columns; FirstName, LastName, Address, City, PostalCode, Country with details; Peter, Tom, 21 claim street, Rivonia, 2001, South Africa}	
	\end{algorithmic}
\end{algorithm}

\begin{lstlisting}[language=SQL,escapechar=@, morekeywords={TRUNCATE},frame=bt, numbers=none, label={lst:label8}, caption={SELECT statement query }]
SELECT * FROM student_information;

SELECT DISTINCT FirstName, LastName 
FROM student_information; 

SELECT * FROM student_information
WHERE FirstName='peter' AND LastName='mark';

SELECT * FROM Customers 
WHERE Country='South Africa' 
OR City = 'Harare';

SELECT * FROM student_information
WHERE FirstName IN ('peter', 'john', 'felix');
\end{lstlisting}

\begin{algorithm}
	\caption{Narration of the SELECT statement}\label{select}
	\begin{algorithmic}[1]
		\State \textit{The query displays all information from the student\_information table}	
		\State \textit{The query displays only the distinct column - FirstName and LastName information from the student\_information table}
		\State \textit{The query displays all the details from the student\_information table where the FirstName is 'peter', and the LastName is 'mark'}
		\State \textit{The query displays all the information from the Customers table where the Country is 'South Africa', or City is Harare}
		\State \textit{The query displays all the information from the student\_information table where the FirstName are either 'peter','john','felix'}
	\end{algorithmic}
\end{algorithm}

\section{Scope and Limitations}
Up to this point, we have presented a tool called \texttt{S-NAR} that uses REs for recognising the constructs of simple DML and DDL queries, some of which had the \texttt{WHERE} clauses. Generating narrations for them was a relatively trivial task. One major limitation to note is that SQL queries sometimes come in more complex forms. For instance, it is possible to have queries and sub-queries, nested to several depths, and cascaded with \textit{balanced parentheses}\footnote{The language of balanced parentheses is not a regular language.} as shown in \autoref{lst:complex_query}. \\

Here, there are two nested \texttt{SELECT} statements. The first one is a simple one that draws values from the Journals table, with the second part specifying the filter. Observe there are two pairs of parentheses in this query. Recognising this language will require more than REs; i.e. a CFG for balanced parentheses can be used for this. This is similar to many programming languages, where REs are only useful for lexical analysis and not building parsers. 

\begin{lstlisting}[language=SQL,escapechar=@, morekeywords={TRUNCATE},frame=bt, numbers=none, label={lst:complex_query}, caption={A non-regular nested SQL query}]
SELECT numcount,Jnls.* 
FROM Journals J 
WHERE numcount<=(SELECT COUNT(*/2) 
                 FROM Journals);
\end{lstlisting}

\section{Chapter Summary}
In this chapter, we have presented \texttt{S-NAR}, a software tool that translates SQL queries into textual description of the implied operations. \texttt{S-NAR} uses REs to first extract and group the tokens of an SQL query into syntactic categories and passes these grouped tokens to a module that uses predefined narration templates for the automatic generation of textual descriptions, referred to as narrations. \texttt{S-NAR} was also tested on 5000 queries scrapped from the Internet and it narrated a subset of these queries (96\%) that do not contain balanced parentheses. This is only recognisable with CFGs or higher classes of formal abstract machines. We have argued that the generated narrations can aid the comprehension of SQL or be used to support teaching, in line with the SFA to programming language pedagogy. \\

\autoref{ch:cfg} presents a CFG used to recognise nested queries cascaded with balanced parentheses. This handles the hitch faced with the current implementation of \texttt{S-NAR}. 
\chapter{Comprehending and Narrating Queries using Context-free Grammars}\label{ch:cfg} 
\lettrine[lines=3,loversize=0.1]{I}\normalsize{n} the previous chapter, a formal approach using REs was used for the recognition of SQL query constructs. A major limitation was identified, where REs were unable to recognise nested SQL queries. This chapter presents an extension to the use of \textit{narrations} for describing nested queries cascaded with balanced parentheses using a \textit{CFG}.

\section{Introduction}
CFGs are more powerful than regular languages (or REs) and have been used to describe nested parenthesis for programming languages~\citep{cereda2017instrumenting,bastani2017synthesizing}. Hence, CFGs are formal notations used for expressing recursive definitions in programming languages. Since nested queries are defined recursively, this work extends the recognition of nested SQL queries using a set of rewriting rules (or production rules). To our knowledge, this appears to be the first time such an approach will be extended to recognise nested SQL queries. The formal definition of CFGs has been provided in \autoref{ch:definitions}.

\begin{figure*}[h]
	\centering
	\includegraphics[width=400px]{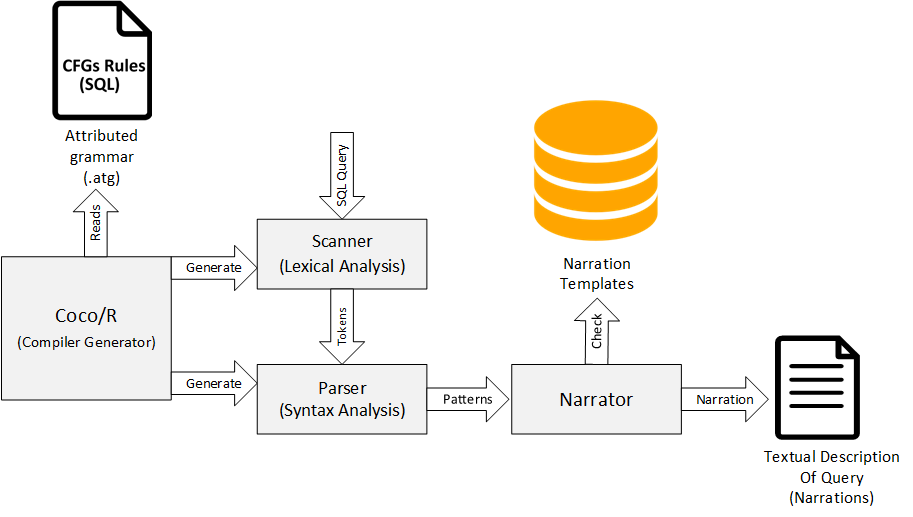}
	\caption{The framework of the SQL Narrator}
	\label{fig:sqlnarr}
\end{figure*}

This chapter presents an SQL Narrator that automatically generates narrations for a nested query. The term ``narrations'' was first coined by \citet{ade2016automatic} and was described as a textual description of programs in plain English and has been shown to aid the comprehension of novice programs \citep{ade2014abstracting,ade2017new}. The authors in \citep{ade2017s} extended the use of narrations for describing simple queries using predefined templates. This was presented in \autoref{ch:regular}. This chapter presents an improved version of the study that describes nested queries with balanced parentheses using CFGs -- a subset of irregular languages to parse and generate narrations from SQL queries. \autoref{fig:sqlnarr} shows how this approach generates a narration for a nested SQL query. First, the query is tokenised and grouped into a syntactic form for recognition, using Coco/R\footnote{http://www.ssw.uni-linz.ac.at/Coco/} that generates a scanner and parser. The sentential forms, derived as patterns, are then available to a narrator that checks these patterns for matches before converting the query into textual descriptions (or narrations). The generated narration is then presented to the learner.

\section{CFG Design for SQL Queries}
This section presents the design of CFGs for the recognition of nested queries. This approach was used in the automatic generation of narrations for SQL queries. In this aspect, CFGs were designed using the compiler generator (Coco/R) that takes an attributed grammar, uses this grammar to generate a scanner and a parser. This is indexed in \autoref{ch:app-cfg-formalism}. These generated elements (parser and scanner) were used to verify syntactic correctness of a query. To design the grammar, we adopt the Ron Savage’s EBNF (Extended Backus–Naur Form) SQL grammar as described in \citep{SQLgrammar}.

\subsection{Building Blocks}
To design the grammar G, we present the lexemes which are sequences of characters matched by a pattern for tokens used to build the production rules, P. These result in a set of terminal symbols $\Sigma$, such as letters, digits, num, etc (Productions \ref{prod:letter} to \ref{prod:log_opr}). In Production \ref{prod:letter} to \ref{prod:ident}, letters are defined and will appear in the list of identifiers (for example, student, lab01, etc). Productions \ref{prod:num} and \ref{prod:digit} show numerical values that may appear. Productions \ref{prod:semi_c} to \ref{prod:period} present semicolon, comma, open and close brackets, open and close quotes and period. Production \ref{prod:all} shows the symbol “all” which is used to display the entire information in a table/database. Production \ref{prod:wspace} shows whitespaces between strings.
 
\setcounter{equation}{0}
\begin{eqnarray}
\texttt{<letter>}^* &\longrightarrow & \texttt{<letter>}(\texttt{<letter>})^* \label{prod:letter}\\
\texttt{<letter>} &\longrightarrow & \texttt{A~|\ldots|~Z| a~|\ldots|~z} \label{prod:letter2}\\
\texttt{<ident>} &\longrightarrow & \texttt{<letter>}^*\texttt{<num>}\label{prod:ident} \\
\texttt{<num>} &\longrightarrow & \texttt{<digit>}\texttt{<digit>}^*\label{prod:num} \\
\texttt{<digit>} &\longrightarrow & \texttt{0~|\ldots|~9} \label{prod:digit} \\
\texttt{<semi\_c>} &\longrightarrow & \texttt{;} \label{prod:semi_c}\\
\texttt{<comma>} &\longrightarrow & \texttt{,} \label{prod:comma}\\
\texttt{<brac\_open>} &\longrightarrow & \texttt{(} \label{prod:braco}\\
\texttt{<brac\_close>} &\longrightarrow & \texttt{)}\label{prod:bracc}\\
\texttt{<open\_q>} &\longrightarrow & \texttt{`} \label{prod:openq}\\
\texttt{<close\_q>} &\longrightarrow & \texttt{'} \label{prod:closeq}\\
\texttt{<period>} &\longrightarrow & \texttt{.} \label{prod:period}\\
\texttt{<all>} &\longrightarrow & ^\texttt{*} \label{prod:all}
\end{eqnarray}

Productions \ref{prod:type} to \ref{prod:bool} show the supported data type. Productions \ref{prod:arith_opr} to \ref{prod:log_opr} present the operators such as arithmetic, relational and logical.

\begin{eqnarray}
\texttt{<wspace>} &\longrightarrow & \texttt{ws} \label{prod:wspace}\\
\texttt{<type>} &\longrightarrow & \texttt{int~|~varchar~|~boolean~|~float} \label{prod:type}\\
\texttt{<boolean>} &\longrightarrow & \texttt{true~|~false} \label{prod:bool}\\
\texttt{<float\_v>} &\longrightarrow & \texttt{<digit><digit><period><digit><digit>} \label{prod:floatvalue}\\
\texttt{<arith\_opr>} &\longrightarrow & \texttt{+~|~-~|~*~|~/~|~\%} \label{prod:arith_opr}\\
\texttt{<rel\_opr>} &\longrightarrow & \texttt{<~|~>~|~=~|~!=~|~!<~|~!>~|~>=~|~<=} \label{prod:rel_opr}\\
\texttt{<log\_opr>} &\longrightarrow & \texttt{OR~|~XOR~|~AND~|~ANY~|~LIKE~|~NOT~|~EXISTS~|}\nonumber\\
&&\texttt{BETWEEN~|~IN~|~IS~NULL~|~UNIQUE} \label{prod:log_opr}
\end{eqnarray}

We proceed to build productions for terms, identifiers and expressions used in Productions \ref{prod:identsepcomma} to \ref{prod:identcondtval}.  Production \ref{prod:valsinquote} is defined recursively allowing the occurrence of either identifiers, numbers or decimal values.

\begin{eqnarray}
\texttt{<ident\_sep\_by\_comma>} &\longrightarrow & \texttt{<ident\_sep\_by\_comma>}(\texttt{<ident>}\texttt{<comma>})^*\nonumber\\
&&\texttt{<ident>~|~<all>} ~\label{prod:identsepcomma}\\
\texttt{<vals\_in\_qts>} &\longrightarrow & \texttt{<open\_q>}\texttt{(<vals\_in\_qts><ident>|} \nonumber\\
&&\texttt{<vals\_in\_qts><num>|<vals\_in\_qts>}\nonumber\\
&&\texttt{<float\_v>)}^*\texttt{<close\_q>}\label{prod:valsinquote}\\
\texttt{<vals\_list\_qts>} &\longrightarrow & \texttt{<brac\_open>}\texttt{<vals\_in\_qts>}\nonumber\\
&&\texttt{<brac\_close>}\label{prod:listofvalsinquote} \\
\texttt{<vals\_list\_sep\_by\_comma>} &\longrightarrow & \texttt{<brac\_open>}\texttt{<vals\_list\_qts>}\nonumber\\
&&\texttt{<brac\_close>}\label{prod:listofvalssepbycomma} \\
\texttt{<ident\_condt\_val>} &\longrightarrow & \texttt{<ident><rel\_opr><vals\_in\_qts>}\label{prod:identcondtval} 
\end{eqnarray}

We have presented the productions for the elements of our grammar as specified from Production \ref{prod:letter} to Production \ref{prod:identcondtval}. Next, we provide productions for the SELECT statement (\texttt{<select\_sub>}) as seen in Production \ref{prod:selectstm} and Production \ref{prod:condtstmt}. The \texttt{<select\_sub>} symbol satisfies the \texttt{SELECT…WHERE…FROM} statement and will be used to build the nested queries.

\begin{eqnarray}
\texttt{<select\_stm>} &\longrightarrow & \texttt{<select\_stm>}\texttt{<select\_sub>}\texttt{<semi\_c>}\label{prod:selectstm}\\
\texttt{<select\_sub>} &\longrightarrow & \texttt{SELECT}\texttt{<wspace>}\texttt{<cols\_list>}\texttt{<wspace>}\texttt{FROM}\texttt{<wspace>}\nonumber\\
&&\texttt{<ident>}\texttt{<wspace>}\texttt{WHERE}\texttt{<wspace>}\texttt{<condt\_stmt>}\label{prod:selectsub}\\
\texttt{<cols\_list>} &\longrightarrow & \texttt{<cols\_list>}\texttt{DISTINCT}\texttt{<wspace>}\texttt{<ident>}~|~\nonumber\\
&&\texttt{<cols\_list>}\texttt{<ident\_sep\_by\_comma>}~|~\nonumber\\
&&\texttt{<cols\_list>}\texttt{<ident>}~|~\nonumber\\
&&\texttt{<cols\_list>}\texttt{<all>}\label{prod:colslist}\\
\texttt{<condt\_stmt>} &\longrightarrow & \texttt{<ident\_condt\_val>}\label{prod:condtstmt}
\end{eqnarray}

\subsection{Nested SQL Queries}
A nested query is essentially a query inside another (\textbf{inner} and \textbf{outer}) query, which is common with the \texttt{SELECT} statements \citep{elhemali2007execution}. These statements are embedded within the \texttt{WHERE} or \texttt{HAVING} clause. In this section, we describe the productions for subqueries in the \texttt{UPDATE}, \texttt{DELETE}, \texttt{INSERT} and \texttt{SELECT} statements.\\

We start by defining the production for subqueries in the \texttt{UPDATE} statement. Production \ref{prod:update_sub_st} describes the \texttt{UPDATE} subquery statement. The symbol \texttt{<select\_sub>} allows the use of the \texttt{SELECT} query within the \texttt{UPDATE} statement to form a subquery. In most cases, this appears within the \texttt{IN} clause.

\begin{eqnarray}
\texttt{<update\_sbqy>} &\longrightarrow & \texttt{UPDATE}\texttt{<wspace>}\texttt{<ident>}\texttt{<wspace>}\texttt{SET}\texttt{<wspace>}\nonumber\\
&&\texttt{<ident>}\texttt{<rel\_opr>}\texttt{<num>}\texttt{<wspace>}\texttt{WHERE}\texttt{<wspace>}\nonumber\\
&&\texttt{<ident>}\texttt{<wspace>}\texttt{IN}\texttt{<brac\_open>}\texttt{<select\_sub>}\nonumber\\
&&\texttt{<brac\_close>}\texttt{<semi\_c>}\label{prod:update_sub_st}
\end{eqnarray}

Production \ref{prod:delete_sub_st} describes the \texttt{DELETE} subquery statement. This includes the \texttt{<select\_sub>} symbol used to present the \texttt{DELETE} subquery statement.

\begin{eqnarray}
\texttt{<delete\_sbqy>} &\longrightarrow & \texttt{DELETE}\texttt{<wspace>}\texttt{FROM}\texttt{<wspace><ident><wspace>WHERE}\nonumber\\
&&\texttt{<wspace><ident><wspace>}\texttt{(<rel\_opr>|<log\_opr>)}\nonumber\\
&&\texttt{<brac\_open><select\_sub><brac\_close><semi\_c>}\label{prod:delete_sub_st}
\end{eqnarray}

Production \ref{prod:insert_sub_st} defines the \texttt{INSERT} subquery statement. In this statement, the \texttt{<select\_sub>} symbol is used within the \texttt{INSERT} query to describe the subquery.

\begin{eqnarray}
\texttt{<insert\_sbqy>} &\longrightarrow & \texttt{INSERT}\texttt{<wspace>}\texttt{INTO}\texttt{<wspace><ident><brac\_open>}\nonumber\\
&&\texttt{(<ident\_sep\_by\_comma>|<ident>)<brac\_close>}\nonumber\\
&&\texttt{<brac\_open><select\_sub><brac\_close><semi\_c>}\label{prod:insert_sub_st}
\end{eqnarray}

In Production \ref{prod:select_sub_st}, the \texttt{SELECT} subquery is described. In this subquery, the \texttt{<select\_sub>} symbol is used within the \texttt{WHERE} clause.

\begin{eqnarray}
\texttt{<select\_sbqy>} &\longrightarrow & \texttt{SELECT<wspace><cols\_list><wspace>FROM<wspace>}\nonumber\\
&&\texttt{<ident><wspace>WHERE<wspace><ident><wspace>IN}\nonumber\\
&&\texttt{<wspace><brac\_open><select\_sub><brac\_close>}\nonumber\\
&&\texttt{<semi\_c>}\label{prod:select_sub_st}
\end{eqnarray}

In conclusion, we define the start symbol $S$$\in$$P$, used to begin the grammar $G$ as described in \autoref{ch:definitions}. This is presented from Production \ref{prod:sql_sub_st1} to \ref{prod:sql_sub_st4}.

\begin{eqnarray}
\texttt{<nested\_qry>} &\longrightarrow & \texttt{<update\_sbqy>~|}\label{prod:sql_sub_st1}\\
&\longrightarrow&\texttt{<delete\_sbqy>~|}\label{prod:sql_sub_st2}\\
&\longrightarrow&~\texttt{<insert\_sbqy>|}\label{prod:sql_sub_st3}\\
&\longrightarrow&~\texttt{<select\_sbqy>}\label{prod:sql_sub_st4}
\end{eqnarray}

\section{Translating Nested Queries into Narrations}
In the previous section, the grammar design for nested queries was described. This section presents how nested queries are translated into textual narrations. Narrations are used to describe programs and they are generally termed \textit{syntax-free textual} algorithms~\citep{ade2014abstracting,ade2017new}. In \autoref{ch:regular}, narrations were applied to describe simple SQL queries. The result showed that REs were not sufficient to generate narrations for complex queries. This section shows how CFGs are able to generate narrations for nested queries. In \autoref{pseudoPSO}, we show how the narration is generated. This recursive function takes a nested query (given as a list of queries and subqueries) and returns a concatenation of the query and all its subqueries.\\

\begin{algorithm}
	\caption{Generating Narrations}\label{pseudoPSO}
	\begin{algorithmic}[1]
		\Function {$getNarration(Q[i])$}{} $returns$ {} $String$ \Comment{$Q[i] = (q_1,q_2,\ldots,q_i)$}
		\If{$(i=1)$}
		\State \textbf{return} $Narrate(q_i)$
		\Else
		\State \textbf{return} $Narrate(q_i) + getNarration(Q[i-1])$\Comment{narration of Q}
		\EndIf
		\EndFunction
	\end{algorithmic}
\end{algorithm}


%
%
%

For nested queries, we present the following narrations. The types of narrations we describe are \textbf{inner to outer} (flow from right to left), \textbf{outer to inner} (left to right) and \textbf{co-joined} (joining the queries together). Hence, the flow of information starts with the first query before ending with the query cascaded inside the balanced parentheses, and vice versa.
In \autoref{lst:label5}, a row is deleted from a table using the nested query, and the corresponding narrations are presented in \autoref{algo:example1}, \autoref{algo:example2} and \autoref{algo:example3}.

\begin{lstlisting}[language=SQL,escapechar=@, morekeywords={DATABASE, SCHEMA},frame=bt,numbers=none, label={lst:label5}, caption= Nested SQL query to delete a row] {}
DELETE FROM country 
WHERE city IN
(SELECT city 
FROM country 
WHERE city = "Pretoria");
\end{lstlisting}

\begin{algorithm}[H]
	\caption{Narration 1: Outer to Inner subquery}\label{algo:example1}
	\textit{This query displays the city information from the country table where the city is equal to Pretoria and removes the entire information from the country table. }
\end{algorithm}

\begin{algorithm}[H]
	\caption{Narration 2: Inner to Outer subquery}\label{algo:example2}
	\textit{This query removes the information from the country table where the city is contained in the values retrieved from the inner query, which gets all the city information that has a city equal to Pretoria}
\end{algorithm}

\begin{algorithm}[H]
	\caption{Narration 3: Co-joined subquery}\label{algo:example3}
	\textit{This nested query contains two queries, where the first query removes the contents from the country table where the city appears in the second query which displays the city information from the country table where the city is equal to Pretoria}
\end{algorithm}

\par This example deletes from the \textit{country} table a subset of rows whose \textit{city} column value satisfies the condition specified in the \texttt{WHERE} clause. In this example, the \texttt{WHERE} ... \texttt{IN} clause specifies which rows to delete returned by the subquery. Hence, only the rows of the \textit{country} table where the \textit{city} is equal to ``Pretoria" is displayed. The next section presents the implementation and results of the \texttt{SQL Narrator}.

\section{Implementation and Results}
\begin{figure*}[h]
	\centering
	\includegraphics[width=400px]{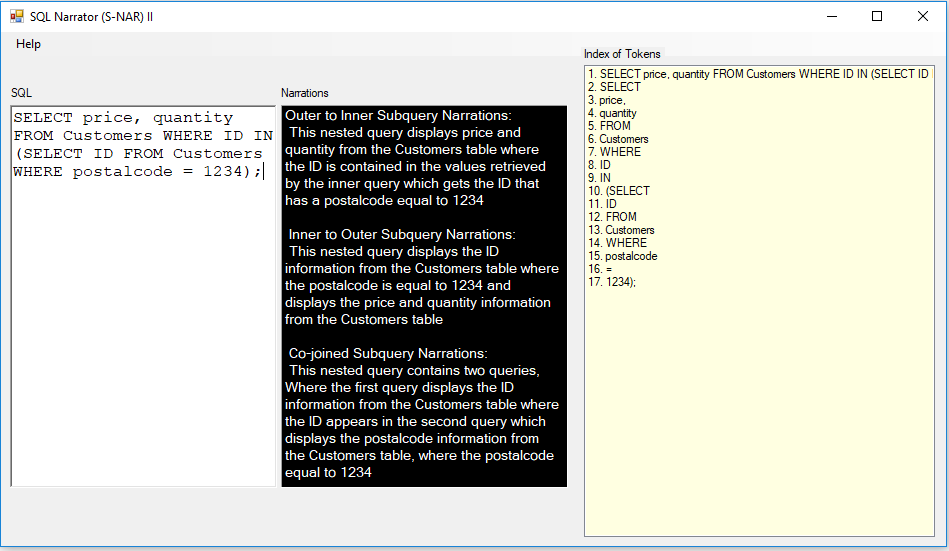}
	\caption{The process of narrating a nested query}
	\label{fig:SNarration}
\end{figure*}

The implementation of the \texttt{SQL Narrator} was carried out using the C\# as the primary language that runs on the .NET Framework. This tool was tested with datasets of nested queries (in \autoref{sql-queries-dataset}) and successfully narrated them. An example of the narration using the \texttt{SQL Narrator} is shown in \autoref{fig:SNarration}.\\


To use the \texttt{SQL Narrator}, a user is expected to enter a nested query into the querybox. The nested query is then converted from characters into tokens and grouped into a sentential form before the narration is displayed to the user. A help file is available to the user. This help file contains a series of steps required to use the narrator. \\

The narration approach that has been presented in this work can be integrated into a SQL pedagogy to assist students in learning nested queries for the first time. We believe that this e-pedagogy will make it easier for students to understand nested queries.

\section{Chapter Summary}
In previous studies within programming pedagogy, it was shown that narrations could be used to enhance learning and assist novices in comprehending programming languages. Another study examined the use of narrations for simple SQL queries. This work presents a new direction of using narrations to assist learner understanding of nested queries. \\

In this chapter, a grammar-based approach that automatically translates nested SQL queries into narrations was presented. This approach used the CFG formalism based on the Coco/R parser generator that takes an attributed grammar and generates a scanner and parser. This approach was implemented in a \texttt{SQL Narrator} based on the C\# language that runs on the .Net Framework. The \texttt{SQL Narrator} was tested with a dataset of nested queries and the narrations were presented.  In \autoref{part3}, we present the synthesis aspect of this study.

\ctparttext{Automatic synthesis of problems into SQL is instrumental to end-users with diverse backgrounds such as business analysts, financial professionals, marketing personnel, etc~\citep{wang2017interactive,yaghmazadeh2017sqlizer}. These users frequently use db applications but lack the technical expertise to write a correct SQL query. Although these users can clearly describe what tasks they intend to perform, they are often faced with how to specify what the intended query should be. Such confusion may increase if they frequently need to engage with technical staff or seek help through online forums just to perform their daily operation. Such process can be time-consuming and frustrating. To mitigate these challenges, we propose different interactive user interfaces that these users can engage with. First, we present a tool that allows a user to specify their query requests in a free-form termed as \emph{narrations}, then we show an interactive visualiser that uses drag and drop interactions to generate a query using icons. Last, we present a speech-query tool that takes a speech input and converts this into a query output.
	
This part contains three chapters. In \autoref{ch:nsql}, we describe the translation of narrations into SQL queries and in \autoref{ch:vsql}, the visualiser for SQL queries is presented. \autoref{ch:ssql} describes the speech synthesiser tool for SQL queries.

}
\part{SQL Synthesis}\label{part3}
\chapter{Narrations to SQL Query}\label{ch:nsql} 
\lettrine[lines=3,loversize=0.1]{I}\normalsize{n} \autoref{ch:cfg} of \autoref{part2}, we presented a grammar-based approach that automatically translates nested SQL queries into narrations. This chapter introduces the synthesis aspect of this thesis. In this chapter, textual \textit{narrations} depicted as natural language descriptions are translated into SQL queries and the resultant feedback is provided to the user. This approach uses a JFA, and was integrated into a tool called \texttt{Narrations-2-SQL}.

\section{Introduction}
JFA is an automata-based algorithm used for processing discontinuous information~\citep{meduna2014regulated,meduna2017modern}. This algorithm has been used in many domains due to its expressive power \citep{meduna2014formal,fernau2015jumping}. Since natural languages are highly ambiguous in nature, we have extended the use of JFA to synthesise SQL queries from natural language specifications. To our knowledge, this is expected to be the first time in which such an approach will be applied for SQL query translation from natural language. Our approach allows users to express their query in a \textit{free-form} -- in natural language to produce the equivalent SQL query \citep{li2014constructing,norouzifard2008using}.  It should be noted that this free-form approach expressed in natural language is regarded as \textit{narrations}, and has shown to improve program comprehension \citep{ade2014abstracting,ade2016automatic}. The formal definition of a JFA has been provided in \autoref{ch:definitions}.

\begin{figure*}[h]
	\centering
	\includegraphics[width=300px]{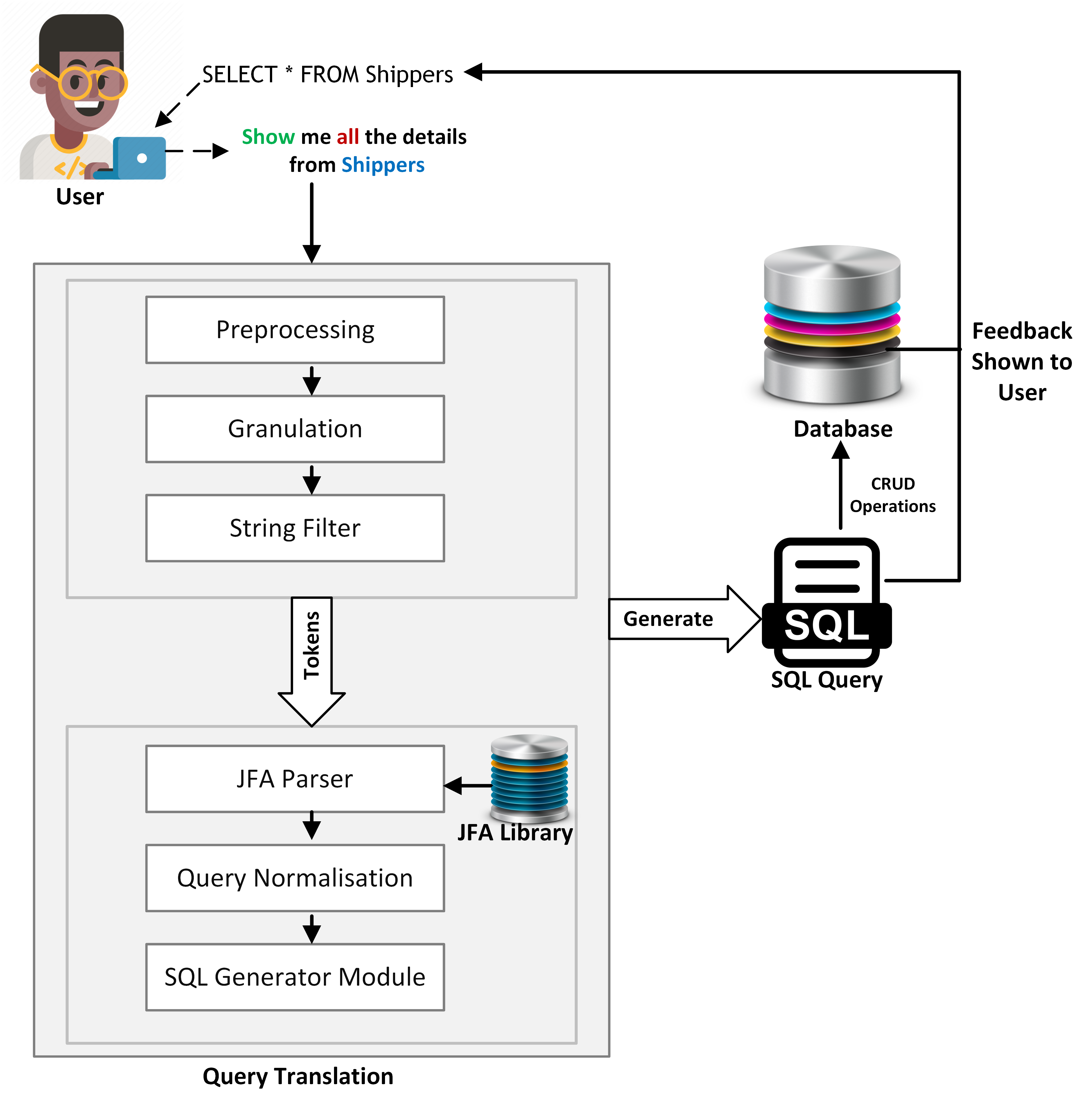}
	\caption{The framework of \texttt{Narrations-2-SQL}}
	\label{fig:frameworknarr}
\end{figure*}

\sloppypar

\par This chapter presents the use of natural language specifications in a tool called \texttt{Narrations-2-SQL}, to aid the understanding of SQL. The \texttt{Narrations-2-SQL} engine uses a JFA -- a type of Finite Machine, to translate natural language specifications into SQL queries, executes the query and provides feedback to a user. \autoref{fig:frameworknarr} shows the framework of \texttt{Narrations-2-SQL}. A user typesets a NLQ, which is then processed by the \texttt{Narrations-2-SQL}. To begin, \texttt{Narrations-2-SQL} preprocesses the typed texts into a stream of tokens, which is then passed to the JFA for matches. At this phase, the tokens are matched and used to construct an SQL query. The equivalent query is used on the \texttt{XNorthwind DB}~\citep{ade2019xnorthwind} and the feedback is provided to the user.

\section{Natural Language to SQL Query}\label{method}
In this section, we show how we have abstracted natural language to a JFA. Our aim at this phase is to remove unwanted features from the dataset and keep the relevant details (tokens) as shown in \autoref{fig:frameworknarr}. To abstract the tokens into a JFA, we use the states, transition process and data representation to understand the abstraction process. The \texttt{XNorthwind} database consists of eight tables and 100000 iterations of datasets, which was used to train our JFA.

\subsection{JFA Design}\label{jfa2}
To abstract natural language specifications into a JFA, we identify the entities with matching colours that make up the alphabet. These entities are query type or ($\sum _\text{QT} = a_{x}$) in \textit{green}, column type or $\sum _\text{CT} = b_{y}$ in \textit{red} and entity or $\sum _\text{ET} = c_{z}$ in \textit{blue}. \autoref{jfa} shows the JFA symbols for the XNorthwind database used for this work. We show an example of a JFA and the language it accepts. \\

\noindent Given a typed request from a user:

\begin{mdframed}[backgroundcolor=cornsilk!20]	
	\begin{enumerate}
		\item Please, help me to \textcolor{ao(english)}{find} \textcolor{bostonuniversityred}{all} the \textcolor{blue}{employees'} information that work for this organisation
	\end{enumerate}
\end{mdframed}  

\noindent The JFA that follows:

\begin{center}
	\textbf{M} = $( \lbrace \textcolor{blue}{R, S, T, U} \rbrace,\lbrace a_5, b_{21}, c_7 \rbrace, R, \textcolor{blue}{R}; \lbrace \textcolor{blue}{U} \rbrace )$\\
\end{center}
$\lbrace \textcolor{blue}{R, S, T, U} \rbrace$ are the states,\\
$\lbrace a_5, b_{21}, c_7 \rbrace,$ are the input alphabets,\\
{R} is the set of rules.\\
$\textcolor{blue}{R}$ is a start state,\\
$\lbrace \textcolor{blue}{U} \rbrace$ is a final state.\\

\noindent \textbf{with}

\begin{center}
	\textbf{R} =  $\lbrace \textcolor{blue}{R}a_5 \rightarrow \textcolor{blue}{S}, \textcolor{blue}{S}b_{21} \rightarrow \textcolor{blue}{T}, \textcolor{blue}{T}c_7 \rightarrow \textcolor{blue}{U}\rbrace$\\
\end{center}

\noindent \textbf{accepts} 

\begin{center}
	L(M) = $ \lbrace w \in \lbrace a_5, b_{21}, c_7 \rbrace$*$: |a_5| = |b_{21}| = |c_7| \rbrace$\\
	i.e. $a_5 =  find ; b_{21} =  {all} ; c_7 = employees$ \\
\end{center}

\begin{align*}
b_{21}a_{5}c_7b_{21}c_7\textcolor{blue}{\underline{R}}a_5  &\curvearrowright b_{21}a_5c_7\textcolor{blue}{\underline{S}}b_{21}c_7 &[\textcolor{blue}{R}a_{5} \rightarrow \textcolor{blue}{S}] \\
&\curvearrowright b_{21}a_5c_7\textcolor{blue}{\underline{T}}c_7 &[\textcolor{blue}{S}b_{21} \rightarrow \textcolor{blue}{T}] \\
&\curvearrowright \textcolor{blue}{\underline{U}}b_{21}a_5c_7 &[\textcolor{blue}{T}c_7 \rightarrow \textcolor{blue}{U}] 
\end{align*}

\begin{figure}[h]
	\begin{center}
		\begin{tikzpicture}[scale=0.20]
		\tikzstyle{every node}+=[inner sep=0pt]
		\draw [black] (14.9,-20.8) circle (3);
		\draw (14.9,-20.8) node {$R$};
		\draw [black] (36.2,-11.3) circle (3);
		\draw (36.2,-11.3) node {$S$};
		\draw [black] (50.8,-22.4) circle (3);
		\draw (50.8,-22.4) node {$T$};
		\draw [black] (29.3,-27) circle (3);
		\draw (29.3,-27) node {$U$};
		\draw [black] (29.3,-27) circle (2.4);
		\draw [black] (7.6,-20.8) -- (11.9,-20.8);
		\fill [black] (11.9,-20.8) -- (11.1,-20.3) -- (11.1,-21.3);
		\draw [black] (16.993,-18.653) arc (132.64752:95.42712:27.811);
		\fill [black] (33.2,-11.42) -- (32.36,-11) -- (32.45,-12);
		\draw (23.58,-13.2) node [above] {$a_{5}$};
		\draw [black] (39.143,-11.863) arc (74.30011:31.21033:17.663);
		\fill [black] (49.47,-19.71) -- (49.48,-18.77) -- (48.63,-19.29);
		\draw (46.06,-14.31) node [above] {$b_{21}$};
		\draw [black] (48.478,-24.295) arc (-54.83409:-101.01279:21.23);
		\fill [black] (32.19,-27.78) -- (32.88,-28.42) -- (33.08,-27.44);
		\draw (41.27,-28.28) node [below] {$c_{7}$};
		\end{tikzpicture}
		\caption{A JFA for Example 1} \label{fig:M1jfa}
	\end{center}
\end{figure}
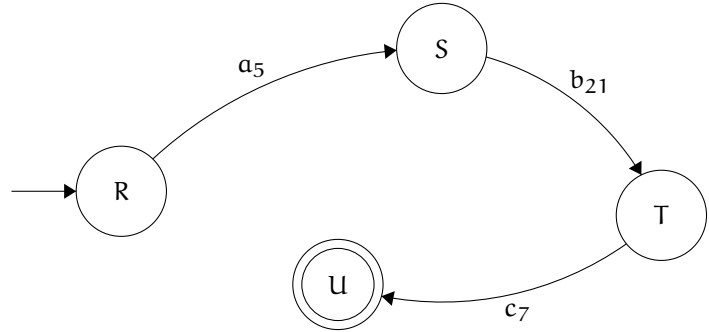
\autoref{fig:M1jfa} shows a JFA with its transitions. It has four states, labeled as \textcolor{blue}{R, S, T, U}. Here, the start state is denoted as \textcolor{blue}{R} and the accepting state, \textcolor{blue}{U}, denoted by the double circle. From the diagram, \textcolor{blue}{R}$a_{5}$ moves to \textcolor{blue}{S}, where the only string found is the \textit{find} keyword. The second transition shows the movement of \textcolor{blue}{S}$b_{21}$ to \textcolor{blue}{T} showing only the \textit{all} keyword. Last, \textcolor{blue}{T}$c_{7}$ moves to \textcolor{blue}{U} showing the \textit{employee} keyword. It is worth noting that this example only shows four states with corresponding transitions. We can have as many states and transitions, depending on the input statement specified from the user. 

\begin{center}
		\captionof{table}{JFA symbols from 1 - 24}
	\label{jfa}
	\begin{tabular}{ |p{2.8cm}|p{3.5cm}|p{2.8cm}|}
		
		\hline
		
		\multicolumn{3}{|c|}{JFA symbols} \\
		\hline
		$a_x$ & $b_y$ & $c_z$\\
		\hline
		$a_0$  \color{ao(english)}{Display}   & $b_0$  \color{bostonuniversityred}{OrderID}    & $c_0$  \color{blue}{Suppliers} \\
		$a_1$  \textcolor{ao(english)}{Select}   & $b_1$  \textcolor{bostonuniversityred}{CompanyName}    & $c_1$  \textcolor{blue}{Products} \\
		$a_2$  \textcolor{ao(english)}{Show}   & $b_2$  \textcolor{bostonuniversityred}{ContactName}    & $c_2$  \textcolor{blue}{OrderDetails} \\
		$a_3$  \textcolor{ao(english)}{List}   & $b_3$  \textcolor{bostonuniversityred}{Supplierscol}    & $c_3$  \textcolor{blue}{Orders} \\
		$a_4$  \textcolor{ao(english)}{Return}   & $b_4$  \textcolor{bostonuniversityred}{ContactTitle}    & $c_4$  \textcolor{blue}{Categories} \\
		$a_5$  \textcolor{ao(english)}{Find}   & $b_5$  \textcolor{bostonuniversityred}{Address}    & $c_5$  \textcolor{blue}{Employees} \\
		$a_6$  \textcolor{ao(english)}{Compute}   & $b_6$  \textcolor{bostonuniversityred}{City}    & $c_6$  \textcolor{blue}{Customers} \\
		$a_7$  \textcolor{ao(english)}{Get}   & $b_7$  \textcolor{bostonuniversityred}{Region}    & $c_7$  \textcolor{blue}{Shippers} \\
		$a_8$  \textcolor{ao(english)}{Remove}   & $b_8$  \textcolor{bostonuniversityred}{PostalCode}    & $-$  \textcolor{blue}{} \\
		$a_9$  \textcolor{ao(english)}{Clear}   & $b_9$  \textcolor{bostonuniversityred}{Country}    & $-$  \textcolor{blue}{} \\
		$a_{10}$  \textcolor{ao(english)}{Delete}   & $b_{10}$  \textcolor{bostonuniversityred}{Phone}    & $-$  \textcolor{blue}{} \\
		$a_{11}$  \textcolor{ao(english)}{Change}   & $b_{11}$  \textcolor{bostonuniversityred}{Fax}    & $-$  \textcolor{blue}{} \\
		$a_{12}$  \textcolor{ao(english)}{Update}   & $b_{12}$  \textcolor{bostonuniversityred}{HomePage}    & $-$ \textcolor{blue}{} \\
		$a_{13}$  \textcolor{ao(english)}{Add}   & $b_{13}$  \textcolor{bostonuniversityred}{ProductID}    & $-$  \textcolor{blue}{} \\
		$a_{14}$  \textcolor{ao(english)}{Give}   & $b_{14}$  \textcolor{bostonuniversityred}{ShippersID}    & $-$  \textcolor{blue}{} \\
		$a_{15}$  \textcolor{ao(english)}{Discontinue}   & $b_{15}$  \textcolor{bostonuniversityred}{CategoryID}    & $-$ \textcolor{blue}{} \\
		$a_{16}$  \textcolor{ao(english)}{Make}   & $b_{16}$  \textcolor{bostonuniversityred}{Quantity}    & $-$ \textcolor{blue}{} \\
		$a_{17}$  \textcolor{ao(english)}{Increase}   & $b_{17}$  \textcolor{bostonuniversityred}{UnitsOnOrder}    & $-$ \textcolor{blue}{} \\
		$a_{18}$  \textcolor{ao(english)}{Create}   & $b_{18}$  \textcolor{bostonuniversityred}{ReorderLevel}    & $-$ \textcolor{blue}{} \\
		$a_{19}$  \textcolor{ao(english)}{Read}   & $b_{19}$  \textcolor{bostonuniversityred}{Discontinued}    & $-$ \textcolor{blue}{} \\
		$a_{20}$  \textcolor{ao(english)}{Insert}   & $b_{20}$  \textcolor{bostonuniversityred}{Productscol}    & $-$  \textcolor{blue}{} \\
		$-$  \textcolor{ao(english)}{}   & $b_{21}$  \textcolor{bostonuniversityred}{All}    & $-$ \textcolor{blue}{} \\
		$-$  \textcolor{ao(english)}{}   & $b_{22}$  \textcolor{bostonuniversityred}{ShipPostalCode}    & $-$ \textcolor{blue}{} \\
		$-$  \textcolor{ao(english)}{}   & $b_{23}$  \textcolor{bostonuniversityred}{HireDate}    & $-$  \textcolor{blue}{} \\
		$-$  \textcolor{ao(english)}{}   & $b_{24}$  \textcolor{bostonuniversityred}{Extension}    & $-$  \textcolor{blue}{} \\	
		\hline
	\end{tabular}
\end{center}

\subsection{Query Normalisation}
This phase takes the keywords extracted from the JFA design to semantically form a proper sequence for the SQL generator module. This approach was further strengthened using WordNet\footnote{https://wordnet.princeton.edu/}. For example, words such as select, choose, pick and display are mapped to SELECT keyword. Words such as insert, add, increase and build are mapped to the INSERT keyword. Also words such as update and amend are mapped to the UPDATE keyword. For the DELETE keyword, words such as remove and delete are mapped. The attributes of the XNorthwind tables are the column details. Once mapped, the information is fed to the query generator.
\subsection{SQL Generator Module}
The query generator transforms the semantic information at the normalisation phase and generates an SQL query. Since this work is limited for a single-relation, this phase is quite straightforward. The generated query is used against the XNorthwind DB and the result is displayed to the user. 
\section{Implementation, Results and Applications}\label{Implement}
\subsection{Implementation and Results}
\begin{figure*}[htb]
	\centering
	\includegraphics[width=400px]{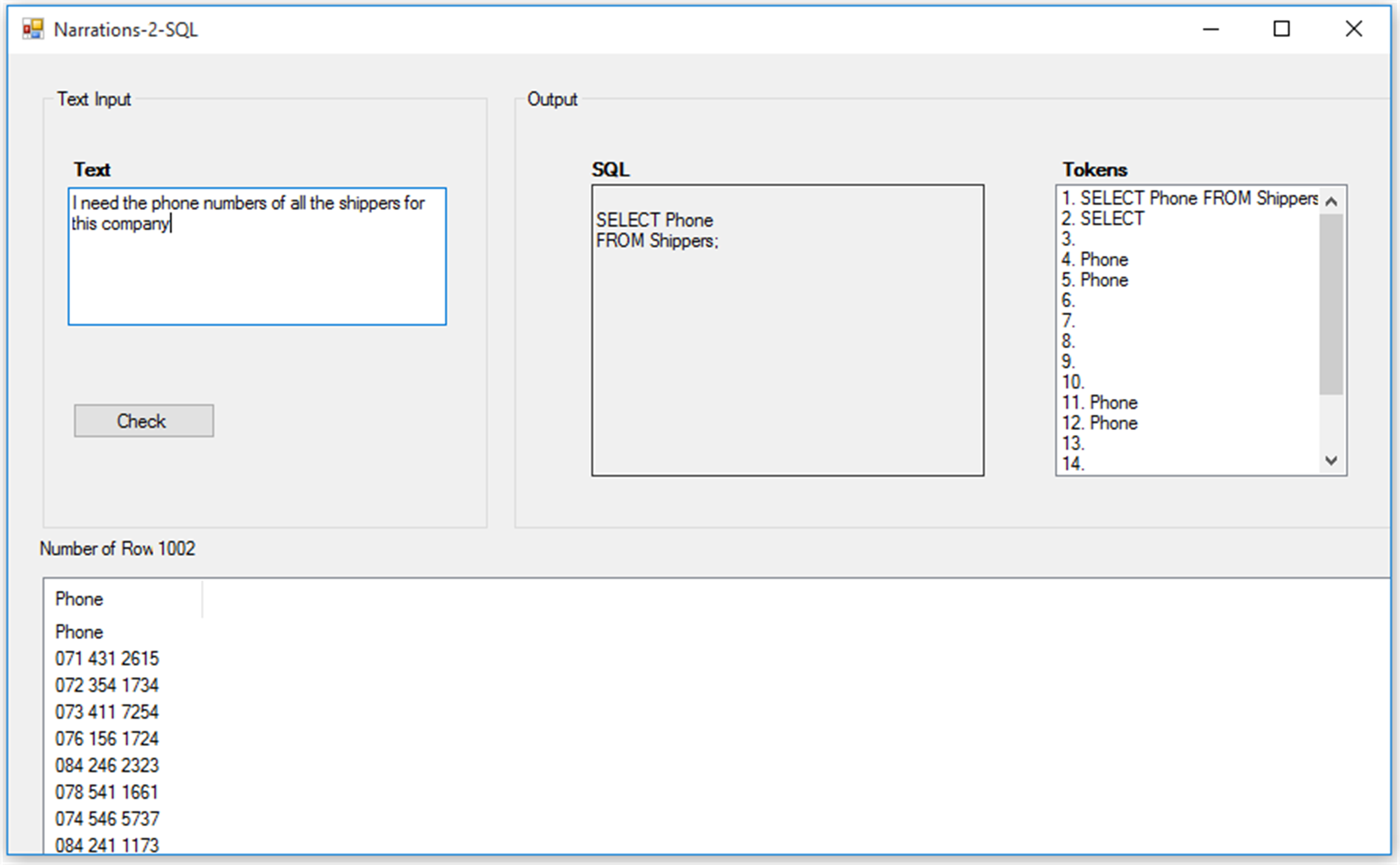}
	\caption{The user interface of \texttt{Narrations-2-SQL}}
	\label{fig:sndone}
\end{figure*}

The JFA technique described in this chapter was implemented into a tool called \texttt{Narrations-2-SQL}. This tool was developed as a C\# application specified using the Microsoft .NET framework. The implemented tool was tested with 204 crowdsourced queries specified in natural language (as presented in \autoref{ch:app-natural-sql}), sourced from the \texttt{XNorthwind DB}. The \texttt{XNorthwind DB} was used in this study, which comprises eight tables and 100000 tuples. The result of the implementation can be found in \autoref{fig:sndone}.

\subsection{Applications of Narrations-2-SQL}
We present possible applications of the tool we have designed for this study. As a tool, \texttt{Narrations-2-SQL} can be used as a:
\begin{enumerate}
	\item QA system,
	\item teaching and learning aid,
	\item tutoring system for SQL,
	\item query tool for complex BI systems, and
	\item natural language interface to query databases.
\end{enumerate} 

\section{Chapter Summary}
This chapter describes a new approach to translating SQL queries from natural language. The technique described in this work uses a JFA, designed into a tool called \texttt{Narrations-2-SQL} for the purpose of SQL query translation from narrations. This work appears to be the first application of abstracting natural language specifications to a JFA, in addition to mapping this to an SQL query. This is a major contribution to the problems faced in the information retrieval (IR) domain. In its framework, \texttt{Narrations-2-SQL} performs operations on the \texttt{XNorthwind} database using simple SQL commands to create, retrieve, modify and delete data. Feedback of the operation is presented to a user. If implemented on a large scale, \texttt{Narrations-2-SQL} will assist end-users in different domains, to specify their queries in natural language, and perform their tasks seamlessly without needing much help from technical users. More so, our evaluation shows that the majority of the users agreed that this approach can be useful in industry.\\

\autoref{ch:vsql} describes the use of visual specifications that explores \textit{drag and drop} interactions of query-like images to generate SQL queries.
\chapter{Visual Specifications to SQL Query}\label{ch:vsql} 
\lettrine[lines=3,loversize=0.1]{I}\normalsize{n} \autoref{ch:nsql}, we described the translation of natural language descriptions, termed \textit{narrations}, into SQL. These narrations are English-like descriptions specified by end-users in natural language forms. This idea was implemented into a software tool called \texttt{Narrations-2-SQL}. The tool is expected to assist end-users in writing correct queries, which was a problem identified in the literature. This chapter presents a tool that uses predefined images that represent SQL commands to generate a query. This study is expected to improve students' comprehension of the SQL query concept.

\section{Introduction}
Visual specifications are symbols used to represent features, which can be used to display some text or program~\citep{rojit2016visual}. In the programming concept, visual specifications have been used to build and demonstrate a programming solution especially with problems faced by students~\citep{roberts2019computer,eden2018round}. This problem is not only limited to program understanding; students struggle to memorise database schema~\citep{kawash2014formulating,cembalo2011savi,garner2015learning}. In addition, writing DML expressions has shown to be problematic for students~\cite{dekeyser2007computer}. In order to provide adequate support to address these problems, there is a need to build interactive platforms, incorporated with either \textit{animation} or \textit{visualisation} aids to support the understanding of SQL. In past decades, a number of tools have been developed to provide support in learning SQL~\citep{cembalo2011savi,folland2016visqlizer}. Some of the existing tools employed interactive visualisations to aid SQL understanding.

\begin{figure*}[h!]
	\centering
	\includegraphics[width=400px]{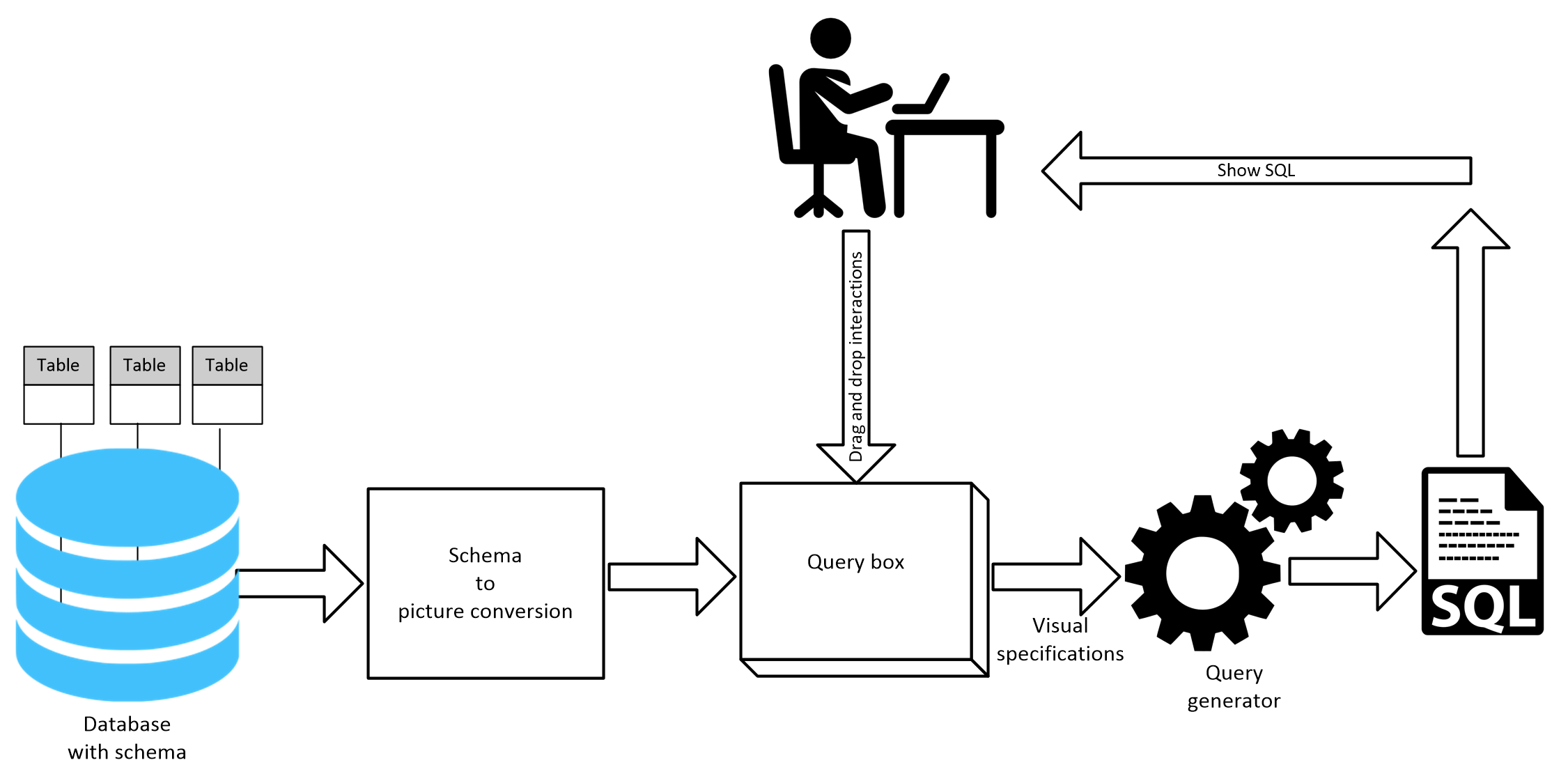}
	\caption{The framework of the SQL query generation}
	\label{fig:framework}
\end{figure*}

\par In this chapter, we propose the use of an interactive visualisation technique to aid the understanding of SQL. The visualisation technique ensures the interaction between visual specifications to build queries and will eliminate the need to memorise database schemas, which is a major problem faced by students learning SQL. In this work, we have developed an interactive visualisation aid called \texttt{SQL visualiser}, which uses visual specifications of the `drag and drop' interactions for generating SQL queries. Although, this approach has found application in the programming languages paradigm such as \texttt{Alice} \citep{dann2008learning}, \texttt{Scratch} \citep{resnick2009scratch} and \texttt{StarLogo TNG} \citep{wang20063d}, it has not (to the best of our knowledge) been applied to SQL queries. \autoref{fig:framework} shows the SQL query generation process. To use the \texttt{SQL visualiser}, images are represented as visual specifications to depict the database model. These images can be moved into the query box. When the moved images correspond to a standard SQL \texttt{SELECT} statement, a query is generated and displayed to the user.

\section{Design of the SQL Visualiser}
One aspect of queries which poses difficulties for students is the SQL \texttt{SELECT} construct \citep{sadiq2004sqlator,qian2012designing}. This type of query is used to extract data from a relational database~\citep{kearns1997teaching}. Hence, our main goal is focused on using visual aids to easily generate queries by means of the SQL \texttt{SELECT} constructs. This idea can be extended to the system of queries.\\

\par It is a common perception that students are better at \textit{recognising} visual constructs rather than at \textit{writing} codes~\citep{dekeyser2007computer}. Thus, this work is motivated by an intention to use visual specifications in order to generate SQL queries. Also, another motive of developing this SQL visualiser is to use it as a teaching and learning aid. We have found that database schemas pose difficulties for students~\citep{dekeyser2007computer}, hence, our intention is to simplify the process of understanding database schemas. We identify three main points to distinguish our visualisation from other approaches.

\begin{description}
	\item[Intuitive] Our visualisation is intuitive to students who are learning SQL queries for the first time. The visualisation uses images to depict each query statement. It helps students better understand SQL queries since it helps them get a glimpse of the behaviour of the image when each image is selected. Hence, students do not require extensive training to understand how to use the visual aid.
	\item[Interactive] The visualisation tool is interactive, which means that students are not required to write any query statement in the application. They can simply click and drag the images across panels. Query statements are generated at the same time.
	\item[Helpful] A help facility is provided before using the visualiser. A user is provided with an instruction of the underlying database schema before using the application. Also, \textit{hints} are provided to the user and are specified using colours (green or red). These colours show whether a query is wrong or correct. In addition, a textual suggestion is offered to the user to ensure that the correct object is selected. 
\end{description}

The \texttt{SQL visualiser} was implemented as a Windows Form Application and was included as part of the .NET framework for the purpose of creating rich client applications \citep{liberty2005programming}. The visualisation tool consists of some components used for the generation of a query. These components include schema, query box and query generator.

\subsection{Schema}
\begin{figure}	
	%
	%
	%
	%
	\centering
	\includegraphics[width=0.9\linewidth]{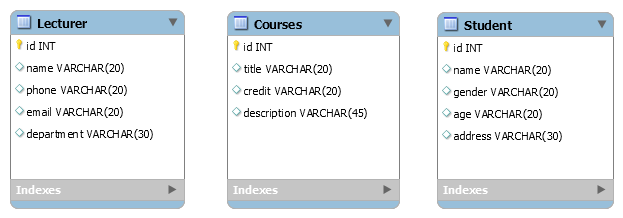}
	\caption{Logical organisation of the data}
	\label{fig:logical}
	
\end{figure}

The schema shows the logical organisation of the data. In the visualiser, the schema consists of tables and associated fields. The SQL query is based on the underlying schema. In addition, if a schema is correct, the generated SQL query is correct, and vice-versa. \autoref{fig:logical} depicts the schema for this model.\\

In the schema, each table is shown by its name displayed at the top and its corresponding attributes shown at the bottom. For example:\\
\tabb \textbf{Lecturer} (\underline{id}, name, phone, email, department)\\
\tabb \textbf{Courses}   (\underline{id}, title, credits, description)\\
\tabb \textbf{Student} (\underline{id}, name, gender, age, address)\\

Each table is linked by a primary key. The unique identifier for the tables is the entity \textit{id}. The visualisation tool relies on the schema to generate the SQL query.

\subsection{Query Box}
The query box is used to specify subsets of the schema that the user is interested in. The query box is also denoted as the building block for the query generator. In the query box, the schema (represented by the images) are extracted into a form used by the query generator to generate the SQL query. Each image is included with a caption for easy identification. \autoref{fig:symbols}, \autoref{fig:lecturer}, \autoref{fig:course} and \autoref{fig:student} represent the pictures and descriptions used to represent the schema. 

\begin{table}[!htb]
	\centering
	\caption{Symbols and description of the SELECT statement}
	\label{fig:symbols}
	\begin{tabular}{|l|l|l|}
		\hline
		\textbf{Symbol}    & \textbf{SQL Block} & \textbf{Description}                   \\ \hline \hline
		\includegraphics[width=21px, height=20px]{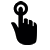} & \multicolumn{1}{c|} {\texttt{SELECT}}      & This icon represents the SELECT statement\\ \hline
	\end{tabular}
\end{table}

\begin{table}[!htb]
	\centering
	\caption {Symbols and descriptions of the Lecturer table}
	\label{fig:lecturer}
	\begin{tabular}{|l|l|l|}
		\hline
		\textbf{Symbol}   & \textbf{SQL Block} & \textbf{Description}                  \\ \hline \hline
		\includegraphics[width=21px, height=20px]{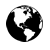} &  \multicolumn{1}{c|} {\texttt{*}} & This denotes ``all'' in rows                          \\ \hline
		\includegraphics[width=21px, height=20px]{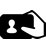}	&      \multicolumn{1}{c|} {\texttt{id}}       &   This icon represents the primary key field                           \\ \hline
		\includegraphics[width=21px, height=23px]{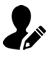}	&      \multicolumn{1}{c|} {\texttt{name}}       &   This icon denotes the name field                            \\ \hline
		\includegraphics[width=21px, height=20px]{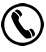}	&      \multicolumn{1}{c|} {\texttt{phone}}       &   A representation for a phone field                          \\ \hline
		\includegraphics[width=21px, height=20px]{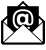}	&     \multicolumn{1}{c|} {\texttt{email}}        &   This symbol denotes an email field                          \\ \hline
		\includegraphics[width=20px, height=24px]{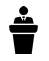}	&      \multicolumn{1}{c|} {\texttt{Lecturer}}       &   A representation for a lecturer field                        \\ \hline
	\end{tabular}
\end{table}

\begin{table}[!htb]
	\centering
	\caption {Symbols and descriptions of the Courses table}
	\label{fig:course}
	\begin{tabular}{|l|l|l|}
		\hline
		\textbf{Symbol}   & \textbf{SQL Block} & \textbf{Description}                  \\ \hline \hline
		\includegraphics[width=21px, height=20px]{gfx/everything} &  \multicolumn{1}{c|} {\texttt{*}} & This denotes ``all'' in row                         \\ \hline
		\includegraphics[width=21px, height=20px]{gfx/id}	&      \multicolumn{1}{c|} {\texttt{id}}       &   This icon represents a primary key field                            \\ \hline
		\includegraphics[width=21px, height=23px]{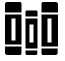}	&      \multicolumn{1}{c|} {\texttt{title}}       &   This icon denotes a title field                           \\ \hline
		\includegraphics[width=21px, height=20px]{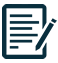}	&      \multicolumn{1}{c|} {\texttt{credit}}       &   A representation for a credit field                           \\ \hline
		\includegraphics[width=21px, height=20px]{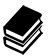}	&      \multicolumn{1}{c|} {\texttt{description}}       &   This symbol represents a description field                           \\ \hline
		\includegraphics[width=20px, height=24px]{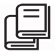}	&      \multicolumn{1}{c|} {\texttt{Courses}}       &   A representation for a course entity                           \\ \hline
	\end{tabular}
\end{table}

\begin{table}[!htb]
	\centering
	\caption {Symbols and descriptions of the Student table}
	\label{fig:student}
	\begin{tabular}{|l|l|l|}
		\hline
		\textbf{Symbol}   & \textbf{SQL Block} & \textbf{Description}                  \\ \hline \hline
		\includegraphics[width=21px, height=20px]{gfx/everything} &  \multicolumn{1}{c|} {\texttt{*}} & This denotes ``all'' in rows                         \\ \hline
		\includegraphics[width=21px, height=20px]{gfx/id}	&      \multicolumn{1}{c|} {\texttt{id}}       &   This icon represents an identity field                            \\ \hline
		\includegraphics[width=21px, height=23px]{gfx/name}	&      \multicolumn{1}{c|} {\texttt{name}}       &   An illustration for a name field                            \\ \hline
		\includegraphics[width=21px, height=20px]{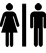}	&      \multicolumn{1}{c|} {\texttt{gender}}       &   A symbol for a gender field                            \\ \hline
		\includegraphics[width=21px, height=20px]{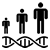}	&      \multicolumn{1}{c|} {\texttt{age}}       &   A representation for age  field                          \\ \hline
		\includegraphics[width=21px, height=20px]{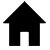}	&     \multicolumn{1}{c|} {\texttt{address}}        &   An icon for an address field                           \\ \hline
		\includegraphics[width=21px, height=20px]{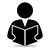}	&   \multicolumn{1}{c|} {\texttt{Student}}            &   This symbol represents a student entity                             \\ \hline
	\end{tabular}
\end{table}

\subsection{Query Generator}
The query generator transforms the images in the query box and presents a query to the user. As more images are added, the query generator also adds the attribute to the query. The generator phase is very straightforward since the scope of this work is limited to a single-relation. The \texttt{SELECT} portion of the query consists of tables and attributes; where the \texttt{FROM} clause defines a table and the \texttt{WHERE} clause is defined by a field attribute and its value. We illustrate this in Example \autoref{lst:visualspecs}.

\begin{exmp}
	\label{lst:visualspecs}
	\textit{Consider a simple database table with the schema: Student (id, name, age). Now, write a simple SQL query to display all information from the student table}.
\end{exmp}


\begin{figure*}[!htb]
	\centering
	\includegraphics[width=0.9\linewidth]{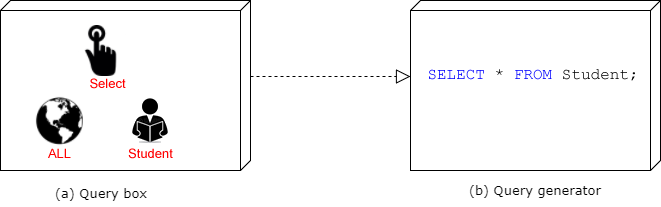}
	\caption{The process of generating an SQL query}
	\label{fig:generator}
\end{figure*}

\par \autoref{fig:generator} presents the process that the query generator uses to present queries. When a user adds the images into the query box (a), the query generator displays the visualisation from the images that were selected into the query box (b). More options used in this chapter are presented in the next section.

\subsection{More Options}
\begin{description}
	\item [Operators and Values] In the SQL visualiser, we have explored some comparison operators, such as the (= or \textit{equal to}, $<$ or \textit{less than} and $>$ or \textit{greater than}). The comparison operators are used within the generated query with values between 0 -- 100 to show the relationship. For each operator, the user can interactively determine which option to select. Further, a user can select the preferred choice of value to use within the query by using the scroll option provided. Once the scroll option is selected, it changes the value within the generated query. \\
	
	\item [Colours] In the HCI interface design specifications, colours have been described to convey information \citep{brown1998human}. Within this specification, the association of colours may be used for many purposes if this is implemented conservatively. Colours have salient features, which are useful in human perception \citep{jost2005assessing}. For example, the colour \textit{red} strongly indicates an error, while \textit{green} indicates a normal or acceptable condition. These colours were explored within the visualiser to indicate either an acceptable condition or to respond criticality to a user's error as presented in \autoref{fig:colour}.   
	
	\begin{figure}[h!]
		\begin{tabular}{| l | r |}
			\hline	
			Red (indicator of an error) & Green (acceptable condition) \\
			\hline
			{\color{red} \textbf{Incorrect query, drag the required field}}  & {\color{green} \textbf{Fantastic! Your query is correct}} \\ \hline
		\end{tabular}
		\centering
		\caption{Colours used to show annotations}
		\label{fig:colour}
	\end{figure} 
	
\end{description}

\section{Results}
\begin{figure*}[h!]
	\centering
	\includegraphics[width=1.0\linewidth]{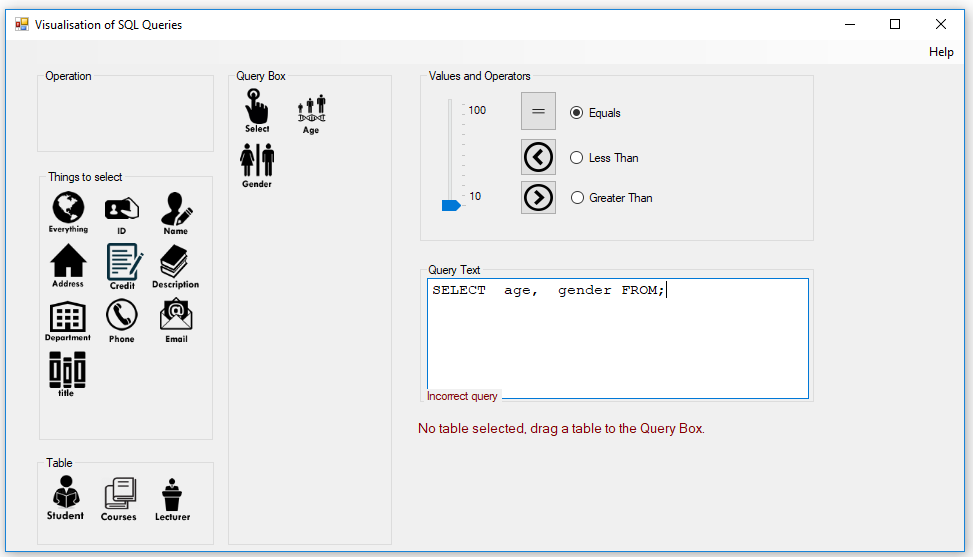}
	\caption{Hints provided to the user}
	\label{fig:hints}
\end{figure*}

\begin{figure*}[h!]
	\centering
	\includegraphics[width=1.0\linewidth]{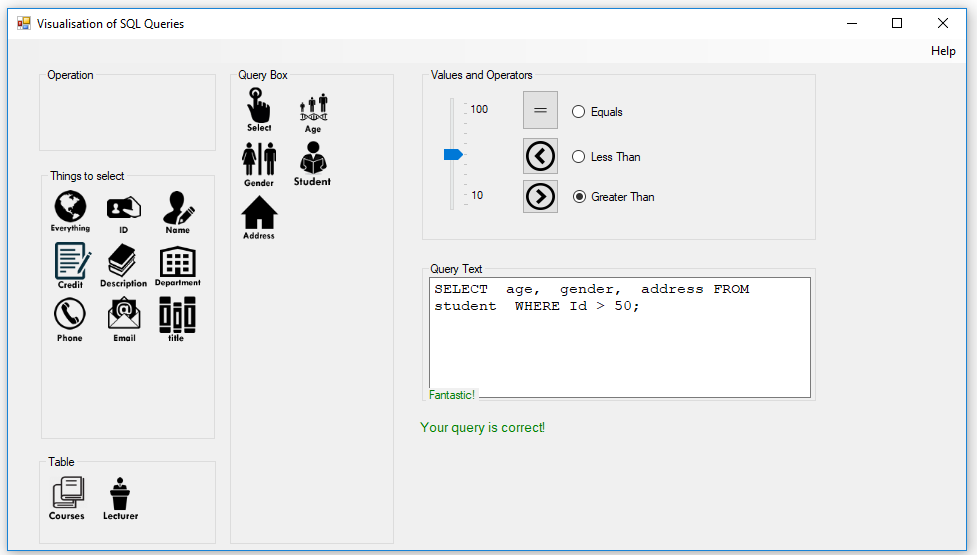}
	\caption{A successfully generated query}
	\label{fig:success}
\end{figure*}

\begin{figure*}[h!]
	\centering
	\includegraphics[width=1.0\linewidth]{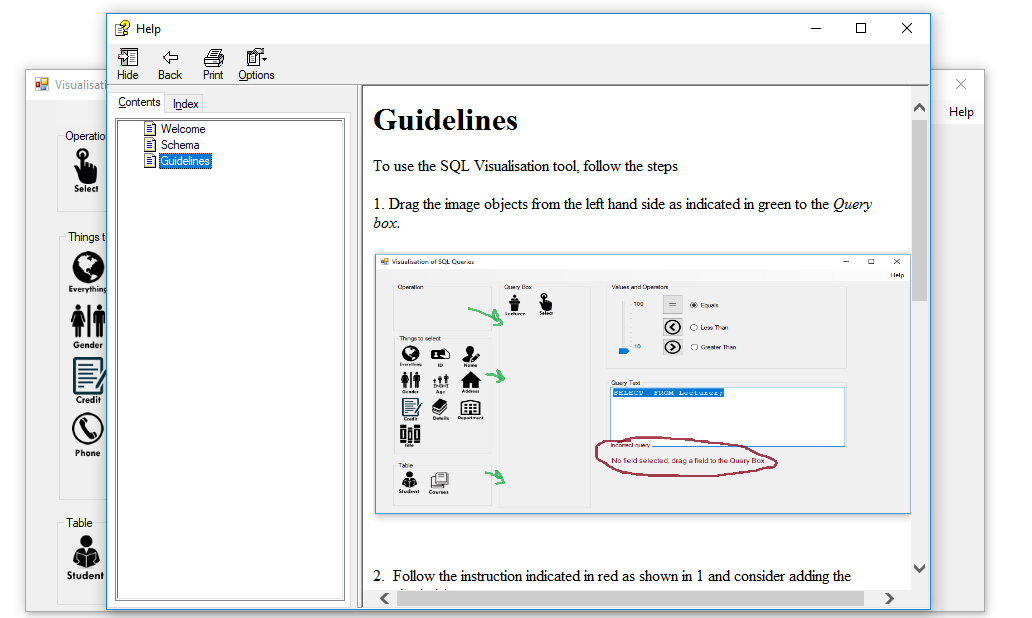}
	\caption{The help feature presents the guidelines for the user}
	\label{fig:help}
\end{figure*}

We present the result from using the visualisation tool. \autoref{fig:hints} shows the feedback received from adding only the \texttt{SELECT} operation into the query box at runtime. The feedback received will assist the user to specify the required visuals before the query can be generated. This example shows that if the user inserts the correct table and its attributes, the query will be successfully generated. The field ``ID'' was chosen as the primary key for each of the table, and the value, ``50'' was selected; See \autoref{fig:success}. The help facility showing the instruction on how to use the \texttt{SQL visualiser} is presented in \autoref{fig:help}. \texttt{SQL visualiser} was compared with a number of SQL visualisation tools. \autoref{fig:review} shows the result of the review. \\

\begin{table*}[h!]
	\scriptsize
	\centering
	\caption{Existing tools versus the SQL Visualiser}
	\label{fig:review}
	\renewcommand{\arraystretch}{1.2}
	\begin{tabular}{|l|c|c|c|c|c|c|c|}
		\hline
		\textbf{Features} & \textbf{eSQL} & \textbf{SAVI} & \textbf{SQlify} & \textbf{sAccess} & \textbf{eledSQL} & \textbf{QueryViz} & \textbf{Our tool (SQL Visualiser)} \\ \hline
		Visualisation of database schema & \ding{55}  & \checkmark & \checkmark & \ding{55} & \ding{55} & \checkmark & \checkmark \\ \hline
		Visualisation of output data & \checkmark  & \checkmark  & \checkmark & \checkmark & \checkmark & \ding{55} & \ding{55} \\ \hline
		SQL query generation & \ding{55} & \ding{55}  & \ding{55} & \ding{55} & \ding{55} & \ding{55} & \checkmark \\ \hline
		Feedback on query semantics & \ding{55} & \ding{55} & \checkmark & \ding{55} & \ding{55} & \checkmark & \checkmark \\ \hline
		Visual object representation & \ding{55} & \ding{55} & \ding{55}  & \ding{55} & \ding{55} & \ding{55} & \checkmark \\ \hline
		\begin{tabular}[c]{@{}l@{}}Ideal for less knowledgeable users \\ (undergraduate students)\end{tabular} & \ding{55} & \ding{55} & \ding{55} & \checkmark & \checkmark & \ding{55} & \checkmark \\ \hline
	\end{tabular}
\end{table*}

We have presented a visualisation tool that applies visual specifications to generate queries. The technique presented in this chapter will find applications in teaching and learning systems. The benefits offered by the visualiser will facilitate human comprehension of SQL queries. This work will particularly aid undergraduate students who are learning SQL queries for the first time. Another application area of this work can be extended to commercial business systems, where the visualiser may be used to assist non-professional users comprehend SQL queries. While such users may be aware of databases, their knowledge of SQL queries may be limited. We believe that our technique's clear communication and visualisation-focus will help users to easily understand SQL queries.\\

\section{Chapter Summary}
\par This chapter has presented an interactive visualisation tool that used visual specifications to build SQL queries. The visualisation tool considered the SQL \texttt{SELECT} constructs in a bid to improve the comprehension process. It is generally agreed that visualisation can encourage active participation and also lead to critical thought processes in students \citep{gray2016visualizing,lye2014review}. Hence, this work is consistent with those that have used visual specifications. In \autoref{ch:ssql}, we will examine another approach that converts speech into SQL queries.

\chapter{Speech to SQL Query}\label{ch:ssql} 
\lettrine[lines=3,loversize=0.1]{T}\normalsize{he} preceding chapter presented the visual specification method that used the \textit{drag and drop} interaction to generate a query. This method was implemented into a tool called the \texttt{SQL Visualiser} that uses images to generate a query. This chapter introduces the verbal specification technique into a speech-based query system named \texttt{TalkSQL} that takes speech inputs from a user, converts these words into SQL queries and returns a feedback to the user. Automatic feedback generation is of immense importance. To achieve this, we have used REs, a representation of regular languages for the recognition of SQL queries and automatically generate feedback using pre-defined templates.

\section{Introduction}
NLP has contributed immensely to the field of HCI, in terms of its theoretical results and practical applications. These applications have led to the emergence of robust speech-enabled user interfaces (VUIs) such as Google's Voice Action, Apple's Siri, Amazon's Alexa, Microsoft's Cortana etc~\citep{feng2017continuous,zhang2018understanding,saha2019framework}. Together, these VUIs have been applied to solve real world problems in Healthcare~\citep{shah2018designing}, Internet of Things (IoT)~\citep{jungbluth2018combining}, Military~\citep{levulis2018effects}, Telecommunication~\citep{kapur2018alterego}, etc. 

\begin{figure*}[hbt!]
	\centering
	\includegraphics[width=410px]{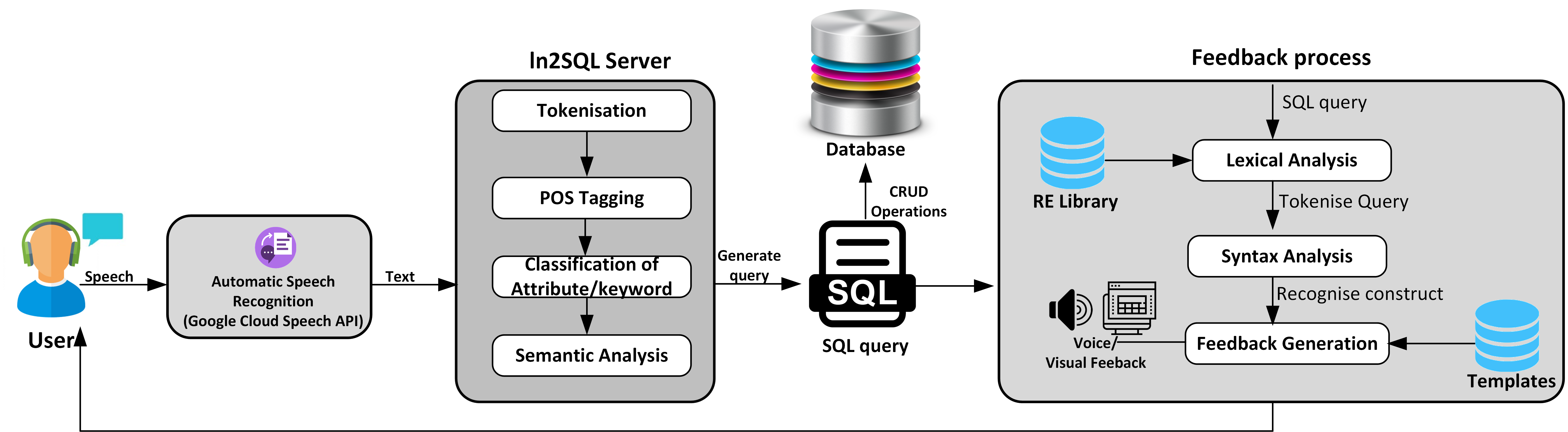}
	\caption{The framework of \texttt{TalkSQL}}
	\label{fig:talksql}
\end{figure*}

This chapter introduces a speech-based, conversational NLIDB system called \texttt{TalkSQL}, used to aid the comprehension of SQL. \texttt{TalkSQL} is an intelligent, conversational tutoring system that translates natural language queries into an executable SQL query to be used on a test database, and presents an output. Similarly, a speech feedback is provided to a learner. \texttt{TalkSQL} applies the CRUD (\texttt{CREATE, SELECT(READ), UPDATE and DELETE}) operations, used against a relational database. \\

In \autoref{fig:talksql}, we show the architecture of \texttt{TalkSQL}. A user initiates a verbal request to the tool, which is then processed and converted into text by the Google Cloud Speech API. The text version is preprocessed by the \texttt{ln2SQL} server that uses the \texttt{spaCy} NLP engine (described in \autoref{ch:Background}) for conversion into a SQL query. The generated query is used on a relational database to request data. Automated feedback is an important aspect of this work. To provide a user with a feedback, the query is tokenised and grouped into syntactic parts. The recognised parts are then matched with a feedback template, and a feedback is generated. The feedback is presented to the user in textual, vocal and visual forms. The \texttt{TalkSQL} engine is similar to that presented in \texttt{ln2SQL}~\citep{couderc2015fr2sql} which considers only the \texttt{SELECT} command.

\section{Feedback Generation}
Feedback has shown to assist novices comprehend their programs and in many cases, applied in textual forms to provide insights~\citep{ade2014abstracting,singh2013automated}. As discussed in the literature (\autoref{ch:Background}), different NLIDB systems have been able to provide feedback in different forms, but to our knowledge, none has undergone our approach. The technique we have developed uses REs to recognise SQL queries before generating a feedback to a user. The feedback is usually concise in a bid to assist learners understand SQL queries. \autoref{algo:example1s} shows spoken words from a user, the query equivalent is presented in \autoref{lst:label5} and the expected feedback is displayed in \autoref{algo:example2s}.

\begin{algorithm}
	\caption{Spoken words from a user (NL)}\label{algo:example1s}
	\textit{Amend the student name to John whose id is equal to 6}	
\end{algorithm}

\begin{lstlisting}[language=SQL,escapechar=@, morekeywords={DATABASE, SCHEMA},frame=bt,numbers=none, label={lst:label5}, caption= SQL query to update a single record] {SQL query to update a single record}
UPDATE Student
SET name = 'John'
Where ID = 6;
\end{lstlisting}

\begin{algorithm}
	\caption{Expected Feedback}\label{algo:example2s}
	\texttt{You have updated a record with a name called John, whose ID number is equal to 6 in the Student table.}
\end{algorithm}

The tool described in this study takes verbal inputs from a user, performs operations on it and generates a query. The feedback from the query generation is similar to that shown in \autoref{algo:example2s}. The next section presents REs for the recognition of the CRUD commands syntaxes in SQL before generating a feedback to the user. 

\section{SQL Constructs Abstraction using Regular Expressions}
\begin{figure}[hbt!]
	\centering
	\includegraphics[width=300px]{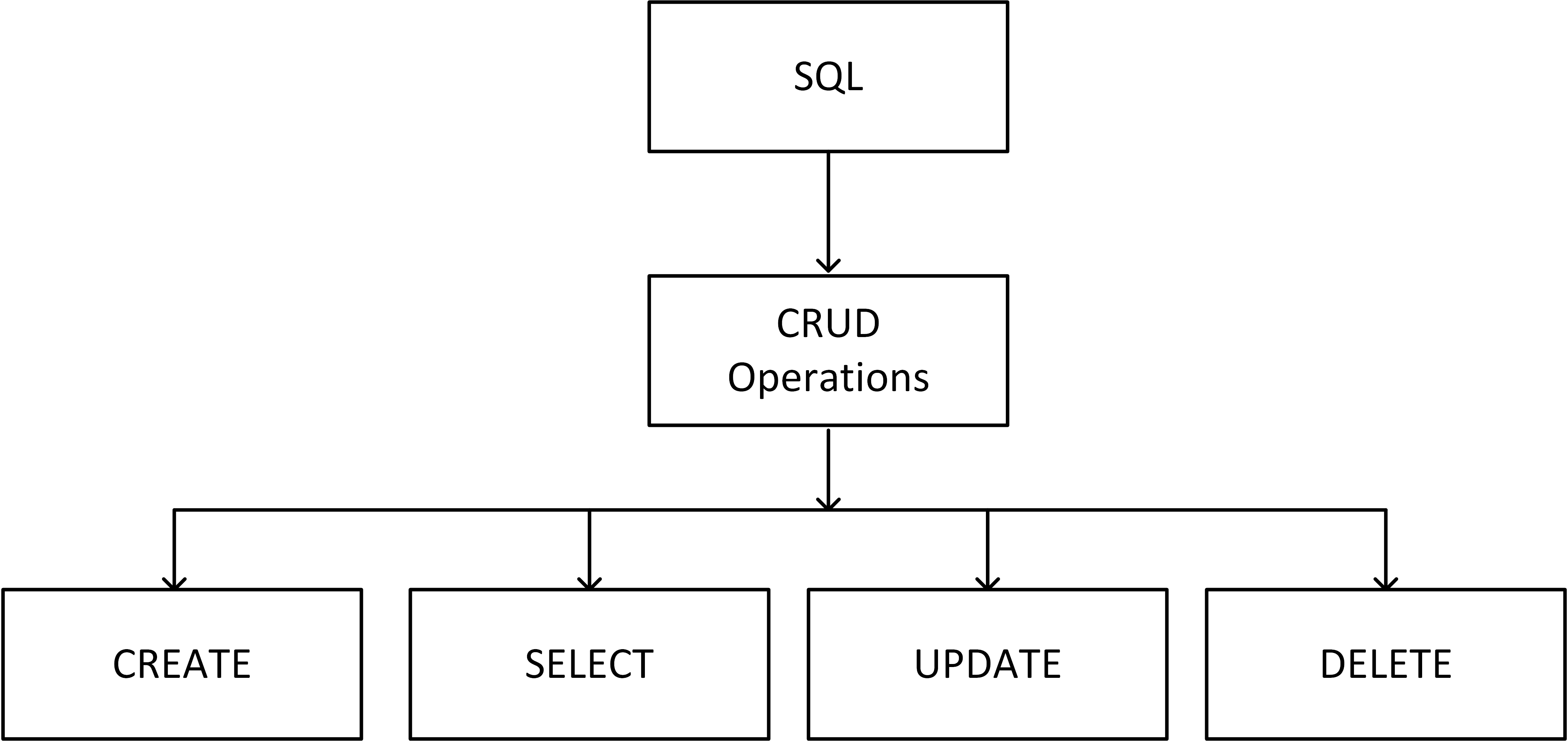}
	\caption{The diagram denoting the CRUD SQL commands}
	\label{fig:crud}
\end{figure}

In this section, we show how we have abstracted the CRUD SQL operations using regular expressions. This stage is regarded as the \textit{Lexical Analysis} phase. This phase shows how the streams of characters (lexemes) that make up the language are grouped into tokens for recognition. To begin, we represent the CRUD operation with a diagram that describes how the building blocks of tokens should be formulated as seen in \autoref{fig:crud}. 
The CRUD commands consists of four statements used to perform operations in a database. These statements are \texttt{Create}, \texttt{Select}, \texttt{Update} and \texttt{Delete}. These statements have been described using regular expressions in~\citep{ade2017s}, as discussed in \autoref{citeresql}.

\section{Introducing TalkSQL}\label{figure:talk}
\begin{figure}[h!]
	\centering
	\includegraphics[width=400px]{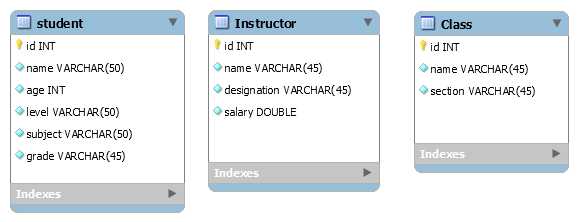}
	\caption{The schema of the test database}
	\label{fig:schema}
\end{figure}

\texttt{TalkSQL} was designed as a C\# Windows Forms Application (WPF) that runs on the .NET framework~\citep{Microsoftregex}. As a VUI, \texttt{TalkSQL} translates natural language specified in verbal inputs into an executable SQL query to be used on a test database and presents an output. \autoref{fig:ui} shows the user interface of TalkSQL. A successful conversion will produce a result to a user in the form of a visualisation. An automatic feedback is also available to a user in textual and speech forms. The idea is to provide a comprehensive feedback to a learner. Errors are also handled by \texttt{TalkSQL} in the form of refinements. \texttt{TalkSQL} informs the user to refine their statement in a conversational manner and produces a result once the statement is understood. The schema for the test database is provided in \autoref{fig:schema}. 

The schema contains three tables (school, instructor, class) with their associated attributes. Each table is linked by a primary key (\textit{id}), which is the unique identifier. \texttt{TalkSQL} uses this schema to perform the query generation. It is worth-noting that \texttt{TalkSQL} can be adapted to any database. We have used this sample database to test the tool that has been developed for this work.

\subsection{Implementation and Results}
\begin{figure*}[hbt!]
	\centering
	\includegraphics[width=1.0\linewidth]{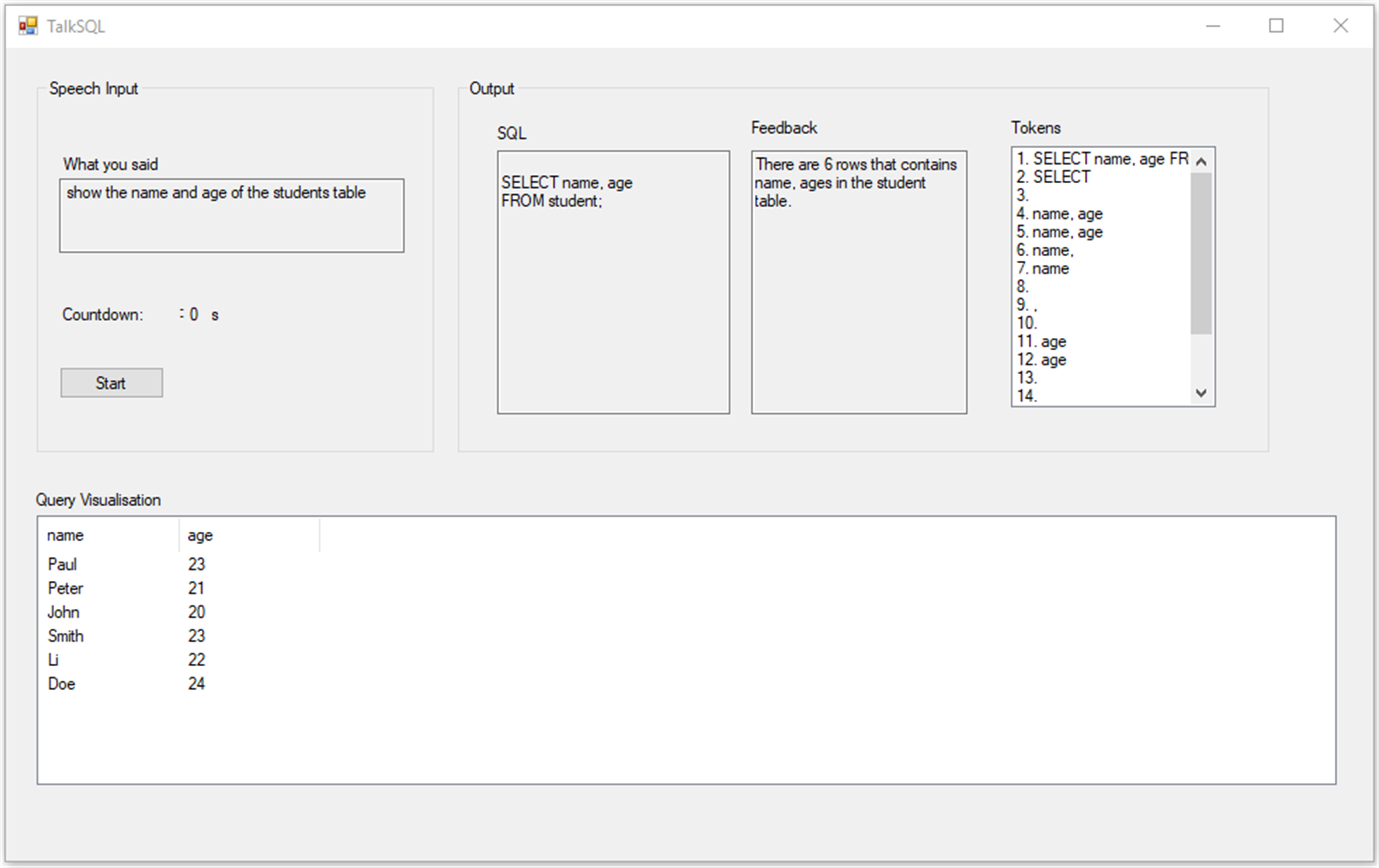}
	\caption{The User Interface of TalkSQL}
	\label{fig:ui}
\end{figure*}

In this section, we show how \texttt{TalkSQL}~(\autoref{fig:ui}) was implemented to translate speech inputs into a SQL query and the feedback is presented to the learner. Next, we describe the error-handling and how it can be refined to construct a query. Last, we present possible applications of \texttt{TalkSQL}.

\subsection{Query Translation}
In this section, we discuss the translation of user inputs from verbal inputs, before a query is generated and feedback is shown to a user. \texttt{TalkSQL} undergoes five phases before a query is generated as presented in \autoref{fig:narration}. In the first phase, words from a user are converted into text by the Google Cloud API service for further processing. This process is known as the speech recognition phase. The second phase takes the recognised text and match it to sequences of tokens. At this phase, whitespaces and noisy features such as comma, etc., are removed. This phase is known as the tokenisation phase. The third stage tags each tokenised sentence to their respective parts-of-speech (also known as POS Tagging). For example, the POS-tagging for the sentence will be:

\begin{center}
	\inline{\textit{Display}}{VB} \inline{\textit{the}}{DT} \inline{\textit{age}}{NN} \inline{\textit{from}}{IN} \inline{\textit{the}}{DT} \inline{\textit{student}}{NN} \inline{\textit{table}}{NN}
\end{center}

In this example, NN stands for a singular noun, DT - determiner, IN - preposition and VB stands for a verb. The fourth phase classifies each word to their syntactic parts. This phase is regarded as the syntactic analysis phase where each word is transformed into structures and shows relations. The fifth phase analyses the structure from the syntactic phase and check for meanings before words are mapped to their equivalent meaning. For example, words such as select, choose, pick and display are mapped to SELECT keyword. Words such as create, produce, make, build are mapped to the CREATE keyword. Also words such as update and amend are mapped to the UPDATE keyword. For the DELETE keyword, words such as remove, delete are mapped. Once attributes and keywords are mapped, the query is generated and used in a database.\\

In this section, we have presented the query translation phase as described in \autoref{fig:narration}. The next section presents the result of the CRUD statement.
\subsection{Results of the CRUD Operations}
\texttt{TalkSQL} was successful in generating feedback for the CRUD statements. We start with the CREATE command as seen in \autoref{algo:nlcreate}. This shows a request from a user. The converted SQL query is seen in \autoref{lst:create} and feedback provided in \autoref{algo:create}.

\begin{algorithm}
	\caption{Create command words from a user (NL)}\label{algo:nlcreate}
	\textit{Make a class table and specify ID as integer, name as alphanumeric entries with at most 45 characters and section as alphanumeric characters of at most 45 characters}
\end{algorithm}
\begin{lstlisting}[language=SQL,escapechar=@, morekeywords={DATABASE, SCHEMA},frame=bt,numbers=none, label={lst:create}, caption= CREATE SQL query] {Create SQL query}
CREATE TABLE class (
ID int,
name varchar(45),
section varchar(45)
);
\end{lstlisting}
\begin{algorithm}
	\caption{Feedback for the CREATE statement}\label{algo:create}
	\texttt{You have created a new table called class with the following information, ID stores integer values, name stores alphanumeric entries that contain 45 characters, section stores alphanumeric entries that contain 45 characters}
\end{algorithm}

The natural language that matches a SELECT command is presented in \autoref{algo:nlselect}. The row count function, count(), is used to retrieve the number of rows in a table. We added this function to the feedback generator to display the number of rows that appear in a table. The converted query version is presented in \autoref{lst:select}. The feedback for this query is presented in \autoref{algo:select}.

\begin{algorithm}
	\caption{Spoken words from a user (NL)}\label{algo:nlselect}
	\textit{Show the name and age of the student table}
\end{algorithm}
\begin{lstlisting}[language=SQL,escapechar=@, morekeywords={DATABASE, SCHEMA},frame=bt,numbers=none, label={lst:select}, caption= SELECT SQL query] {SELECT SQL query}
SELECT name, age
FROM student;
\end{lstlisting}
\begin{algorithm}
	\caption{Feedback for the SELECT statement}\label{algo:select}
	\texttt{There are 6 rows that contains name and ages in the student table}
\end{algorithm}

\autoref{algo:nlupdate} shows the update keyword specified in natural language spoken by the user to update a table. The NL query is converted into SQL queries in \autoref{lst:update}. The corresponding feedback is presented in \autoref{algo:update}.

\begin{algorithm}
	\caption{Spoken words from a user (NL)}\label{algo:nlupdate}
	\textit{Amend the student name to John whose id is equal to 6}
\end{algorithm}
\begin{lstlisting}[language=SQL,escapechar=@, morekeywords={DATABASE, SCHEMA},frame=bt,numbers=none, label={lst:update}, caption= UPDATE SQL query] {Update SQL query}
UPDATE student
SET name = 'John'
Where ID = 6;
\end{lstlisting}
\begin{algorithm}
	\caption{Feedback for the Update statement}\label{algo:update}
	\texttt{You have updated a record with a name called John, whose ID number is equal to 6 in the Student table.}
\end{algorithm}

\autoref{algo:nldelete} shows the delete keyword specified in NL, spoken by a user to delete a record from a table. \autoref{lst:delete} shows the converted SQL query. The feedback is presented in \autoref{algo:delete}.

\begin{algorithm}
	\caption{Spoken words from a user (NL)}\label{algo:nldelete}
	\textit{Remove a record from the lecturer table where the name is John}
\end{algorithm}

\begin{lstlisting}[language=SQL,escapechar=@, morekeywords={DATABASE, SCHEMA},frame=bt,numbers=none, label={lst:delete}, caption= DELETE SQL query] {DELETE SQL query}
DELETE FROM lecturer
Where name = 'John';
\end{lstlisting}

\begin{algorithm}
	\caption{Feedback for the Update statement}\label{algo:delete}
	\texttt{You have deleted 1 record from the lecturer table where the name is John}
\end{algorithm}

\subsection{Error-handling and Refinement}
To handle errors, \texttt{TalkSQL} uses two phases to resolve any ambiguity. The first phase requests a user to supply a missing attribute and table information, while the second phase finds missing keywords from a user statement.

\subsubsection{Table and Attribute Ambiguity}
\texttt{TalkSQL} handles errors that may occur in a conversational manner. If some information is required before a query can be generated, \texttt{TalkSQL} notifies the user to complete the statement with the missing information using a speech-based conversation. For example, if the user says: ``Show me the names in a table". It can be seen that the \textit{name} attribute appears in the three tables as specified in the schema in \autoref{fig:schema}. \texttt{TalkSQL} uses this template to notify the user:
\begin{center}
	\textit{Do you mean the \{attribute\} in the \texttt{Table 1 $|$ Table 2 $|$ \dots $|$ Table N}} table ?
\end{center}

In this case, the \texttt{TalkSQL} would ask the user: ``Do you mean the name in Lecturer, Student or Class table?''. Once the user clarifies the missing table. \texttt{TalkSQL} proceeds with generating the SQL query and the feedback is provided to the user. Similarly, if a user makes a request such as ``Display all details from the table" for a query to be generated and does not supply a table name, we use the template to request a user to supply the required table: 
\begin{center}
	\textit{Do you mean the \texttt{Table 1 $|$ Table 2 $|$ \dots $|$ Table N}} table ?
\end{center}

As specified in the template, the request would be: ``Do you mean the Lecturer, Student or Class table ?''. Here, \texttt{TalkSQL} would ask the user to complete the statement and once provided, a query will be generated.

\subsubsection{Keywords}
In situations where the keywords such as table and attributes information are missing, \texttt{TalkSQL} requests a user to provide additional information. For example, if a user requests: ``Show the information from the table''. \texttt{TalkSQL} uses this template to request the user to provide additional details before generating a query: 

\begin{center}
	\textit{Could you specify which \{Table\}, and its \{Attributes\}} ?
\end{center}

Once the information required is provided by the user, \texttt{TalkSQL} generates the query. We compared \texttt{TalkSQL} with a number of NLIDB systems as presented in \autoref{fig:reviewvsql}.

\begin{table*}[]
	\centering
	\caption{Existing tools versus \texttt{TalkSQL}}
	\label{fig:reviewvsql}
	\renewcommand{\arraystretch}{1.0}
	\resizebox{\textwidth}{!}	{\begin{tabular}{l|l|l|l|l|l|l|l|l|l|l|}
			\cline{2-10}
			& \multicolumn{1}{c|}{Focus} & Purpose & \multicolumn{4}{c|}{SQL Commands} & \multicolumn{3}{c|}{Feedback}  \\ \hline
			\multicolumn{1}{|l|}{}  & \begin{tabular}[c]{@{}l@{}}Target\\ (Novice)\end{tabular} & \begin{tabular}[c]{@{}l@{}}Teaching\\ and\\ Learning\\ aid\end{tabular}& CREATE    & SELECT   & UPDATE   & DELETE   & \multicolumn{1}{c|}{Speech} & \multicolumn{1}{c|}{Text} & Visuals \\ \hline
			\multicolumn{1}{|l|}{LUNAR} &\checkmark  &\checkmark  &\ding{55} &\checkmark  &\ding{55}  &\ding{55}  &\ding{55}  &\ding{55}  &\checkmark   \\ \hline
			\multicolumn{1}{|l|}{MaNaLa}& \checkmark &\checkmark  &\ding{55}   &\checkmark &\ding{55}   &\ding{55}      &\ding{55}  &\ding{55}   &\checkmark  \\ \hline
			\multicolumn{1}{|l|}{EchoQuery}  & \checkmark  &\checkmark  &\ding{55}  &\checkmark &\ding{55}  &\ding{55} &\checkmark  &\ding{55}   &\checkmark  \\ \hline
			\multicolumn{1}{|l|}{SpeakQL} &\checkmark &\checkmark  &\ding{55} &\checkmark  &\ding{55} &\ding{55} &\checkmark &\ding{55}  &\checkmark  \\ \hline
			\multicolumn{1}{|l|}{Cyrus}   &\checkmark  &\checkmark &\ding{55} &\checkmark &\ding{55} &\ding{55} &\checkmark &\ding{55} &\checkmark    \\ \hline
			\multicolumn{1}{|l|}{Ln2QL}  &\checkmark &\checkmark &\ding{55} &\checkmark &\ding{55} &\ding{55} &\ding{55} &\ding{55} &\checkmark   \\ \hline
			\multicolumn{1}{|l|}{\begin{tabular}[c]{@{}l@{}}TalkSQL\\ (Our tool)\end{tabular}} &\checkmark  &\checkmark  &\checkmark  &\checkmark   &\checkmark  &\checkmark  &\checkmark  &\checkmark  & \checkmark \\ \hline
	\end{tabular}}
\end{table*}

\subsection{Applications of TalkSQL}
In this section, we present possible applications of \texttt{TalkSQL} that were introduced in \autoref{figure:talk}. \texttt{TalkSQL} may find applications in:

\begin{enumerate}
	\item QA systems: As a tool, it may be used to provide a solution to a user's question specified in natural language.
	\item learning aids: It may be applied to assist users understand and improve their cognition of the SQL concepts.
	\item ITSs: \texttt{TalkSQL} may be used to provide immediate feedback to a learner without relying solely on an instructor. In addition, it can be used as a practice aid to assist learners understand SQL.
	\item assistive technologies: For a visually impaired learner, \texttt{TalkSQL} may be used to enhance SQL learning, since it is hands-free and provides feedback in speech forms.
	\item improving SQL comprehension: Generally, as a tool, it may be used by non-technical end-users in industry and students in higher institutions of learning to improve their SQL understanding.
\end{enumerate}

\section{Scope and Limitations}
In this section, we present the scope and limitations of \texttt{TalkSQL}. Up until this point, \texttt{TalkSQL} can perform error-checking and provide feedback for simple CRUD operations. The tool uses REs for the recognition of the simple CRUD operations in SQL. A noticeable feature that \texttt{TalkSQL} cannot handle is generating feedback for queries in complex forms -- especially queries enclosed with balanced parentheses (See \autoref{lst:label10}). 

\begin{lstlisting}[language=SQL,escapechar=@, morekeywords={TRUNCATE},frame=bt, numbers=none, label={lst:label10}, caption={A non-regular nested SQL query statement}]
SELECT firstname
FROM   employee 
WHERE  empid IN (
                SELECT DISTINCT
                empid
                FROM 
                employee);
\end{lstlisting}

The example in \autoref{lst:label10} contains two nested SELECT statements. The inner query within the parentheses displays only the empid information. The outer query displays the firstname from the employee table. These types of queries can only be recognised by a CFG -- a type of irregular language. It is interesting to note that REs are only used for lexical analysis tasks rather than syntactic parsing. A parser generator such as Coco/R~\citep{mossenbock20051} will fix this issue. 

\section{Chapter Summary}
In this chapter, a speech-based tool called \texttt{TalkSQL} was introduced, that takes speech inputs from a user, converts these words into SQL queries, and then returns a feedback to a user. We improved an existing NLIDB framework to accommodate the CRUD operations which allows a user to create, retrieve, modify and delete data from a database. \texttt{TalkSQL} uses REs to recognise these SQL operations and generate a feedback to be shown to a user. \texttt{TalkSQL} was able to recognise simple query commands that do not contain balanced parentheses. As a tool, it may find applications in real-world scenarios and improve SQL learning.\\

\par \autoref{part4} contains two chapters. The first chapter, \autoref{ch:eproto}, contains the evaluation results of the prototypes developed from \autoref{ch:regular} to \autoref{ch:ssql}. The conclusion of this thesis is presented in \autoref{ch:conclusion}.

\ctparttext{Throughout this study, we have presented a number of algorithms together with some software prototypes that assist end-users comprehend SQL queries. These prototypes are (i) \texttt{S-NAR}, a narrator for explaining SQL queries using predefined templates; (ii) \texttt{SQL-Narrator}, an improved SQL narrator for nested queries; (iii) \texttt{Narrations-2-SQL}, a tool that translates a narration into SQL queries; (iv) \texttt{SQL Visualiser}, a visualisation tool that uses the drag and drop interaction for query generation; and (v) \texttt{TalkSQL}, a speech-based query tool, that converts speech into SQL queries. For each prototype, we examine end-users' perceptions and form a basis for future studies. Many evaluations were conducted using an online means, and where necessary, we undertook some performance evaluation.
	
This part consists of two chapters. In \autoref{ch:eproto}, we present the evaluation results for the prototypes designed in this study and provide the conclusion with discussions for future work in \autoref{ch:conclusion}.

}
\part{Evaluations, Conclusions and Future Work}\label{part4}
\chapter{Evaluation of Prototypes}\label{ch:eproto} 
\section{Introduction}
\sloppypar
\lettrine[lines=3,loversize=0.1]{T}\normalsize{hroughout} this study, we have developed a number of prototypes from \autoref{ch:regular} to \autoref{ch:ssql}.  These tools have contributed immensely to the problem of SQL understanding and synthesis. In this chapter, we present the result of the evaluation that was conducted for each of the studies. Majority of the evaluation was conducted through an online survey designed as questionnaires. These questionnaires were designed using Google Forms. A link was sent to all participants at the university for their contribution. The questions used for these surveys are provided in \autoref{ch:app-question-results}.

\section{Evaluation of S-NAR}\label{prot1}
This section presents the accuracy of the \texttt{S-NAR} tool. \texttt{S-NAR} was tested on 5000 queries scrapped from the Internet (indexed in \autoref{sql-queries-dataset}), and it successfully narrated a subset of these queries (96\%). This is presented in \autoref{equations}.

\begin{equation}\label{equations}
\begin{aligned}
\text{Accuracy}=\frac{4824}{5000} = 96.48\%
\end{aligned}
\end{equation}

We noticed that the remaining 4\% of the queries were nested queries and had balanced parentheses in them. This could only be recognisable by a CFG or higher classes of formal abstract machines. This was already addressed in \autoref{ch:cfg}.

\section{Evaluation of the SQL Narrator}\label{prot2}
In this section, we present the result of the evaluation carried out using an online survey from 161 students at the University of the Witwatersrand. The participants were mostly undergraduate CS students at the University of the Witwatersrand whom had already been taught simple and nested queries. Participation was strictly non-mandatory and participants' profiles were kept anonymous. The survey is available in the link: \texttt{https://goo.gl/CQcVZN}.

\subsection{Result of the Survey}
\begin{figure}[!htb]
	\centering
	\begin{minipage}{.5\textwidth}
		\centering
		\includegraphics[width=200px]{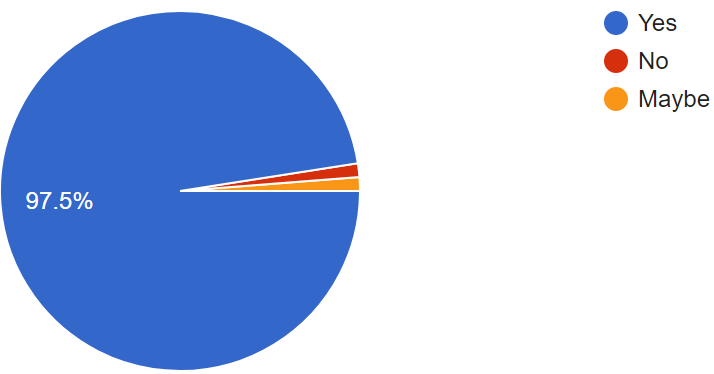}
		\caption[Knowledge of simple queries]{Knowledge of simple queries\\}
		\label{fig:ex3-a}
	\end{minipage}%
	\begin{minipage}{0.5\textwidth}
		\centering
		\includegraphics[width=200px]{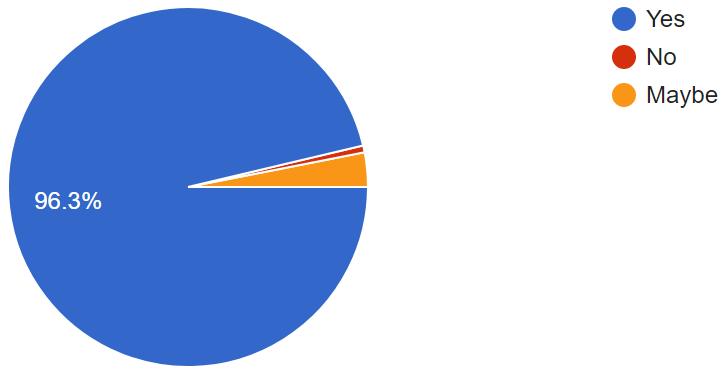}
		\caption[Are nested queries difficult?]{Are nested queries difficult? \\}
		\label{fig:ex3-b}
	\end{minipage}
\end{figure}

Out of the 161 responses received, 97.5\% claimed they are familiar with SQL queries, 1.2\% indicated no familiarity and 1.2\% were unsure about their response - this is presented in \autoref{fig:ex3-a}. In addition, we asked the participants if they think nested queries are difficult and 96.3\% agreed that nested queries are difficult, 8.5\% claimed they do not think nested queries are difficult, while 3.4\% were unsure about their responses (see \autoref{fig:ex3-b}).\\

\begin{figure}[!htb]
	\centering
	\begin{minipage}{.5\textwidth}
		\centering
		\includegraphics[width=200px]{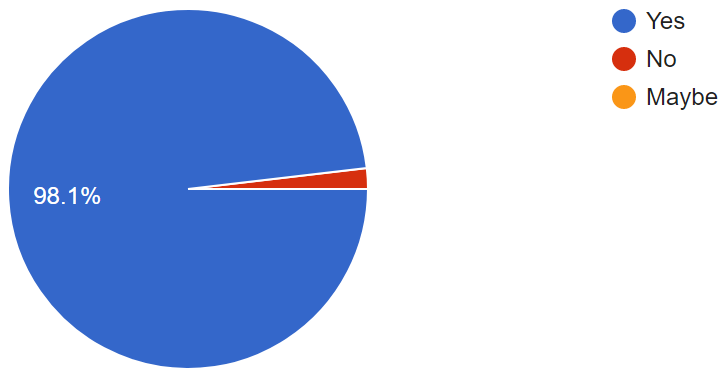}
		\caption[Comprehend nested queries using the SQL Narrator]{Comprehend nested queries \\ using the SQL Narrator\\}
		\label{fig:ex3-c}
	\end{minipage}%
	\begin{minipage}{0.5\textwidth}
		\centering
		\includegraphics[width=270px]{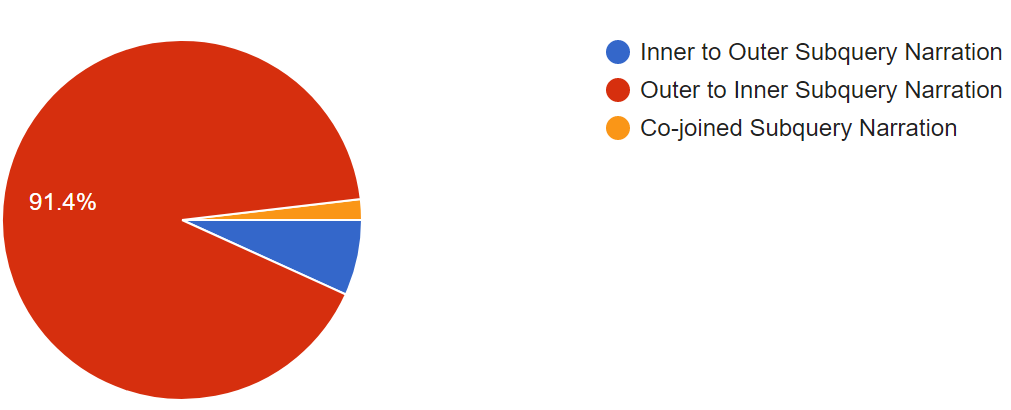}
		\caption[Comprehend which of the generated narrations]{Comprehend which of the \\ generated narrations \\}
		\label{fig:ex3-d}
	\end{minipage}
\end{figure}

\begin{figure*}[h!]
	\centering
	\includegraphics[width=430px]{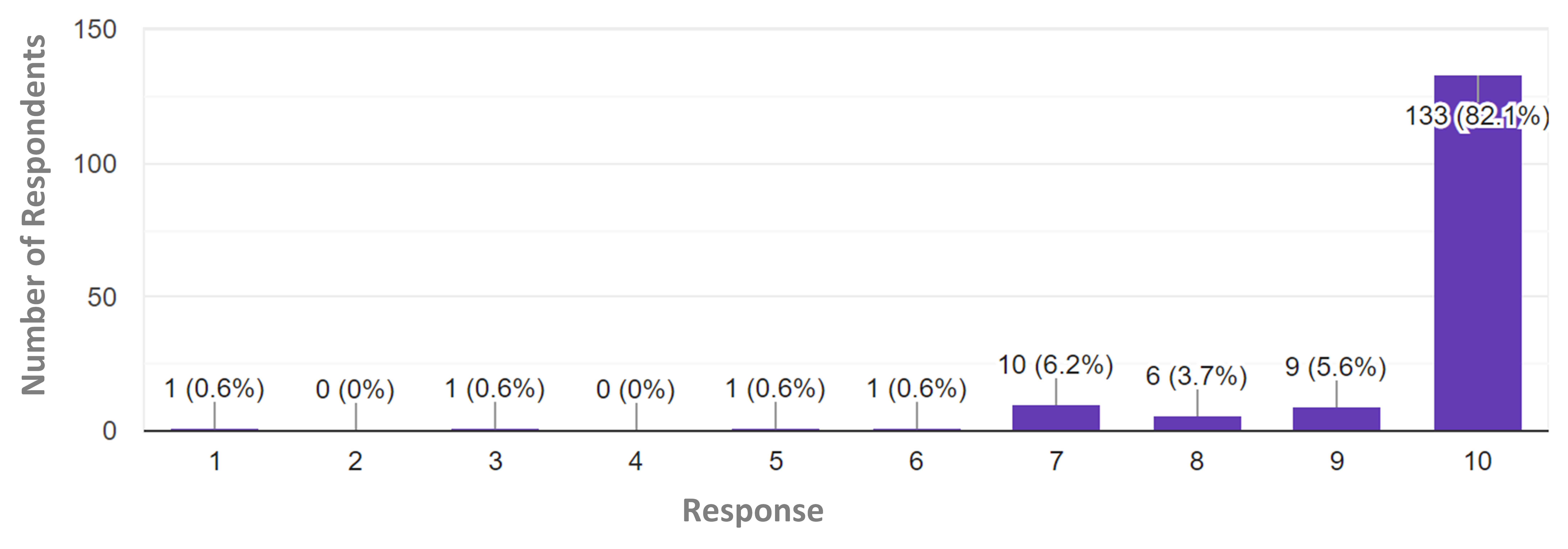}
	\caption[Rate the SQL Narrator]{Rate the SQL Narrator}
	\label{fig:ratenarr}
\end{figure*}

A total of 98.1\% agreed that they were able to comprehend nested queries using the \texttt{SQL Narrator} and 1.9\% claimed that they could not comprehend the nested queries using the narrator (in \autoref{fig:ex3-c}). Furthermore, we asked the participants to answer which of the generated narrations they were able to comprehend (see \autoref{fig:ex3-d}). About 91.4\% strongly agreed that the outer to inner subquery narration was easier to comprehend while 1.9\% chose the co-joined subquery narration, and 6.8\% agreed with the inner to outer subquery narration. It is interesting to note that the majority of the participants agreed with the outer to inner narration for the subquery. Perhaps, they were comfortable with the chronological flow of the explanation of the subquery. In addition, we asked the participants to rate the \texttt{SQL Narrator} on a scale of 1- 10; it was seen that majority agreed that the tool was useful to them (see \autoref{fig:ratenarr}). 

\subsection{Respondent Feedback}
The last part of the questionnaire was open-ended. We asked the respondent to provide any recommendation they might have for us to improve the tool. Some of the recommendations provided by the participants are given in this section. With the result of the evaluation, and the recommendation provided, we can conclude that adopting this tool will improve student comprehension of nested queries. Some comments that were anonymously given by the participants are as follows.

\begin{flushright}{\slshape    
	    ``\texttt{I would not suggest anything, as the tool makes nested queries very easy to understand.}''} \\ \medskip
\end{flushright}

\begin{flushright}{\slshape    
		``\texttt{This tool is easy to understand compared to other narration tools that I have used.}''} \\ \medskip
\end{flushright}

\begin{flushright}{\slshape    
		``\texttt{It's much nicer to see what the queries are, and explanations of each of them were provided.}''} \\ \medskip
\end{flushright}

\begin{flushright}{\slshape    
		``\texttt{It was easier to read the sentences and follow it through. I am sure the narrator will be more useful with more complicated examples, this will benefit a novice programmer altogether.}''} \\ \medskip
\end{flushright}

\begin{flushright}{\slshape    
		``\texttt{Nothing to suggest. It seems to do exactly what it's meant to and does it really well. I liked the fact that the narrations are well structured sentences and aren't complicated explanations.}''} \\ \medskip
\end{flushright}

\section{Evaluation of Narrations-2-SQL}\label{prot3}
The evaluation was carried out in a two-fold manner: (1) Using the crowdsourced XNorthwind dataset indexed in \autoref{ch:app-natural-sql}, we show the accuracy of the \texttt{Narrations-2-SQL} tool. (2) Using human subjects, we show the users' perceptions of the tool we have designed for this study. The survey can be accessed through \texttt{https://bit.ly/2m62guw}.

\subsection{Accuracy of Narrations-2-SQL}
We used the crowdsourced XNorthwind dataset to train our tool. The end-users were asked to test \texttt{Narrations-2-SQL} with their narrations. We discovered that about 180 narrations from the end-users were able to successfully match an SQL query. To determine the accuracy, we take:  \\

\begin{equation}\label{equation}
\begin{aligned}
\text{Accuracy}=\frac{180}{204} = 88\%
\end{aligned}
\end{equation} 

It is worth noting that the decrease in the accuracy was due to some of the narrations being outside are out of our JFA training data. In future work, we will improve our JFA to recognise more queries by adding more keywords and semantic rules. 
\subsection{Survey Design} 
The survey was carried out through an online means and feedback was received from 162 participants. The results of the survey in this section and was strictly anonymous. The survey can be accessed via \texttt{https://shorturl.at/aJMN8}.

\subsubsection{Result of the Survey}
\begin{figure}[!htb]
	\centering
	\begin{minipage}{.5\textwidth}
		\centering
		\includegraphics[width=200px]{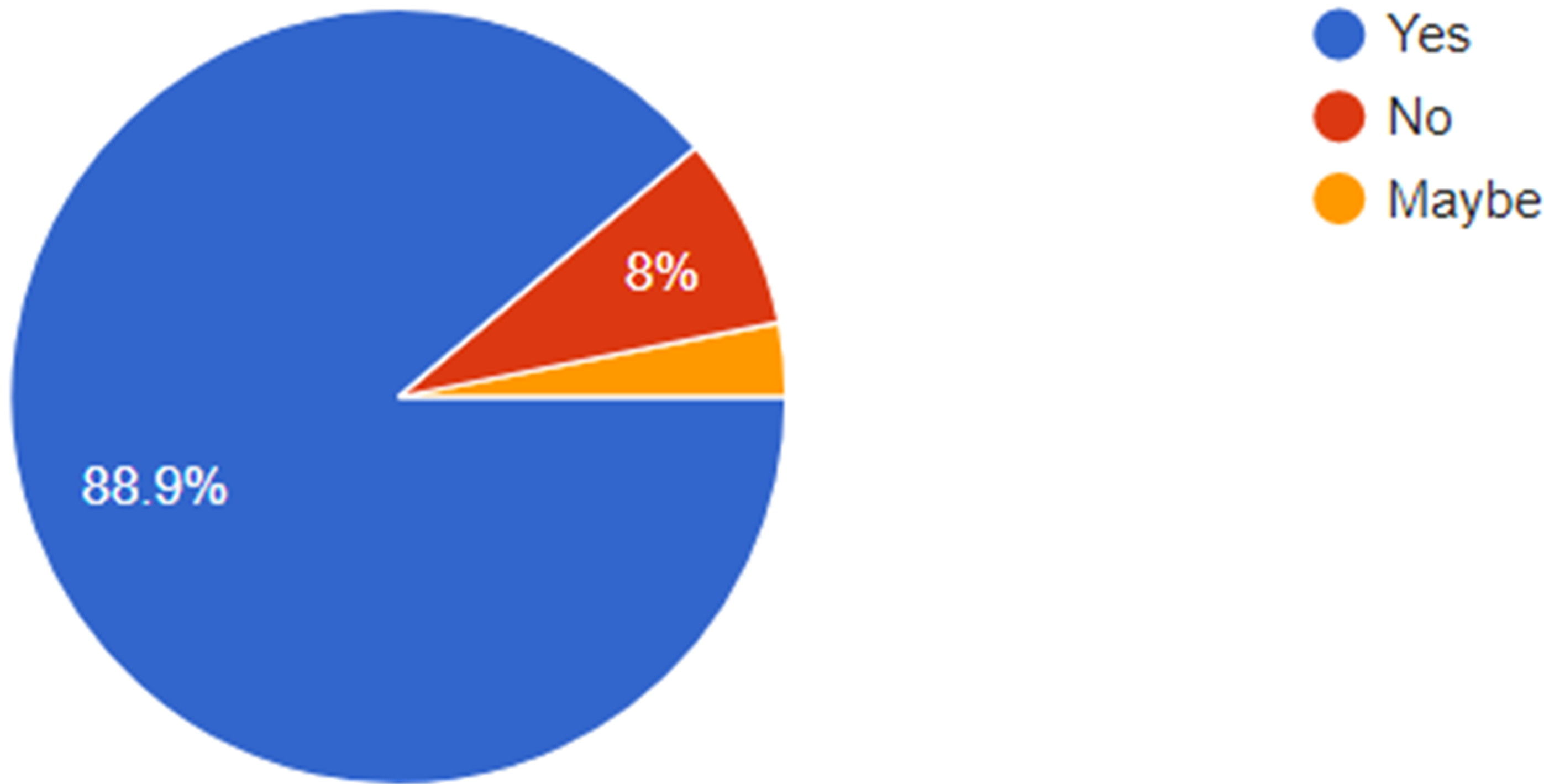}
		\caption[Familiar with SQL]{Familiar with SQL\\}
		\label{fig:proto4-1}
	\end{minipage}%
	\begin{minipage}{0.5\textwidth}
		\centering
		\includegraphics[width=200px]{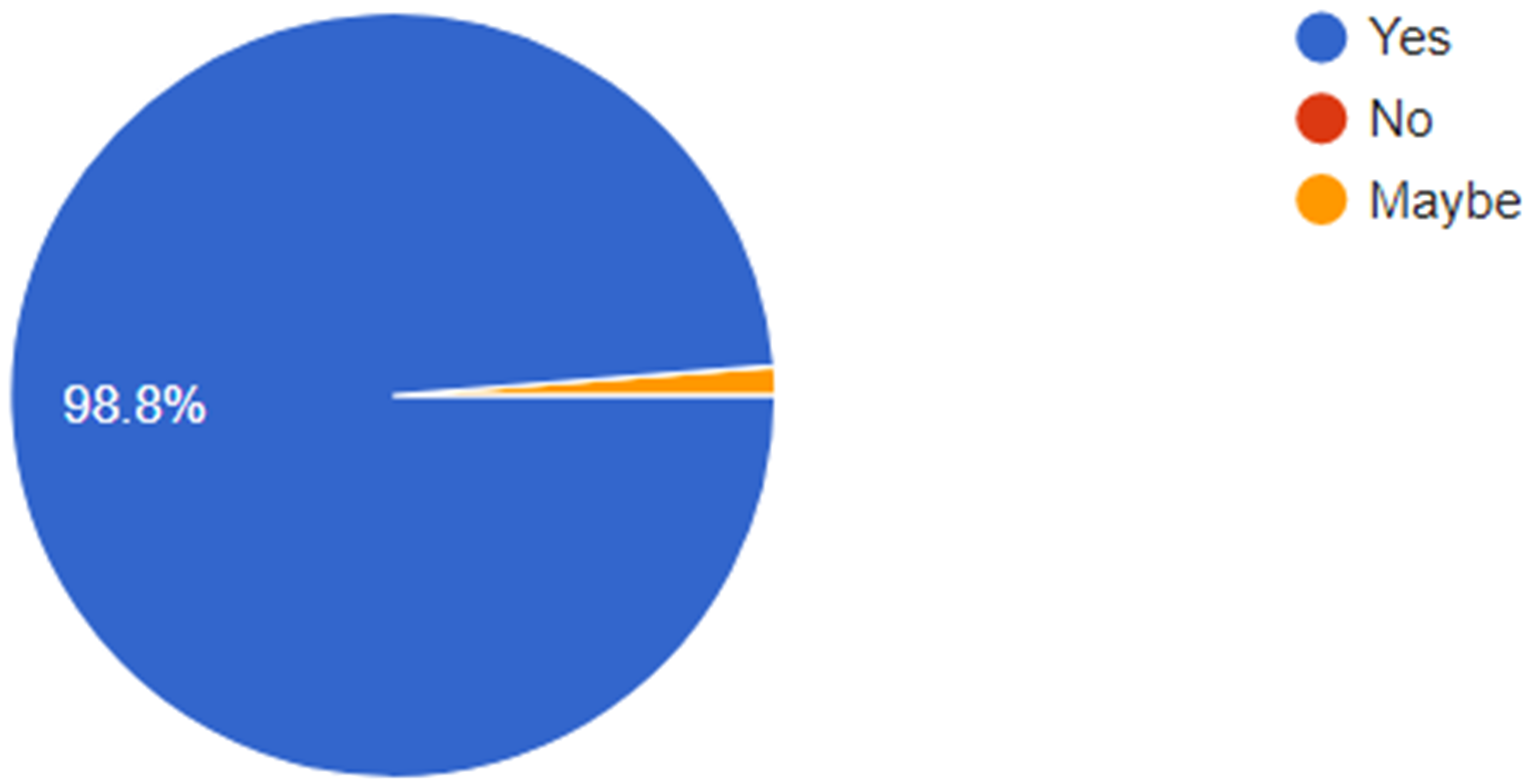}
		\caption[The user friendliness of Narrations-2-SQL]{The user friendliness of Narrations-2-SQL \\}
		\label{fig:proto4-2}
	\end{minipage}
\end{figure}

\begin{figure}[!htb]
	\centering
	\begin{minipage}{.5\textwidth}
		\centering
		\includegraphics[width=200px]{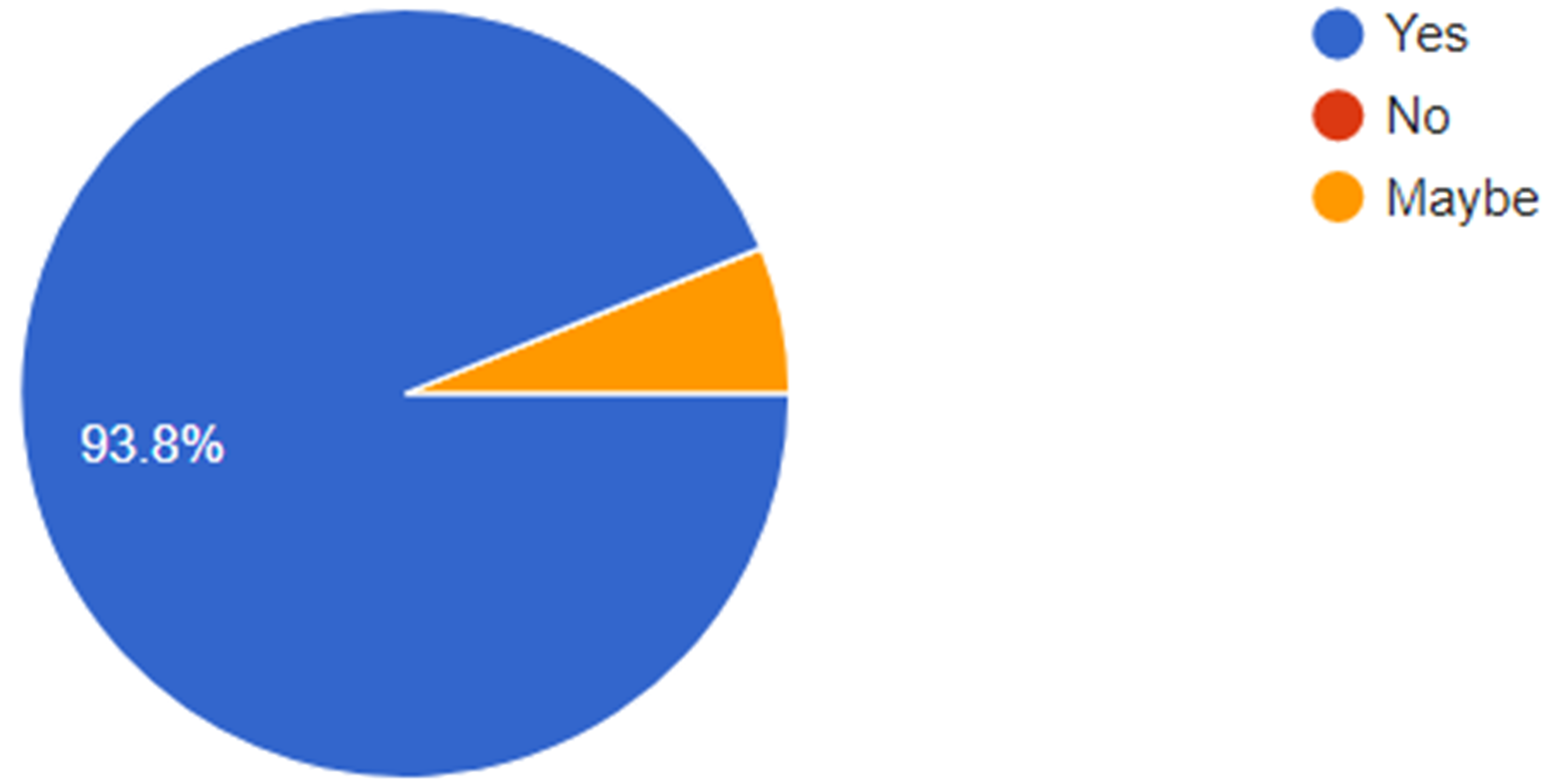}
		\caption[The generated SQL query a correct translation of narrations]{The generated SQL query a\\ correct translation of narrations\\}
		\label{fig:proto4-3}
	\end{minipage}%
	\begin{minipage}{0.5\textwidth}
		\centering
		\includegraphics[width=200px]{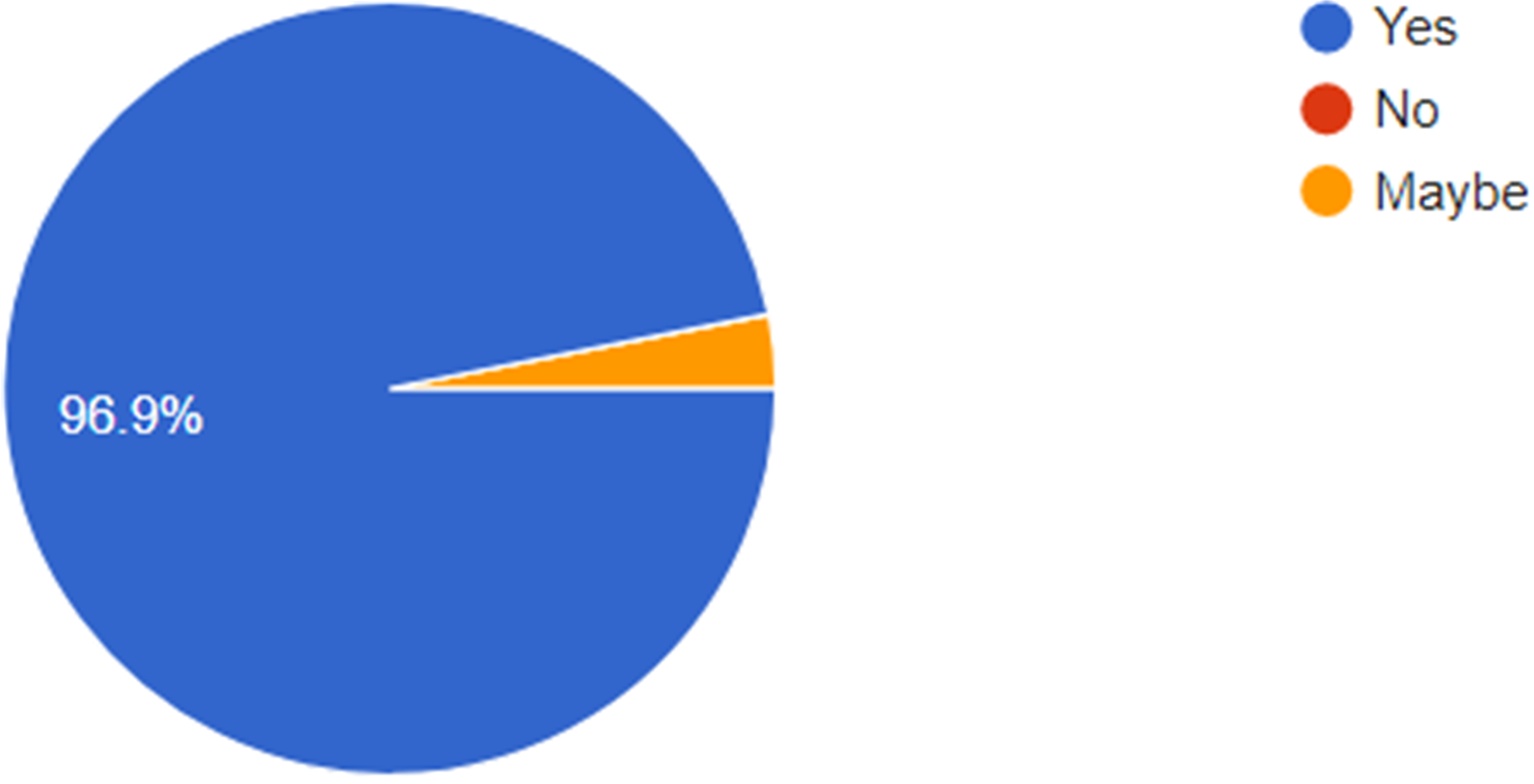}
		\caption[The tool will help end-users in industry with no knowledge of SQL]{The tool will help end-users in\\ industry with no knowledge of SQL \\}
		\label{fig:proto4-4}
	\end{minipage}
\end{figure}

A total of 162 responses were received, and 88.9\% agreed that they were familiar with SQL, 8.0\% admitted no familiarity of SQL and 3.1\% were unsure (see \autoref{fig:proto4-1}). In addition, the participants were asked if they thought \texttt{Narrations-2-SQL} is user-friendly; 98.8\% agreed that the tool was user-friendly and 1.2\% were unsure about their responses (see \autoref{fig:proto4-2}). They were also asked if they thought the generated SQL query was a correct translation of their narrations; about 93.8\% admitted yes and 6.2\% were unsure about their responses (see \autoref{fig:proto4-3}). \\ 

\begin{figure*}[h!]
	\centering
	\includegraphics[width=430px]{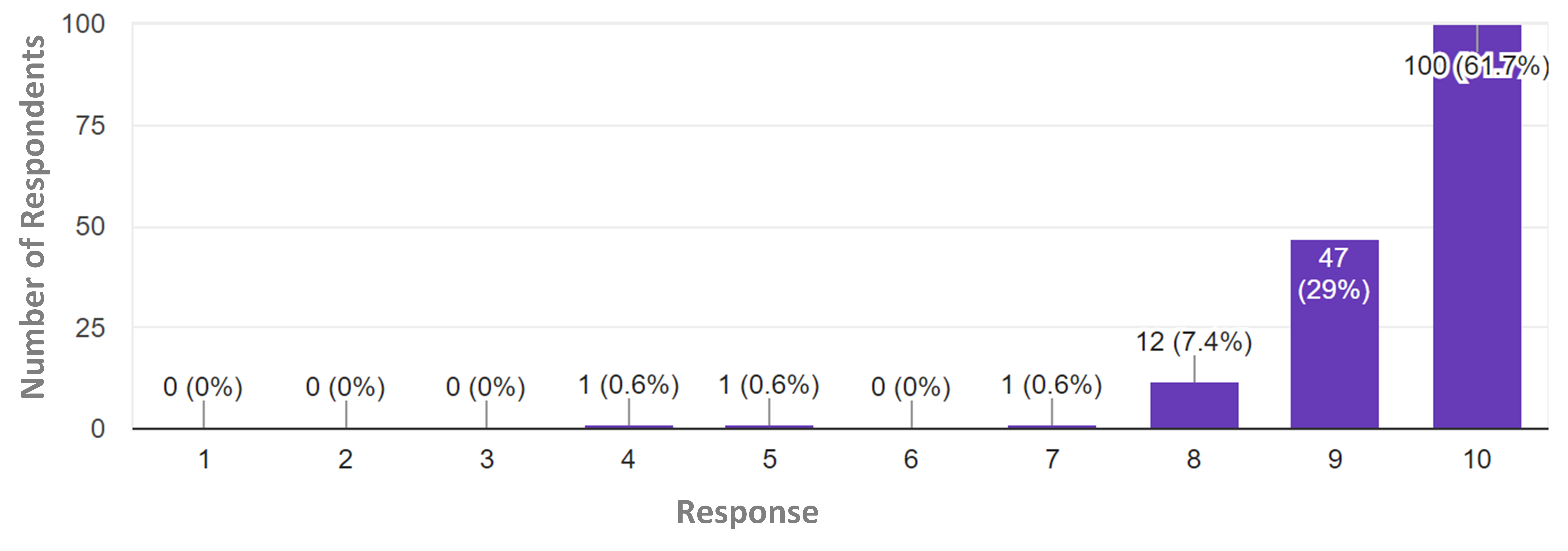}
	\caption[Rate the Narrations-2-SQL]{Rate the Narrations-2-SQL}
	\label{fig:ratenarr3}
\end{figure*}

When the users were asked if the tool will help end-users in industry with no knowledge of SQL, 96.9\% of them agreed that the tool will be helpful to industry users and 3.1\% were unsure (in \autoref{fig:proto4-4}). In addition, we asked the participants to rate the tool that we have developed on a scale of 1-10. The result is presented in \autoref{fig:ratenarr3}.

\subsubsection{Respondent Feedback}
In this section, we capture participants' comments about the \texttt{Narrations-2-SQL}. Open-ended questions were asked to get participants to get their insights about what they thought of \texttt{Narrations-2-SQL}. It was worth noting that the feedback provided was interesting and would enable us improve \texttt{Narrations-2-SQL} for future research. Some comments from the participants are as follows.

\begin{flushright}{\slshape    
		``\texttt{This is a very helpful tool and will improve a lot of successful my SQL code creation skills.}''} \\ \medskip
\end{flushright}

\begin{flushright}{\slshape    
		``\texttt{Great to use and very simple. It will be easy for people that do not even know much about SQL}''} \\ \medskip
\end{flushright}

\begin{flushright}{\slshape    
		``\texttt{I think this tool will help eliminate a lot of errors in SQL coding}''} \\ \medskip
\end{flushright}

\begin{flushright}{\slshape    
		``\texttt{I would honestly recommend this tool to anyone who intends to queries in SQL}''} \\ \medskip
\end{flushright}

\begin{flushright}{\slshape    
		``\texttt{I have two observations: Could you include voice prompt so that it will be easier to use, and could you increase the font size of the query result.}''} \\ \medskip
\end{flushright}

With these results, we conclude that automating narrations into SQL will be beneficial for end-users.

\section{Evaluation of SQL Visualiser}\label{prot4}
The evaluation of the \texttt{SQL Visualiser} was carried out using an online survey from 121 students from the University of the Witwatersrand. The respondents were mostly undergraduate CS students and majority of them had knowledge of SQL. The questionnaire is split into two parts: the first required the students to answer general questions about their knowledge of visualisers, while the second focused on their perception of the \texttt{SQL Visualiser} we have designed in this research. In addition, the students were asked to provide feedback on ways to improve the \texttt{SQL Visualiser}. Constructive feedback was received. The survey is available in the link: \texttt{https://bit.ly/2m1Pf4Y}.

\subsection{Result of the Survey}
\begin{figure}[!htb]
	\centering
	\begin{minipage}{.5\textwidth}
		\centering
		\includegraphics[width=200px]{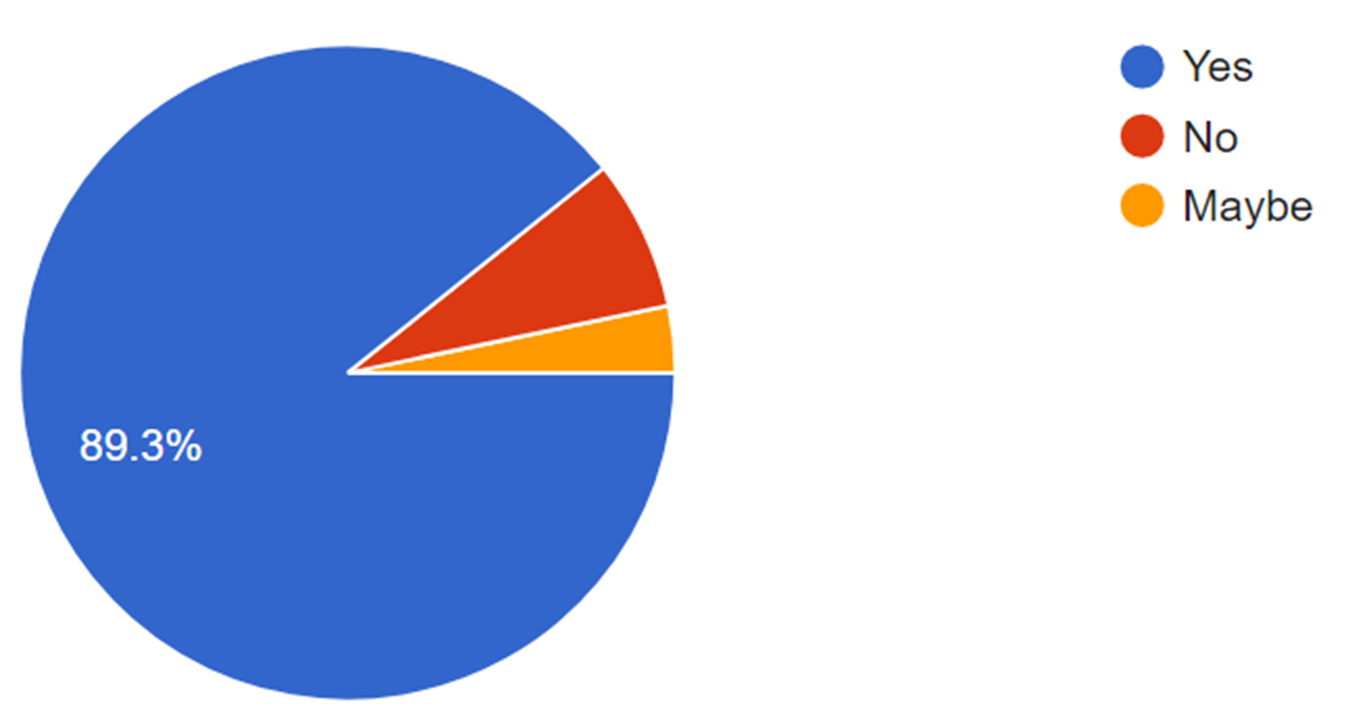}
		\caption[Knowledge of SQL]{Knowledge of SQL\\}
		\label{fig:proto5-1}
	\end{minipage}%
	\begin{minipage}{0.5\textwidth}
		\centering
		\includegraphics[width=200px]{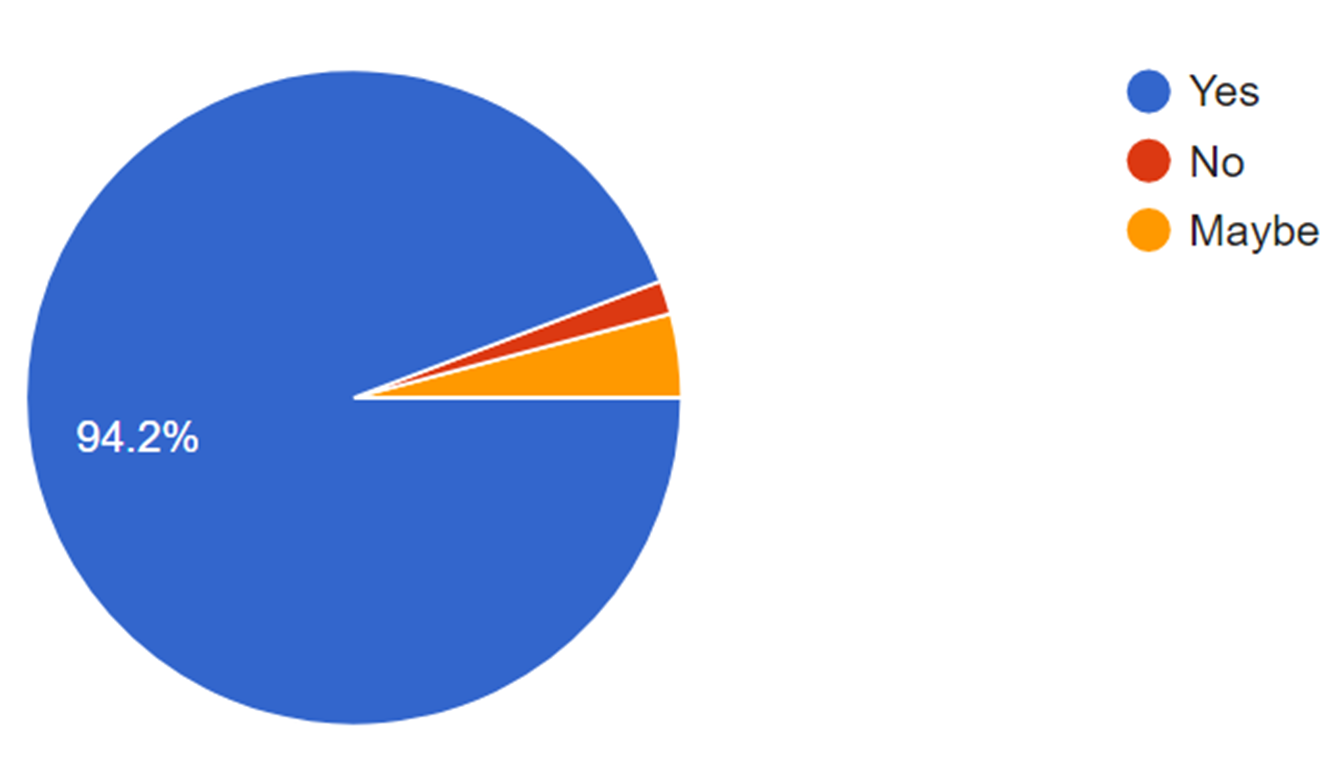}
		\caption[The user friendliness of the SQL Visualiser]{The user friendliness of the SQL Visualiser \\}
		\label{fig:proto5-2}
	\end{minipage}
\end{figure}

Out of the 121 responses, 89.3\% admitted to have knowledge of SQL, 7.4\% affirmed no knowledge of SQL and 3.3\% were unsure about their responses -- this is presented in \autoref{fig:proto5-1}. Of the participants, 94.2\% agreed that the \texttt{SQL Visualiser} was user-friendly, 4.1\% admitted that the visualiser was not user-friendly, and 1.7\% were unsure about their responses (see \autoref{fig:proto5-2}). \\

\begin{figure}[!htb]
	\centering
	\begin{minipage}{.5\textwidth}
		\centering
		\includegraphics[width=200px]{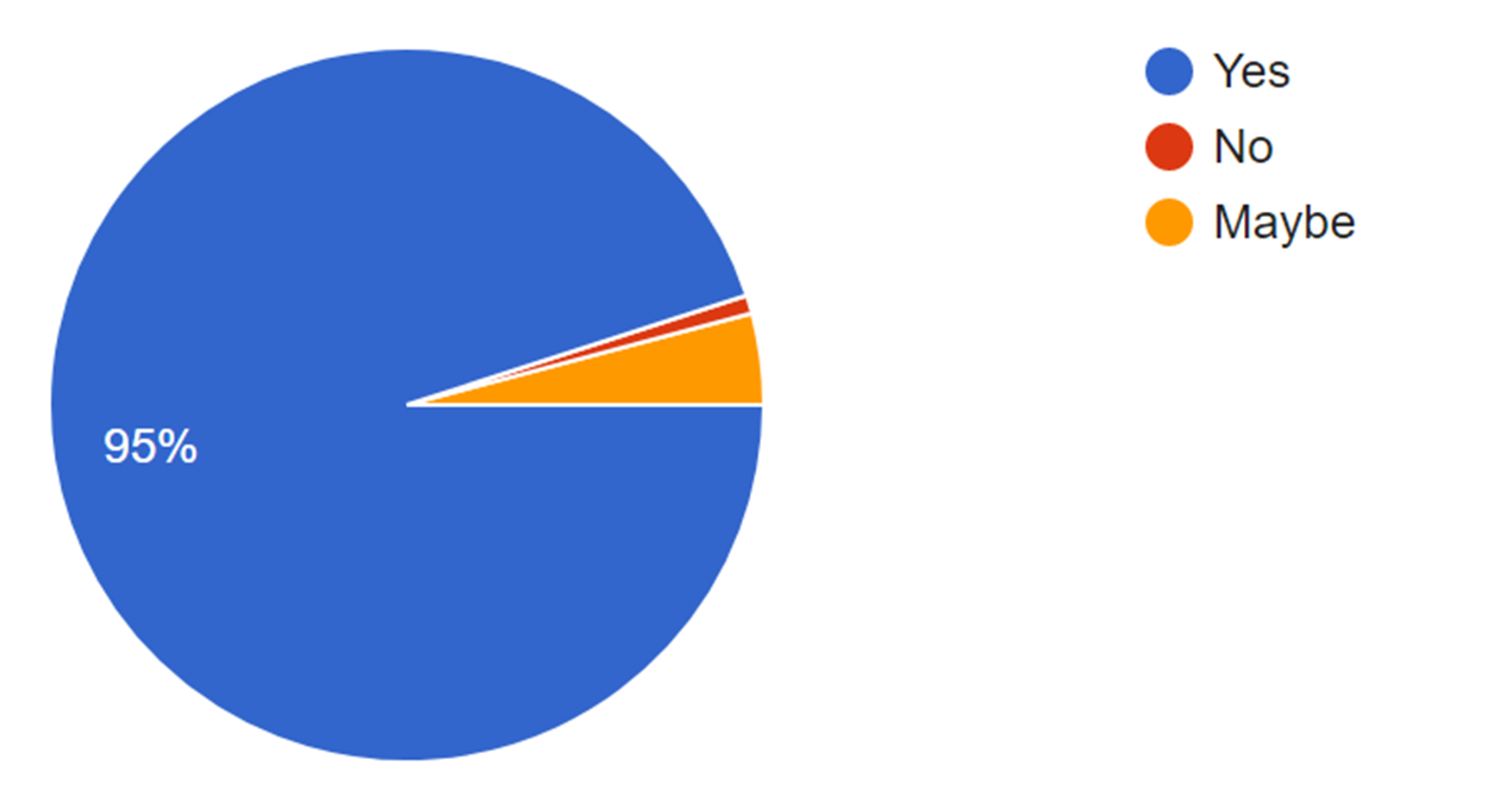}
		\caption[Ability to synthesise basic SQL queries using the visualiser]{Ability to synthesise basic\\ SQL queries using the visualiser\\}
		\label{fig:proto5-3}
	\end{minipage}%
	\begin{minipage}{0.5\textwidth}
		\centering
		\includegraphics[width=200px]{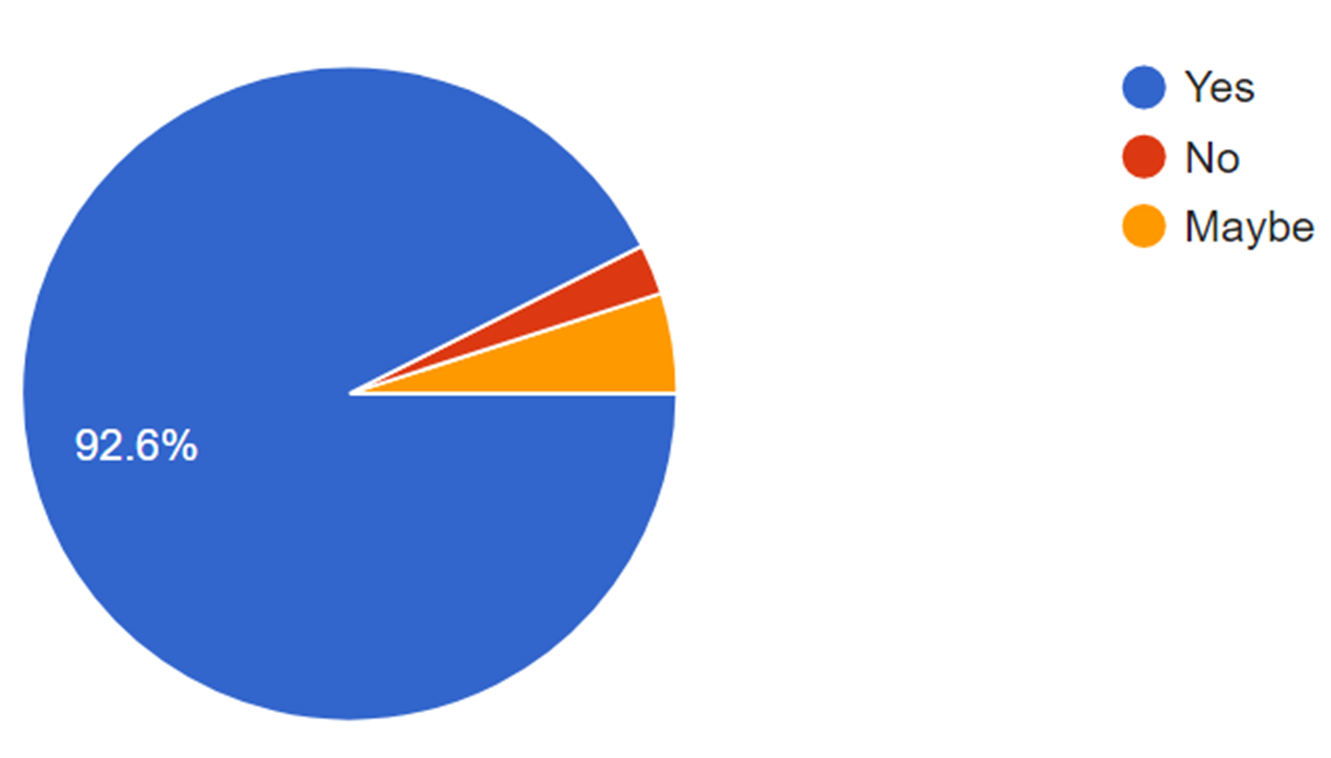}
		\caption[Visual specifications helped comprehend SQL]{Visual specifications helped \\comprehend SQL\\}
		\label{fig:proto5-4}
	\end{minipage}
\end{figure}

\begin{figure*}[h!]
	\centering
	\includegraphics[width=420px]{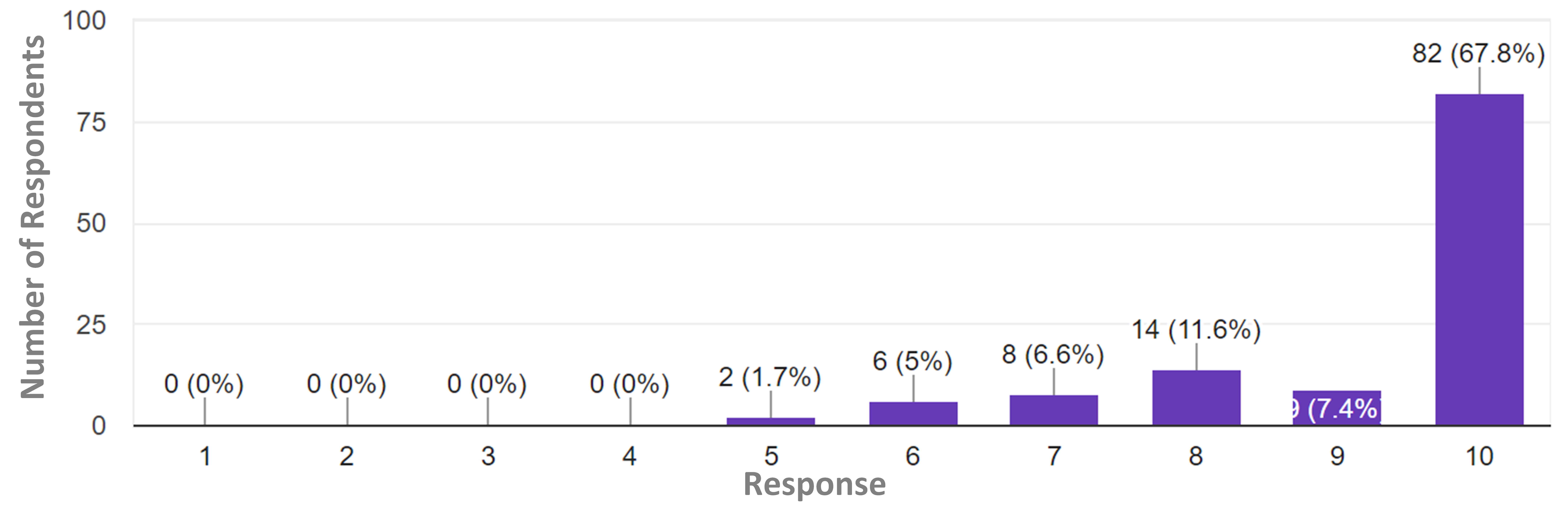}
	\caption[Rate the SQL Visualiser]{Rate the SQL Visualiser}
	\label{fig:ratenarr4}
\end{figure*}

Furthermore, the students were asked if they were able to synthesise basic SQL queries using the visualiser (in \autoref{fig:proto5-3}). About 95\% agreed that the visualiser helped them comprehend SQL queries, 4\% admitted that they find the visualiser difficult to use and 1\% stayed indifferent. In addition, 92.6\% admitted that visual specifications helped them understand the syntax of the SQL queries, 5\% did not agree and 2.4\% stayed indifferent (see \autoref{fig:proto5-4}). We asked the participants to rate the \texttt{SQL Visualiser}; their responses are captured in \autoref{fig:ratenarr4}.

\subsection{Respondent Feedback}
The feedback the respondents made was helpful and some respondents highlighted some limitations of the \texttt{SQL Visualiser}. Examples of the positive comments made were:

\begin{flushright}{\slshape    
		``\texttt{The visualiser was extremely helpful. I liked the way the pictures assisted in displaying the query. As a learner, I think it will help other students understand SQL queries better and improve our knowledge of the SQL concept.}''} \\ \medskip
\end{flushright}

\begin{flushright}{\slshape    
		``\texttt{Icons were a good idea and helpful. It is not boring.}''} \\ \medskip
\end{flushright}

\begin{flushright}{\slshape    
		``\texttt{Far much better visualiser l have used so far.}''} \\ \medskip 
\end{flushright}

\begin{flushright}{\slshape    
		``\texttt{It seems simple and straight-forward, the user does not have to try hard to understand its functionality and operation.}''} \\ \medskip
\end{flushright}

\setlength{\parindent}{10ex}
\noindent Some of the limitations mentioned by the respondents include:

\begin{flushright}{\slshape    
		``\texttt{There should be explanations, the use of comments in the query text box would be extremely useful. I have used SQL before, therefore this is mostly targeted at novice users. Other methods should be integrated to cater for expert users.}''} \\ \medskip
\end{flushright}

\begin{flushright}{\slshape    
		``\texttt{The icons colour choice is boring.}''} \\ \medskip 
\end{flushright}

\begin{flushright}{\slshape    
		``\texttt{Possibly increase text sizing for the visually impaired user.}''} \\ \medskip
\end{flushright}

\begin{flushright}{\slshape    
		``\texttt{This tool should allow users to choose their own icons (e.g. students could be a different icon).}''} \\ \medskip
\end{flushright}

\noindent The majority of the students commended the use of visual specifications to aid their comprehension. These results are consistent with the evaluation carried out on a visualiser \citep{satyanarayan2014lyra} for program comprehension, where users perceptions supported the usefulness and importance of visual specifications. We believe that adopting this tool in higher institutions of learning will improve students' comprehension of SQL. 

\section{Evaluation of TalkSQL}\label{prot5}
The survey was carried out online and the feedback was received from 113 participants. The majority of the participants were undergraduate CS students taking a database course and most of them were familiar with SQL. The survey can be accessed through \texttt{https://bit.ly/2ktYNVX}. The result of the survey is presented in the next section.

\subsection{Result of the Survey}
\begin{figure}[!htb]
	\centering
	\begin{minipage}{.5\textwidth}
		\centering
		\includegraphics[width=200px]{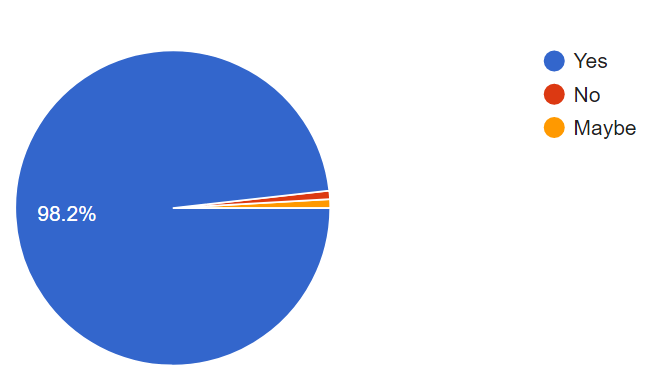}
		\caption[Knowledge of SQL]{Knowledge of SQL\\}
		\label{fig:proto6-1}
	\end{minipage}%
	\begin{minipage}{0.5\textwidth}
		\centering
		\includegraphics[width=200px]{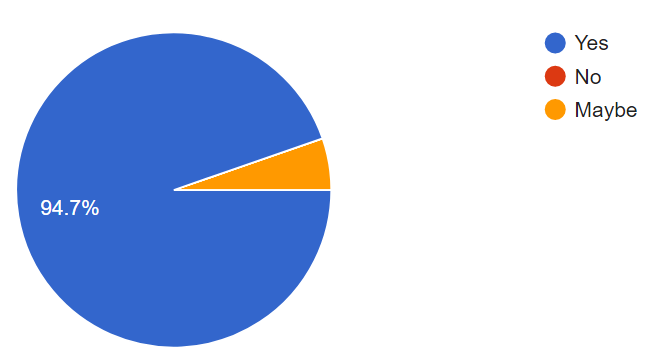}
		\caption[Ease of use of TalkSQL]{Ease of use of TalkSQL\\}
		\label{fig:proto6-2}
	\end{minipage}
\end{figure}

\begin{figure}[!htb]
	\centering
	\begin{minipage}{.5\textwidth}
		\centering
		\includegraphics[width=200px]{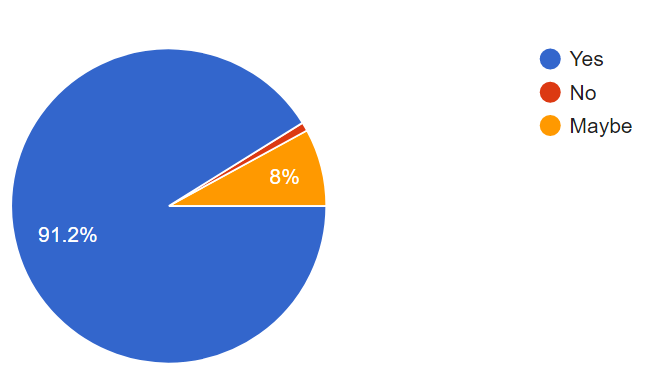}
		\caption[Able to understand CRUD operations using TalkSQL]{Able to understand CRUD operations\\ using TalkSQL\\}
		\label{fig:proto6-3}
	\end{minipage}%
	\begin{minipage}{0.5\textwidth}
		\centering
		\includegraphics[width=200px]{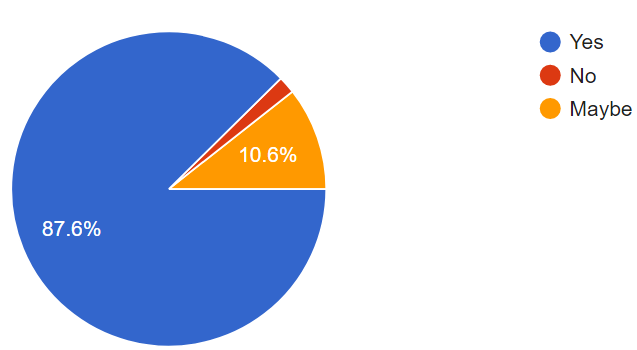}
		\caption[Able to assist the visually impaired understand SQL]{Able to assist the visually impaired\\ understand SQL\\}
		\label{fig:proto6-4}
	\end{minipage}
\end{figure}

\begin{figure*}[h!]
	\centering
	\includegraphics[width=420px]{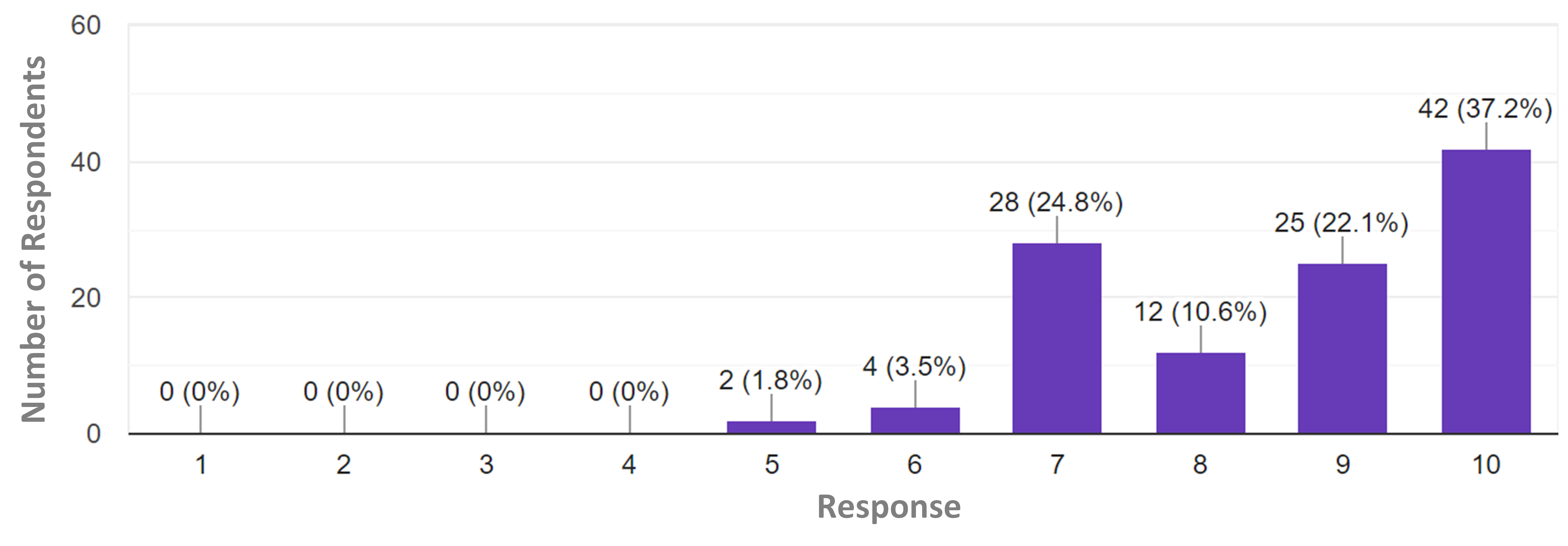}
	\caption[Rate the TalkSQL tool]{Rate the TalkSQL tool}
	\label{fig:ratenarr5}
\end{figure*}

We received a total of 113 responses, out of which 98.2\% agreed that they had knowledge of SQL, 0.9\% were unsure and 0.9\% indicated no knowledge of SQL -- this is shown in \autoref{fig:proto6-1}. Furthermore, we asked the participants if the \texttt{TalkSQL} tool was easy to use; 98.7\% claimed that the tool was easy to use and 5.3\% were not sure of their responses (see \autoref{fig:proto6-2}). About 91.2\% claimed that they were able to understand the CRUD commands, 8\% were indifferent and 0.9\% claimed they could not understand the CRUD operation using \texttt{TalkSQL} (see \autoref{fig:proto6-3}). Next, we asked the participants if they think a visually impaired learner could understand SQL with \texttt{TalkSQL}, a total of 87.6\% agreed, 10.6\% were unsure and 1.8\% do not agree that this category of learners would understand SQL query using \texttt{TalkSQL} (in \autoref{fig:proto6-4}). We asked the participants to rate TalkSQL, their responses are indicated in \autoref{fig:ratenarr5}.

\subsubsection{Respondent Feedback}
Open-ended questions were asked from the participants to indicate what we could do to improve \texttt{TalkSQL}. The feedback received was quite helpful, while some highlighted a couple of limitations. Some positive feedback received are as follows:

\begin{flushright}{\slshape    
		``\texttt{It is really nice and very helpful. It will help learners who are colour-blinded to learn SQL.}''} \\ \medskip
\end{flushright}

\begin{flushright}{\slshape    
		``\texttt{I liked the part where a voice feedback was read back to me, no need to type!.}''} \\ \medskip
\end{flushright}

\begin{flushright}{\slshape    
		``\texttt{Very interesting tool! I could rephrase my words over again. Nice!.}''} \\ \medskip
\end{flushright}

\begin{flushright}{\slshape    
		``\texttt{No suggestions. TalkSQL works perfectly.}''} \\ \medskip
\end{flushright}

\noindent Some limitations mentioned were:

\begin{flushright}{\slshape    
		``\texttt{Please, increase the font size!}''} \\ \medskip
\end{flushright}

\begin{flushright}{\slshape    
		``\texttt{What if my voice is strained, I think you should add typing features.}''} \\ \medskip
\end{flushright}

\begin{flushright}{\slshape    
		``\texttt{This tool should cater for nested SQL queries as well.}''} \\ \medskip
\end{flushright}

\noindent Indeed, the feedback received will help us improve \texttt{TalkSQL} and make it accessible for learners that require its services. These results are consistent with the survey by Wilson~\citep{wilson2010may} where the participants agreed that verbal specifications can assist learners understand SQL. We are very positive that adopting this tool will stimulate students' interest of SQL in higher institutions of learning.

\section{Chapter Summary}
This chapter presented the evaluation of the prototypes that were designed in this research. In the first prototype, we measured the accuracy and presented some results. Next, we conducted online surveys for the remainder of the prototypes, as seen from \autoref{prot2} to \autoref{prot5}. Overall, the feedback received on each survey indicated that the participants agreed that the tools stimulated their interests in the SQL concepts.\\ 

\noindent \autoref{ch:conclusion} concludes this research and provides ideas for future research.

\chapter{Conclusion and Future Work}\label{ch:conclusion} 
\section{Conclusion}
\setlength{\parindent}{2em}
\lettrine[lines=3,loversize=0.1]{L}\normalsize{earning} and writing correct SQL queries have shown to be significant problems in the academic environment as well as in industry. These problems have a seemingly unending impact on students and non-technical end-users alike. Even when there are highly skilled instructors, who are experts in the field, their solutions may not address the issue adequately. In most universities, a class may be large and an instructor might be less responsive to every student’s needs. Similarly, this may occur in the industry whereby a developer who is saddled with all the technical intricacies of a company may leave the organisation, leaving the less-skilled user to scramble for solutions to their query needs. Consequently, such users might opt for online forums as a last resort, which may offer little or no help to their immediate query needs. Therefore, the overall problem is two-fold: 
\begin{enumerate}
	\item A student may be willing to learn SQL, which is paramount to passing a course and it is of equal importance when it comes to employability, but lacks sufficient knowledge to do so.
	\item A non-technical end-user may have an urgent need to write queries as part of a routine task, but lacks the skill.
\end{enumerate}

This thesis presented \textit{SQL Comprehension and Synthesis} to address the aforementioned challenges. Additionally, it suggests interactive learning aids in an attempt to assist these choices of users to enable them to understand and write correct queries. Since the ultimate goal of learning aids is to enhance the teaching and learning process, this research is aimed at the improvement, and the comprehension process utilised by interactive aids. Ultimately, as stated in this research, interactive aids aim to improve the understanding of SQL queries, an area that students and non-technical end-users often struggle to comprehend. If implemented on a large-scale, the tools may be applied in numerous applications in real world scenarios and to improve SQL learning. The following prototypes have been presented in this thesis:

\begin{description}
\item[S-NAR.] \autoref{ch:regular} introduced \texttt{S-NAR}, a tool that used REs in its engine to recognise SQL queries for the purpose of improving SQL comprehension. The tool translated the recognised SQL queries into textual explanations called narrations. This information is presented to a learner in real-time. In addition, the tool could serve as a learning aid by students or non-technical end-users that require explanations to queries written by technical staff. Similarly, the tool could be used to support teaching in line with the SFA to programming language pedagogy~\citep{fincher1999we,ade2014abstracting}. Another major advantage seen through the use of \texttt{S-NAR} is that it provides immediate feedback about the correctness of a query. \texttt{S-NAR} was tested with 5000 queries scrapped from the Internet, and it successfully reported an accuracy of 96\%, which was a major success. In its current formation, the tool is unable to recognise nested queries enclosed in balanced parentheses. However, this may only be addressed by using an irregular language such as CFGs or higher classes of formal abstract machines, as presented in \autoref{ch:cfg}. \texttt{S-NAR} has appeared in \citet{ade2017s}.

\item[SQL Narrator.] In \autoref{ch:cfg}, a CFG was designed for the automatic generation of narrations for nested SQL queries. This was implemented using Coco/R, a compiler generator that takes an attributed grammar and generates a scanner and a parser. These generated elements, both the scanner and the parser, were used to verify the correctness of a nested query. The designed grammar was implemented into a tool called \texttt{SQL Narrator} based on the C\# language. This further runs on the .Net framework. \texttt{SQL Narrator} was tested with the remaining 4\% of queries that could not recognised by \texttt{S-NAR} in the previous study in \autoref{ch:regular}. The tool successfully translated these queries. This idea has appeared in \citet{obaido2019generating}. The idea of using a CFG in this study was extended to the synthesis of 100000 hypothetical datasets that are similar to the Northwind database. The resulting database was referred to as XNorthwind~(Extended Northwind) and has appeared in an extended study in \citet{ade2019xnorthwind}. This database was used to train the tool, discussed in \autoref{ch:nsql}.

\item[Narrations-2-SQL.] \autoref{ch:nsql} proposed an approach that uses a JFA – a type of Finite Machine for translating natural language descriptions into SQL queries, which then further executes the queries, as well as provides feedback to a user. This technique was implemented into a tool called \texttt{Narrations-2-SQL}. This idea has appeared in \citet{obaido2019narrations} An experimental evaluation was performed on 204 crowdsourced queries in natural language from the XNorthwind DB. The result thereof reported an accuracy of 88\%. This report revealed that there is room for improvement. To our knowledge, this is expected to be the first time in which such an approach would be applied for SQL query translation from natural language. Since a natural language is context-sensitive, the JFA approach has shown to be an effective technique to the problem of query translation from natural language. If implemented on a large scale, \texttt{Narrations-2-SQL} may assist end-users in different domains to specify queries in natural language and perform tasks seamlessly without requiring much help from technical users.

\item[SQL Visualiser.] \autoref{ch:vsql} presented an approach which made use of images that depict SQL commands to generate a query. This was designed into a tool called the \texttt{SQL Visualiser}. The visualisation technique ensured the interaction between visual specifications to build queries. This is expected to eliminate the need to memorise database schemas, which is a major problem faced by students while learning SQL. \texttt{SQL Visualiser} used visual specifications for ‘drag and drop’ interactions for generating SQL queries. So far, the tool is only able to generate queries within the \texttt{SELECT} command. An extended visualiser is anticipated to recognise more commands to support the JOIN, ORDER BY, GROUP BY and aggregate functions. \texttt{SQL Visualiser} has appeared in \citet{obaido2018generating}. 

\item[TalkSQL.] In \autoref{ch:ssql}, a speech-based query system called \texttt{TalkSQL} was designed to assist end-users to specify queries using speech inputs. \texttt{TalkSQL} has appeared in \citet{obaido2019talksql}. This tool relied on an existing framework which makes use of the \texttt{spaCy} NLP engine to recognise SQL commands. For speech translation, \texttt{TalkSQL} uses the Google Speech API that incorporates a Deep Neural Network, alongside the HMM to transcribe speech. This speech engine was chosen due to its WER of 9\% which outperformed other automatic speech recognition engines as discussed in \autoref{deeplearn}. For query explanation, \texttt{TalkSQL} uses the narration technique that automatically generates using regular expressions as described in \autoref{ch:regular}. Currently, the tool is unable to provide explanations for queries in nested forms. We anticipate a narrator engine developed using Coco/R in \autoref{ch:cfg}, which could be used to fix this hitch.
\end{description}

Finally, an evaluation of the prototypes was provided in \autoref{ch:eproto}. During the first study, only an experimental evaluation was performed to determine the accuracy of the tool. This showed an accuracy of 96\%. The second study reported that out of 161 participants, 98.1\%  agreed that the tool enabled them understand nested queries. In the third study, an accuracy of 88\% was reported as experimental evaluation and 96.9\% out of 162 participants agreed that the tool would be helpful to industry users. The fourth study showed that out of 121 responses, 92.16\% indicated that the tool aided their understanding of the SQL syntax. In the fifth study, out of 113 responses, 87.6\% acknowledged that the tool would certainly help visually impaired learners to correctly write queries using voice inputs. The next section presents discussions for future directions.

\section{Future Work}
The prototypes developed in this thesis provide a basis for future research in several areas. At least five such areas can be identified. These areas include:

\begin{enumerate}
	\item Extending \texttt{narrations} to other query forms.
	\item Improving the \texttt{SQL Visualiser} to accommodate other SELECT statements such as JOIN, ORDER BY and other aggregate functions. 
	\item Extending the \texttt{Narrations-2-SQL} tool to work with multiple different databases.
	\item Developing a practical quiz engine for SQL quiz grading.
\end{enumerate}

\noindent The following sections elaborate the areas that have been highlighted in more detail.

\subsection{Extending Narrations}
This work has used narrations for aiding the comprehension of SQL queries. An area of exploration might be to abstract these explanations into a form that is free from language keywords. For example, we have used `alphanumeric' to represent datatypes. It may be worth noting that some end-users may not clearly understand what the term `alphanumeric' means. One way of capturing such a detail might be to use terms such as: `letters and numbers'. These terms are well-defined in an end-user’s vocabulary. Another interesting area of application of narrations to might be to explain queries that appear in other forms such as XML, SPARQL, JSON, etc. These will promote a clearer understanding of these queries.

\subsection{Improving the Visualiser}
The developed tool was able to easily generate simple queries using images. The next plan of action is to extend the visualiser to support other \texttt{SELECT} operation statements such as \texttt{JOIN, ORDER BY, GROUP BY} and aggregate functions. It will be interesting to see how learners could use these images to generate queries that contain these kinds of statements. Another desirable application of the visualiser is to generate nested SQL queries. These types of queries have shown to be problematic for students. Such solutions will help users to learn nested queries using such a visualiser.  Other areas of exploration are to allow students to define their schema. This will provide a richer learning experience for the more advanced learner.  It will be interesting to see how the extended visualiser can be made to generate queries to manipulate data in a database, and then produce a result. This will promote a more ‘realistic’ use of the tool. We have seen that the \texttt{Narrations-2-SQL} and \texttt{TalkSQL} use this approach on a live database and produce an output, which students find very useful.

\subsection{Extending the Narration-2-SQL Tool}
We have used the \texttt{Narrations-2-SQL} tool to work with the XNorthwind DB, which shows a good accuracy. It will be worthwhile to investigate the tool to work with multiple different databases such as Geo880, Academic, British National Corpus and IMDB to determine its accuracy. This may be an interesting problem to investigate. Recent studies by \citet{baik2019bridging,cai2017encoder,yaghmazadeh2017sqlizer} have tested their approaches on multiple different databases. It will be interesting to compare our results with these studies. Additionally, it will be beneficial to use the JFA technique to other domains that
require natural language specifications to generate code. For example, JFA can be applied to map query stored in JSON, XML, SPARQL and NoSQL formats.

\subsection{Developing a Quiz Game}
REs are useful for pattern matching and have been demonstrated in this research. It will be worth investigating this approach for a  game-based scenario for learning SQL. Such an application may be useful to test learners' query skills. 

\appendix
\cleardoublepage

\ctparttext{This appendix contains materials that are supplementary to contents that have been discussed in this thesis. This part has been organised as follows: the questionnaires used for the survey design are presented in \autoref{ch:app-question-results}, \autoref{ch:app-regularexpression} contains the REs library written in the Microsoft .NET framework, the atg file used for the grammar design by the Coco/R parser generator is presented in \autoref{ch:app-cfg-formalism}, the crowdsourced natural language descriptions dataset used to train the \texttt{Narrations-2-SQL} is highlighted in \autoref{ch:app-natural-sql} and the 5000 queries tested by \texttt{S-NAR} is provided in \autoref{sql-queries-dataset}.

}
\part{Appendix}

\chapter{Questionnaires}\label{ch:app-question-results}
This appendix contains the questionnaires that were used in this study. 



# Generating Narrations from SQL Queries using Context-Free Grammars

Narrations provide explanations of a concept in a domain. For example, a program narrator explains details about lines of code. In this study, narrations are described as a textual explanation of a query. We have developed a tool called SQL Narrator that generates narrations for nested SQL queries. This section contains general questions about SQL queries. The next section presents specific questions on the SQL Narrator.

* Required

1. **Have you used a narrator before?** *
   *Mark only one oval.*
   - ◯ Yes
   - ◯ No
   - ◯ Maybe

2. **What was the narrator's task?**

   _____
   _____
   _____
   _____

3. **How would you rate the narrator on a scale from 1 to 10 (1 for not at all useful and 10 for very useful)** *
   *Mark only one oval.*

   |  | 1 | 2 | 3 | 4 | 5 | 6 | 7 | 8 | 9 | 10 |  |
   |---|---|---|---|---|---|---|---|---|---|---|---|
   | Not at all useful | ◯ | ◯ | ◯ | ◯ | ◯ | ◯ | ◯ | ◯ | ◯ | ◯ | Very useful |

4. **Are you familiar with simple SQL queries?** *
   *Mark only one oval.*
   - ◯ Yes
   - ◯ No
   - ◯ Maybe

5. **Are nested queries too difficult?** *

   *Mark only one oval.*

   ◯ Yes

   ◯ No

   ◯ Maybe

6. **Have you used a SQL-based Narrator before?** *

   *Mark only one oval.*

   ◯ Yes

   ◯ No

   ◯ Maybe

7. **Would you be interested in using a SQL-based narrator?** *

   *Mark only one oval.*

   ◯ Yes

   ◯ No

   ◯ Maybe

# SQL Narrator

Specific questions on the SQL Narrator tool. SQL Narrator generates narrations from nested SQL queries. Answer the following questions.

## The SQL Narrator tool generates narrations from a nested SQL query

8. **Was the SQL Narrator easy to use?** *

   *Mark only one oval.*

   ( ) Yes
   ( ) No
   ( ) Maybe

9. **Were you able to comprehend the nested queries using SQL Narrator?** *

   *Mark only one oval.*

   ( ) Yes
   ( ) No
   ( ) Maybe

10. **Out of the three narrations generated, which one were you able to comprehend?** *

    *Mark only one oval.*

    ( ) Inner to Outer Subquery Narration
    ( ) Outer to Inner Subquery Narration
    ( ) Co-joined Subquery Narration

11. **How would you rate the narrator on a scale of 1 to 10 (1 for not at all useful and 10 for very useful)?** *

    *Mark only one oval.*

    |  | 1 | 2 | 3 | 4 | 5 | 6 | 7 | 8 | 9 | 10 |  |
    |---|---|---|---|---|---|---|---|---|---|---|---|
    | Less useful | ( ) | ( ) | ( ) | ( ) | ( ) | ( ) | ( ) | ( ) | ( ) | ( ) | Very useful |

12. **Any suggestions/comments to improve the SQL Narrator to aid your cognitive process?**

    _____
    _____
    _____
    _____
    _____

Powered by

![Google Forms]

# Synthesis of SQL Queries from Narrations

In this study, narrations are translated into SQL Queries. We have developed a tool that translates a natural language specification into SQL queries called narration-2-SQL. This section contains general questions on SQL queries. The next section presents specific questions about the narrations-2-SQL tool.

* Required

1. **Are you familiar with SQL?** *
   *Mark only one oval.*
   - ◯ Yes
   - ◯ No
   - ◯ Maybe

2. **Rate your familiarity with SQL on a scale of 1 to 10?** *
   *Mark only one oval.*

   |  | 1 | 2 | 3 | 4 | 5 | 6 | 7 | 8 | 9 | 10 |  |
   |---|---|---|---|---|---|---|---|---|---|---|---|
   | Not at all familiar | ◯ | ◯ | ◯ | ◯ | ◯ | ◯ | ◯ | ◯ | ◯ | ◯ | Very familiar |

3. **Are you familiar with the CREATE, SELECT, UPDATE and DELETE operations?** *
   *Mark only one oval.*
   - ◯ Yes
   - ◯ No
   - ◯ Maybe

4. **Do you know of any tool that helps you to transform a natural language specification into SQL query before?** *
   *Mark only one oval.*
   - ◯ Yes
   - ◯ No
   - ◯ Maybe

5. **If yes, what is the name of such a tool?**

   _____

## Narrations-2-SQL tool

Specific questions on the Narrators to SQL tool. This tool translates a natural language specification into SQL. Answer the following questions.

# The tool translates natural language specifications into SQL queries

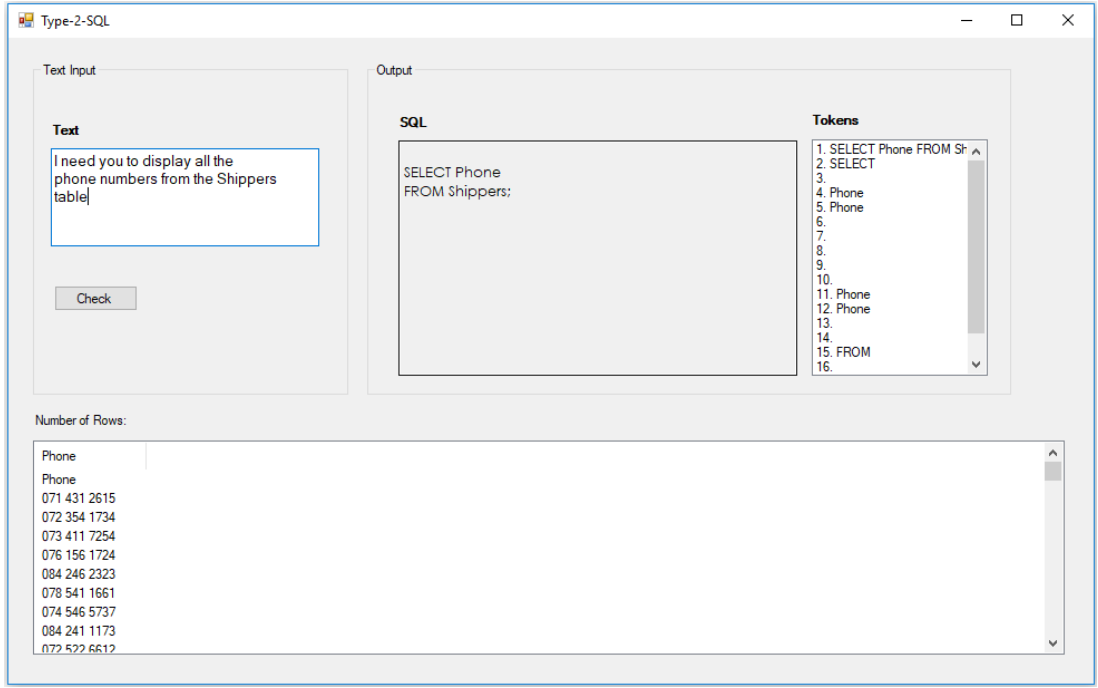

6. **Was Narration-2-SQL easy to use? ***

   *Mark only one oval.*

   ( ) Yes

   ( ) No

   ( ) Maybe

7. **Was the generated SQL query a correct translation of your narration? ***

   *Mark only one oval.*

   ( ) Yes

   ( ) No

   ( ) Maybe

8. **Do you think this tool will assist end users in the industry who has no knowledge of SQL to work with SQL? ***

   *Mark only one oval.*

   ( ) Yes

   ( ) No

   ( ) Maybe

9. **How would you rate Narration-2-SQL (1 for not at all useful and 10 for very useful)? ***
   *Mark only one oval.*

   |  | 1 | 2 | 3 | 4 | 5 | 6 | 7 | 8 | 9 | 10 |  |
   |---|---|---|---|---|---|---|---|---|---|---|---|
   | Less useful | ◯ | ◯ | ◯ | ◯ | ◯ | ◯ | ◯ | ◯ | ◯ | ◯ | Very useful |

10. **Any suggestions to improve Narration-2-SQL to aid your understanding?**

    _____
    _____
    _____
    _____

Powered by
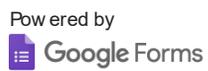
Google Forms

# Generating SQL Queries from Visual Specifications

This study proposes the use of interactive visualisation technique to aid the understanding of SQL. We have developed a tool that uses images that depicts SQL operation to generate SQL queries. This section contains general questions about the use of visualisers. The next section presents specific questions about the SQL visualiser that was developed for this study.

* Required

1. **Have you used a visualiser before? ***
   *Mark only one oval.*

   ◯ Yes
   ◯ No
   ◯ Maybe

2. **What was the visualiser task? ***

   _____
   _____
   _____
   _____
   _____

3. **How would you rate the visualiser on a scale from 1 to 10 ***
   *Mark only one oval.*

   |  | 1 | 2 | 3 | 4 | 5 | 6 | 7 | 8 | 9 | 10 |  |
   |---|---|---|---|---|---|---|---|---|---|---|---|
   | Not at all useful | ◯ | ◯ | ◯ | ◯ | ◯ | ◯ | ◯ | ◯ | ◯ | ◯ | Very useful |

4. **Are you familiar with SQL ***
   *Mark only one oval.*

   ◯ Yes
   ◯ No
   ◯ Maybe

5. **Rate your familiarity with SQL ***
   *Mark only one oval.*

   |  | 1 | 2 | 3 | 4 | 5 | 6 | 7 | 8 | 9 | 10 |  |
   |---|---|---|---|---|---|---|---|---|---|---|---|
   | Less familiar | ◯ | ◯ | ◯ | ◯ | ◯ | ◯ | ◯ | ◯ | ◯ | ◯ | Very familiar |

6. **Have you used a SQL-based visualiser** *

   *Mark only one oval.*

   ◯ Yes

   ◯ No

   ◯ Maybe

## SQL Visualiser

Specific questions on SQL Visualiser. The SQL visualiser uses the drag and drop method to generate SQL queries.

## The SQL Visualiser, generating a correct successful query

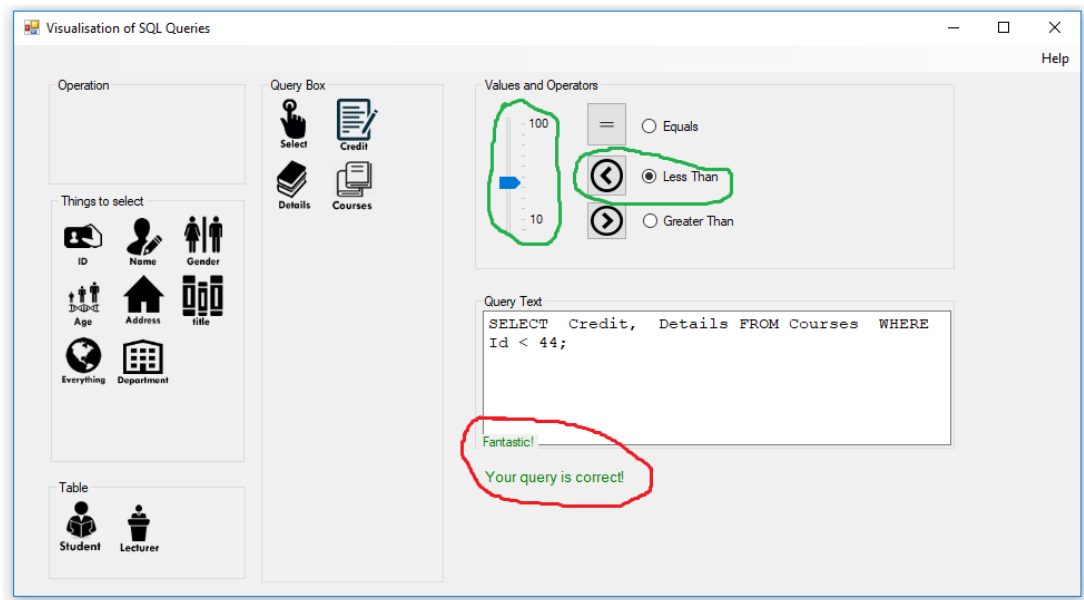

## The SQL visualiser generating a wrong query

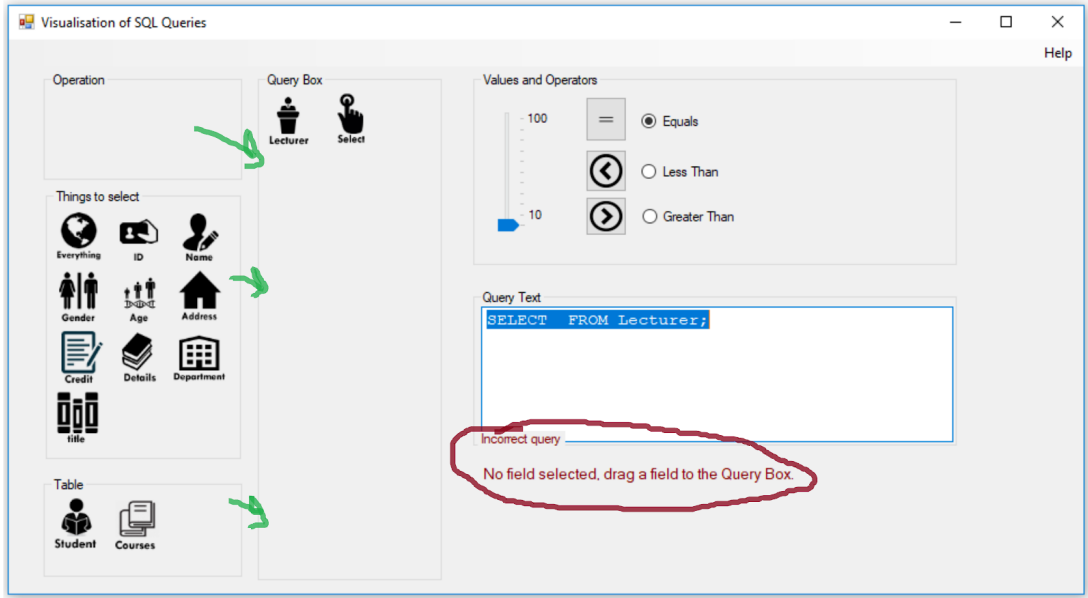

7. **Is the visualiser easy to use?** *

   *Mark only one oval.*

   ◯ Yes

   ◯ No

   ◯ Maybe

8. **Were you able to synthesize the basic SQL queries?** *

   *Mark only one oval.*

   ◯ Yes

   ◯ No

   ◯ Maybe

9. **Do icons help you analyse the syntax of SQL queries** *

   *Mark only one oval.*

   ◯ Yes

   ◯ No

   ◯ Maybe

10. **How would you rate the visualiser on a scale of 1 to 10?** *

    *Mark only one oval.*

    |  | 1 | 2 | 3 | 4 | 5 | 6 | 7 | 8 | 9 | 10 |  |
    |---|---|---|---|---|---|---|---|---|---|---|---|
    | Less useful | ◯ | ◯ | ◯ | ◯ | ◯ | ◯ | ◯ | ◯ | ◯ | ◯ | Very useful |

11. **Any suggestions/comments to improve the visualiser to aid your cognitive process?** *

---

Powered by
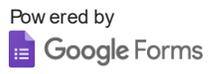

# Synthesis of SQL Queries from Verbal Specifications

In this study, a speech to SQL method was proposed. We have developed a tool called TalkSQL, that takes voice inputs from a user, converts these words into SQL queries and returns feedback to the user. This section contains general questions on users' familiarity with SQL and voice-based NLIDBs. The next section presents specific questions about the TalkSQL tool.

* Required

1. **Are you familiar with simple SQL queries? ***

   *Mark only one oval.*

   ◯ Yes

   ◯ No

   ◯ Maybe

2. **Are you familiar with the CRUD (Create, Read, Update, Delete) operations? ***

   *Mark only one oval.*

   ◯ Yes

   ◯ No

   ◯ Maybe

3. **Have you used a voice-based natural language interface to databases (NLIDB) before? ***

   *Mark only one oval.*

   ◯ Yes

   ◯ No

   ◯ Maybe

4. **Would you be interested in a voice-based NLIDB? ***

   *Mark only one oval.*

   ◯ Yes

   ◯ No

   ◯ Maybe

## TalkSQL

TalkSQL takes voice inputs from a user, convert these words into SQL queries and returns feedback to a user

## The TalkSQL tool takes voice inputs from a user and generates a query.

**5. Was TalkSQL easy to use?** *

*Mark only one oval.*

( ) Yes

( ) No

( ) Maybe

**6. Were you able to understand the CRUD command using TalkSQL?** *

*Mark only one oval.*

( ) Yes

( ) No

( ) Maybe

**7. Do you think this might help visually impaired learners understand SQL?** *

*Mark only one oval.*

( ) Yes

( ) No

( ) Maybe

**8. Was the feedback you received comprehensive?** *

*Mark only one oval.*

( ) Yes

( ) No

( ) Maybe

9. **How would you rate TalkSQL (1 for not at all useful and 10 for very useful)?** *

    *Mark only one oval.*

    |  | 1 | 2 | 3 | 4 | 5 | 6 | 7 | 8 | 9 | 10 |  |
    |---|---|---|---|---|---|---|---|---|---|---|---|
    | Less useful | ○ | ○ | ○ | ○ | ○ | ○ | ○ | ○ | ○ | ○ | Very useful |

10. **Any suggestions to improve TalkSQL to aid your understanding?**

    _____
    _____
    _____
    _____

Powered by

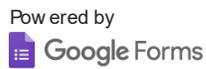


\chapter{RE Library}\label{ch:app-regularexpression}
This section contains the REs library that have been specified in the .NET framework, written in VB.NET. This was used to recognise simple SQL query constructs as seen in Listing \ref{lst:sqlvbconstructs}.

\begin{lstlisting}[style=A,caption={REs defined in .NET for SQL query constructs},label=lst:sqlvbconstructs]%please try style=B
'----------------- Lowest granularity level ----------------------------'
'for all letters including hyphens
letter As String = "([A-Za-z-])" 
s_spc As String = "(\s)"
n_spc As String = "(\s+)"
spc As String = "(\s*)"
all As String = "(\*)"
comma As String = "(\,)"
bra_open As String = "(\()"
bra_close As String = "(\))"
ass_sym As String = "(\=)"
greater_than As String = "(\>)"
less_than As String = "(\<)"
not_equal_to As String = "(\!\=)"
less_than_equal As String = "(\<\=)"
greater_than_equal As String = "(\>\=)"
not_greater_than As String = "(\!\>)"
not_less_than As String = "(\!\<)"
or_greater_or_less As String = "(\<\>)"
ident As String = "([A-Za-z_][A-Za-z0-9_]+)"
number As String = "[1-9][0-9]*"
semi_colon As String = "(\;)"
float_number As String = "(\d\d?\.\d\d?)"

val_in_quote As String = "((\')" & "(" & _ident & "|" & _number & "|" & _float_number & "|" & _n_spc & "|" & _comma & ")+" & "(\'))"

list_of_vals_in_quote As String = "(" & "(" & _val_in_quote & _spc & _comma & _spc & ")*(" & _val_in_quote & "))"

ident_sep_by_comma As String = "(" & "(" & _ident & _spc & _comma & _spc & ")*(" & _ident & "))"

list_of_values_sep_by_comma As String = _bra_open & _spc & _list_of_vals_in_quote & _spc & _bra_close

comp_op As String = "(" & _ass_sym & "|" & _greater_than & "|" & _less_than & "|" & _not_equal_to & "|" & _less_than_equal & "|" & _greater_than_equal & "|" & _not_less_than & "|" & _or_greater_or_less & ")"


'----------------- SQL (INSERT) ----------------------------
insert_suffix_into As String = _ident & _spc & _bra_open & _ident_sep_by_comma & _bra_close
insert_suffix_end As String = _bra_open & _list_of_values_sep_by_comma & _bra_close
insert_command As String = "(INSERT)" & _n_spc & "(INTO)" & _n_spc & _insert_suffix_into & _spc & "(VALUES)" & _spc & _insert_suffix_end & _semi_colon


'----------------- SQL (CREATE_TABLE) ----------------------------
create_suffix_into As String = _ident & _spc & _bra_open & _spc & _ident_sep_by_comma & _spc & _bra_close
create_command As String = "(CREATE)" & _n_spc & "(TABLE)" & _n_spc & _create_suffix_into & _semi_colon

'----------------- SQL (CREATE_DB) ---------------------------------------
create_database As String = "(CREATE)" & _n_spc & "(DATABASE)" & _n_spc & ("IF") & _n_spc & ("NOT") & _n_spc & ("EXISTS") & _n_spc & _ident & _semi_colon


'----------------- SQL (DROP) ----------------------------
drop_suffix_into As String = _ident_sep_by_comma
drop_command As String = "(DROP)" & _n_spc & "(DATABASE)" & _n_spc & "(IF)" & _spc & "(EXISTS)" & _spc & _drop_suffix_into & _semi_colon
drop_command_table As String = "(DROP)" & _n_spc & "(TABLE)" & _n_spc & "(IF)" & _spc & "(EXISTS)" & _spc & _drop_suffix_into & _semi_colon


'----------------- SQL (RENAME) ----------------------------
rename_suffix_into As String = _ident & _n_spc & "(TO)" & _n_spc & _ident
rename_command As String = "(RENAME)" & _n_spc & "(TABLE)" & _n_spc & _rename_suffix_into & _semi_colon

'----------------- SQL (TRUNCATE) ----------------------------
truncate_suffix_into As String = _ident
truncate_command As String = "(TRUNCATE)" & _n_spc & "(TABLE)" & _n_spc & _truncate_suffix_into & _semi_colon

'----------------- SQL (ALTER) ----------------------------
alter_suffix_into As String = _ident & _n_spc & "(RENAME)" & _n_spc & "(TO)" & _n_spc & _ident
alter_command As String = "(ALTER)" & _n_spc & "(TABLE)" & _n_spc & _alter_suffix_into & _semi_colon

'----------------- SQL (DELETE) ----------------------------
delete_suffix_into As String = _ident & _n_spc & "(WHERE)" & _n_spc & _ident & _comp_op & _val_in_quote
delete_command As String = "(DELETE)" & _n_spc & "(FROM)" & _n_spc & _delete_suffix_into & _semi_colon

'----------------- SQL (SELECT) ----------------------------
select_suffix_all As String = _all
select_suffix_one_or_more As String = "(" & _ident_sep_by_comma & "|" & "(\*)" & ")+"
select_numb_or_string_in_quote As String = "(" & "((\')" & _ident & "(\'))" & "|" & _number & ")+"
select_command_all As String = "(SELECT)" & _n_spc & _select_suffix_all & _n_spc & "(FROM)" & _n_spc & _ident & _semi_colon
select_command_all_more As String = "(SELECT)" & _n_spc & _select_suffix_one_or_more & _n_spc & "(FROM)" & _n_spc & _ident & _semi_colon
select_command_distinct As String = "(SELECT)" & _n_spc & "(DISTINCT)" & _n_spc & _select_suffix_one_or_more & _n_spc & "(FROM)" & _n_spc & _ident & _semi_colon
select_command_where As String = "(SELECT)" & _n_spc & _select_suffix_one_or_more & _n_spc & "(FROM)" & _n_spc & _ident & _n_spc & "(WHERE)" & _n_spc & _select_suffix_one_or_more & _comp_op & _select_numb_or_string_in_quote & _semi_colon
select_command_where_and As String = "(SELECT)" & _n_spc & _select_suffix_one_or_more & _n_spc & "(FROM)" & _n_spc & _ident & _n_spc & "(WHERE)" & _n_spc & _select_suffix_one_or_more & _comp_op & _select_numb_or_string_in_quote & _n_spc & "(AND)" & _n_spc & _select_suffix_one_or_more & _comp_op & _select_numb_or_string_in_quote & _semi_colon
select_command_where_or As String = "(SELECT)" & _n_spc & _select_suffix_one_or_more & _n_spc & "(FROM)" & _n_spc & _ident & _n_spc & "(WHERE)" & _n_spc & _select_suffix_one_or_more & _comp_op & _select_numb_or_string_in_quote & _n_spc & "(OR)" & _n_spc & _select_suffix_one_or_more & _comp_op & _select_numb_or_string_in_quote & _semi_colon
select_command_where_not As String = "(SELECT)" & _n_spc & _select_suffix_one_or_more & _n_spc & "(FROM)" & _n_spc & _ident & _n_spc & "(WHERE)" & _n_spc & "(NOT)" & _n_spc & _select_suffix_one_or_more & _comp_op & _select_numb_or_string_in_quote & _semi_colon
select_command_where_in As String = "(SELECT)" & _n_spc & _select_suffix_one_or_more & _n_spc & "(FROM)" & _n_spc & _ident & _n_spc & "(WHERE)" & _n_spc & _ident & _n_spc & "(IN)" & _n_spc & _list_of_values_sep_by_comma & _semi_colon
select_command_where_between As String = "(SELECT)" & _n_spc & _select_suffix_one_or_more & _n_spc & "(FROM)" & _n_spc & _ident & _n_spc & "(WHERE)" & _n_spc & _ident & _n_spc & "(BETWEEN)" & _n_spc & _select_numb_or_string_in_quote & _n_spc & "(AND)" & _n_spc & _select_numb_or_string_in_quote & _semi_colon
select_command_where_like As String = "(SELECT)" & _n_spc & _select_suffix_one_or_more & _n_spc & "(FROM)" & _n_spc & _ident & _n_spc & "(WHERE)" & _n_spc & _ident & _n_spc & "(LIKE)" & _n_spc & _select_numb_or_string_in_quote & _semi_colon
select_command_where_groupby As String = "(SELECT)" & _n_spc & _select_suffix_one_or_more & _n_spc & "(FROM)" & _n_spc & _ident & _n_spc & "(GROUP BY)" & _n_spc & _ident & _semi_colon
select_command_where_orderby_asc As String = "(SELECT)" & _n_spc & _select_suffix_one_or_more & _n_spc & "(FROM)" & _n_spc & _ident & _n_spc & "(ORDER BY)" & _n_spc & _ident & _n_spc & "(ASC)" & _semi_colon
select_command_where_orderby_desc As String = "(SELECT)" & _n_spc & _select_suffix_one_or_more & _n_spc & "(FROM)" & _n_spc & _ident & _n_spc & "(ORDER BY)" & _n_spc & _ident & _n_spc & "(DESC)" & _semi_colon
select_command_count As String = "(SELECT)" & _n_spc & "(COUNT)" & _bra_open & _ident & _bra_close & _n_spc & "(FROM)" & _n_spc & _ident & _semi_colon
\end{lstlisting}

\chapter{CFG Rules}\label{ch:app-cfg-formalism}
The following contains the attributed grammar (atg) of the SQL source language used by Coco/R to generate a scanner and parser for the language. This ideas were taken from the EBNF grammar defined by Ron Savage~\citep{SQLgrammar}. Listing \ref{lst:cocor} shows the atg file used by the Coco/R engine.

\begin{lstlisting}[style=A,caption={Attributed grammar design using Coco/R},label=lst:cocor]
COMPILER SqlGrammar
public Narrator narrator;

CHARACTERS
letter    = "ABCDEFGHIJKLMNOPQRSTUVWXYZabcdefghijklmnopqrstuvwxyz_".
digit     = "0123456789".

TOKENS
ident			=	letter { letter | digit | '_' }.
number			=	digit { digit }.
semi_colon		=	';'.

IGNORE '\r' + '\n' + '\t'

PRODUCTIONS

SqlGrammar												(. narrator = new Narrator(); .) 
=
TruncateCommand | DeleteCommand | SelectCommand .

TruncateCommand
=														(. string tableName; .) 
"TRUNCATE" "TABLE" StringVal<out tableName> semi_colon	(. narrator.NarrateTruncate(tableName); .).

DeleteCommand											(. ArrayList list = null; .)
=
"DELETE" { "*" | StringList<out list > } "FROM" StringVal<out string tableName> semi_colon
(. narrator.NarrateDelete(tableName, list); .).


SelectCommand											(.	bool nested = false;
bool noConditions = false;
string comparisonOp = null;
string conditionColumn = null;
object criteriaValue = null;
string logicalOp = null; 
ArrayList innerOptions = new ArrayList(); 
string innerTableName = null;
string innerConditionValue = null;
string innerComparisonOp = null; .)
= 
"SELECT" StringList<out ArrayList options> "FROM" StringVal<out string tableName> 
(. noConditions = true; .)
[														
"WHERE" StringVal<out conditionColumn>					
(
LogicalOp<out logicalOp> 
"(" 
"SELECT" StringList<out innerOptions> "FROM" StringVal<out innerTableName>   
"WHERE" StringVal<out innerConditionValue> ComparisonOp<out innerComparisonOp> AnyValue<out criteriaValue> 
")"												(. nested = true; noConditions = false;.)

|

ComparisonOp<out comparisonOp> AnyValue<out criteriaValue>
(. noConditions = false; .)
) 
] semi_colon											(. narrator.NarrateSelect(	
nested,
noConditions,
options, 
tableName,
comparisonOp,
conditionColumn, 
innerOptions, 
innerTableName, 
innerConditionValue, 
innerComparisonOp, 
criteriaValue); .).


StringVal<out String s> 
=														(. s = null;  .)
ident													(. s = t.val;  .).

StringList<out ArrayList list >
=
StringVal< out string s>								(. list = new ArrayList{s}; .)
{"," StringVal< out s>									(. list.Add(s); .)
}.

LogicalOp<out string op>								(. op = null; .)
=
"AND"													(. op = t.val; .)
|
"OR"													(. op = t.val; .)
|
"IN"													(. op = t.val; .)
|
"NOT"													(. op = t.val; .)
|
"SOME"													(. op = t.val; .)
|
"ALL"													(. op = t.val; .).

ComparisonOp<out string op>								(. op = null; .)
=
"="														(. op = "equal to"; .)
|
">"														(. op = "greater than"; .)
|
"<"														(. op = "less than"; .)
|
">="													(. op = "igreater than or equal to"; .)
|
"<="													(. op = "less than or equal to"; .)
|
"<>"													(. op = "not "; .).


AnyValue<out object value>								(. value = null; .)
=
ident													(. value = "'" + t.val + "'"; .)
|
number													(. value = t.val; .).




//quote = "'" .
//string_num = quote | number.
//ident_list		=	ident  {"," ident}.
//logicalop		=	"AND" | "OR" | "IN" | "NOT" | "SOME" | "ALL".
END SqlGrammar.
\end{lstlisting}
\chapter{Natural Language Query to SQL TRANSLATION}\label{ch:app-natural-sql}
This following shows the JSON file that contains 204 translations of natural language descriptions into SQL queries (as shown in \autoref{lst:narr2sqljson}). This dataset was used to train the \texttt{narrations-2-SQL} tool as discussed in \autoref{ch:nsql}. This is broken into \textit{Item}, denoting the numbering, \textit{Narrations} described as natural language descriptions, and the equivalent SQL queries. We have only presented 20 items here, the complete list can be accessed via: \url{http://tiny.cc/fe2oiz}.

\begin{lstlisting}[language=json,firstnumber=1,caption={JSON file containing natural language descriptions of SQL queries},label=lst:narr2sqljson]
[
{
"Item": 1,
"Narrations": "Please, show me all the information from the customers table.",
"SQL Queries": "SELECT * FROM Customers;"
},
{
"Item": 2,
"Narrations": "Retrieve all the order details information",
"SQL Queries": "SELECT * FROM order_details;"
},
{
"Item": 3,
"Narrations": "Display the orders information",
"SQL Queries": "SELECT * FROM orders;"
},
{
"Item": 4,
"Narrations": "Display all the products details",
"SQL Queries": "SELECT * FROM products;"
},
{
"Item": 5,
"Narrations": "Display all employee records",
"SQL Queries": "SELECT * FROM employee;"
},
{
"Item": 6,
"Narrations": "Display all the categories information",
"SQL Queries": "SELECT * FROM Categories;"
},
{
"Item": 7,
"Narrations": "Please can you show me all the shippers details from the table",
"SQL Queries": "SELECT * FROM shippers;"
},
{
"Item": 8,
"Narrations": "I need you to select all the suppliers data",
"SQL Queries": "SELECT * FROM suppliers;"
},
{
"Item": 9,
"Narrations": "Show all the employee cities.",
"SQL Queries": "SELECT cities FROM employees;"
},
{
"Item": 10,
"Narrations": "Show me only the employee countries.",
"SQL Queries": "SELECT country FROM employees;"
},
{
"Item": 11,
"Narrations": "Show all the employeeID.",
"SQL Queries": "SELECT employeeID FROM employees;"
},
{
"Item": 12,
"Narrations": "Select all ids from the customer table.",
"SQL Queries": "SELECT * FROM Customerdemographics;"
},
{
"Item": 13,
"Narrations": "List all customers from South Africa or USA",
"SQL Queries": "SELECT Id, FirstName, LastName, City, Country FROM Customers WHERE Country = 'South Africa' OR Country = 'USA';"
},
{
"Item": 14,
"Narrations": "select the Customer Name and company Name",
"SQL Queries": "SELECT ContactName, CompanyName FROM Customers;"
},
{
"Item": 15,
"Narrations": "select all columns from customer table  where the Country column has South Africa for its value",
"SQL Queries": "SELECT * FROM Customers WHERE Country='South Africa';"
},
{
"Item": 16,
"Narrations": "return only the Customer contact name and phone number where country is equal to South Africa",
"SQL Queries": "SELECT phone, ContactName FROM Customers WHERE Country='Sout-Africa';"
},
{
"Item": 17,
"Narrations": "select the First name and title from customers",
"SQL Queries": "SELECT ContactName, ContactTitle FROM Customers;"
},
{
"Item": 18,
"Narrations": "List the first name, Phone, and city of all customers",
"SQL Queries": "SELECT ContactName, phone, city FROM Customers;"
},
{
"Item": 19,
"Narrations": "List the order id, order date and shipped date for all orders.",
"SQL Queries": "SELECT orderID, orderDate and shippedDate FROM orders;"
},
{
"Item": 20,
"Narrations": "List the customers in Sweden",
"SQL Queries": "SELECT * FROM Customer WHERE Country = 'Sweden';"
}
]
\end{lstlisting}
\chapter{Dataset of SQL Queries}\label{sql-queries-dataset}
The following contains SQL queries scrapped from the Internet. In total, 5000 queries were scrapped from the Internet. We have only showed 44 queries here. The entire file can be accessed via \url{http://tiny.cc/qs1adz}. \autoref{allquery} shows some of queries scrapped from Internet.

\begin{lstlisting}[language=SQL,escapechar=@, morekeywords={WHERE, INSERT, INTO, UPDATE, DELETE, ORDER, DESC, MIN, DISTINCT, SUM, AVG,COUNT,JOIN}, upquote=true, showstringspaces=false, stringstyle=\color{violet}, basicstyle=\ttfamily, caption={Dataset of SQL queries scrapped from the Internet}, label={allquery}, captionpos=t]
INSERT INTO Student (SELECT * FROM LateralStudent);
INSERT INTO Student (ROLL_NO,NAME,Age) SELECT ROLL_NO, NAME, Age FROM LateralStudent;
INSERT INTO Student SELECT * FROM LateralStudent WHERE Age = 18;
INSERT INTO categories(category_id, category_name)VALUES(150, 'Miscellaneous');
INSERT INTO customers(customer_id, last_name, first_name)SELECT employee_number AS customer_id, last_name, first_name FROM employees WHERE employee_number < 1003;
SELECT name(s)FROM student WHERE name = 'peter' AND name = 'doe';
SELECT name AS 'Alias' FROM student;
SELECT AVG(name)FROM student;
SELECT name(s)FROM student WHERE name BETWEEN 'peter' AND 'doe';
SELECT name,CASE WHEN condition THEN 'Result_1'WHEN condition THEN 'Result_2'ELSE 'Result_3'END FROM student;
SELECT COUNT(name)FROM student;
SELECT B.FirstName AS FirstName1, B.LastName AS LastName1, A.FirstName AS FirstName2, A.LastName AS LastName2, B.City, B.Country FROM Customer A, Customer B WHERE A.Id <> B.Id AND A.City = B.City AND A.Country = B.Country ORDER BY A.Country;
SELECT column-names FROM table-name UNION SELECT column-names FROM table-name;
SELECT 'Customer' As Type,FirstName + ' ' + LastName AS ContactName,City, Country, Phone FROM Customer UNION SELECT 'Supplier', ContactName, City, Country, Phone FROM Supplier;
SELECT column-names FROM table-name1 WHERE value IN (SELECT column-name FROM table-name2 WHERE condition);
SELECT column1 = (SELECT column-name FROM table-name WHERE condition),column-names FROM table-name WEHRE condition;
SELECT ProductName FROM Product WHERE Id IN (SELECT ProductId FROM OrderItem WHERE Quantity > 100);
SELECT FirstName, LastName, OrderCount = (SELECT COUNT(O.Id) FROM [Order] O WHERE O.CustomerId = C.Id) FROM Customer C ;
SELECT column-names FROM table-name WHERE column-name operator ANY (SELECT column-name FROM table-name WHERE condition);
SELECT column-names FROM table-name WHERE column-name operator ALL(SELECT column-name FROM table-name WHERE condition);
SELECT ProductName FROM Product WHERE Id = ANY(SELECT ProductId FROM OrderItem WHERE Quantity = 1);
SELECT DISTINCT FirstName + ' ' + LastName as CustomerName FROM Customer, [Order]WHERE Customer.Id = [Order].CustomerId AND TotalAmount > ALL (SELECT AVG(TotalAmount)FROM [Order]GROUP BY CustomerId);
SELECT column-names FROM table-name WHERE EXISTS (SELECT column-name FROM table-name WHERE condition);
SELECT CompanyName FROM Supplier WHERE EXISTS(SELECT ProductName FROM Product WHERE SupplierId = Supplier.Id AND UnitPrice > 100)	;
SELECT column-names INTO new-table-name FROM table-name WHERE EXISTS(SELECT column-name FROM table-name WHERE condition);
SELECT * INTO SupplierUSA FROM Supplier WHERE Country = 'USA';
INSERT INTO table-name (column-names)SELECT column-names FROM table-name WHERE condition;
INSERT INTO Customer (FirstName, LastName, City, Country, Phone)SELECT LEFT(ContactName, CHARINDEX(' ',ContactName) - 1) AS FirstName,SUBSTRING(ContactName, CHARINDEX(' ',ContactName) + 1, 100) AS LastName,City, Country, Phone FROM Supplier WHERE Country = 'Canada';
SELECT column_list FROM table-name [WHERE Clause][GROUP BY clause][HAVING clause][ORDER BY clause];
SELECT first_name FROM student_details;
SELECT first_name, last_name FROM student_details;;
SELECT first_name + ' ' + last_name AS emp_name FROM employee;
SELECT * FROM EMPLOYEE_TBL;
SELECT EMP_ID FROM EMPLOYEE_TBL;
SELECT EMP_ID FROM EMPLOYEE_TBL;
SELECT EMP_ID, LAST_NAME FROM EMPLOYEE_TBL;
SELECT EMP_ID, LAST_NAME FROM EMPLOYEE_TBL WHERE EMP_ID = '333333333';
SELECT EMP_ID, LAST_NAME FROM EMPLOYEE_TBL WHERE CITY = 'INDIANAPOLIS' ORDER BY EMP_ID;
SELECT EMP_ID, LAST_NAME FROM EMPLOYEE_TBL WHERE CITY = 'INDIANAPOLIS' ORDER BY EMP_ID, LAST_NAME DESC;
SELECT EMP_ID, LAST_NAME FROM EMPLOYEE_TBL WHERE CITY = 'INDIANAPOLIS' ORDER BY 1;
INSERT INTO CUSTOMER(CustomerName,ContactName, Address, City, PostalCode, Country)('Cardinal','Tom B','Erichsen','Sagen 21','Stavanger','4006','Norway');
INSERT INTO CATEGORIES(Category_id, Category_Name)(150,'Miscellaneous');
INSERT INTO PRODUCT (ProductID, ProductName, Price, ProductDescription)(1,'Clamp',12.48,'Workbench clamp');
INSERT INTO CUSTOMER(FirstName,LastName,PhoneNumber,EmailAddress,priority,CreatedDate)('Jonah','Hook','0114022558','Jonah@neverdull.com',1,'2011-09-01');
\end{lstlisting}


\nocite{*}
\bibliography{references}\addcontentsline{toc}{chapter}{Bibliography}

\cleardoublepage\pagestyle{empty}

\hfill

\vfill

\pdfbookmark[0]{Colophon}{colophon}
\section*{Colophon}

\lettrine[lines=3,loversize=0.1]{\color{darkgray}T}\normalsize{his} work is based on a research supported by the Council for Scientific and Industrial Research of South Africa (CSIR). Any opinion, findings and conclusions or recommendations expressed in this material are those of the author and therefore, the CSIR does not accept liability in regard thereto.

\bigskip
\bigskip

\noindent This document was typeset using the typographical look-and-feel \texttt{classicthesis} developed by Andr\'e Miede. 
The style was inspired by Robert Bringhurst's seminal book on typography ``\emph{The Elements of Typographic Style}''. 
\texttt{classicthesis} is available for both \LaTeX\ and \mLyX at: \url{https://bitbucket.org/amiede/classicthesis/}. The author of this thesis has made many modifications to the \texttt{classicthesis} template to arrive at this final version

\bigskip
\noindent\rule{\linewidth}{0.4pt}
\begin{center}
	\emph{\myTitle}	
	
	\bigskip
	
	\textbf{George Rabeshi Obaido | \emph{2020} }\\
	\url{rabeshi.george@gmail.com} \\
	
	\bigskip
	
	Copyright~\textcopyright~University of the Witwatersrand, Johannesburg, South Africa.
	
\end{center}

\end{document}